\documentclass[a4paper,11pt]{article}
\pdfoutput=1 

\usepackage{jheppub} 

\usepackage[T1]{fontenc} 
\usepackage{comment}

\newcommand{\bi}{\begin{itemize}}
	\newcommand{\ei}{\end{itemize}}
\newcommand{\bea}{\begin{eqnarray}}
	\newcommand{\eea}{\end{eqnarray}}
\newcommand{\bt}{\begin{tabular}}
	\newcommand{\et}{\end{tabular}}
\newcommand{\bc}{\begin{center}}
	\newcommand{\ec}{\end{center}}

\newcommand{\be}{\begin{equation}}
	\newcommand{\ee}{\end{equation}}
\newcommand{\ba}{\begin{array}}
	\newcommand{\ea}{\end{array}}
\newcommand{\p}[1]{(\ref{#1})}
\newcommand{\lb}[1]{\label{#1}}

\def\bbox{{\,\lower0.9pt\vbox{\hrule \hbox{\vrule height 0.2 cm
				\hskip 0.2 cm \vrule height 0.2 cm}\hrule}\,}}
\newcommand{\dsl}{\pa \kern-0.5em /}

\newcommand{\nn}{\nonumber \\}

\makeatletter \@addtoreset{equation}{section} \makeatother

\def\slashchar#1{\setbox0=\hbox{$#1$}           
	\dimen0=\wd0                                 
	\setbox1=\hbox{/} \dimen1=\wd1               
	\ifdim\dimen0>\dimen1                        
	\rlap{\hbox to \dimen0{\hfil/\hfil}}      
	#1                                        
	\else                                        
	\rlap{\hbox to \dimen1{\hfil$#1$\hfil}}   
	/                                         
	\fi}

\pdfoutput=1

\title{\boldmath $\mathcal{N}=2$ higher spins:  superfield equations of motion, the hypermultiplet
supercurrents, \break and the component structure}

\author[a,b,c]{Ioseph Buchbinder,}
\author[b,d]{Evgeny~Ivanov,}
\author[b,d]{Nikita~Zaigraev}

\affiliation[a]{Center of Theoretical Physics, Tomsk State
Pedagogical University,\\ 634061, Tomsk,  Russia}

\affiliation[b]{Bogoliubov Laboratory of Theoretical Physics,
JINR,\\141980 Dubna, Moscow region, Russia}

\affiliation[c]{National Research Tomsk State University, \\634050,
Tomsk, Russia}

\affiliation[d]{Moscow Institute of Physics and Technology,\\ 141700
Dolgoprudny, Moscow region, Russia}

\emailAdd{joseph@tspu.edu.ru}
\emailAdd{ eivanov@theor.jinr.ru}
\emailAdd{nikita.zaigraev@phystech.edu}

\abstract{As a continuation of our previous papers \href{https://arxiv.org/abs/2109.07639}{arXiv:2109.07639}
and \href{https://arxiv.org/abs/2202.08196}{arXiv:2202.08196}, we study the linearized structure of the manifestly
$4D,\,\mathcal{N}=2$ supersymmetric theory of the cubic couplings of the higher spin gauge
superfields to the matter hypermultiplets.
We consider in detail the superfield equations
of motion, construct the conserved hypermultiplet superfield currents, explore
their component structure (basically in the bosonic sector) and compare it with
the corresponding currents in the conventional higher-spin bosonic theory.
We thoroughly study the $\mathcal{N}=2$ spin $\mathbf{2}$ and
$\mathbf{3}$ models as instructive examples.
}

\begin{document}
\maketitle
\flushbottom

\section{Introduction}
\label{sec:intro}

Higher spin theory raises a lot of conceptual and challenging
questions concerning the general structure of classical and
quantum field theory (see, e.g., the reviews
\cite{Bekaert:2022poo}). The most important issue is how to define
consistent field theories involving spin 2 together with higher
spins $s >2$. Free theories of higher spins are well known
\cite{FronsdalInteg, FronsdalHalfint}.
Naive attempts to build interacting Lagrangians for higher spins
(beyond cubic vertices) face a set of various no-go theorems
\cite{Bekaert:2010hw}. So in order to formulate consistent higher
spin interacting theories, the new approaches are of need. A
significant progress towards this goal has been achieved in Vasiliev
theory which embodies the interactions of higher spins on (A)dS$_d$
backgrounds \cite{Vasiliev:1990en, Vasiliev:1992av, Vasiliev:2003ev,
VAS} (for a review see \cite{V1, V2, V3, Bekaert:2004qos,
Didenko:2014dwa}). There still remains the problem of the Lagrangian
description of these interactions.

It is a natural task to study supersymmetric generalizations of
higher spin theories. The origin of interest in such extensions
roots in the existence of close relationship between string
theory and higher-spin fields which inevitably appear in the string
spectrum. So the field theory of higher spins must describe some
limit of superstring theory and, because of supersymmetry
of the latter, should also be supersymmetric.

An adequate framework to deal with supersymmetric theories is the
superspace formalism. The $\mathcal{N}=1$ superfield methods are
well known \cite{Gates:1983nr, Wess:1992cp, BK} and they have been
applied with great success for the construction of free massless
$\mathcal{N}=1$ superfield higher-spin theories \cite{Kuz1, Kuz2},
on both flat and AdS \cite{Kuz3} backgrounds, as well as  of
$\mathcal{N}=1$ free massive higher spin theories \cite{K}. Also
$\mathcal{N}=1$ superspace approach was used to develop the
Lagrangian description of the free massless $\mathcal{N}=2$ higher
spin multiplets \cite{KSG1, KSG2}. Various $\mathcal{N}=1$
supersymmetric cubic vertices in terms of $\mathcal{N}=1$
superfields have been constructed in \cite{Kuzenko:2017ujh, Buchbinder:2017nuc,
Buchbinder:2018gle, Buchbinder:2018nkp, Buchbinder:2018wwg,
Buchbinder:2018wzq,  HK1, HK2, Hutomo:2020wca, 
Gates:2019cnl}.
Also $\mathcal{N}$-extended cubic interactions were built by using
light-cone momentum superspace \cite{Metsaev:2019aig}.

The most appropriate superfield approach  to theories with $\mathcal{N}=2$ supersymmetry is $\mathcal{N}=2$ harmonic superspace (HSS) \cite{18, HSS, HSS1}.
Its main advantages are the possibility of unconstrained formulation of the hypermultiplet,
as well as geometric formulations of $\mathcal{N}=2$ supersymmetric Yang Mills theory and supergravity
in terms of unconstrained gauge potentials and supervielbeins.
The pivotal feature of the HSS approach is the presence of an $\mathcal{N}=2$ analytic subspace in $\mathcal{N}=2$ HSS,
on which the basic geometric quantities of all $\mathcal{N}=2$ supersymmetric theories of interest are defined as unconstrained superfields.

In ref. \cite{Buchbinder:2021ite} we have elaborated $\mathcal{N}=2$
extension of Fronsdal theory for the integer highest spin
$\mathbf{s}$\footnote{We use bold $\mathbf{s}$ to denote
$\mathcal{N}=2$ gauge supermultiplet with the higher spin s. On
shell such a supermultiplet contains the bosonic fields carrying
spins $s$ and $s-1$  and a doublet of the spin $s-\tfrac{1}{2}$
fermionic fields. In our notation, the spin $\mathbf{s} =
\mathbf{\tfrac{1}{2}}$ supermultiplet is the matter hypermultiplet.
It involves the physical fields of spins 0 and $\tfrac{1}{2}$.}
supermultiplets by heavily using the HSS approach. It is worth to
point out that all nice geometric features of the harmonic approach
have their manifestations in the $\mathcal{N}=2$ higher spin theory
constructed. It is quite possible that the clear geometry which
underlies the $\mathcal{N}=2$ higher spin theories in HSS
formulation can also provide some new insights into the general
geometric structure of higher spin theories.

The next natural step in  $\mathcal{N}=2$ higher spin theory was to allow for interactions.
In ref. \cite{Buchbinder:2022kzl}, we have presented $\mathcal{N}=2$ off-shell cubic vertices for the interaction
of a matter hypermultiplet with $\mathcal{N}=2$ massless higher spins gauge supermultiplets in the HSS approach.
This construction starts with identifying proper rigid symmetries of the free hypermultiplet action.
Gauging these symmetries by analytic superfield parameters  gives rise to superfield Noether cubic couplings. Demanding manifest rigid $\mathcal{N}=2$ supersymmetry imposes
rather restrictive  constraints on the possible form of cubic vertices. All these interactions have a universal form: they
are obtained by covariantizing the analyticity-preserving harmonic derivative through adding some higher-derivative supersymmetric differential operators
which involve analytic potentials of $\mathcal{N}=2$ higher spins theories and commute with rigid $\mathcal{N}=2$ supersymmetry.
The construction of the off-shell hypermultiplet couplings to $\mathcal{N}=2$ higher spins made use of  the whole arsenal of the harmonic methods and
demonstrated once more the crucial role of the HSS approach in revealing the intrinsic geometries of $\mathcal{N}=2$ supersymmetric theories,
including the higher spin ones.

In this article, we focus on a number of additional issues as a natural followup to the consideration
in \cite{Buchbinder:2022kzl, Buchbinder:2021ite}:

\begin{itemize}

    \item \textbf{Equations of motion}

    Superfield equations of motion are of interest from various points of view. The most natural motivation
    to study them is that they incorporate in a compact way the set of field equations for each spin in the supermultiplet \cite{Ogievetsky:1976qb}.
    A deeper motivation is to study their relation to the irreducible representations of $\mathcal{N}=2$ extended Poincar\'e superalgebra \cite{Rittenberg:1981cp}.
    So far, the realization of irreducible representations of $\mathcal{N}=2$ supersymmetry algebra on harmonic superfields has not been
    investigated in full generality. Moreover, the  study of such equations could give hints to the structure of massive
    $\mathcal{N}=2$ supersymmetric higher-spin theories.

    We shall derive $\mathcal{N}=2$ superfield equations corresponding to the theories that have been
    built in \cite{Buchbinder:2021ite}.
    We shall accomplish this in two ways: by varying with respect to the basic unconstrained
    analytic potentials and by varying
    with respect to the unconstrained generalized
    Mezincescu-type prepotentials\footnote{The spin $\mathbf{1}$
    Mezincescu prepotential for abelian gauge ${\cal N}=2$ gauge theory
    was defined in \cite{Mezincescu:1979af},
    and the analogous linearized spin $\mathbf{2}$ prepotentials for ${\cal N}=2$ supergravity were
    found in \cite{Gates:1981qq} and \cite{Zupnik:1998td}.}.

    \item \textbf{Hypermultiplet supercurrents}

    If some theory, in addition to the sector of the gauge fields, involves their interaction with some other fields, the latter induce sources in the equations
    for gauge fields.
    In this article, we study such sources for the $\mathcal{N}=2$ supersymmetric interaction of a
    hypermultiplet with the higher spin multiplets \cite{Buchbinder:2022kzl}.
    The relevant superfield sources are bilinear in the hypermultiplet superfields and are obtained
    by varying the unconstrained analytic potentials in the corresponding
    cubic vertices.
    The component conserved currents encoded in these superfield currents are related to the global symmetries of the component hypermultiplet action via Noether theorem.
   The higher-spin hypermultiplet superfield currents on their own can also be derived using a superfield version of the Noether theorem for global symmetries of the free superfield
   hypermultiplet action. This is in full agreement with the standard Noether construction of cubic vertices.

    Given the superfield currents just mentioned, we study their component structure, which allows us
    to compare them with the known component results \cite{Berends:1985xx, Gelfond:2006be} and
    to display  a number of restrictions imposed by rigid $\mathcal{N}=2$ supersymmetry.
    In addition, we reveal the structure behind all the obtained $\mathcal{N}=2$ superfield currents
    which is universal for all $\mathcal{N}=2$ supersymmetric
    theories in harmonic superspace, and not specific just for a free hypermultiplet.
    We pay special attention to the hypermultiplet  $\mathcal{N}=2$ supercurrent (spin $\mathbf{2}$ current superfield), which couples to the linearized $\mathcal{N}=2$ Einstein
    supergravity multiplet (spin $\mathbf{2}$ gauge supermultiplet) and encompasses the conserved doublet of the fermionic currents corresponding to the invariance
    of the  supersymmetric theory under $\mathcal{N}=2$ global supersymmetry, together with the
    energy-momentum tensor (coupled to the spin 2 gauge field) and
    the $R$-symmetry vector current (coupled to the spin 1 gauge field)\footnote{All these component currents constitute the hypermultiplet part of the total ${\cal N}=2$ supercurrent involving also
    a part depending on the components of the gauge ${\cal N}=2$ spin ${\bf 2}$ multiplet. At the free
    field level, the total supercurrent is just a sum of these two parts.}.

     Note that in ref. \cite{Kuzenko:2021pqm} there were constructed $\mathcal{N}=2$ higher spin conformal supercurrents for the on-shell massless hypermultiplet
     in the description through the conventional constrained ${\cal N}=2$ superfields.
     We start from the off-shell analytic hypermultiplet and construct the hypermultiplet supercurrents which make sense both off and on shell.
     In addition to exploring the superfield expressions, we also study the component structure of these supercurrents.

    \item \textbf{Component contents and $\xi$-interaction}

    As an important task we consider the component structure of the interactions built in ref. \cite{Buchbinder:2022kzl}.
    In particular, the role of the $\xi$-freedom present in the hypermultiplet couplings to odd higher ${\cal N}=2$ spins is of special interest.
     The question is as to whether it is possible to get rid of the $\xi$-interaction after the elimination of the auxiliary fields and/or through a field redefinition.
     We demonstrate that at the component level in the bosonic sector  the $\xi$-interaction in the $({3}, {0}, {0})$ vertex
     can be discarded on shell.

    The structure of the bosonic cubic $({s},0,0)$ interaction was described in \cite{Fotopoulos:2008ka, Bekaert:2009ud,  Khabarov:2020bgr, Zinoviev:2010cr}.
    The component structure of $\mathcal{N}=1$ supersymmetric vertices was studied in
    \cite{Khabarov:2020deh, Buchbinder:2021qkt}. We establish the relationship with these general
    results and find the agreement with them.

\end{itemize}

The plan of the paper is as follows. Section \ref{Harmonic superspace} gives a concise presentation of the basic elements of the HSS approach and introduces
the notations we make use of. In Section \ref{N=2 higher spin action in harmonic superspace}
we briefly review the  spin $\mathbf{s}$ analytic potentials $h^{++\dots}(\zeta)$
and the HSS formulation of $\mathcal{N}=2$ higher spin theories. Section \ref{equations of motion} and the related Appendix \ref{Component reduction of bosonic gauge sector}
are devoted to deriving the superfield equations of motion
for $\mathcal{N}=2$ higher spin gauge supermultiplets and inspecting their component field contents.
 We derive these equations in the two equivalent ways: using unconstrained analytic potentials
 $h^{++\dots}(\zeta)$ and unconstrained generalized
 Mezincescu-type prepotentials $\Psi^{-\dots}(\zeta, \theta^-)$ introduced in Appendix 
 \ref{Pre-prepotentials}. We study the component field contents of the superfield equations of motion in the bosonic sector using the relations of
 Appendix \ref{Component reduction of bosonic gauge sector} (where all the necessary elements for the component analysis of the bosonic sectors
of $\mathcal{N}=2$ spins $\mathbf{2}$ and $\mathbf{3}$ supermultiplet Lagrangians are collected).
The cubic superfield $(\mathbf{3}, \mathbf{1/2}, \mathbf{1/2} )$ couplings are reviewed in Section \ref{Hyper-current-superfields}, where we use them for deducing sources
in the superfield higher spin equations of motions. Also we give a detailed description of the component contents of the current superfields
in the spin $\mathbf{2}$ case and exhibit the most characteristic features of the spin $\mathbf{3}$ and general spin $\mathbf{s}$ cases.
We also analyze the component contents of the superfield equations in the presence of the hypermultiplet sources in the spin $\mathbf{2}$ and the spin $\mathbf{3}$ cases. Section \ref{Hypermultiplet equations of motion and
component reduction of cubic couplings} is devoted to the derivation of the hypermultiplet equations of motion in the presence of cubic coupling to the general spin $\mathbf{s}$ gauge supermultiplet. We explicitly present
the cubic couplings in terms of the free on-shell hypermultiplet superfield, as well as the component vertices implied by these superfield couplings. Summary and outlook are the contents of Section \ref{Summary and outlook}.
Appendix \ref{current} is devoted to the description of analytic current superfields and their general structure. In Appendix \ref{Noether} we explain the relation of these current superfields to global symmetries of the
free hypermultiplet action via Noether theorem. Appendix \ref{Fields redefinitions} contains various field redefinition for the $(2,0,0)$ and $(3,0,0)$ bosonic couplings, which were used in the component analysis  of the
corresponding superfield cubic couplings. In Appendix \ref{Gauge fixing} we sketch some details of gauge-fixing which can be useful for further studying the group theoretical aspects of the superfield equations.

\section{Harmonic superspace}
\label{Harmonic superspace}

We deal with ${\cal N}=2$ harmonic superspace (HSS) in the
analytic basis parametrized by the following set of coordinates
\cite{HSS,18, HSS1}
\be
Z = \big(x^m, \theta^{+\mu},
\bar\theta^{+\dot\mu}, u^\pm_i, \theta^{-\mu},
\bar\theta^{-\dot\mu}\big) \equiv \big(\zeta, \theta^{-\mu},
\bar\theta^{-\dot\mu}\big), \lb{HSS}
\ee
where the standard notation of ref. \cite{18} is used. In addition to the standard superspace
coordinates $(x, \theta)$ this superspace involves harmonic
variables. Harmonic variables $u^\pm_i$ parametrize the internal
sphere $S^2 \simeq {\rm SU}(2)/{\rm U}(1)$, $u^{+i}u^-_i =1$.  The
indices $\pm$ denote the harmonic ${\rm U}(1)$ charges of various
quantities and $i = 1, 2$ is the doublet index of the automorphism
group ${\rm SU}(2)_{aut}$  acting only on the harmonic variables.
The set \p{HSS} is closed under the rigid ${\cal N}=2$ supersymmetry
transformations
\be
\delta_\epsilon x^{\alpha\dot\alpha} = -2i
\big(\epsilon^{- \alpha}\bar\theta^{+ \dot\alpha} + \theta^{+
\alpha}\bar\epsilon^{- \dot\alpha}\big), \quad
\delta_\epsilon\theta^{\pm \hat \mu} = \epsilon^{\pm \hat \mu}\,,
\quad \delta_\epsilon u^{\pm}_i = 0\,, \quad \epsilon^{\pm \hat \mu}
= \epsilon^{\hat \mu i } u^\pm_i\,, \lb{N2SUSY}
\ee
where we employed the condensed notation, $\hat\mu = (\mu, \dot\mu)$. These
transformations also leave intact the harmonic analytic subspace of
\p{HSS},
\be
\zeta := \big(x^m, \theta^{+\mu},
\bar\theta^{+\dot\mu}, u^\pm_i\big). \lb{AHSS}
\ee

Moreover, analytic subspace is invariant under the generalized tilde-conjugation, which acts on the coordinates as:
\begin{equation}\label{tilde}
    \widetilde{x^m} = x^m\,, \qquad \widetilde{\theta^\pm_\alpha} = \bar{\theta}^\pm_{\dot{\alpha}} \,, \qquad
    \widetilde{\bar{\theta}^\pm_{\dot{\alpha}}} = - \theta^\pm_\alpha \,, \qquad
    \widetilde{u^{\pm i}} = - u_i^\pm\,, \qquad
    \widetilde{u^\pm_i} = u^{\pm i}\,.
\end{equation}
The tilde-conjugation acts on arbitrary function with $SU(2)$ indices as ordinary complex conjugation:
\begin{equation}\label{tilde-func}
    \widetilde{f^{i_1\dots i_n}_{}(x)} =    \overline{f^{i_1\dots i_n}_{}(x)} = \bar{f}_{i_1\dots i_n}^{}(x)\,.
\end{equation}
The complex conjugation of $SU(2)$ doublet gives $\overline{f^i} (x) = \bar{f}_i (x) = \epsilon_{ij} \bar{f}^{j}(x)$  and it satisfies the
property $\overline{\left(\overline{f^i}\right)}=\overline{\left(\bar{f}_i\right)} = f^i$.

The existence of the analytic superspace allows one to describe the hypermultiplet and $\mathcal{N}=2$ gauge supermultiplets by unconstrained analytic superfields.
It is very crucial for our further consideration. In particular, in order to derive the equations of motion from the action, we need to vary it with respect to unconstrained analytic superfields.

The HSS formulation of ${\cal N}=2$ gauge theories coupled to the hypermultiplet
matter uses an extension
of the HSS \p{HSS} by a fifth coordinate $x^5$,
\be
Z \Longrightarrow (Z, x^5)\,\supset \, (\zeta, x^5)\,, \lb{Ext5}
\ee
with the following
analyticity-preserving transformation law under ${\cal N}=2$
supersymmetry,
\be
\delta_\epsilon x^5 = 2i\big(\epsilon^-\theta^+ -
\bar\epsilon^-\bar\theta^+ \big). \lb{Tranfifth}
\ee
This coordinate
can be interpreted as associated with a central charge in ${\cal N}=2$ Poincar\'e superalgebra. All superfields except for the hypermultiplet
are assumed to be $x^5$-independent, while the action of $\partial_5$
on the hypermultiplet is identified with the generator of some ${\rm U}(1)$ isometry of the free hypermultiplet action (see Section \ref{sec:hyper}).

An important ingredient of the HSS formalism is the harmonic
derivatives $\mathcal{D}^{++}$ and $\mathcal{D}^{--}$ which have the following form in
the analytic basis \footnote{Hereafter, we use the notations
    $\hat{\mu} \equiv (\mu, \dot{\mu})$, $\partial^\pm_{\hat{\mu}} =
    \partial/ \partial \theta^{\mp \hat{\mu}}$, $(\theta^{\hat{+}})^2
    \equiv (\theta^+)^2 - (\bar{\theta}^+)^2$ and
    $\partial_{\alpha\dot\alpha} =
    \sigma^m_{\alpha\dot\alpha}\partial_m$. The summation rules are
    $\psi\chi = \psi^\alpha \chi_\alpha, \bar\psi\bar\chi =
    \bar\psi_{\dot\alpha} \bar\chi^{\dot\alpha}$, Minkowski metric is
    ${\rm diag} (1, -1, -1, -1)$ and $\Box = \partial^m\partial_m =
    \frac12 \partial^{\alpha\dot\alpha}\partial_{\alpha\dot\alpha}$.}
\begin{eqnarray}
    && \mathcal{D}^{++} = \partial^{++} - 2i \theta^{+\rho} \bar{\theta}^{+\dot{\rho}} \partial_{\rho\dot{\rho}} + \theta^{+\hat{\mu}} \partial^{+}_{\hat{\mu}}
    +
    i (\theta^{\hat{+}})^2 \partial_5\,,  \lb{Dflat+} \\
    && \mathcal{D}^{--} = \partial^{--}- 2i \theta^{-\rho} \bar{\theta}^{-\dot{\rho}} \partial_{\rho\dot{\rho}} + \theta^{-\hat{\mu}} \partial^{-}_{\hat{\mu}}
    +
    i (\theta^{\hat{-}})^2 \partial_5\,, \lb{Dflat} \\
    && [\mathcal{D}^{++}, \mathcal{D}^{--}] = D^0\,, \quad D^0 = \partial^0 + \theta^{+ \hat\mu}\partial^-_{\hat\mu}
    -\theta^{- \hat\mu}\partial^+_{\hat\mu}\,,\lb{Flatness}
\end{eqnarray}
where we have used the notations
\begin{equation}
    \partial^{++} = u^{+i} \frac{\partial}{\partial u^{-i}}\,,\qquad
    \partial^{--} = u^{-i} \frac{\partial}{\partial u^{+i}}\,,\qquad
    \partial^0 = u^{+ i}\frac{\partial}{\partial u^{+i}} - u^{- i}\frac{\partial}{\partial u^{-i}}\,.
\end{equation}
The crucial difference between the derivatives $\mathcal{D}^{++}$
and $\mathcal{D}^{--}$ is that $\mathcal{D}^{++}$ preserves
analyticity, while $\mathcal{D}^{--}$ does not. In what follows we
will also need the explicit form of the spinor derivatives
$D^{+\alpha}, \bar{D}^{+\dot\alpha}$ in the analytic basis:
\be
D^{+}_{\alpha} = \frac{\partial}{\partial \theta^{-\alpha}}\,, \quad
\bar{D}^{+}_{\dot\alpha}= \frac{\partial}{\partial
\bar\theta^{-\dot{\alpha}}}\,. \label{DbarD}
\ee
The spinor derivatives$D^{-}_{\alpha}, \bar{D}^{-}_{\dot\alpha}$ can be obtained by
commuting ${\cal D}^{--}$ defined in \eqref{Dflat} with \eqref{DbarD}
\begin{eqnarray}
 &&   D^-_{\mu} = [{\cal D}^{--}, D^+_{\mu}] = - \partial_\mu^- + 2i \bar{\theta}^{-\dot{\mu}} \partial_{\mu\dot{\mu}} - 2i \theta^-_{\mu} \partial_5\,, \nonumber \\
  &&  \bar{D}^-_{\dot{\mu}} = [{\cal D}^{--}, \bar{D}^+_{\dot{\mu}}] = -\partial_{\dot{\mu}}^- - 2i \theta^{-\mu} \partial_{\mu\dot{\mu}}
  - 2i \bar{\theta}^-_{\dot{\mu}} \partial_5\,.\label{D-}
\end{eqnarray}

\section{$\mathcal{N}=2$ higher spin action in harmonic superspace}
\label{N=2 higher spin action in harmonic superspace}

The free $\mathcal{N}=2$ higher spin $\bf{s}$ theory is described by the following gauge-invariant action in harmonic
superspace \cite{Buchbinder:2021ite}\footnote{In action \eqref{Action} we assume that $s\geq 2$. The action for the spin 1 case has the similar form,
see section \ref{Maxwell}. The measure of integration over the total harmonic superspace is defined as:
$$
\int d^4x d^8\theta du = \int d^4x du\; (D^+)^4 (D^-)^4\,,\quad (D^{\pm})^4 = \frac{1}{16} (\bar{D}^\pm)^2 (D^\pm)^2\,, \;\;
(D^\pm)^4 (\theta^{\mp})^2 (\bar{\theta}^\mp)^2 = 1\,.$$}.
\begin{equation}\label{Action}
    \begin{split}
    S^{(s)} = (-1)^{s+1} \int d^4x d^8\theta du\; \Bigr\{ &G^{++\alpha(s-1)\dot{\alpha}(s-1)} G^{--}_{\alpha(s-1)\dot{\alpha}(s-1)}
\\&\;\;\;+ G^{++5\alpha(s-2)\dot{\alpha}(s-2)} G^{--5}_{\alpha(s-2)\dot{\alpha}(s-2)} \Bigr\}\,.
    \end{split}
\end{equation}

Here $G^{++}$-superfields are constructed out of the analytic bosonic prepotentials $h^{++\alpha(s-1)\dot{\alpha}(s-1)}$, $h^{++\alpha(s-2)\dot{\alpha}(s-2)}$ and
analytic mutually conjugated fermionic prepotentials $h^{++\alpha(s-1)\dot{\alpha}(s-2)+}$, $h^{++\alpha(s-2)\dot{\alpha}(s-1)+}$ as
\begin{equation}\label{s-1}
    \begin{split}
    G^{++\alpha(s-1)\dot\alpha(s-1)} =&
    h^{++\alpha(s-1)\dot\alpha(s-1)}
    \\&+2i
    \big[h^{++\alpha(s-1)(\dot\alpha(s-2)+}\bar\theta^{-\dot\alpha_{s-1})}
    -
    h^{++\dot\alpha(s-1)(\alpha(s-2)+}\,\theta^{-\alpha_{s-1})}
    \big],
        \end{split}
\end{equation}
\begin{equation}\label{s-2}
    \begin{split}
    G^{++5\alpha(s-2)\dot\alpha(s-2)} &=
    h^{++5\alpha(s-2)\dot\alpha(s-2)} \\&- 2i
    \big[h^{++(\alpha(s-2)\alpha_{s-1})
        \dot\alpha(s-2)+}\theta^{-}_{\alpha_{s-1}} +
    h^{++\alpha(s-2)(\dot\alpha(s-2)\dot\alpha_{s-1})+}\,\bar\theta^{-}_{\dot{\alpha}_{s-1}}
    \big]\,.
        \end{split}
\end{equation}

$G^{--}$-superfields satisfy the zero-curvature equations:
\begin{equation}\label{zero-curv}
    \mathcal{D}^{--} G^{++\alpha(s-1)\dot\alpha(s-1)} = \mathcal{D}^{++} G^{--\alpha(s-1)\dot\alpha(s-1)}\,,
\end{equation}
\begin{equation}\label{zero-curv2}
    \mathcal{D}^{--} G^{++5\alpha(s-2)\dot\alpha(s-2)} = \mathcal{D}^{++} G^{--5\alpha(s-2)\dot\alpha(s-2)}\,.
\end{equation}
They can be expressed in terms of $G^{++}$-superfields as unique solutions of these equations.

The action \eqref{Action} is invariant under the gauge transformations:
\begin{equation}\label{Gauge_s}
    \begin{split}
        \delta_\lambda h^{++\alpha(s-1)\dot\alpha(s-1)} =\;& \mathcal{D}^{++} \lambda^{\alpha(s-1)\dot\alpha(s-1)} \\&+
        2i \big[\lambda^{+\alpha(s-1)(\dot\alpha(s-2)}\bar\theta^{+\dot\alpha_{s-1})} + \theta^{+(\alpha_{s-1}} \bar\lambda^{+\alpha(s-2))\dot\alpha(s-1)} \big], \\
        \delta_\lambda h^{++5\alpha(s-2)\dot\alpha(s-2)} =\;& \mathcal{D}^{++} \lambda^{5\alpha(s-2)\dot\alpha(s-2)} \\&-
        2i\,\big[\lambda^{+(\alpha(s-2)\alpha_{s-1})\dot\alpha(s-2)} \theta^+_{\alpha_{s-1}} +
        \bar\lambda^{+(\dot\alpha(s-2)\dot\alpha_{s-1})\alpha(s-2)} \bar\theta^+_{\dot\alpha_{s-1}} \big], \\
        \delta_\lambda  h^{++\alpha(s-1)\dot\alpha(s-2)+} =\;& \mathcal{D}^{++}\lambda^{+\alpha(s-1)\dot\alpha(s-2)}\,, \\
        \delta_\lambda h^{++\dot\alpha(s-1)\alpha(s-2)+} =\;&
        \mathcal{D}^{++}\tilde{\lambda}^{+\dot\alpha(s-1)\alpha(s-2)}\,.
    \end{split}
\end{equation}
Here the gauge parameters $\lambda^{\dots} (x, \theta^+, u)$ are arbitrary analytic functions. These transformations
give rise to the following gauge transformations of the superfields $G^{++\ldots}$ and $G^{--\dots}$:
\begin{subequations}\label{Lambda gauge}
\begin{equation}
    \delta_\lambda G^{\pm\pm\alpha(s-1)\dot{\alpha}(s-1)}
    =
    \mathcal{D}^{\pm\pm} \Lambda^{\alpha(s-1)\dot{\alpha}(s-1)}\,,
\end{equation}
\begin{equation}
    \delta_\lambda G^{\pm\pm5\alpha(s-2)\dot{\alpha}(s-2)}
    =
    \mathcal{D}^{\pm\pm} \Lambda^{5\alpha(s-2)\dot{\alpha}(s-2)}\,,
\end{equation}
\end{subequations}
where the gauge $\Lambda$-parameters are defined as
\begin{subequations}\label{Lambda}
\begin{equation}
    \begin{split}
    \Lambda^{\alpha(s-1)\dot\alpha(s-1)} =&
    \lambda^{\alpha(s-1)\dot\alpha(s-1)} \\&+
    2i\big[\lambda^{+\alpha(s-1)(\dot\alpha(s-2)}\bar\theta^{-\dot\alpha_{s-1})}
    - \bar\lambda^{+\dot\alpha(s-1)(\alpha(s-2)} \theta^{-\alpha_{s-1})}
    \big],
        \end{split}
\end{equation}
\begin{equation}
    \begin{split}
    \Lambda^{5\alpha(s-2)\dot\alpha(s-2)} =&
    \lambda^{5\alpha(s-2)\dot\alpha(s-2)}
    \\&-2i\big[\lambda^{+(\alpha(s-2)\alpha_{s-1})\dot\alpha(s-2)}\theta^{-}_{\alpha_{s-1}}
    -\,
    \bar\theta^{-}_{\dot\alpha_{s-1}}
    \bar\lambda^{+(\dot\alpha(s-2)\dot\alpha_{s-1})\alpha(s-2)}\big]\,.
    \end{split}
\end{equation}
\end{subequations}

The analytic potentials $h^{++}(x, \theta^+, u)$ contain infinitely many terms in their harmonic expansions, but most of them are pure gauge degrees of freedom.
Using the gauge freedom \eqref{Gauge_s}, one can impose Wess-Zumino-type gauge by gauging out all these redundant degrees:
\begin{equation}\label{WZ3}
    \begin{split}
        &h^{++\alpha(s-1)\dot{\alpha}(s-1)}
        =
        -2i \theta^{+\rho} \bar{\theta}^{+\dot{\rho}} \Phi^{\alpha(s-1)\dot{\alpha}(s-1)}_{\rho\dot{\rho}}
        +  (\bar{\theta}^+)^2 \theta^+ \psi^{\alpha(s-1)\dot{\alpha}(s-1)i}u^-_i  \\
        & \;\; \;\;\;\; \;\;\;\; \;\;\;\; \;\;\; \;\;\;\; \;\;\;\;\;+\, (\theta^+)^2 \bar{\theta}^+ \bar{\psi}^{\alpha(s-1)\dot{\alpha}(s-1)i}u_i^-
        +  (\theta^+)^2 (\bar{\theta}^+)^2 V^{\alpha(s-1)\dot{\alpha}(s-1)(ij)}u^-_iu^-_j\,, \\
        &  h^{++5\alpha(s-2)\dot{\alpha}(s-2)} =
        -2i \theta^{+\rho} \bar{\theta}^{+\dot{\rho}} C^{\alpha(s-2)\dot{\alpha}(s-2)}_{\rho\dot{\rho}}
        + (\bar{\theta}^+)^2 \theta^+ \rho^{\alpha(s-2)\dot{\alpha}(s-2)i}u^-_i
        \\
        & \;\; \;\;\;\; \;\;\;\; \;\;\;\; \;\;\; \;\;\;\; \;\;\;\;\;+\,
        (\theta^+)^2 \bar{\theta}^{+} \bar{\rho}^{\alpha(s-2)\dot{\alpha}(s-2)i}u_i^-
        + (\theta^+)^2 (\bar{\theta}^+)^2 S^{\alpha(s-2)\dot{\alpha}(s-2)(ij)}u^-_iu^-_j\,, \\
        &  h^{++\alpha(s-1)\dot{\alpha}(s-2)+} = (\theta^+)^2 \bar{\theta}^+_{\dot{\mu}} P^{\alpha(s-1)\dot{\alpha}(s-2)\dot{\mu}}
        \\
        & \;\; \;\;\;\; \;\;\;\; \;\;\;\; \;\;\; \;\;\;\; \;\;\;\;\;
        +  \left(\bar{\theta}^+\right)^2 \theta^+_\nu \left[\varepsilon^{\nu(\alpha} M^{\alpha(s-2))\dot{\alpha}(s-2)} + T^{\dot{\alpha}(s-2)(\alpha(s-1)\nu)}\right]
        \\
        & \;\; \;\;\;\; \;\;\;\; \;\;\;\; \;\;\; \;\;\;\; \;\;\;\;\;+\,
        (\theta^+)^2 (\bar{\theta}^+)^2 \chi^{\alpha(s-1)\dot{\alpha}(s-2)i}u^-_i\,, \\
        &  h^{++\dot{\alpha}(s-1)\alpha(s-2)+} = \widetilde{\left(h^{++\alpha(s-1)\dot{\alpha}(s-2)+}\right)}\,.
    \end{split}
\end{equation}
Here the fields
\begin{equation}\label{physical}
    \Phi^{\alpha(s-1)\dot{\alpha}(s-1)}_{\rho\dot{\rho}},\;\;\;\;\;
    \psi^{\alpha(s-1)\dot{\alpha}(s-1)i}_\mu          , \;\;\;\;\;
    \bar{\psi}^{\alpha(s-1)\dot{\alpha}(s-1)i}_{\dot\mu}, \;\;\;\;\;
    C^{\alpha(s-2)\dot{\alpha}(s-2)}_{\rho\dot{\rho}}
\end{equation}
are physical and stand for the bosonic integer spin $s$ gauge field, a doublet of the fermionic $s-\tfrac{1}{2}$ spin gauge fields and the bosonic spin $s-1$ gauge field.
All other fields are auxiliary and obey algebraic equations of motion in the theory with action \eqref{Action}. So the
analytical superfields $h^{++\alpha(s-2)\dot{\alpha}(s-2)M}$ ($M = (\alpha\dot{\alpha}, \hat{\alpha}+,5)$  with gauge transformations \eqref{Gauge_s} represent
 $\mathcal{N}=2$ gauge supermultiplet with the highest spin $\mathbf{s}$. The residual gauge freedom in the Wess-Zumino gauge coincides with the gauge freedom  of
 the higher spin theories by Fronsdal \cite{FronsdalInteg} and by Fang-Fronsdal \cite{FronsdalHalfint}. It is worth noting that in what follows we shall deal with the maximally reduced bosonic
gauge fields in which the linearized local Lorentz invariance and its higher spin analogs are  completely fixed, leaving the highest and lowest rank irreducible components
of these fields $\big(\Phi^{\alpha(s)\dot\alpha(s)}, \;\Phi^{\alpha(s-2)\dot\alpha(s-2)}\,(s\geq 2) \big)$ and $\big(C^{\alpha(s-1)\dot\alpha(s-1)}, \; C^{\alpha(s-3)\dot\alpha(s-3)}\, (s\geq 3) \big)$ as independent.
For further details see \cite{Buchbinder:2021ite}.

It is instructive to note that all the physical fields \eqref{physical} originally sit in the analytic bosonic potentials $h^{++\alpha(s-1)\dot{\alpha}(s-1)}$
and $h^{++5\alpha(s-2)\dot{\alpha}(s-2)}$. The spinor potentials contain only the auxiliary degrees of freedom.
However, these components possess non-trivial transformation laws under the residual gauge transformations and, in order to make them inert,  one needs to perform
some field redefinitions giving
rise to the appearance of effective dependence on physical fields in spinor potentials \cite{Buchbinder:2021ite}. So the spinor  potentials play a crucial
role in the harmonic superspace construction of higher spin theories and cannot be naively
eliminated \footnote{The role of spinor  potentials is somewhat similar to the auxiliary fields in the
frame-like formalism of higher spins, see, e.g., \cite{Zinoviev:2010cr}.}.

The bosonic analytic potentials
have non-standard transformation laws under $\mathcal{N}=2$ global supersymmetry
\footnote{$\delta_\epsilon$ denotes the  ``passive'' transformation, {\it i.e.}, $\delta_\epsilon \Phi(Z) = \Phi^\prime(Z^\prime) - \Phi(Z)$, which differs from the ``active'' transformation
$\delta_\epsilon^*$ defined as $\delta^*_\epsilon \Phi(Z) = \Phi^\prime(Z) - \Phi(Z)$,
    by the ``transport term'', $\delta^*_\epsilon = \delta_\epsilon -\delta_\epsilon Z^M \partial_M$.}:
\begin{subequations}\label{susy}
    \begin{equation}\label{susy1}
        \begin{split}
            \delta_\epsilon h^{++\alpha(s-1)\dot\alpha(s-1)} = -2i\big[h^{++\alpha(s-1)(\dot\alpha(s-2)+}\bar\epsilon^{-\dot\alpha_{s-1})}-
            h^{++\dot\alpha(s-1)(\alpha(s-2)+}\,\epsilon^{-\alpha_{s-1})}
            \big]\,, \\
            \delta_\epsilon h^{++5\alpha(s-2)\dot\alpha(s-2)} =2i\big[h^{++(\alpha(s-2)\alpha_{s-1})
                \dot\alpha(s-2)+}\epsilon^{-}_{\alpha_{s-1}} +
            h^{++\alpha(s-2)(\dot\alpha(s-2)\dot\alpha_{s-1})+}\,\bar\epsilon^{-}_{\dot{\alpha}_{s-1}}
            \big] \,.
        \end{split}
    \end{equation}
   The  spinor potentials have the standard $\mathcal{N}=2$ superfield transformation rules\footnote{In the full nonlinear case of
        hypothetical ``higher-spin ${\cal N}=2$ supergravity'' the rigid ${\cal N}=2$ supersymmetry is expected to be included in the general gauge transformations.}:
    \begin{equation}\label{susy2}
        \delta_\epsilon h^{++\alpha(s-1)\dot\alpha(s-2)+} = 0,
        \;\;\;\;\;\;\;\;
        \delta_\epsilon h^{++\dot\alpha(s-1)\alpha(s-2)+} = 0\,.
    \end{equation}
\end{subequations}
It is straightforward to be convinced that $\delta_{\epsilon}G^{++\cdots} = 0$ and, as a consequence of zero-curvature conditions, $\delta_{\epsilon}G^{--\cdots} = 0$.
So the action \eqref{Action} is manifestly invariant under $\mathcal{N}=2$ supersymmetry.

The structure of supersymmetry transformations \eqref{susy} implies that the analytic potentials $h^{++M}$ form not fully reducible representation
of $\mathcal{N}=2$ supersymmetry. The spinor potentials $h^{++\alpha(s-1)\dot\alpha(s-2)+}$ and $h^{++\dot\alpha(s-1)\alpha(s-2)+}$
are closed under $\mathcal{N}=2$ supersymmetry, while the bosonic potentials transform through them.
In the subsequent Sections we shall show how this subtlety affects the structure of $\mathcal{N}=2$ supercurrents. Note that the unconstrained prepotentials
$\Psi^{-\alpha(s-2)\dot{\alpha}(s-2)\hat{\mu}}$ introduced  in Appendix \ref{Pre-prepotentials} are scalars under rigid ${\cal N}=2$ supersymmetry, see \eqref{susy PSI}.
However, they are not analytic and display a larger gauge freedom of non-geometric nature, which makes them not too appropriate for constructing interactions with the
analytic hypermultiplet superfields.

Since in  what follows we will be basically interested in the component structure of $\mathbf{s}=\mathbf{2}$ and $\mathbf{s}=\mathbf{3}$ higher-spin theories,
it is useful to give the explicit form of the Wess-Zumino gauge for these particular cases.

The off-shell $\mathcal{N}=2$, $\mathbf{s}=\mathbf{2}$ multiplet (linearized multiplet of $\mathcal{N}=2$ minimal Einstein supergravity \cite{Fradkin:1979as, Fradkin:1979cw, deWit:1979xpv}) is spanned
by the following fields:
\begin{equation}\label{Spin 2 multiplet}
    \left\{ \Phi^{\alpha\dot{\alpha}}_{\beta\dot{\beta}} , \psi^{\alpha\dot{\alpha}i}_\mu,
    \bar{\psi}^{\alpha\dot{\alpha}i}_{\dot{\mu}}, C_{\alpha\dot{\alpha}}  \Bigr|  V^{\alpha\dot{\alpha}(ij)}, \rho_\alpha^i, \bar{\rho}^i_{\dot{\alpha}}, S^{(ij) }, P^{\alpha\dot{\alpha}}, \bar{P}^{\alpha\dot{\alpha}},  M, T^{(\alpha\beta)}, \bar{T}^{(\dot{\alpha}\dot{\beta})}, \chi^{\alpha i}, \bar{\chi}^{\dot{\alpha i }}   \right\}.
\end{equation}
Here $4D$ fields $\left\{\Phi^{\alpha\dot{\alpha}}_{\beta\dot{\beta}} , \psi^{\alpha\dot{\alpha}}_\mu,
\bar{\psi}^{\alpha\dot{\alpha}}_{\dot{\mu}}, C_{\alpha\dot{\alpha}}\right\}$ are physical and, after going on shell, form  $\mathcal{N}=2$ spin $\mathbf{2}$ multiplet (the spin 2 field, a doublet of
the spin $\tfrac{3}{2}$ fields and
the spin 1 field). All other off-shell fields are auxiliary. The multiplet \eqref{Spin 2 multiplet} is accommodated by the analytic potentials:
\begin{equation}\label{WZ3 s=2}
    \begin{split}
        &h^{++\alpha\dot{\alpha}}
        =
        -2i \theta^{+\rho} \bar{\theta}^{+\dot{\rho}} \Phi^{\alpha\dot{\alpha}}_{\rho\dot{\rho}}
        +  (\bar{\theta}^+)^2 \theta^+ \psi^{\alpha\dot{\alpha}i}u^-_i  \\
        & \;\; \;\;\;\; \;\;\;\; +\, (\theta^+)^2 \bar{\theta}^+ \bar{\psi}^{\alpha\dot{\alpha}i}u_i^-
        +  (\theta^+)^2 (\bar{\theta}^+)^2 V^{\alpha\dot{\alpha}(ij)}u^-_iu^-_j\,, \\
        &  h^{++5} =
        -2i \theta^{+\rho} \bar{\theta}^{+\dot{\rho}} C_{\rho\dot{\rho}}
        + (\bar{\theta}^+)^2 \theta^+ \rho^{i}u^-_i+\,
        (\theta^+)^2 \bar{\theta}^{+} \bar{\rho}^{i}u_i^-
        + (\theta^+)^2 (\bar{\theta}^+)^2 S^{(ij)}u^-_iu^-_j\,, \\
        &  h^{++\alpha+} = (\theta^+)^2 \bar{\theta}^+_{\dot{\mu}} P^{\alpha\dot{\mu}}
        +  \left(\bar{\theta}^+\right)^2 \theta^+_\nu \left[\varepsilon^{\nu(\alpha} M^{} + T^{(\alpha\nu)}\right]+\,
        (\theta^+)^2 (\bar{\theta}^+)^2 \chi^{\alpha i}u^-_i\,, \\
        &  h^{++\dot{\alpha}+} = \widetilde{\left(h^{++\alpha+}\right)}.
    \end{split}
\end{equation}
The residual gauge freedom amounts to the standard redundancy of the free spin ${2}$, spin $3/2$ and spin $1$ gauge fields (in the ``frame-like'' formalism).

Analogously, the $\mathcal{N}=2$ $\mathbf{s}=\mathbf{3}$ off-shell multiplet,
\begin{equation}\label{Spin 3 multiplet}
    \begin{split}
    \Bigr\{ &\Phi^{(\alpha\beta)(\dot{\alpha}\dot{\beta})}_{\rho\dot{\rho}} , \psi^{(\alpha\beta)(\dot{\alpha}\dot{\beta})i}_\mu,
    \bar{\psi}^{(\alpha\beta)(\dot{\alpha}\dot{\beta})i}_{\dot{\mu}}, C^{\alpha\dot{\alpha}}_{\beta\dot{\beta}}  \;\Bigr| \; V^{(\alpha\beta)(\dot{\alpha}\dot{\beta})(ij)},
\rho_\alpha^{\beta\dot{\beta}i}, \bar{\rho}^{\beta\dot{\beta}i}_{\dot{\alpha}}, S^{\alpha\dot{\alpha}(ij) }, \\&P^{(\alpha\beta)\dot{\alpha}\dot{\beta}}, \bar{P}^{\alpha\beta(\dot{\alpha}\dot{\beta})}, M^{\alpha\dot{\alpha}},
T^{\dot{\alpha}(\alpha\beta\nu)}, \bar{T}^{\alpha(\dot{\alpha}\dot{\beta}\dot{\nu})}, \chi^{(\alpha\beta)\dot{\alpha} i}, \bar{\chi}^{(\dot{\alpha}\dot{\beta})\alpha i }   \Bigr\},
    \end{split}
\end{equation}
involves the physical fields   $\left\{ \Phi^{(\alpha\beta)(\dot{\alpha}\dot{\beta})}_{\rho\dot{\rho}} , \psi^{(\alpha\beta)(\dot{\alpha}\dot{\beta})i}_\mu,
\bar{\psi}^{(\alpha\beta)(\dot{\alpha}\dot{\beta})i}_{\dot{\mu}}, C^{\alpha\dot{\alpha}}_{\beta\dot{\beta}} \right\}$ (the spin 3 gauge field, a doublet of the spin $\tfrac{5}{2}$ fields
and the spin 2 field)
which are going to form  $\mathcal{N}=2$ spin ${\bf 3}$ multiplet on shell. All other fields are
auxiliary. The multiplet \eqref{Spin 3 multiplet}
is accommodated by the  analytic potentials:
\begin{equation}\label{WZ3 s=3}
    \begin{split}
        &h^{++(\alpha\beta)(\dot{\alpha}\dot{\beta})}
        =
        -2i \theta^{+\rho} \bar{\theta}^{+\dot{\rho}} \Phi^{(\alpha\beta)(\dot{\alpha}\dot{\beta})}_{\rho\dot{\rho}}
        +  (\bar{\theta}^+)^2 \theta^+ \psi^{(\alpha\beta)(\dot{\alpha}\dot{\beta})i}u^-_i  \\
        & \;\; \;\;\;\; \;\;\;\; \;\;\;\; \;\;\; +\, (\theta^+)^2 \bar{\theta}^+ \bar{\psi}^{(\alpha\beta)(\dot{\alpha}\dot{\beta})i}u_i^-
        +  (\theta^+)^2 (\bar{\theta}^+)^2 V^{(\alpha\beta)(\dot{\alpha}\dot{\beta})(ij)}u^-_iu^-_j\,, \\
        &  h^{++5\alpha\dot{\alpha}} =
        -2i \theta^{+\rho} \bar{\theta}^{+\dot{\rho}} C^{\alpha\dot{\alpha}}_{\rho\dot{\rho}}
        + (\bar{\theta}^+)^2 \theta^+ \rho^{\alpha\dot{\alpha}i}u^-_i
        \\
        & \;\; \;\;\;\; \;\;\;\; \;\;\;\; \;\;\; +\,
        (\theta^+)^2 \bar{\theta}^{+} \bar{\rho}^{\alpha\dot{\alpha}i}u_i^-
        + (\theta^+)^2 (\bar{\theta}^+)^2 S^{\alpha\dot{\alpha}(ij)}u^-_iu^-_j\,, \\
        &  h^{++(\alpha\beta)\dot{\alpha}+} = (\theta^+)^2 \bar{\theta}^+_{\dot{\mu}} P^{(\alpha\beta)\dot{\alpha}\dot{\mu}}
        +  \left(\bar{\theta}^+\right)^2 \theta^+_\nu \left[\varepsilon^{\nu(\alpha} M^{\beta)\dot{\alpha}} + T^{\dot{\alpha}(\alpha\beta\nu)}\right]
        \\
        & \;\; \;\;\;\; \;\;\;\; \;\;\;\; \;\;\; +\,
        (\theta^+)^2 (\bar{\theta}^+)^2 \chi^{(\alpha\beta)\dot{\alpha}i}u^-_i\,, \\
        &  h^{++(\dot{\alpha}\dot{\beta})\alpha+} = \widetilde{\left(h^{++(\alpha\beta)\dot{\alpha}+}\right)}\,.
    \end{split}
\end{equation}
The residual gauge freedom corresponds to the standard frame-like redundancy of spin 3, spin $\tfrac{5}{2}$ and spin 2 fields.

Recall that in our further analysis the bosonic gauge fields will be taken in the ``physical gauge'', with only the highest and lowest rank irreducible components
being non-vanishing. All the intermediate rank components are assumed to be gauged away by fixing gauges with respect to the linearized local Lorentz symmetry and its
spin $3$ analog \cite{Buchbinder:2021ite}.

\section{Superfield equations of motion} \label{equations of motion}

In order to derive the equations of motion from some action it is necessary, first of all, to identify unconstrained (super)fields.
In our theories, the equations of motion can be derived from the action \eqref{Action} in the two equivalent ways, by varying it with respect to:
\begin{itemize}
    \item   unconstrained analytic potentials $h^{++\alpha(s-2)\dot{\alpha}(s-2)M}(\zeta)$\,;
    \item  unconstrained Mezincescu-type non-analytic prepotentials $\Psi^{-\alpha(s-2)\dot{\alpha}(s-2)\hat{\mu}}(Z)$ introduced in Appendix \ref{Pre-prepotentials}.
\end{itemize}
     Of course, the final results must be identical.

     Firstly, we will analyze the simplest example of $\mathcal{N}=2$ Maxwell theory, deriving its superfield equations of motion in the two ways mentioned above.
     This example provides hints as to how the procedure of deriving equations of motion  works in more complicated cases.
     Further  we will proceed to deriving the equations of motion for an arbitrary $\mathcal{N}=2$ higher spin supermultiplet.
     Initially, we use unconstrained analytic potentials and then reproduce the same results  in an alternative way, by
     using the Mezincescu-type prepotentials. In the end of this Section, we
     shall consider the component structure of the  superfield equations of motion for some simplest cases.

\subsection{$\mathcal{N}=2$ supersymmetric spin $\mathbf{1}$ theory}\label{Maxwell}

We begin with the simplest example of $\mathcal{N}=2$ supersymmetric gauge theory, $\mathcal{N}=2$ Maxwell theory \cite{18,HSS,HSS1}.
Its action can be treated as a degenerate case of the general $\mathcal{N}=2$ higher spin action \eqref{Action}:
\begin{equation}\label{Maxwell action}
    S_{(s=1)} =  \int d^4x d^8\theta du\; h^{++}h^{--}\,.
\end{equation}
Here $h^{++}(\zeta)$ is an analytic potential and $h^{--}$ is the solution of zero-curvature condition:
\begin{equation}\label{zero-M}
    \mathcal{D}^{--} h^{++} = \mathcal{D}^{++} h^{--}\,.
\end{equation}

The action and zero-curvature conditions are invariant under abelian gauge transformations with an arbitrary analytic gauge parameter $\lambda(\zeta)$:
\begin{equation}
    \delta h^{++} = \mathcal{D}^{++} \lambda\,, \qquad  \delta h^{--} = \mathcal{D}^{--} \lambda\,.
\end{equation}

Using these gauge transformations, one can choose the Wess-Zumino gauge for $h^{++}$,
\begin{equation}\label{Maxwell WZ}
    \begin{split}
    h^{++}_{WZ} =& -2i \theta^+\sigma^m \bar{\theta}^+ A_m - i (\theta^+)^2 \bar{\phi} + i (\bar{\theta}^+)^2 \phi \\&+ 4 (\bar{\theta}^+)^2 \theta^{+\alpha} \psi^i_\alpha u^-_i - 4 (\theta^+)^2 \bar{\theta}^+_{\dot{\alpha}} \bar{\psi}^{\dot{\alpha}i} u^-_i + (\theta^+)^2 (\bar{\theta}^+)^2 D^{ij} u^-_i u^-_j\;,
        \end{split}
\end{equation}
which yields just the off-shell field content of massless $\mathcal{N}=2$ spin $\mathbf{1}$ multiplet involving a complex scalar, a doublet of gaugini,
Maxwell gauge field and a real triplet of auxiliary fields:
\begin{equation}
    \phi, \quad \psi^i_\alpha, \quad A_m\,,\quad D^{(ij)}\,.
\end{equation}

Now we derive the superfield equations of motion corresponding to the action \eqref{zero-M} by two equivalent methods.

\medskip
\underline{\textit{\textbf{By varying with respect to the analytic potential}}}
\medskip

$\mathcal{N}=2$ Maxwell action can be rewritten only in terms of the analytic potential $h^{++}$ using harmonic distributions\footnote{For a pedagogical discussion of harmonic distributions see section 4.4 of \cite{18}.}:
\begin{equation}\label{analytic}
    S_{(s=1)} =  \int \frac{d^4x d^8\theta du du_1}{(u^+u_1^+)^2}\; h^{++} (x, \theta, u) h^{++} (x, \theta, u_1)\,.
\end{equation}
It is important that all analytic superfields here are written in the central basis,
\begin{equation}\label{cetral basis}
     h^{++} (x, \theta, u) =  h^{++} (x^m - 2i \theta^{(i} \sigma^m \bar{\theta}^{j)} u^+_i u^-_j, \theta^i_{\hat{\alpha}}u^+_i, u^\pm)\,,
\end{equation}
and the integral is taken over the product $\mathbb{R}^{4|8}\times S^2 \times S^2$.

To rewrite $\mathcal{N}=2$ Maxwell action \eqref{Maxwell action} in such a form, we use the solution of zero-curvature equation \eqref{zero-M} involving the harmonic distributions:
\begin{equation}\label{solution ZC}
    h^{--}(u) = \int\frac{du_1}{(u^+u_1^+)^2 } h^{++}(u_1)\,.
\end{equation}
Here $h^{++}$ is taken in the central basis \eqref{cetral basis}.
Using the property
\begin{equation}
    D^{++} \frac{1}{(u^+u_1^+)^2} = D^{--} \delta^{(2,-2)} (u, u_1)\,, \label{IdentDdelta}
\end{equation}
one can easily check that \eqref{solution ZC} satisfies the zero-curvature equation \eqref{zero-M}.

Now one can vary the action, using the representation \eqref{analytic},
\begin{equation}
    \delta S_{(s=1)} = 2 \int \frac{d^4x d^8\theta du du_1}{(u^+u_1^+)^2}\; \delta h^{++} (x, \theta, u)\, h^{++} (x, \theta, u_1)\,.
\end{equation}
With the help of \eqref{solution ZC} and \eqref{IdentDdelta} one can rewrite the variation of the action as an integral over harmonic superspace,
\begin{equation}
    \delta S_{(s=1)} = 2 \int d^4x d^8\theta du \; \delta h^{++} (\zeta)\, h^{--} (\zeta, \theta^-)\,.
\end{equation}
Then one can pass to the integral over the analytic superspace:
\begin{equation}
    \delta S_{(s=1)} = 2 \int  d\zeta^{(-4)}\; \delta h^{++} (\zeta)\, (D^+)^4 h^{--} (\zeta, \theta^-)\,,
\end{equation}
where
\be
d\zeta^{(-4)} =  d^4x du (D^-)^4
\ee
is the analytic superspace integration measure. Finally, keeping in mind that $\delta h^{++} (\zeta)$ is an arbitrary unconstrained analytic superfield,
we derive the sought equation of motion from the minimal action principle $\delta S_{(s=1)} = 0$,
\begin{equation}\label{Maxwell eom}
    (D^+)^4 h^{--} = 0\,.
\end{equation}

\medskip
\underline{\textit{\textbf{By varying with respect to Mezincescu's prepotential}}}
\medskip

The $\mathcal{N}=2$ analytic Maxwell potential $h^{++}(\zeta)$ and its gauge parameter $\lambda(\zeta)$ can be expressed in terms of general harmonic superfields \cite{18}:
\begin{equation}\label{M-superfield}
    h^{++} (\zeta) = (D^+)^4 M^{--} (Z)\,,
\end{equation}
\begin{equation}
    \lambda(\zeta) = (D^+)^4 \rho^{(-4)} (Z)\,.
\end{equation}
The superfields $M^{--}$ and $\rho^{(-4)}$ are unconstrained harmonic superfields. One can choose a supersymmetric gauge with a finite set of fields:
\begin{equation}\label{M-gauge}
    M^{--}(Z) = M^{(ij)} u^-_i u^-_j \qquad \Leftrightarrow \qquad (D^{++})^3 h^{++} = 0\,,
\end{equation}
where $M^{(ij)}$ is some ordinary harmonic-independent ${\cal N}=2$ superfield. The residual gauge transformations of $M^{(ij)}$ are:
\begin{equation}
    \delta M^{(ij)} = D^\alpha_k \chi_\alpha^{(ijk)} + \bar{D}_{\dot{\alpha}k} \bar{\chi}^{(ijk)\dot{\alpha}}\,.
\end{equation}
This coincides with the result of Mezincescu \cite{Mezincescu:1979af} obtained by directly solving the superfield constraints of $\mathcal{N}=2$ Maxwell theory.
We will call the Mezincescu prepotential also the harmonic superfield $M^{--}(Z)$ defined in \eqref{M-superfield} (equally as its higher-spin counterparts).

We now derive the equations of motion of $\mathcal{N}=2$ Maxwell theory, using the representation \eqref{M-superfield}.
Varying the action \eqref{analytic} with respect to $M^{--}(Z)$, we obtain:
\begin{equation}
    \begin{split}
    \delta S_{(s=1)} &= 2 \int \frac{d^4x d^8\theta du du_1}{(u^+u_1^+)^2}\; \left((D^+)^4 \delta M^{--}(Z)\right) \; h^{++} (x, \theta, u_1)
    \\& = 2 \int d^4 x d^8\theta du \; \delta M^{--}(Z) \; (D^+)^4 h^{--} (x, \theta, u) \,.
    \end{split}
\end{equation}
While passing to the second line, we used \eqref{solution ZC} and integrated by parts.
Since $\delta M^{--}(Z)$ is an unconstrained harmonic superfield,
we once again can apply the minimal action principle $\delta S_{(s=1)}=0$ to come back to the same equation of motion \eqref{Maxwell eom}.

\medskip
\underline{\textit{\textbf{Comments}}}
\medskip

\begin{enumerate}
    \item
Before turning to the study of the component field content of eq. \eqref{Maxwell eom}, it is important to note
that adding of the superfield current (source) to it implies the essential constraints on this current.
First of all, the superfield current must be analytic and possess the harmonic charge $+2$.
Moreover, it must satisfy the harmonic conservation equation:
\begin{equation}
    (D^+)^4 h^{--} = J^{++}\qquad \Rightarrow \qquad\mathcal{D}^{++}J^{++} =0 \,.
\end{equation}
The general component structure of such analytic current superfields is discussed in Appendix \ref{N=2 current superfield}. The current
corresponding to the minimal interaction of the hypermultiplet with $\mathcal{N}=2$ Maxwell multiplet is described in Section \ref{sec:cubic Maxwell}.

\item The superfield equation \eqref{Maxwell eom} involves the basic analytic potential $h^{++}(\zeta)$ only implicitly, through the zero-curvature condition.
In fact, eq. \eqref{Maxwell eom}, taken on its own, can be treated as restricting the component expansion of the superfield $h^{--}$. E.g., the gauge invariance of this equation
can be checked without resorting to the zero-curvature condition:
\begin{equation}
    \delta \left[ (D^+)^4 h^{--} \right] = (D^+)^4 \mathcal{D}^{--} \lambda = 0\,.
\end{equation}

\item It is straightforward to use the Wess-Zumino gauge \eqref{Maxwell WZ} in order to explicitly solve the zero-curvature equations
and to find the component equations of motion.

As an illustration, consider how this works for the gauge field $A_m$. To obtain the relevant piece of $h^{--}$, one needs to solve the zero-curvature equation \eqref{zero-M} for
\begin{equation}
    h^{++}_A = -2i \theta^+ \sigma^m \bar{\theta}^+ A_m\,.
\end{equation}
This part of the zero-curvature equation is solved by
\begin{equation}
    \begin{split}
    h_A^{--} &= -2i\theta^- \sigma^m \bar{\theta}^- A_m
    +
    2(\theta^-)^2 \bar{\theta}^{-(\dot{\rho}}\bar{\theta}^{+\dot{\beta})} \partial_{\dot{\rho}}^\beta A_{\beta\dot{\beta}}
    -
    2 (\bar{\theta}^-)^2 \theta^{-(\rho} \theta^{+\beta)} \partial_\rho^{\dot{\beta}} A_{\beta\dot{\beta}}
      \\&\qquad+ 2i (\theta^-)^2 (\bar{\theta}^-)^2 \theta^+ \sigma^m \bar{\theta}^+ \left(\Box A_m - \partial_m \partial^n A_n \right)\,.
        \end{split}
\end{equation}
The last term can be written in terms of Maxwell  field strength:
\begin{equation}
    \Box A_m - \partial_m \partial^n A_n  = \partial^n F_{nm}\,,\qquad F_{nm} = \partial_n A_m - \partial_m A_n\,.
\end{equation}
Thus, in the gauge field sector, we obtain the ordinary Maxwell equation:
\begin{equation}
    (D^+)^4 h^{--}_A =2i \theta^+ \sigma^m \bar{\theta}^+ \partial^n F_{nm} = 0 \qquad \Rightarrow \qquad  \partial^n F_{nm} = 0\,.
\end{equation}

\end{enumerate}

In the next Sections, we shall generalize this consideration to the case of arbitrary integer spin $\mathcal{N}=2$ supermultiplets.

\subsection{Arbitrary $\mathcal{N}=2$ higher spin: analytic potentials}

Now we turn to the derivation of the equations of motion for $\mathcal{N}=2$ higher spin ${\bf s}$ multiplet. The procedure is similar to that for $\mathcal{N}=2$ Maxwell multiplet,
though there are some important distinctions.

To obtain the equations of motion it is necessary to rewrite the action \eqref{Action} in terms of the analytic potentials $h^{++\dots}(\zeta)$.
One can do this at cost of appearance of non-localities in harmonics:
\begin{multline}\label{action++++}
    S_{(s)} = (-1)^{s-1} \int \frac{d^4x d^8\theta du du_1}{(u^+u^+_1)^2}\; \Bigr\{G^{++\alpha(s-1) \dot{\alpha}(s-1) }(u)G^{++}_{\alpha(s-1) \dot{\alpha}(s-1)}(u_1) \\
    +  G^{++5\alpha(s-2) \dot{\alpha}(s-2)}(u)G^{++5}_{\alpha(s-2) \dot{\alpha}(s-2)}(u_1)  \Bigr\}\,.
\end{multline}

Here we made use of the solution of zero-curvature equations \eqref{zero-curv}:

\begin{equation}\label{zero-1}
    G^{--\alpha(s-1) \dot{\alpha}(s-1) }(u) = \int\frac{du_1}{(u^+u_1^+)^2 } G^{++\alpha(s-1) \dot{\alpha}(s-1) }(u_1)\,,
\end{equation}
\begin{equation}\label{zero-2}
    G^{--5\alpha(s-2) \dot{\alpha}(s-2) }(u) = \int\frac{du_1}{(u^+u_1^+)^2 } G^{++5\alpha(s-2 ) \dot{\alpha}(s-2) }(u_1)\,.
\end{equation}
All involved superfields are in the central basis, as in the previous Section, recall eq. \eqref{cetral basis}.

Using this representation, we can directly vary the action \eqref{action++++} with respect to the unconstrained analytic potentials $h^{++\dots}$, taking into account how
the non-analytical potentials $G^{--\dots}$ are expressed through them (eqs. \eqref{s-1} and \eqref{s-2}). Since there are four independent analytic gauge potentials,
we obtain four different superfield equations of motion.

Let us consider independent equations separately.

\medskip

\textbf{\textit{\underline{Varying with respect to $h^{++\alpha(s-1) \dot{\alpha}(s-1) }$}}}

\medskip

Using the relations
\begin{equation}
    \delta G^{++\alpha(s-1)\dot\alpha(s-1)} =
    \delta h^{++\alpha(s-1)\dot\alpha(s-1)} \,,
\end{equation}
\begin{equation}
    \delta G^{++\alpha(s-2)\dot\alpha(s-2)} = 0\,,
\end{equation}
we immediately obtain the relevant variation of the action \eqref{action++++}:
\begin{equation}
    \delta S_{(s)} = (-1)^{s-1}  \int \frac{d^4x d^8\theta du du_1}{(u^+u^+_1)^2}\; \Bigr\{ 2 \delta h^{++\alpha(s-1) \dot{\alpha}(s-1) }(u)G^{++}_{\alpha(s-1) \dot{\alpha}(s-1)}(u_1)   \Bigr\}\,.
\end{equation}

Using the representation \eqref{zero-1} and passing to the analytic integration measure, we have
\begin{equation}
    \delta S_{(s)} = (-1)^{s-1}  \int d\zeta^{(-4)} \; \Bigr\{ 2 \delta h^{++\alpha(s-1) \dot{\alpha}(s-1) }(D^+)^4 G^{--}_{\alpha(s-1) \dot{\alpha}(s-1)}   \Bigr\}\,.
\end{equation}

The variation $\delta h^{++\alpha(s-1) \dot{\alpha}(s-1) }$ is an arbitrary unconstrained analytic superfield.
So, from the minimal action principle one obtains  the sought equation of motion as
\begin{equation}\label{eq s-1}
    \boxed{(D^+)^4 G^{--\alpha(s-1) \dot{\alpha}(s-1) }= 0 }\,.
\end{equation}

\medskip

\textbf{\textit{\underline{Varying with respect to $h^{++\alpha(s-2) \dot{\alpha}(s-2) }$}}}

\medskip

The variation with respect to the analytic potential $h^{++5\alpha(s-2) \dot{\alpha}(s-2) }$ can be found in the same way,
leading to the superfield equation:
\begin{equation}\label{eq s-2}
        \boxed{ (D^+)^4 G^{--5\alpha(s-2) \dot{\alpha}(s-2) } = 0 } \,.
\end{equation}

\medskip

\textbf{\textit{\underline{Varying with respect to $h^{++\alpha(s-2) \dot{\alpha}(s-1) +}$}}}

\medskip

Variations of $G^{++}$-superfields in this case read:

\begin{equation}\label{s-1-spinor}
        \delta G^{++\alpha(s-1)\dot\alpha(s-1)} =
        -
        2i
        \delta
        h^{++\dot\alpha(s-1)(\alpha(s-2)+}\,\theta^{-\alpha_{s-1})}\,,
\end{equation}
\begin{equation}\label{s-2-spinor}
    \begin{split}
        \delta G^{++\alpha(s-2)\dot\alpha(s-2)} = - 2i
        \delta h^{++\alpha(s-2)(\dot\alpha(s-2)\dot\alpha_{s-1})+}\,\bar\theta^{-}_{\dot{\alpha}_{s-1}} \,.
    \end{split}
\end{equation}

The relevant variation of the action \eqref{action++++} is:
\begin{multline}
    \delta S_{(s)} = - 4i (-1)^{s-1} \int \frac{d^4x d^8\theta du du_1}{(u^+u^+_1)^2}\; \Bigr\{
    \delta
    h^{++\dot\alpha(s-1)\alpha(s-2)+}(u) \,\theta^{-\alpha_{s-1}}\, G^{++}_{\alpha(s-1) \dot{\alpha}(s-1)}(u_1) \\
      +
    \delta h^{++\alpha(s-2)(\dot\alpha(s-2)\dot\alpha_{s-1})+} (u)\,\bar\theta^{-}_{\dot{\alpha}_{s-1}} G^{++5}_{\alpha(s-2) \dot{\alpha}(s-2)}(u_1)  \Bigr\}\,.
\end{multline}

We then use \eqref{zero-1} and \eqref{zero-2} in order to obtain:
\begin{multline}
    \delta S_{(s)} = - 4i (-1)^{s-1} \int d^4x d^8\theta du \; \Bigr\{
    \delta
    h^{++\dot\alpha(s-1)\alpha(s-2)+} \,\theta^{-\alpha_{s-1}}\, G^{--}_{\alpha(s-1) \dot{\alpha}(s-1)} \\
    +
    \delta h^{++\alpha(s-2)(\dot\alpha(s-2)\dot\alpha_{s-1})+} \,\bar\theta^{-}_{\dot{\alpha}_{s-1}} G^{--5}_{\alpha(s-2) \dot{\alpha}(s-2)}  \Bigr\}\,.
\end{multline}

This leads us to the equation:
\begin{eqnarray}\label{ComplEq}
    (D^+)^4 \left[ \theta^{-\alpha_{s-1}}\, G^{--}_{\alpha(s-1) \dot{\alpha}(s-1)} + \bar\theta^{-}_{(\dot{\alpha}_{s-1}} G^{--5}_{\alpha(s-2) \dot{\alpha}(s-2))}  \right]
    = 0\,.
\end{eqnarray}
It can be identically rewritten as
\begin{eqnarray}
&& \frac{1}{8}(\bar{D}^+)^2 D^{+\alpha_{s-1}} G^{--}_{\alpha(s-1)
\dot{\alpha}(s-1)}
    -
    \frac{1}{8}
    (D^+)^2 \bar{D}^+_{(\dot{\alpha}_{s-1}} G^{--5}_{\alpha(s-2)
    \dot{\alpha}(s-2))} \nonumber \\
    &&+\, \theta^{-\alpha_{s-1}}\, (D^+)^4 G^{--}_{\alpha(s-1)
    \dot{\alpha}(s-1)} + \bar\theta^{-}_{(\dot{\alpha}_{s-1}} (D^+)^4 G^{--5}_{\alpha(s-2)
    \dot{\alpha}(s-2))} = 0\,.\label{ComplEq2}
\end{eqnarray}
Then, using \eqref{eq s-1} and \eqref{eq s-2}, we can bring it in
the following final form:
\begin{equation}\label{spinor1}
        \boxed{\frac{1}{8}(\bar{D}^+)^2 D^{+\alpha_{s-1}} G^{--}_{\alpha(s-1) \dot{\alpha}(s-1)}
    -
    \frac{1}{8}
    (D^+)^2 \bar{D}^+_{(\dot{\alpha}_{s-1}} G^{--5}_{\alpha(s-2) \dot{\alpha}(s-2))}\;
    = 0 }\,.
\end{equation}

\medskip

\textbf{\textit{\underline{Varying with respect to $h^{++ \alpha(s-1) \dot{\alpha}(s-2)+}$ }}}

\medskip

The same procedure immediately leads to the superfield equation
which is the tilde-conjugate of \eqref{spinor1}:
\begin{equation}\label{spinor2}
        \boxed{ \frac{1}{8}(D^+)^2  \bar{D}^{+\dot{\alpha}_{s-1}} G^{--}_{\alpha(s-1) \dot{\alpha}(s-1)}
    +
    \frac{1}{8}(\bar{D}^+)^2 D^+_{(\alpha_{s-1}} G^{--5}_{\alpha(s-2)) \dot{\alpha}(s-2)} \;
    =
    0 }\,.
\end{equation}

Note that the equations obtained by varying with respect to bosonic
potentials, {\it i.e.}  eqs.  \eqref{eq s-1} and \eqref{eq s-2}, are
secondary, in the sense that they can be obtained as a consequence
of eqs. \eqref{spinor1} and \eqref{spinor2}. Indeed, to recover them
it suffices to act by the spinor derivative $\bar{D}^+_{\dot\alpha}$ or $D^+_{\alpha}$  on \eqref{spinor1} or \eqref{spinor2}.

\subsection{Summary and component contents of superfield equations}
\label{component content of superfield equations}

So the complete system of equations for an arbitrary $\mathcal{N}=2$ integer higher spin $\mathbf{s}$ supermultiplet with the action \eqref{Action} is as follows:

\begin{equation}\label{EOM}
    \begin{cases}
        (D^+)^4 G^{--\alpha(s-1) \dot{\alpha}(s-1) }= 0\,,
        \\
            (D^+)^4 G^{--5\alpha(s-2) \dot{\alpha}(s-2) } = 0\,,
            \\
                (\bar{D}^+)^2 D^{+\alpha_{s-1}} G^{--}_{\alpha(s-1) \dot{\alpha}(s-1)}
            -
            (D^+)^2 \bar{D}^+_{(\dot{\alpha}} G^{--5}_{\alpha(s-2) \dot{\alpha}(s-2))} \;
            = 0\,,
            \\
            (D^+)^2  \bar{D}^{+\dot{\alpha}_{s-1}} G^{--}_{\alpha(s-1) \dot{\alpha}(s-1)}
            +
            (\bar{D}^+)^2 D^+_{(\alpha} G^{--5}_{\alpha(s-2)) \dot{\alpha}(s-2)} \;
            =
            0\,.
    \end{cases}
\end{equation}
These equations manifestly reveal ${\cal N}=2$ supersymmetry because of invariance of the $G$-superfields.
It is also easy to check that these equations are invariant with respect to the gauge transformations \eqref{Lambda gauge}.
Recall that the $G^{--\dots}$ superfields are solutions of the zero-curvature equations:
\begin{equation}
    \mathcal{D}^{++}  G^{--\alpha(s-1) \dot{\alpha}(s-1)}  - \mathcal{D}^{--}  G^{++\alpha(s-1) \dot{\alpha}(s-1)} = 0\,,
\end{equation}
\begin{equation}
    \mathcal{D}^{++}  G^{--5\alpha(s-2) \dot{\alpha}(s-2)}  - \mathcal{D}^{--}  G^{++5\alpha(s-2) \dot{\alpha}(s-2)} = 0\,,
\end{equation}
where the $G^{++\dots}$ superfields are composed out of the analytic potentials $h^{++\dots}(\zeta)$ according to eqs. \eqref{s-1} and \eqref{s-2}.

It is instructive to reveal the  meaning of each equation in the set \eqref{EOM} from the component approach viewpoint. We will restrict our brief analysis to the simplest examples of
the purely bosonic sector  of the spin $\bf{2}$ and spin $\bf{3}$ theories.

\subsubsection{${\cal N}=2$ spin $\mathbf{2}$} \label{spin 2 equations}
In this case, the equations take the simplest form:
\begin{equation}\label{EOM spin 2}
    \begin{cases}
        (D^+)^4 G^{--\alpha \dot{\alpha} }= 0\,,
        \\
        (D^+)^4 G^{--5} = 0\,,
        \\
        (\bar{D}^+)^2 D^{+\alpha} G^{--}_{\alpha \dot{\alpha}}
        -
        (D^+)^2 \bar{D}^+_{\dot{\alpha}} G^{--5}_{} \;
        = 0\,,
        \\
        (D^+)^2  \bar{D}^{+\dot{\alpha}} G^{--}_{\alpha \dot{\alpha}}
        +
        (\bar{D}^+)^2 D^+_{\alpha} G^{--5}_{} \;
        =
        0\,.
    \end{cases}
\end{equation}

In our analysis, we will make use of the component structure of the superfields $G^{--\dots}$ displayed in  Appendix \ref{spin 2 theory -- components}.

\medskip
\textit{\underline{\textbf{${\cal N}=2$ spin 2 theory: component contents of the bosonic spin 2 sector}}}
\medskip

Here we shall deal with the irreducible spin 2 components $\Phi_{(\alpha\beta)(\dot\alpha\dot\beta)}, \Phi$ defined in \eqref{Spin2Dec}.

The first equation in \eqref{EOM spin 2} in its bosonic sector yields Pauli-Fierz equations (linearized vacuum Einstein equations) augmented with some combinations of the auxiliary fields:
\begin{equation}\label{Ein-linear}
    \begin{split}
    &(D^+)^4 G_{(\Phi, P)}^{--\alpha \dot{\alpha} } = 4i  \left[- \theta^{+\rho}\bar{\theta}^{+\dot{\rho}} \mathcal{G}^{\alpha\dot{\alpha}}_{\rho\dot{\rho}} + \theta^{+\rho} \bar{\theta}^{+(\dot{\alpha}} \partial^{\dot{\rho})}_\rho \mathcal{P}^\alpha_{\dot{\rho}}
    -
    \bar{\theta}^{+\dot{\rho}} \theta^{+(\alpha} \partial^{\rho)}_{\dot{\rho}} \bar{\mathcal{P}}^{\dot{\alpha}}_{\rho}\right]
    = 0\,, \\&\qquad
    \Rightarrow
    \qquad  \mathcal{G}_{\rho\dot{\rho}}^{\alpha\dot{\alpha}}
    -
    \frac{1}{2} \partial^{\dot{\alpha}}_\rho \mathcal{P}^\alpha_{\dot{\rho}}
    -
    \frac{1}{2} \partial^\alpha_{\dot{\rho}} \bar{\mathcal{P}}^{\dot{\alpha}}_{\rho}
    -
    \frac{1}{2} \delta^{\dot{\alpha}}_{\dot{\rho}} \partial^{\dot{\mu}}_\rho \mathcal{P}^\alpha_{\dot{\mu}}
    -
    \frac{1}{2} \delta^\alpha_\rho \partial^\mu_{\dot{\rho}} \bar{\mathcal{P}}_\mu^{\dot{\alpha}}
    = 0\,.
        \end{split}
\end{equation}
Here we used eq. \eqref{spin 2 highest component}. Various
quantities appearing in \eqref{Ein-linear} are defined in eqs.
\eqref{p-field} - \eqref{scalar curvature}.

The component form of the second equation in \eqref{EOM spin 2} is
obtained  after substituting $G^{--5}_{(\Phi, P)}$ from \eqref{G--5
P}:
\begin{equation}\label{Ein-linear-2}
    \begin{split}
    (D^+)^4 G^{--5}_{(\Phi, P)} &= 2i (\theta^+)^2 \partial_{\rho\dot{\rho}} P^{\rho\dot{\rho}} - 2i (\bar{\theta}^+)^2 \partial_{\rho\dot{\rho}} \bar{P}^{\rho\dot{\rho}}
    \\&= 2i (\theta^+)^2 \left[ -2 \mathcal{R} + \partial_{\rho\dot{\rho}} \mathcal{P}^{\rho\dot{\rho}} \right] - 2i (\bar{\theta}^+)^2 \left[ -2 \mathcal{R} + \partial_{\rho\dot{\rho}} \bar{\mathcal{P}}^{\rho\dot{\rho}} \right] = 0\,,
    \\ \Rightarrow \qquad & \begin{cases}
        -2 \mathcal{R} + \partial_{\rho\dot{\rho}} \mathcal{P}^{\rho\dot{\rho}} = 0\,,\\
        -2 \mathcal{R} + \partial_{\rho\dot{\rho}} \bar{\mathcal{P}}^{\rho\dot{\rho}} = 0\,.
        \end{cases}
    \end{split}
\end{equation}
It is straightforward to see that \eqref{Ein-linear-2} follow from
\eqref{Ein-linear} by contracting the spinorial indices in the latter. In particular, the definition \eqref{lin-ein} implies that $\mathcal{G}_{\alpha\dot{\alpha}}^{\alpha\dot{\alpha}} = - 4\mathcal{R}$.
{Thus we conclude that, in the $(\Phi, P)$ sector, the second equation  does
not result in anything new.

The highest components of the third and fourth equations in
\eqref{EOM spin 2} are also trivially satisfied as a consequence of
eqs. \eqref{Ein-linear} and \eqref{Ein-linear-2}, and the only role
of these superfield equations is to nullify the auxiliary fields:
\begin{equation}
    \begin{split}
    &\frac{1}{8}(\bar{D}^+)^2 D^{+\alpha} G^{--}_{(\Phi,P)\alpha \dot{\alpha}}
    -
    \frac{1}{8}
    (D^+)^2 \bar{D}^+_{\dot{\alpha}} G^{--5}_{(\Phi,P)}
    \\&\qquad\qquad=
    \theta^+_{\mu} \left(\mathcal{P}^\mu_{\dot{\alpha}} - 3 \bar{\mathcal{P}}^\mu_{\dot{\alpha}} \right)
    -
    \theta^{-\alpha} (D^+)^4 G^{--}_{(\Phi, P)\alpha\dot{\alpha}}
    -
    \bar{\theta}^-_{\dot{\alpha}} (D^+)^4 G^{--5}_{(\Phi, P)}
    = 0\,,
    \\&\quad\quad\; \Rightarrow \qquad \mathcal{P}^\mu_{\dot{\alpha}} - 3 \bar{\mathcal{P}}^\mu_{\dot{\alpha}} =
    0\,,
    \end{split}
\end{equation}
\begin{equation}
    \begin{split}
    &\frac{1}{8}(D^+)^2  \bar{D}^{+\dot{\alpha}} G^{--}_{(\Phi,P)\alpha \dot{\alpha}}
    +
    \frac{1}{8}
    (\bar{D}^+)^2 D^+_{\alpha} G^{--5}_{(\Phi, P)}
    \\&\qquad\qquad =
    - \bar{\theta}^{+}_{\dot{\rho}} \left( \bar{\mathcal{P}}^{\dot{\rho}}_\alpha - 3 \mathcal{P}^{\dot{\beta}}_\alpha\right)
    + \bar{\theta}^{-\dot{\alpha}} (D^+)^4 G^{--}_{\alpha\dot{\alpha}}
    -
    \theta^-_{\alpha} (D^+)^4 G^{--5}_{(\Phi, P)}
    =
    0\,,
    \\&\quad\quad \Rightarrow \qquad \bar{\mathcal{P}}^{\dot{\rho}}_\alpha - 3 \mathcal{P}^{\dot{\beta}}_\alpha = 0\,.
    \end{split}
\end{equation}
The algebraic equations for the auxiliary fields indeed imply
\begin{equation}
    \mathcal{P}_{\alpha\dot{\alpha}} = \bar{\mathcal{P}}_{\alpha\dot{\alpha}} =
    0\,.\label{AuxZero}
\end{equation}

Finally, substituting  \eqref{AuxZero} in \eqref{Ein-linear}, we
conclude that in the spin 2 sector the whole on-shell dynamics amounts to the
single equation:
\begin{equation}\label{BasEq}
    \mathcal{G}_{\rho\dot{\rho}}^{\alpha\dot{\alpha}}  = 0\,.
\end{equation}
With taking into account the definitions \eqref{lin-ein},
\eqref{Ricchi-lin} and \eqref{scalar curvature}, eq. \eqref{BasEq}
is recognized as the equation of motion for the spin 2
fields $\Phi^{(\alpha\beta)(\dot{\alpha}\dot{\beta})}$ and $\Phi$:
\begin{eqnarray}
       && \eqref{BasEq} \; \Rightarrow \;
        \frac{1}{2} \partial_{(\alpha(\dot{\alpha}} \partial^{\sigma\dot{\sigma}} \Phi_{\beta)\sigma\dot{\beta})\dot{\sigma}}
    -
    \frac{1}{2} \Box \Phi_{(\alpha\beta)(\dot{\alpha}\dot{\beta})}
    -
    \frac{1}{2} \partial_{(\alpha(\dot{\alpha}} \partial_{\beta)\dot{\beta})} \Phi
    \nonumber \\
&& \qquad\quad  -\, \varepsilon_{\alpha\beta} \varepsilon_{\dot{\alpha}\dot{\beta}}
    \left\{\frac{1}{8} \partial^{\rho\dot{\rho}} \partial^{\sigma\dot{\sigma}}  \Phi_{(\rho\sigma) (\dot{\rho}\dot{\sigma})}
    - \frac{3}{4} \Box \Phi \right\} = 0\,,\label{Spintwo}
\end{eqnarray}
where the field in the first term is assumed to be the irreducible one $ \Phi_{(\beta\sigma)(\dot{\beta}\dot{\sigma})}$. This is the linearized Einstein equations. Equivalently, these
equations (symmetric and antisymmetric parts of eq. \eqref{Spintwo})
can be directly derived from the component action
\eqref{Pauli_Fiertz}.

\medskip
\textit{\underline{\textbf{${\cal N}=2$ spin 2 theory: component contents of the spin 1 sector}}}
\medskip

The spin 1 sector is described by the superfields
$G^{--\alpha\dot\alpha}_{(C, T)}$ with the component contents
exhibited in eqs. \eqref{spin 1-1} and \eqref{spin 1-2}, $C_{\alpha\dot\alpha}$ being the spin 1 gauge field.
Substitution of these expressions in the first superfield equation
in the set \eqref{EOM spin 2} gives:

\begin{equation}\label{spin 1 sf 1}
        (D^+)^4G^{--\alpha\dot{\alpha}}_{(C, T)}
    =
    (\bar{\theta}^+)^2 \partial^{\dot{\alpha}}_\nu T^{(\alpha\nu)}
    -
     (\theta^+)^2  \partial^\alpha_{\dot{\nu}} \bar{T}^{(\dot{\alpha}\dot{\nu})}
     =
     0\,,
     \qquad
     \Rightarrow
     \qquad\begin{cases}
        \partial^{\dot{\alpha}}_\nu T^{(\alpha\nu)} =0 \,,\\
        \partial^\alpha_{\dot{\nu}} \bar{T}^{(\dot{\alpha}\dot{\nu})}=0\,.
     \end{cases}
\end{equation}

The second equation in \eqref{EOM spin 2} leads to:
\begin{equation}\label{spin 1 sf 2}
    \begin{split}
        &(D^+)^4 G^{--5}_{(C, T)}
        =
    2i  \theta^{+\rho} \bar{\theta}^{+\dot{\rho}} \left[ \Box C_{\rho\dot{\rho}} - \frac{1}{2} \partial_{\rho\dot{\rho}} \partial^{\beta\dot{\beta}} C_{\beta\dot{\beta}} \right]
    \\&\qquad\qquad\qquad\quad+
    2 \bar{\theta}^{+\dot{\rho}} \theta^+_\rho \partial_{\mu\dot{\rho}} T^{(\mu\rho)}
    +
    2
    \theta^{+\rho} \bar{\theta}^+_{\dot{\rho}} \partial_{\rho\dot{\mu}} \bar{T}^{(\dot{\mu}\dot{\rho})} = 0\,,
    \\& \Rightarrow \qquad i \left[ \Box C^{\rho\dot{\rho}} - \frac{1}{2} \partial^{\rho\dot{\rho}} \partial^{\beta\dot{\beta}} C_{\beta\dot{\beta}} \right]
    -
    \partial_{\mu}^{\dot{\rho}} T^{(\mu\rho)}
    -
    \partial^{\rho}_{\dot{\mu}} \bar{T}^{(\dot{\mu}\dot{\rho})} = 0\,.
\end{split}
\end{equation}
As a consequence of \eqref{spin 1 sf 1}, the terms involving tensors
$T$ can be omitted in \eqref{spin 1 sf 2}. So we recover Maxwell
equation:
\begin{equation}
     \Box C^{\rho\dot{\rho}} - \frac{1}{2} \partial^{\rho\dot{\rho}} \partial^{\beta\dot{\beta}} C_{\beta\dot{\beta}}
     =
     0\,.
\end{equation}

Eqs. \eqref{spin 1 sf 1} and \eqref{spin 1 sf 2} are just the
$\theta^-$ components of the third and fourth equations in
\eqref{EOM spin 2}. So we are left with the following independent
component equations:
\begin{equation}
    \begin{split}
 &(\bar{D}^+)^2 D^{+\alpha} G^{--}_{(C, T)\alpha \dot{\alpha}}
    -
    (D^+)^2 \bar{D}^+_{\dot{\alpha}} G^{--5}_{(C, T)} \Bigr|_{\theta^- = 0}
    \\&
    =
    8 \bar{\theta}^{+\dot{\beta}} \partial_{(\dot{\alpha}}^\beta C_{\beta\dot{\beta})} - 8i \bar{\theta}^{+\dot{\beta}} \bar{T}_{(\dot{\alpha}\dot{\beta})}
    = 0
    \,\qquad \Rightarrow \qquad
    \bar{T}_{(\dot{\alpha}\dot{\beta})}  = -i \partial_{(\dot{\alpha}}^\beta C_{\beta\dot{\beta})}\,.
    \end{split}
\end{equation}
\begin{equation}
    \begin{split}
    &(D^+)^2  \bar{D}^{+\dot{\alpha}} G^{--}_{(C, T)\alpha \dot{\alpha}}
    +
    (\bar{D}^+)^2 D^+_{\alpha} G^{--5}_{(C, T)} \Bigr|_{\theta^- = 0}
    \\&=
    8 \theta^{+\beta} \partial^{\dot{\beta}}_{(\alpha} C_{\beta)\dot{\beta}}
    +
    8i \theta^{+\beta} T_{(\alpha\beta)}
    =
    0\,\qquad \Rightarrow \qquad
    T_{(\alpha\beta)} = i \partial^{\dot{\beta}}_{(\alpha} C_{\beta)\dot{\beta}}\,.
    \end{split}
\end{equation}
Thus $T_{(\alpha\beta)}$ and $\bar{T}_{(\dot{\alpha}\dot{\beta})}$
are self-dual $(1,0)$ and anti-self-dual $(0,1)$ irreducible parts
of the Maxwell field strength and eqs. \eqref{spin 1 sf 1} are just
the corresponding irreducible parts of Maxwell equation.
Alternatively, these equations can be directly deduced from the
action \eqref{Maxwell action -- spin2}.

\bigskip

\subsubsection{${\cal N}=2$ spin $\mathbf{3}$}

The superfield equations of motion for the spin $\mathbf{3}$ ${\cal N}=2$ model read:
\begin{equation}\label{EOM spin 3}
    \begin{cases}
        (D^+)^4 G^{--(\alpha\beta) (\dot{\alpha}\dot{\beta})}= 0\,,
        \\
        (D^+)^4 G^{--5\alpha\dot{\alpha} } = 0\,,
        \\
        (\bar{D}^+)^2 D^{+\alpha} G^{--}_{(\alpha\beta) (\dot{\alpha}\dot{\beta})}
        -
        (D^+)^2 \bar{D}^+_{(\dot{\alpha}} G^{--5}_{\beta \dot{\beta})}\;
         = 0\,,
        \\
        (D^+)^2  \bar{D}^{+\dot{\alpha}} G^{--}_{(\alpha\beta) (\dot{\alpha}\dot{\beta})}
        +
        (\bar{D}^+)^2 D^+_{(\alpha} G^{--5}_{\beta) \dot{\beta}}\;
        =
        0\,.
    \end{cases}
\end{equation}
The relevant component fields are defined in Appendix \ref{spin 3 bosonic sector}, the basic gauge fields being given in \eqref{spin 3 field} and \eqref{spin 3-2 sector}.
Using these definitions, we find, like in the spin $\mathbf{2}$
case, that the third and fourth equations set the auxiliary fields
equal to zero:
\begin{equation}
    \mathcal{P}^{(\alpha\beta)\dot{\alpha}\dot{\beta}} = \bar{\mathcal{P}}^{(\dot{\alpha}\dot{\beta})\alpha\beta}
    = 0\,.
\end{equation}

With these relations in mind, we obtain from the first equation:
\begin{equation}
    \begin{split}
    (D^+)^4 G^{--(\alpha\beta) (\dot{\alpha}\dot{\beta})}
        =
    &- 3i \theta^{+\rho} \bar{\theta}^{+\dot{\rho}}  \mathcal{G}_{\rho\dot{\rho}}^{(\alpha\beta)(\dot{\alpha}\dot{\beta})}
    \\&+
        i (\bar{\theta}^+)^2  \partial_{\rho}^{(\dot{\alpha}} H^{\dot{\beta})\rho(\alpha\beta)}
    -
    i (\theta^+)^2 \partial^{(\alpha}_{\dot{\rho}} H^{\beta)\dot{\rho}(\dot{\alpha}\dot{\beta})}
    = 0\,,
    \end{split}
\end{equation}
where we also used the solutions \eqref{G--31} and \eqref{G-- 321}
of the relevant zero-curvature conditions. As the result, we arrive
at the system:
\begin{equation}\label{sys 1}
    \begin{cases}
        \mathcal{G}_{(\alpha\beta)\rho(\dot{\alpha}\dot{\beta})\dot{\rho}}^{} = 0\,,\\
        \partial_{\rho}^{(\dot{\alpha}} H^{\dot{\beta})\rho(\alpha\beta)} = 0 \,\Rightarrow \,   \mathcal{R}^{(\alpha\beta)(\dot{\alpha}\dot{\beta})} = 0\,,\\
        \partial^{(\alpha}_{\dot{\rho}} H^{\beta)\dot{\rho}(\dot{\alpha}\dot{\beta})}= 0 \,\Rightarrow \,
        \mathcal{R}^{(\alpha\beta)(\dot{\alpha}\dot{\beta})} = 0\,.
    \end{cases}
\end{equation}
Here we made use of the notation \eqref{Ricchi-lin} for the
linearized curvature  constructed out of the
spin 2 field \eqref{spin 3-2 sector}, as well as  the relations \eqref{HR}.

The second equation in \eqref{EOM spin 3} yields:
\begin{equation}
    (D^+)^4 G^{--5\alpha\dot{\alpha}}
    =
    4i \theta^{+}_{\rho} \bar{\theta}^{+}_{\dot{\rho}}
    \mathcal{G}^{\alpha\rho\dot{\alpha}\dot{\rho}}
    +
    i (\theta^+)^2  \partial_{\rho\dot{\rho}} B^{(\rho\alpha)\dot{\rho}\dot{\alpha}}
    -
    i (\bar{\theta}^+)^2  \partial_{\rho\dot{\rho}} \bar{B}^{\rho\alpha(\dot{\rho}\dot{\alpha})}
    =
    0\,,
\end{equation}
where we used the solutions \eqref{G--32} and \eqref{spin 3-2
solution} of the relevant zero-curvature conditions. Thus we arrive
at the set:
\begin{equation}\label{sys 2}
    \begin{cases}
        \mathcal{G}_{\alpha\rho\dot{\alpha}\dot{\rho}} = 0\,,\\
         \partial_{\rho\dot{\rho}} B^{(\rho\alpha)\dot{\rho}\dot{\alpha}}  =  0 \,\Rightarrow \, \mathcal{R}^{\alpha\dot{\alpha}} =  0\,,\\
         \partial_{\rho\dot{\rho}} \bar{B}^{\rho\alpha(\dot{\rho}\dot{\alpha})} = 0 \, \Rightarrow \, \mathcal{R}^{\alpha\dot{\alpha}} = 0\,.
    \end{cases}
\end{equation}
Here we have used the definitions \eqref{A34}, \eqref{A35}, the relations \eqref{BR} and
the expression \eqref{lin-ein} for the linearized
Einstein tensor constructed from the spin 2 field \eqref{spin 3-2 sector}.

So in the bosonic sector of the systems \eqref{sys 1} and
\eqref{sys 2} there actually remain only two independent equations,
\bea \label{BasicTwo}
\mathcal{G}_{(\alpha\beta)\rho(\dot{\alpha}\dot{\beta})\dot{\rho}}^{} = 0\,, \qquad \mathcal{G}_{\alpha\rho\dot{\alpha}\dot{\rho}} = 0\,.
\eea

The first one is the equation of motion for the spin 3 field \eqref{spin 3 field}:
\begin{equation}
    \mathcal{G}_{(\alpha_1\alpha_2)\alpha_3 (\dot{\alpha}_1  \dot{\alpha}_2) \dot{\alpha}_3}
    =
    \mathcal{R}_{(\alpha_1\alpha_2\alpha_3) (\dot{\alpha}_1  \dot{\alpha}_2 \dot{\alpha}_3)}
    - \frac{4}{9}
    \mathcal{R}_{(\alpha_1 (\dot{\alpha}_1} \epsilon_{\dot{\alpha}_2) \dot{\beta}} \epsilon_{\alpha_2) \beta} = 0\,.
\end{equation}
It can be split into the two irreducible equations:
\begin{eqnarray}
    \mathcal{R}_{(\alpha_1\alpha_2\alpha_3)( \dot{\alpha}_1\dot{\alpha}_2\dot{\alpha}_3)}
    &=&
    \partial_{(\alpha_1(\dot{\alpha}_1}\partial^{\rho\dot{\rho}} \Phi_{\alpha_2\alpha_3)\rho \dot{\alpha}_2\dot{\alpha}_3)\dot{\rho}}
    -
    \frac{2}{3}
    \Box \Phi_{(\alpha_1\alpha_2\alpha_3)( \dot{\alpha}_1\dot{\alpha}_2\dot{\alpha}_3)} \nn
    &&-\,
    \partial_{(\alpha_1(\dot{\alpha}_1}
    \partial_{\alpha_2\dot{\alpha}_2}
    \Phi_{\alpha_3)\dot{\alpha}_3)} = 0\,, \label{SymmSpin3}
\end{eqnarray}
\begin{equation}\label{R alpha dot alpha}
    \mathcal{R}_{\alpha_1\dot{\alpha}_1}
    =
    \partial^{\alpha_2\dot{\alpha}_2} \partial^{\alpha_3\dot{\alpha}_3} \Phi_{(\alpha_1\alpha_2\alpha_3)( \dot{\alpha}_1\dot{\alpha}_2\dot{\alpha}_3)}
    -
    \frac{1}{4}\partial_{\alpha_1\dot{\alpha}_1} \partial^{\alpha_2\dot{\alpha}_2} \Phi_{\alpha_2\dot{\alpha}_2}
    -
    \frac{5}{2} \Box \Phi_{\alpha_1\dot{\alpha}_1}
    =
    0\,,
\end{equation}
where in the first term in eq. \eqref{SymmSpin3} we omitted, for simplicity, symmetrization of indices of the involved field. These are just Fronsdal equations for the
spin 3 in the spinor formalism. They can  also be
directly derived  from  the action \eqref{spin 3 action}.

The second equation in the set \eqref{BasicTwo} amounts to the linearized Einstein equations for the spin 2 field \eqref{spin 3-2 sector}:
\begin{eqnarray}
&&        {\cal G}_{\alpha\beta\dot{\alpha}\dot{\beta}} =
        \frac{1}{2} \partial_{(\alpha(\dot{\alpha}} \partial^{\sigma\dot{\sigma}} C_{\beta)\sigma\dot{\beta})\dot{\sigma}}
        -
        \frac{1}{2} \Box C_{(\alpha\beta)(\dot{\alpha}\dot{\beta})}
        -
        \frac{1}{2} \partial_{(\alpha(\dot{\alpha}} \partial_{\beta)\dot{\beta})} C
       \nonumber  \\
      &&\qquad\qquad  -\, \varepsilon_{\alpha\beta} \varepsilon_{\dot{\alpha}\dot{\beta}}
        \left\{\frac{1}{8} \partial^{\rho\dot{\rho}} \partial^{\sigma\dot{\sigma}}  C_{(\rho\sigma) (\dot{\rho}\dot{\sigma})}
        - \frac{3}{4} \Box C \right\} = 0\,,\label{Cspin3}
\end{eqnarray}
where, once again, we omitted symmetrization of indices of $C_{(\beta\sigma)(\dot\beta\dot\sigma)}$ in the first term. Eq. \eqref{Cspin3} also admits splitting into the two irreducible ones which
can be independently derived from the action \eqref{spin 2 action -- spin 3}.

\subsubsection{General ${\cal N}=2$ spin $\mathbf{s}$}

It is not difficult to solve the zero-curvature equations in the general case of an arbitrary spin $\mathbf{s}$ ${\cal N}=2$ multiplet
and to obtain the component equivalents of the corresponding superfield equations. The whole procedure can be carried out in a complete analogy
with the ${\cal N}=2$ spin $\mathbf{2}$ and spin $\mathbf{3}$ examples  considered above.
Here we do not present all the technical details and only say that in the case of the spin $\mathbf{s}$ supermultiplet
the resulting superfield equations in the bosonic sector
give rise to the standard free Fronsdal equations for the gauge fields of the spins $s$ and $s-1$. Also, they set equal to zero all the bosonic auxiliary fields.

\section{Hypermultiplet current superfields}\label{Hyper-current-superfields}

Hypermultiplet is the basic $\mathcal{N}=2$ matter  supermultiplet. All other $\mathcal{N}=2$ matter supermultiplets are related to it by various
duality transformations \cite{Dulities}. The off-shell formulation of the hypermultiplet requires infinitely many auxiliary fields.
It was given in harmonic superspace in terms of an unconstrained analytic harmonic superfield $q^+(\zeta)$ in which an infinite set of auxiliary fields
is naturally accommodated by the harmonic expansion on the internal sphere $S^2 \sim {\rm SU}(2)_{aut}/{\rm U}(1)_{aut}$.
This formulation allowed one to construct most general self-interaction of hypermultiplets yielding most general hyper-K\"ahler sigma model
in the bosonic sector and to make manifest other nice features of extended supersymmetry, such as non-renormalization theorems \cite{Buchbinder:1997ib}.
The harmonic superspace description of hypermultiplet proved to be of pivotal importance also for constructing
manifestly $\mathcal{N}=2$ supersymmetric interactions with $\mathcal{N}=2$ higher spins \cite{Buchbinder:2022kzl}.

In this Section we start by reviewing the main features of the harmonic superspace formulation of the  hypermultiplet \cite{18}
and setting up the cubic couplings of the hypermultiplet to higher spins \cite{Buchbinder:2022kzl}. Then we shall construct
conserved current superfields associated with the higher spin rigid symmetries of the free hypermultiplet.
These current superfields  determine the cubic Noether couplings of the hypermultiplet to $\mathcal{N}=2$ higher spin supermultiplets.
In Appendix \ref{current} we shall discuss some general features and a universal structure of these supercurrents, which
are not peculiar to just free hypermultiplet theory but are in fact inherent in any $\mathcal{N}=2$ theory.
Also in this Section we shall exhibit the component structure of some simplest current superfields.

\subsection{Hypermultiplet in harmonic superspace} \label{sec:hyper}

The free  hypermultiplet action in $\mathcal{N}=2$ harmonic superspace is written as an integral over the analytic harmonic superspace \cite{18, HSS}:
\begin{equation}\label{hyper}
    S_{free}
    = -\int d\zeta^{(-4)}  \; \frac{1}{2} q^{+a} \mathcal{D}^{++} q^+_a = -\int d\zeta^{(-4)}  \; \tilde{q}^+ \mathcal{D}^{++} q^+ \,.
\end{equation}

The first representation explicitly displays the Pauli-G\"ursey ${\rm SU}(2)_{PG}$ global symmetry acting on the doublet indices $a$.
It is related to the second description, with only ${\rm U}(1)_{PG} \subset {\rm SU}(2)_{PG}$ manifest, as :
\begin{equation}
    q^+_a = (q^+, - \tilde{q}^+),
    \;\;\;\;\;\;\;\;\;\;
    \widetilde{q^+_a}\equiv q^{+a} = \epsilon^{ab} q^+_b = (\tilde{q}^+, q^+)\,.
\end{equation}
The formulation with manifest Pauli-G\"ursey ${\rm SU}(2)$ symmetry allows one to write more compactly the interactions with higher spin fields.

The hypermultiplet described by the action \eqref{hyper} can be either massive or massless.
In the case of massive $\mathcal{N}=2$ hypermultiplet (without gauge fields) the minimal mechanism of generating
mass is through a non-zero central charge in ${\cal N}=2$ superalgebra, with mass being the eigenvalue
of the central charge generator.
Such a massive hypermultiplet describes on shell a short $\mathcal{N}=2$ supermultiplet.

The most convenient way to introduce the central charge is to add an extra fictitious coordinate $x^5$,
as was already announced in \eqref{Ext5}. The hypermultiplet is assumed to depend on $x^5$ almost trivially:
\begin{equation}
    q^+(x, \theta^+, u, x^5) = e^{im x^5}  q^+(x, \theta^+, u) \quad \Leftrightarrow \quad  \partial_5 q^+ =  i m q^+\,,
\end{equation}
or, with ${\rm SU}(2)_{PG}$ indices taken manifest,
\begin{equation}\label{global-1}
    \qquad\partial_5 q^{+a} := -im (\tau^3)^a_{\;b} q^{+b},
\end{equation}
where
\begin{equation}\label{tau3}
    (\tau_3)^a_{\;b}
    =
    {\scriptstyle   \begin{pmatrix}
            1 & 0 \\
            0 & -1
    \end{pmatrix}}
    ,
    \;\;\;\;\;\;\;\;\;
    (\tau_3)_{ab} = \epsilon_{ac}   (\tau_3)^c_{\;b}
    =
    {\scriptstyle   \begin{pmatrix}
            0 & -1 \\
            -1 & 0
    \end{pmatrix}}.
\end{equation}
For avoiding any integration over $x^5$ in the hypermultiplet action \eqref{hyper},
we impose the standard Scherk-Schwarz condition that
$\partial_5 q^{+a}$ coincides with the action
of some ${\rm U}(1)_{PG} \subset {\rm SU}(2)_{PG}$. The parameter $m$ is a mass of the hypermultiplet,
so we deal with the massive hypermultiplet in the general case. It is evident that
\be
\partial_5 (q^{+a}\mathcal{D}^{++} q^+_a) = 0\,. \label{x5independence}
\ee
The action \eqref{hyper} is invariant under the rigid $\mathcal{N}=2$ supersymmetry,
\begin{equation}\label{sup-hyp}
    \delta^*_{\epsilon} q^{+a} = -\delta_\epsilon Z^M\partial_M q^{+a}\,,
\end{equation}
where we need to take into account also the transformation of $x^5$ \eqref{Tranfifth} for $m\neq 0$. Because of the property
\eqref{x5independence}, introducing a mass does not affect $\mathcal{N}=2$ supersymmetry
transformation of the Lagrangian, though modifies  $\mathcal{N}=2$ transformation of $q^{+a}$. The internal symmetry
of the single hypermultiplet action is ${\rm SU}(2)_{PG} \times{\rm SU}(2)_{aut}$ in the massless case
($\partial_5 q^{+a} =0$) and ${\rm U}(1)_{PG} \times{\rm SU}(2)_{aut}$ in the massive case ($\partial_5 q^{+a} \neq 0$).

The equation of motion for the free hypermultiplet reads:
\begin{equation}
    \mathcal{D}^{++} q^{+a} = 0\,.\label{hyperEq}
\end{equation}
On shell, using the non-dynamical part of \eqref{hyperEq} (terms up to the second order in $\theta^{+\alpha}, \bar{\theta}^{+\dot\alpha}$),
the analytic harmonic superfields $q^{+},\, \tilde{q}^+$ are reduced to:
\begin{eqnarray}
&&    q^+(\zeta) = f^i u^+_i
    + \theta^{+\alpha} \psi_\alpha
    + \bar{\theta}^+_{\dot{\alpha}} \bar{\chi}^{\dot{\alpha}}
    + m(\theta^{\hat{+}})^2 f^i u^-_i
    + 2i \theta^+ \sigma^n \bar{\theta}^+ \partial_n f^i u_i^-\,,\label{hyper q} \\
&& \tilde{q}^+ (\zeta) =
    -
    \bar{f}^{i} u^{+}_{i}
    +
    \bar{\theta}^{+}_{\dot{\alpha}} \bar{\psi}^{\dot{\alpha}}
    -
    \theta^{+\alpha} \chi_\alpha
    +
    m (\hat{\theta}^+)^2
    \bar{f}^{i} u^{-}_{i}
    -
    2i \theta^+ \sigma^n \bar{\theta}^+ \partial_n
    \bar{f}^{i} u_i^-\,.\label{hyper q tilde}
\end{eqnarray}

As is seen from these expressions, an infinite set of original auxiliary fields vanishes on the equations of motion
and there remain only the physical fields forming the standard on-shell $\mathcal{N}=2$ hypermultiplet. The latter
consists of a doublet of complex scalar fields $f^i$ and Weyl fermions $\psi_\alpha$ and $\kappa_\alpha$.
Action of the on-shell $\mathcal{N}=2$ supersymmetry on these fields can be deduced directly from the superfield transformation law \eqref{sup-hyp}:
\begin{equation}\label{susy hyp}
    \begin{cases}
    \delta^*_\epsilon f^i = - \epsilon^{\mu i} \psi_\mu - \bar{\epsilon}^i_{\dot{\mu}} \bar{\chi}^{\dot{\mu}}\,,\\
    \delta^*_\epsilon \bar{f}_i = -\epsilon_i^\mu \chi_\mu - \bar{\epsilon}_{\dot{\mu} i} \bar{\psi}^{\dot{\mu}}\,,\\
    \delta^*_\epsilon \psi_\alpha = -2i \bar{\epsilon}^{\dot{\alpha}i} \partial_{\alpha\dot{\alpha}} f_i + 2m \epsilon^i_\alpha f_i\,,\\
    \delta^*_\epsilon \bar{\psi}_{\dot{\alpha}} = -2i \epsilon^{\alpha i} \partial_{\alpha\dot{\alpha}} \bar{f}_i
    +
    2m \bar{\epsilon}_{\dot{\alpha}}^i \bar{f}_i
    \,,\\
    \delta^*_\epsilon \chi_\alpha = 2i \bar{\epsilon}^{\dot{\alpha}i} \partial_{\alpha\dot{\alpha}} \bar{f}_i - 2m \epsilon^{\alpha i} \bar{f}_i\,,\\
    \delta^*_\epsilon \bar{\chi}_{\dot{\alpha}} =  2i \epsilon^{\alpha i} \partial_{\alpha\dot{\alpha}} f_i - 2m \bar{\epsilon}^i_{\dot{\alpha}} f_i\,.
    \end{cases}
\end{equation}
Eq. \eqref{hyperEq} (terms $\sim \theta^3, \theta^4$ in it) also implies the massive equations of motion for the physical fields:
\begin{equation}\label{hyper-equations}
    \begin{cases}
    \left(\Box + m^2\right) f^i = 0\,,\\
    i \partial_{\alpha\dot{\alpha}} \bar{\chi}^{\dot{\alpha}} + m \psi_\alpha = 0\,,\\
    i \partial_{\alpha\dot{\alpha}} \psi^\alpha - m \bar{\chi}_{\dot{\alpha}} =0 \,.
    \end{cases}
\end{equation}

Using \eqref{hyper q} and \eqref{hyper q tilde} one can deduce the component on-shell hypermultiplet action from the off-shell one \eqref{hyper}:
\begin{equation}\label{hyper on shell}
    \begin{split}
    S_{on-shell} = \int d^4x\; \Bigr[ &\partial_n f^i \partial^n \bar{f}_i - m^2 f^i \bar{f}_i
    \\&+
    \frac{i}{2} \bar{\psi}_{\dot{\alpha}} \partial^{\dot{\alpha}\alpha} \psi_\alpha
        +
    \frac{i}{2} \bar{\chi}_{\dot{\alpha}} \partial^{\dot{\alpha}\alpha} \chi_\alpha
    +
    \frac{m}{2} \bar{\psi}_{\dot{\alpha}} \bar{\chi}^{\dot{\alpha}}
    +
    \frac{m}{2} \chi^\alpha \psi_\alpha
    \Bigr]\,.
    \end{split}
\end{equation}
This action is invariant under $\mathcal{N}=2$ supersymmetry transformations \eqref{susy hyp}. The doublet of  conserved Noether fermionic
currents associated with the latter is given by the expression:
\begin{equation}\label{N=2 supercurrents}
    \begin{split}
    S^{\alpha\dot{\alpha}}_{\mu\, k}
    =&
    \frac{1}{2} \left(\partial^{\alpha\dot{\alpha}} \psi_\mu \bar{f}_k -  \psi_\mu \partial^{\alpha\dot{\alpha}}\bar{f}_k  \right)
    -
    \frac{1}{2}
    \left(\partial^{\alpha\dot{\alpha}} \chi_\mu f_k - \chi_\mu \partial^{\alpha\dot{\alpha}} f_k \right)
    \\
    &
    +
    \delta_\mu^\alpha \left[ f_k \left(\partial^{\dot{\alpha}\beta}\chi_\beta + im \bar{\psi}^{\dot{\alpha}}\right)
    -
    \bar{f}_k \left( \partial^{\dot{\alpha}\beta}\psi_\beta + im \bar{\chi}^{\dot{\alpha}} \right)
     \right]\,,
    \end{split}
\end{equation}
These currents are conserved on the equations of motion \eqref{hyper-equations} :
\begin{equation}
    \partial_{\alpha\dot{\alpha}} S^{\alpha\dot{\alpha}}_{\mu k} = 0\,.
\end{equation}

It is useful to present another form of the superfield equation of motion \eqref{hyperEq}:
\begin{equation}
    \mathcal{D}^{++} q^+ = 0 \qquad \Leftrightarrow \qquad \left(\mathcal{D}^{--}\right)^2 q^+ = 0\,.\label{hyperEq2}
\end{equation}
Hitting the second equation by derivatives $\left(D^+\right)^4$ immediately yields the Klein-Gordon equation:
\begin{equation}\label{superfield-KG}
    \left( \Box + m^2 \right) q^+ = 0\,.
\end{equation}
So on shell all the components of the hypermultiplet satisfy the massive Klein-Gordon equation, in full agreement with the
component equations \eqref{hyper-equations}.

Another feature which will be important for our further consideration is the presence of an infinite number of rigid symmetries
in the hypermultiplet action \eqref{hyper}.  All currents to be explored  in this Section are associated with these higher-derivative rigid symmetries,
as explained in Appendix \ref{Noether}.

\subsection{Cubic coupling of hypermultiplet to Maxwell spin $\mathbf{1}$ multiplet; the spin $\mathbf{1}$ current superfield}\label{sec:cubic Maxwell}
We begin the description of cubic vertices with the simplest example of spin $\mathbf{1}$ $\mathcal{N}=2$ supermultiplet.

Coupling of hypermultiplet to $\mathcal{N}=2$ Maxwell theory \cite{18} is accomplished through covariantization of the flat harmonic derivative $\mathcal{D}^{++}$
in the free hypermultiplet action \eqref{hyper} by the analytic gauge connection $h^{++}(\zeta)$ defined in section \ref{Maxwell},
$\mathcal{D}^{++} \to \mathcal{D}^{++} + i\kappa_1 h^{++}\,,$
\begin{equation}
    \begin{split}
    S^{(s=1)}_{gauge} =& -\int d\zeta^{(-4)}  \; \frac{1}{2} q^{+a} \left( \mathcal{D}^{++} \delta_a^b + i \kappa_1 h^{++}(\tau_3)^b_{\;a} \right) q^+_b
    \\=& - \int d\zeta^{-4} \; \tilde{q}^+ \left(\mathcal{D}^{++} + i \kappa_1 h^{++}\right) q^+\,.
        \end{split}
\end{equation}
Here $\kappa_1$ is a dimensionful ($[\kappa] =1$) coupling constant.

The corresponding current superfield is obtained by varying the action with respect to the analytical potential $h^{++}(\zeta)$:
\begin{equation}\label{electric current}
    J^{++} \equiv \frac{1}{\kappa_1}\frac{\delta S^{(s=1)}_{gauge} }{ \delta h^{++}} = - i\frac{1}{2} q^{+a} (\tau_3)^b_{\;a} q^+_b  = - i \tilde{q}^+ q^+\,.
\end{equation}
The current superfield is real with respect to the tilde-conjugation \eqref{tilde}, $\widetilde{J^{++}} = J^{++}$. Here we have used the property $\widetilde{(\tilde{q}^+)} = - q^+$.

The current $J^{++}$ provides a source in the Maxwell superfield  equation \eqref{Maxwell eom}:
\begin{equation}
    (D^+)^4 h^{--} + \frac12\kappa_1 J^{++} = 0 \,.
\end{equation}
The equation of motion for the hypermultiplet in the  presence of cubic coupling reads:
\begin{equation}\label{interacting Max eom}
        \left(\mathcal{D}^{++} + i \kappa_1 h^{++}\right) q^+ = 0\,.
\end{equation}

The free equations of motion of the hypermultiplet \eqref{hyperEq} ensures the ``conservation'' of $J^{++}$
\footnote{On the complete equations of motion \eqref{interacting Max eom} the current superfield is also conserved:
\begin{equation*}
    \mathcal{D}^{++} J^{++}
    =
    - i \left(\mathcal{D}^{++} \tilde{q}^+\right) q^+
    - i \tilde{q}^+ \left(\mathcal{D}^{++} q^+ \right)
    =
    \kappa_1 h^{++} \tilde{q}^+ q^+
    -
    \kappa_1 h^{++} \tilde{q}^+ q^+
    =
    0\,.
\end{equation*}
Treating the spin ${\bf 1}$ system as a toy model for higher spin ${\cal N}=2$ systems, we limit our attention to the conservation of current superfield on the free equations of motion, i.e. to the first order  in $\kappa_1$. }
\begin{equation}
    \mathcal{D}^{++} J^{++} = 0\,.
\end{equation}
Thus the components of $J^{++}$ constitute the tensor (or ``linear'')  ${\cal N}=2$ multiplet \cite{deWit:1979xpv, Wess:1975ns, Siegel:1984bm, 18}.

In accord with this identification  one can check that $J^{++}$ has the following general structure:
\begin{equation}\label{electric supercurrent components}
    \begin{split}
        J^{++}(x, \theta^+, u) =& j^{(ij)}(x) u^+_i u^+_j  + \theta^{+\alpha} j^{i }_{\alpha} (x) u^+_i
        +
        \bar{\theta}^{+\dot{\alpha}} \bar{j}^{i }_{\dot{\alpha}} (x) u^+_i
        \\&+
        (\theta^{+})^2 j(x) +
        (\bar{\theta}^+)^2 \bar{j}(x)
        \\& +
        2i \theta^+ \sigma^n \bar{\theta}^+ \left[ j_n (x) + \partial_n j^{(ij) }(x) u^+_i u^-_j \right]
        \\&
        - i
        (\bar{\theta}^+)^2\theta^{+\alpha} \partial^{\dot{\alpha}}_\alpha j^{i }_{\dot{\alpha}} u^-_i
        - i
        (\theta^+)^2 \bar{\theta}^{+\dot{\alpha}} \partial_{\dot{\alpha}}^\alpha \bar{j}^{i}_{\alpha} u^-_i
        \\&
        -
        (\theta^+)^2 (\bar{\theta}^+)^2 \Box j^{(ij)} u^-_i u^-_j\,, \qquad \partial^n j_n (x) = 0\,.
    \end{split}
\end{equation}
As explained in Appendix \ref{current}, this structure is in fact universal for all current superfields.

Now, for $J^{++}= - i\tilde{q}^+ q^+$, we can explicitly express all the components in \eqref{electric supercurrent components}
in terms of the on-shell hypermultiplet fields \eqref{hyper q}:

\begin{itemize}
\item The  lowest component is the composite ${\rm SU}(2)$ triplet operator:
\begin{equation}\label{SU(2)}
    j^{(ij)}(x) = i \bar{f}^{(i} f^{j)}\,.
\end{equation}

\item The coefficients of $\theta^+$ and $\bar\theta^+$ in $J^{++}$ are composite fermionic operators:
\begin{equation}\label{pre S current}
    j_\alpha^i = i \left(  \bar{f}^{i} \psi_\alpha +  f^i \chi_\alpha  \right),\qquad  \bar{j}^i_{\dot{\alpha}} = i\left(  f^i \bar{\psi}_{\dot{\alpha}} -  \bar{f}^{i} \bar{\chi}_{\dot{\alpha}}  \right).
\end{equation}

\item The coefficients of $\left(\theta^+\right)^2$ and $\left(\bar{\theta}^+\right)^2$ are the operators:
\begin{equation}
    j(x) =  -im f^i \bar{f}_{i} - \frac{i}{2} \psi^\beta \chi_\beta\,,\qquad \bar{j}(x) =  im
    f^i \bar{f}_{i} - \frac{i}{2} \bar{\chi}^{\dot{\alpha}} \bar{\psi}_{\dot{\alpha}}\,.
\end{equation}
\item The independent component in the coefficient of $\theta^+ \sigma^n \bar{\theta}^+$ is the ${\rm U}(1)$ current

\begin{equation}\label{el current}
    j_m^{el} = \frac{i}{2} \left( f^i \partial_m \bar{f}_i - \partial_m f^i \bar{f}_i  \right) + \frac{1}{2} \psi \sigma_m \bar{\psi} - \frac{1}{2} \chi \sigma_m \bar{\chi}\,.
\end{equation}
It is easy to verify that it is indeed conserved on the hypermultiplet equations of motion \eqref{hyper-equations},
\begin{equation}
    \partial^m  j_m^{el} = 0\,,
\end{equation}
in accord with the transversality constraint in \eqref{electric supercurrent components}.

The ${\rm SU}(2)$ triplet part of the coefficient of $\theta^+ \sigma^n \bar{\theta}^+$, in agreement with the general structure \eqref{electric supercurrent components},
is the derivative of the ${\rm SU}(2)$ composite operator \eqref{SU(2)}:
\begin{equation}
    j_n^{(ij)} = i \left( \bar{f}^{(i} \partial_n f^{j)} + f^{(i} \partial_n \bar{f}^{j)}\right)
    =
     i \partial_n \left(\bar{f}^{(i} f^{j)}\right)
    =
     \partial_n j^{(ij)}\,;
\end{equation}

\item The higher-order components in the current superfield are derivatives of the lower-order
components, in accordance with the general expansion \eqref{electric supercurrent components}.

\end{itemize}

It is easy to explicitly verify that the above composite operators are transformed through each other under the supersymmetry transformations \eqref{susy hyp}. In particular,
\begin{equation}
    \delta^*_\epsilon j^{(ij)}(x)
    =
     i \delta^*_\epsilon \left( \bar{f}^{(i} f^{j)} \right)
    =
     - i \epsilon^{\mu (i} \underbrace{\left( \bar{f}^{j)} \psi_\mu + f^{j)} \kappa_\mu\right)}_{-i j^j_\mu}
     +\,
     ...\,.
\end{equation}

Thus we see that the superfield current \eqref{electric current} involves the conserved ${\rm U}(1)$ current and its $\mathcal{N}=2$ supersymmetry partners.
We now turn to the general case of cubic vertex for an arbitrary $\mathcal{N}=2$ spin $\mathbf{s}$ supermultiplet.

\subsection{Cubic couplings of hypermultiplet to $\mathcal{N}=2$ higher spins} \label{hs-couplings}

In \cite{Buchbinder:2022kzl} we have constructed gauge-invariant cubic coupling of hypermultiplet with $\mathcal{N}=2$ higher  spins.
Coupling to $\mathcal{N}=2$ spin $\bf{s}$ multiplet  has the universal form:
\begin{equation}\label{action-final}
    S^{(s)}_{gauge}
    =
    - \frac{1}{2} \int d\zeta^{(-4)}\;
    q^{+a} \left(\mathcal{D}^{++} + \kappa_s\hat{\mathcal{H}}^{++}_{(s)}  (J)^{P(s)} + \kappa_s \xi \Gamma^{++}_{(s)} (J)^{P(s)} \right) q^+_a\,.
\end{equation}
The action \eqref{action-final} is obtained by covariantization of harmonic derivative $\mathcal{D}^{++}$ under higher spin gauge group by $\mathcal{N}=2$
invariant differential operator $\hat{\mathcal{H}}^{++}_{(s)}$ and the superfield $\Gamma^{++}_{(s)}$ defined as
\begin{equation}\label{diff operator}
    \hat{\mathcal{H}}^{++}_{(s)} :=  \hat{\mathcal{H}}^{++\alpha(s-2)\dot{\alpha}(s-2)} \partial^{s-2}_{\alpha(s-2)\dot{\alpha}(s-2)}\,,
\end{equation}
with
\begin{equation}\label{operator-in}
    \hat{\mathcal{H}}^{++\alpha(s-2)\dot{\alpha}(s-2)} : = h^{++\alpha(s-2)\dot{\alpha}(s-2)M} \partial_M\,,
\end{equation}
and
\begin{equation}\label{Gamma-main}
    \Gamma^{++}_{(s)} =  \left(\partial^{s-2}_{\alpha(s-2)\dot\alpha(s-2)}    \Gamma^{++\alpha(s-2)\dot{\alpha}(s-2)}\right),
\end{equation}
with
\begin{eqnarray}
    && \Gamma^{++\alpha(s-2)\dot{\alpha}(s-2)} = (-1)^{P(M)} \partial_M h^{++\alpha(s-2)\dot{\alpha}(s-2) M} \nonumber
    \\
    && =  \partial_{\beta\dot{\beta}}h^{++(\alpha(s-2)\beta)(\dot{\alpha}(s-2)\dot{\beta})}-
    \partial^-_{\beta}h^{++(\alpha(s-2)\beta)\dot{\alpha}(s-2)+}
    -  \partial^-_{\dot{\beta}} h^{++\alpha(s-2)(\dot{\alpha}(s-2)\dot{\beta})+} \label{PreciseInd}  \,,
\end{eqnarray}
where $\partial^{s}_{\alpha(s)\dot\alpha(s)} := \partial_{(\alpha_1 \dot{\alpha}_1} \ldots  \partial_{\alpha_s) \dot{\alpha}_s}$.
Both quantities, $\hat{\mathcal{H}}^{++}_{(s)}$ and $\Gamma^{++}_{(s)}$, manifestly preserve the analyticity. To avoid a possible confusion, let
us point out that the undotted and dotted indices hidden in the multi-index $M$ in \eqref{operator-in} are assumed to be properly
symmetrized with the same type of indices of the relevant analytic potentials, as is explicitly given in \eqref{PreciseInd}.

The most general cubic coupling \eqref{action-final} involves the matrix operator $J$:
\begin{eqnarray}\label{J operator}
&&     J q^{+ a} = i  (\tau^3)^{a}_{\;b} q^{+ b}\,,
    \quad\quad\quad  (\tau_3)^a_{\;b}
    =
    {\scriptstyle  \begin{pmatrix}
        1 & 0 \\
        0 & -1
    \end{pmatrix}}\,, \nonumber \\
    &&J q^{+ } = -iq^{+ }\,, \quad J \tilde{q}^{+ } =  i\tilde{q}^{+ }\,.
\end{eqnarray}
This operator  necessarily enters the interaction of odd higher spin multiplets with the hypermultiplet and does not appear for even spins.
So the couplings in \p{action-final} are different for odd and even spins. They are distinguished by the projection operator $P(s)$
\begin{equation}
     P(s) = \frac{1 + (-1)^{s+1}}{2}\,,
\end{equation}
which is zero for even spins and equals $1$ for odd spins. Note that for even spins $\bf{s}$ the $\xi$-term in \eqref{action-final}
disappears due to the kinematical relation $q^{+a}q^+_a = 0$. Thus only odd spin couplings admit non-trivial $\xi$-terms.

The superfield $\Gamma^{++}_{(s)}$ is invariant under  $\mathcal{N}=2$ supersymmetry:
\begin{equation}
    \delta_\epsilon \Gamma^{++}_{(s)} = 0\,.
\end{equation}
It also possesses the simple gauge transformation law with respect to \eqref{Gauge_s}:
\begin{equation}\label{Gamma}
    \delta_\lambda \Gamma^{++}_{(s)} =  \mathcal{D}^{++}\Omega_{(s)}\,, \qquad \Omega_{(s)} := \big(\partial^{s-2}_{\alpha(s-2)\dot{\alpha}(s-2)}\Omega^{\alpha(s-2)\dot{\alpha}(s-2)}\big),
\end{equation}
where
\be \label{Omega}
\Omega^{\alpha(s-2)\dot{\alpha}(s-2)} := (-1)^{P(M)} \big(\partial_M \lambda^{\alpha(s-2)\dot{\alpha}(s-2) M}\big).
\ee

In the first order in $\kappa_s$, the coupling \eqref{action-final} is gauge invariant, provided that
the gauge transformations are realized on the hypermultiplet $q^{+a}$ via higher derivatives:
\begin{equation}\label{transformations}
    \begin{split}
        \delta^{(s)}_{\lambda} q^{+a}
        =& - \frac{\kappa_s}{2}\left\{\hat{\Lambda}^{\alpha(s-2)\dot{\alpha}(s-2)}, \partial^{s-2}_{\alpha(s-2)\dot{\alpha}(s-2)}\right\} (J)^{P(s)} q^{+a}
        \\&-\frac{\kappa_s}{2}
        \partial^{s-2}_{\alpha(s-2)\dot{\alpha}(s-2)} \Omega^{\alpha(s-2)\dot{\alpha}(s-2)} (J)^{P(s)}  q^{+a}\,.
    \end{split}
\end{equation}
Here we used the notation \eqref{LAMBDA s-2} for the differential operator.
The derivatives act freely on all objects to the right. In the presence of $\xi$-term the action respects some extra gauge transformations:
\begin{equation}\label{xi-s-f}
    \delta^{(s)}_{\xi} q^{+a} = -   P(s) \kappa_s \xi \left( \partial^{s-2}_{\alpha(s-2)\dot{\alpha}(s-2)} \Omega^{\alpha(s-2)\dot{\alpha}(s-2)} \right)  J q^{+a}\,.
\end{equation}
Gauge transformations of gauge supermultiplets have the form \eqref{Gauge_s} and, as a consequence, all terms in the action \eqref{action-final}
transform according to \eqref{transf-H} and \eqref{Gamma}.

\medskip

There are several important peculiarities of the cubic interactions \eqref{action-final}:
\begin{itemize}
    \item For odd higher spins the cubic coupling explicitly breaks ${\rm SU}(2)_{PG}$;
    \item Also, for odd higher spin $\bf{s}$, the coupling to massive hypermultiplet ($\partial_5 q^{+a} \neq 0$) can be rewritten through $\partial_5$ derivative \eqref{global-1},
    \begin{equation}\label{partial 5 and mass}
        \partial_5 q^{+a} = - m    J q^{+a}\,,
    \end{equation}
    keeping in mind that in this case $\partial_5$ coincides with $J$ modulo a numerical factor. For a massless hypermultiplet ($\partial_5 q^{+a} = 0, m=0$), the generator $J$ should still be retained in all previous formulas;
    \item For even higher spin multiplets the couplings with both massive and massless hypermultiplet preserve ${\rm SU}(2)_{PG}$ and so do not  require introducing $J$,
    as is obvious from the action \eqref{action-final} with $P(s)=0$;

    \item The interactions of massive and massless hypermultiplets differ from each other  due to the presence of $\partial_5$ in the differential operator \eqref{diff operator}.

    \item We ascribe canonical dimensions to superfields, $[h^{++\alpha(s-1)\dot{\alpha}(s-1)}]=
    [h^{++5\alpha(s-2)\dot{\alpha}(s-2)}]$ $= 0\,,\; [h^{++\alpha(s-1)\dot{\alpha}(s-2)+}] = [h^{++\alpha(s-2)\dot{\alpha}(s-1)+}] = \frac{1}{2}$,
    $[q^+]=1$,  so the coupling constants have the dimensions $[\kappa_s] = - (s-1)$;

    \item At the superfield level, there are 2 types of interactions for odd spins, with and without the parameter $\xi$. The distinction  between these interactions
    after elimination of the auxiliary fields in the spin $\mathbf{3}$ case will be explained in Section \ref{spin 3 coupling}.
\end{itemize}

The couplings \eqref{action-final} will lead to the appearance of matter sources $J^{++\dots}$ in the higher spin equations of motion \eqref{EOM}.
Our next goals will be  to derive the superfield form of these sources and to explore their component structure.

\subsection{Construction of current superfields}\label{Construction of current superfields}

The current superfields $J^{++\dots}(\zeta)$ are analytic superfields appearing as sources in $\mathcal{N}=2$ higher spin gauge superfield
equations of motion (described in Section \ref{equations of motion}) in the presence of matter couplings \eqref{action-final}:
\begin{equation}
    \delta_h S^{(s)}_{gauge} := \kappa_s \int d\zeta^{(-4)} \; \delta h^{++\alpha(s-2)\dot{\alpha}(s-2) M} J^{++}_{\alpha(s-2)\dot{\alpha}(s-2) M}\,.
\end{equation}

$\mathcal{N}=2$ spin $\mathbf{s}$ multiplets are generically described by a set of four analytic prepotentials:
$$
h^{++\alpha(s-1)\dot{\alpha}(s-1)}\,, \quad
h^{++\alpha(s-1)\dot{\alpha}(s-2)}\,, \quad
h^{++\alpha(s-2)\dot{\alpha}(s-1)}\,, \quad
h^{++\alpha(s-2)\dot{\alpha}(s-2)}\,.
$$
There exist four current superfields associated with these prepotentials.
In this Section we shall study these four kinds of the superfield sources.

\medskip

\textbf{\textit{\underline{Varying $h^{++\alpha(s-1) \dot{\alpha}(s-1) }$}}}

\medskip
In this case, the variation of the action \eqref{action-final} is expressed as:
\begin{equation}
    \begin{split}
    \delta_h S^{(s)}_{gauge}
    &=
    - \frac{\kappa_s}{2}  \int d\zeta^{(-4)}\;
    q^{+a} \Bigr[ \delta h^{++\alpha(s-1) \dot{\alpha}(s-1) } \partial^{s-1}_{\alpha(s-1) \dot{\alpha}(s-1)} (J)^{P(s)}
    \\
    &\quad\quad+
      \xi \left( \partial^{s-1}_{\alpha(s-1) \dot{\alpha}(s-1)} \delta h^{++\alpha(s-1) \dot{\alpha}(s-1) } \right)  (J)^{P(s)} \Bigr] q^+_a\,.
        \end{split}
\end{equation}
After integration by parts one obtains
\begin{equation}\label{Supercurrent1}
    \boxed{J^{++}_{\alpha(s-1)\dot{\alpha}(s-1)} =
    - \frac{1}{2}
    q^{+a}  \partial^{s-1}_{\alpha(s-1) \dot{\alpha}(s-1)} (J)^{P(s)} q^+_a
    -
    \frac{1}{2} \xi  \, \partial^{s-1}_{\alpha(s-1) \dot{\alpha}(s-1)} \left[q^{+a}  (J)^{P(s)}  q^+_a \right]}.
\end{equation}
For even $\bf{s}$ the $\xi$-term in this current superfield vanishes (because of the identity $q^{+a}q^+_a = 0$).

It is straightforward to verify that the current superfield \eqref{Supercurrent1} is conserved on the free hypermultiplet equations of motion \eqref{hyperEq}:
\begin{equation}\label{cons 1}
    \mathcal{D}^{++} q^{+a} = 0\, \;\;\;\;\; \Rightarrow \;\;\;\;\;     \mathcal{D}^{++} J^{++}_{\alpha(s-1)\dot{\alpha}(s-1)} = 0 \,.
\end{equation}

\medskip

\textbf{\textit{\underline{Varying $h^{++\alpha(s-2) \dot{\alpha}(s-2) }$}}}

\medskip

In this case, the variation takes a simpler form, since $h^{++\alpha(s-2) \dot{\alpha}(s-2) }$ is not present in $\Gamma^{++}_{(s)}$:
\begin{equation}
    \begin{split}
        \delta_h S^{(s)}_{gauge}
        &=
        - \frac{\kappa_s}{2} \int d\zeta^{(-4)}\;
        q^{+a} \Bigr( \delta h^{++\alpha(s-2) \dot{\alpha}(s-2) } \partial^{s-2}_{\alpha(s-2) \dot{\alpha}(s-2)} (J)^{P(s)}   \partial_5
         \Bigr) q^+_a\,.
    \end{split}
\end{equation}

Thus the relevant current superfield reads:
\begin{equation}\label{Supercurrent2}
    \boxed{ J^{++}_{\alpha(s-2)\dot{\alpha}(s-2)} = - \frac{1}{2}
    q^{+a}  \partial^{s-2}_{\alpha(s-2) \dot{\alpha}(s-2)} (J)^{P(s)}   \partial_5
     q^+_a}\,.
\end{equation}
It is important, that it is non-vanishing only for the massive hypermultiplet ($\partial_5 q^+_a \neq 0$). Also, \eqref{Supercurrent2} is $\xi$-independent.
The free hypermultiplet equations of motion ensure the conservation of  \eqref{Supercurrent2}:
\begin{equation}\label{cons 2}
    \mathcal{D}^{++} J^{++}_{\alpha(s-2)\dot{\alpha}(s-2)} = 0\,.
\end{equation}

\medskip

\textbf{\textit{\underline{Varying $h^{++\alpha(s-2) \dot{\alpha}(s-1)+ }$}}}

\medskip

The variation of the action \eqref{action-final} with respect to the spinor potentials has a similar form, though there are some distinctions
related to different index structures:
\begin{eqnarray}
&& \delta_h S^{(s)}_{gauge} =
        - \frac{\kappa_s}{2} \int d\zeta^{(-4)}\;
        q^{+a} \Bigr[ \delta h^{++\alpha(s-2) \dot{\alpha}(s-1)+ } \partial^{s-2}_{\alpha(s-1)  \dot{\alpha}(s-1)} (J)^{P(s)} \partial^-_{\dot{\alpha}}
        \nonumber \\
&&\qquad\qquad\qquad-\,
         \xi \left( \partial^{s-2}_{\alpha(s-2) \dot{\alpha}(s-2)} \partial^-_{\dot{\alpha}} \delta h^{++\alpha(s-2) \dot{\alpha}(s-1)+ } \right)  (J)^{P(s)} \Bigr] q^+_a\,. \label{Var54}
\end{eqnarray}
Then,  the corresponding superfield current reads:
\begin{equation}\label{Supercurrent3}
    \boxed{J^{+}_{\alpha(s-2)\dot{\alpha}(s-1)} = -\frac{1}{2} q^{+a}  \partial^{s-2}_{\alpha(s-2)  (\dot{\alpha}(s-2)}  \partial^-_{\dot{\alpha})} (J)^{P(s)} q^+_a
    -
    \frac{1}{2} \xi \partial^{s-2}_{\alpha(s-2)(\dot{\alpha}(s-2)} \partial^-_{\dot{\alpha})} \left[q^{+a}  (J)^{P(s)}  q^+_a \right]}.
\end{equation}
Here the $\xi$-term also survives only for odd spins. Unlike the bosonic currents, the fermionic current does not respect manifest supersymmetry due to the presence
of spinor derivative\footnote{The standard lore is that the gauge (super)fields couple to conserved (super)currents. And indeed the supercurrents coupled to bosonic gauge prepotentials are conserved.
The reason why the spinor currents obey non-standard conservation laws is the non-trivial transformation laws
of bosonic prepotentials \eqref{Gauge_s}, which leads to extra terms containing the spinor parameter.}.
Since the derivatives $\mathcal{D}^{++}$ and $\partial^-_{\dot{\alpha}}$ do not commute,
\begin{equation}
    [\mathcal{D}^{++}, \partial^-_{\dot{\alpha}}] = - 2i \theta^{+\rho} \partial_{\rho\dot{\alpha}} - \partial^+_{\dot{\alpha}} - 2i \theta^+_{\dot{\alpha}} \partial_5\,,
\end{equation}
the superfield \eqref{Supercurrent3}, in contrast to the bosonic current superfields, satisfies a more complicated ``conservation'' law:
\begin{equation}\label{non-cons}
    \begin{split}
    \mathcal{D}^{++} J^{+}_{\alpha(s-2)\dot{\alpha}(s-1)}
    =&
    i q^{+a} \partial^{s-2}_{\alpha(s-2)(\dot{\alpha}(s-2)}
    \left(\theta^{+\rho}\partial_{\rho\dot{\alpha})} + \bar{\theta}^+_{\dot{\alpha})} \partial_5\right) (J)^{P(s)} q^+_a\\
    &+
    i\xi \partial^{s-2}_{\alpha(s-2)(\dot{\alpha}(s-2)} \theta^{+\rho}\partial_{\rho\dot{\alpha})}  \left[q^{+a}  (J)^{P(s)}  q^+_a \right],
\end{split}
\end{equation}
where we have used the property $\partial_5 [q^{+a}(J)^{P(s)}  q^+_a ] = 0$. Using the bosonic supercurrents \eqref{Supercurrent1} and \eqref{Supercurrent2}, one
can rewrite the generalized conservation condition \eqref{non-cons} as:
\begin{equation}
    \mathcal{D}^{++} J^{+}_{\alpha(s-2)\dot{\alpha}(s-1)} =
    -2i \theta^{+\rho} J^{++}_{(\alpha(s-2)\rho)\dot{\alpha}(s-1)}
    -
    2i \bar{\theta}^+_{(\dot{\alpha}} J^{++}_{\alpha(s-2)\dot{\alpha}(s-2))}\,. \label{ModConserv}
\end{equation}
As a consequence of the relations \eqref{cons 1} and \eqref{cons 2}, the superfield \eqref{Supercurrent3} satisfies the modified conservation law:
\begin{equation}
    \left(\mathcal{D}^{++}\right)^2 J^{+}_{\alpha(s-2)\dot{\alpha}(s-1)} = 0\,.
\end{equation}

The form \eqref{ModConserv} of the conservation law suggests that one can define  the modified spinor superfield current as
 \begin{equation}\label{covariant current superfield}
    \begin{split}
 \mathcal{J}^{+}_{\alpha(s-2)\dot{\alpha}(s-1)} = &\;\frac{1}{2} q^{+a}  \partial^{s-2}_{\alpha(s-2)  (\dot{\alpha}(s-2)}  D^-_{\dot{\alpha})} (J)^{P(s)} q^+_a
 \\&\;+
 \frac{1}{2} \xi \partial^{s-2}_{\alpha(s-2)(\dot{\alpha}(s-2)} D^-_{\dot{\alpha})} \left[q^{+a}  (J)^{P(s)}  q^+_a \right].
    \end{split}
  \end{equation}
Here we used the covariant spinor derivatives \eqref{D-}. The newly defined current is conserved in the standard way (due to $[\mathcal{D}^{++}, D^-_{\hat{\mu}}] = D^+_{\hat{\mu}}$ and analyticity of hypermultiplet),
\bea
\mathcal{D}^{++}\mathcal{J}^{+}_{\alpha(s-2)\dot{\alpha}(s-1)} = 0\,.
\eea
It is invariant with respect to ${\cal N}=2$ supersymmetry, but is not analytic. It is also useful to rewrite the newly defined current \eqref{covariant current superfield} as:
\begin{equation}\label{covariant fermionic 1}
     \mathcal{J}^{+}_{\alpha(s-2)\dot{\alpha}(s-1)} =
     J^{+}_{\alpha(s-2)\dot{\alpha}(s-1)}
     +
     2i \theta^{-\mu} J^{++}_{(\alpha(s-2)\mu)\dot{\alpha}(s-1)}
     +
     2i
     \bar{\theta}^-_{(\dot{\alpha}} J^{++}_{\alpha(s-2)\dot{\alpha}(s-2))}\,.
\end{equation}

\medskip

\textbf{\textit{\underline{Varying  $h^{++ \alpha(s-1) \dot{\alpha}(s-2) + }$}}}

\medskip

In complete analogy with the previous variation, we derive:
\begin{equation}\label{Supercurrent4}
    \boxed{J^{+}_{\alpha(s-1)\dot{\alpha}(s-2)} = -\frac{1}{2} q^{+a}  \partial^{s-2}_{(\alpha(s-2)  \dot{\alpha}(s-2)}  \partial^-_{\alpha)} (J)^{P(s)} q^+_a
    -
    \frac{1}{2} \xi \partial^{s-2}_{(\alpha(s-2)\dot{\alpha}(s-2)} \partial^-_{\alpha)} \left[q^{+a}  (J)^{P(s)}  q^+_a \right]}.
\end{equation}
This current superfield is also not invariant under supersymmetry. As in the  previous case, due to the non-trivial commutation relation,
\begin{equation}
    [\mathcal{D}^{++}, \partial^-_\alpha] =
    2i \bar{\theta}^{+\dot{\rho}} \partial_{\alpha\dot{\rho}} - \partial^+_{\alpha} -2i \theta^+_\alpha \partial_5\,,
\end{equation}
the superfield \eqref{Supercurrent4} is not conserved in the conventional sense,
\begin{equation}\label{ModConser2}
    \mathcal{D}^{++} J^{+}_{\alpha(s-1)\dot{\alpha}(s-2)}
    =
    2i \bar{\theta}^{+\dot{\rho}} J^{++}_{\alpha(s-1)(\dot{\alpha}(s-2)\dot{\rho})}
    -
    2i
    \theta^+_{(\alpha} J^{++}_{\alpha(s-2))\dot{\alpha}(s-2)}\,.
\end{equation}

In full analogy with the construction of the current superfield \eqref{covariant current superfield},
we can construct the supersymmetry-invariant and conserved non-analytic current superfield:
\begin{equation}\label{covariant current superfield 2}
    \mathcal{J}^{+}_{\alpha(s-1)\dot{\alpha}(s-2)} = \frac{1}{2} q^{+a}  \partial^{s-2}_{(\alpha(s-2)  \dot{\alpha}(s-2)} D^-_{\alpha)} (J)^{P(s)} q^+_a
        +
        \frac{1}{2} \xi \partial^{s-2}_{(\alpha(s-2)\dot{\alpha}(s-2)} D^-_{\alpha)} \left[q^{+a}  (J)^{P(s)}  q^+_a \right].
\end{equation}

This superfield can also be rewritten through the analytic superfields:
\begin{equation}\label{covariant fermionic 2}
        \mathcal{J}^{+}_{\alpha(s-1)\dot{\alpha}(s-2)}
        =
        J^{+}_{\alpha(s-1)\dot{\alpha}(s-2)}
        -
        2i \bar{\theta}^{-\dot{\mu}} J^{++}_{\alpha(s-1)(\dot{\alpha}(s-2)\dot{\mu})}
        +
        2i \theta^-_{(\alpha} J^{++}_{\alpha(s-2))\dot{\alpha}(s-2)}\,.
\end{equation}
The corresponding conservation law reads:
\begin{equation}
    \mathcal{D}^{++}  \mathcal{J}^{+}_{\alpha(s-1)\dot{\alpha}(s-2)} = 0\,.
\end{equation}

\medskip

Using the obtained expressions for analytic current superfields, one can cast the  cubic
vertices \eqref{action-final} in the Noether form
\begin{equation}\label{current coupling}
    S_{gauge}^{(s)} = \kappa_s \int d\zeta^{(-4)}\; h^{++\alpha(s-2)\dot{\alpha}(s-2)M} J^{++}_{\alpha(s-2)\dot{\alpha}(s-2)M}\,.
\end{equation}
Requiring the coupling to be invariant under the gauge transformations of analytic potentials \eqref{Gauge_s} leads to the on-shell conservation laws \eqref{cons 1}, \eqref{cons 2}
and the modified conservation laws  \eqref{ModConserv}, \eqref{ModConser2} (see also Appendix
\ref{Noether}).

By construction, the cubic coupling
    \eqref{current coupling} is manifestly invariant under $\mathcal{N}=2$ supersymmetry. So, because
    of nontrivial supersymmetry transformations of the bosonic potentials \eqref{susy1},
the current superfields should also possess non-trivial transformation laws
:
\begin{subequations}\label{SUSY currents}
\begin{equation}\label{SUSY fermioic currents 1}
    \delta_\epsilon J^{+}_{(\dot{\alpha}(s-2)\dot{\beta})\alpha(s-2)}  = -2i \epsilon^{-\beta} J^{++}_{(\alpha(s-2)\beta)\dot{\alpha}(s-1)}
    - 2i \bar{\epsilon}^-_{(\dot{\beta}} J^{++}_{\dot{\alpha}(s-2))\alpha(s-2)}\,,
\end{equation}
\begin{equation}
    \label{SUSY fermioic currents 2}
    \delta_\epsilon J^{+}_{(\alpha(s-2)\beta)\dot{\alpha}(s-2)} = 2i \bar{\epsilon}^{-\dot{\beta}} J^{++}_{(\dot{\alpha}(s-2)\dot{\beta})\alpha(s-2)}
    -
    2i \epsilon^-_{(\beta} J^{++}_{\alpha(s-2)) \dot{\alpha}(s-2)}\,,
\end{equation}
\begin{equation}\label{SUSY bosonic currents}
    \delta_\epsilon J^{++}_{\alpha(s-1)\dot{\alpha}(s-1)} = 0\,,
    \qquad\qquad
    \delta_\epsilon J^{++}_{\alpha(s-2)\dot{\alpha}(s-2)} = 0 \,.
\end{equation}
\end{subequations}
These transformation laws indicate that the  analytic supercurrents constructed constitute not fully reducible representation of ${\cal N}=2$ supersymmetry, reflecting the similar property
of the analytical potentials \eqref{susy}. The supercurrents $\mathcal{J}^{+}_{\alpha(s-2)\dot{\alpha}(s-2)\hat{\mu}}$ can be found by varying with respect
to unconstrained non-analytic Mezincescu-type prepotentials (see Appendix \ref{non analytical supercurrents}) and they possess a manifest $\mathcal{N}=2$ supersymmetry,
albeit at cost of the loss of analyticity.

According to the general discussion in Appendix \ref{N=2 current superfield}, the even current superfields $J^{++}_{\alpha(s-1)\dot{\alpha}(s-1)}$ and
$J^{++}_{\alpha(s-2)\dot{\alpha}(s-2)}$ involve
    the conserved bosonic currents which can be identified as the spin $s$ and spin $s-1$  Noether currents for rigid bosonic symmetries of the hypermultiplet action.
    The spin $s-\tfrac{1}{2}$ odd currents associated with the fermionic spin $s-\tfrac{3}{2}$ symmetries of the free hypermultiplet
    are not directly contained in the fermionic current superfields $J^+_{\alpha(s-1)\dot{\alpha}(s-2)}$, $J^+_{\alpha(s-2)\dot{\alpha}(s-1)}$ and are sums of some
     components of these current superfields and of the bosonic current superfields
    $J^{++\alpha(s-1)\dot{\alpha}(s-1)}$ and $J^{++\alpha(s-2)\dot{\alpha}(s-2)}$ (see Appendix \ref{5/2 symmetry}
    for the explicit formulas in the  case of the spin $\tfrac{3}{2}$ symmetry, with $s=3$).
    The conditions \eqref{ModConserv} and \eqref{ModConser2}
    imply conservation of such sums of the constituent fermionic currents, as explained in
    Appendix \ref{N=2 current superfield}.

The bosonic current superfields transform as scalars under $\mathcal{N}=2$ rigid supersymmetry \eqref{SUSY bosonic currents} so $\mathcal{N}=2$ supermultiplets of conserved currents of integer spin $s$
are  embodied by analytic superfields $J^{++}_{\alpha(s-1)\dot{\alpha}(s-1)}$.
As follows from \eqref{SUSY currents}, the analytic fermionic current superfields $J^{+}_{\alpha(s-1)\dot{\alpha}(s-2)}$ and  $J^{+}_{\alpha(s-2)\dot{\alpha}(s-1)}$ are
not $\mathcal{N}=2$ scalars, so the supermultiplets of conserved half-integer spin currents cannot be described by such analytic fermionic superfields alone. However,
such current supermultiplets can be incorporated  in non-analytic $\mathcal{N}=2$ scalar fermionic superfields  $\mathcal{J}^{+}_{\alpha(s-2)\dot{\alpha}(s-1)}$
and $\mathcal{J}^{+}_{\alpha(s-1)\dot{\alpha}(s-2)}$ defined in \eqref{covariant fermionic 1} and \eqref{covariant fermionic 2}. Since these current superfields
are expressed in terms of analytic current superfields, it suffices to deal only with the latter ones, despite their unusual ${\cal N}=2$ transformation laws.

After elimination of the auxiliary fields and performing the Berezin integration, the cubic coupling \eqref{current coupling}
in the leading order in  $\kappa_s$  will produce the component couplings of  the generic Noether form\footnote{There could appear also $\kappa_s^2$ contributions as a result of eliminating the auxiliary fields.
Presumably, in this way it is possible to generate component vertices with a larger number of higher spin gauge fields (for the spin ${\bf 1}$
example see footnote 15). We hope to address this interesting question elsewhere.}:
\begin{equation}\label{Noether form interaction}
    \begin{split}
    S_{gauge}^{(s)} =\, &\kappa_s \int d^4x \;     \Big(\Phi^{\alpha(s-1)\dot{\alpha}(s-1) \beta\dot{\beta}} Y_{\alpha(s-1)\dot{\alpha}(s-1) \beta\dot{\beta}}
    +
    \psi^{\alpha(s-1)\dot{\alpha}(s-1)\beta i} Y_{\alpha(s-1)\dot{\alpha}(s-1) \beta i}
    \\
    & \qquad
    +
    \bar{\psi}^{\alpha(s-1)\dot{\alpha}(s-1)\dot{\beta} i } \bar{Y}_{\alpha(s-1)\dot{\alpha}(s-1)\dot{\beta} i}
    +
    C^{\alpha(s-2)\dot{\alpha}(s-2)\beta\dot{\beta}} Y_{\alpha(s-2)\dot{\alpha}(s-2)\beta\dot{\beta}}
    \Big) + \mathcal{O}(\kappa_s^2).
    \end{split}
\end{equation}
Here all $Y$ currents  are constructed from the physical fields of hypermultiplet and satisfy the appropriate on-shell conservation laws as
a guarantee of the gauge invariance in the linearized approximation.
Specific examples of such component $Y$ currents for $\mathcal{N}=2$ spin $\mathbf{2}$ and
spin $\mathbf{3}$ couplings  are given
in Sections \ref{spin 2 coupling} and \ref{spin 3 coupling}.

\subsection{Superfield higher spin equations of motion with sources}
\label{Superfield higher spins equations of motion with sources}

Using the results of Section \ref{Construction of current superfields}, one can immediately find how the free equations \eqref{EOM}
are modified in the presence of $\mathcal{N}=2$ cubic coupling \eqref{action-final}.

The modification of eqs. \eqref{eq s-1} and \eqref{eq s-2} is evident:
\begin{subequations}\label{bosonic equations currents}
\begin{equation}
    (D^+)^4 G^{--\alpha(s-1) \dot{\alpha}(s-1) }= \frac{1}{2} (-1)^{s} \kappa_s J^{++\alpha(s-1) \dot{\alpha}(s-1)}\,,
\end{equation}
\begin{equation}
        (D^+)^4 G^{--5\alpha(s-2) \dot{\alpha}(s-2) } = \frac{1}{2} (-1)^{s} \kappa_s J^{++\alpha(s-2) \dot{\alpha}(s-2)}\,.
\end{equation}
\end{subequations}
Their fermionic counterpart has the form:
\begin{equation}\label{FermSFeq}
    \begin{split}
        (D^+)^4 \left[ \theta^{-\alpha_{s-1}}\, G^{--}_{\alpha(s-1) \dot{\alpha}(s-1)} + \bar\theta^{-}_{(\dot{\alpha}_{s-1}} G^{--5}_{\alpha(s-2) \dot{\alpha}(s-2))}  \right]
        +
         \frac{1}{4i} (-1)^{s+1}\kappa_s J^+_{\alpha(s-2)\dot{\alpha}(s-1)}
        = 0\,.
        \end{split}
\end{equation}

Using eqs. \eqref{bosonic equations currents} and transforming the first term in \eqref{FermSFeq} as in passing from eq. \eqref{ComplEq} to eq. \eqref{ComplEq2},
one  can bring \eqref{FermSFeq} to the form:
\begin{equation}\label{FermSFeq2}
(\bar{D}^+)^2 D^{+\beta} G^{--}_{(\beta\alpha(s-2)) \dot{\alpha}(s-1)}
-
(D^+)^2 \bar{D}^+_{(\dot{\alpha}} G^{--5}_{\alpha(s-2) \dot{\alpha}(s-2))}
= 2i \kappa_s (-1)^{s} \mathcal{J}^+_{\alpha(s-2) \dot{\alpha}(s-1)}\,,
\end{equation}
where the current superfield $\mathcal{J}^+_{\alpha(s-2) \dot{\alpha}(s-1)}$ was defined in \eqref{covariant fermionic 1}. It is scalar with respect to ${\cal N}=2$ supersymmetry but
non-analytic, just as the left hand side of eq. \eqref{FermSFeq2}. The conjugated equation can be obtained directly:
\begin{equation}
    (D^+)^2  \bar{D}^{+\dot{\beta}} G^{--}_{\alpha(s-1) (\dot{\beta}\dot{\alpha}(s-2))}
    +
    (\bar{D}^+)^2 D^+_{(\alpha} G^{--5}_{\alpha(s-2)) \dot{\alpha}(s-2)} \;
    =
    2i \kappa_s (-1)^{s} \mathcal{J}^+_{\alpha(s-1) \dot{\alpha}(s-2)}\,.
\end{equation}
The corresponding current superfield is defined in \eqref{covariant fermionic 2}.

Finally,  we are left with the system of superfield equations:
\begin{equation}\label{EOM with currents}
    \begin{cases}
        (D^+)^4 G^{--\alpha(s-1) \dot{\alpha}(s-1) }= \frac{1}{2} (-1)^{s} \kappa_s J^{++\alpha(s-1) \dot{\alpha}(s-1)}\,,
        \\
        (D^+)^4 G^{--5\alpha(s-2) \dot{\alpha}(s-2) } = \frac{1}{2} (-1)^{s} \kappa_s J^{++\alpha(s-2) \dot{\alpha}(s-2)}\,,
        \\
        (\bar{D}^+)^2 D^{+\beta} G^{--}_{(\beta\alpha(s-2)) \dot{\alpha}(s-1)}
        -
        (D^+)^2 \bar{D}^+_{(\dot{\alpha}} G^{--5}_{\alpha(s-2) \dot{\alpha}(s-2))}
        = 2i \kappa_s (-1)^{s} \mathcal{J}^+_{\alpha(s-2) \dot{\alpha}(s-1)}\,,
        \\
        (D^+)^2  \bar{D}^{+\dot{\beta}} G^{--}_{\alpha(s-1) (\dot{\beta}\dot{\alpha}(s-2))}
        +
        (\bar{D}^+)^2 D^+_{(\alpha} G^{--5}_{\alpha(s-2)) \dot{\alpha}(s-2)} \;
        =
        2i \kappa_s (-1)^{s} \mathcal{J}^+_{\alpha(s-1) \dot{\alpha}(s-2)}\,.
    \end{cases}
\end{equation}
The explicit form of the hypermultiplet current superfields was given in \eqref{Supercurrent1}, \eqref{Supercurrent2}, \eqref{covariant fermionic 1} and \eqref{covariant fermionic 2}.
One can alternatively derive these equations by varying with respect to the unconstrained Mezincescu-type $\Psi^{-\dots}$ prepotentials (see Appendix \ref{Pre-prepotentials}).
Like in the sourceless case, the first two equations in \eqref{EOM with currents} can be obtained from the remaining ones, using the relations:
\begin{equation}
    D^+_\rho \mathcal{J}^+_{\alpha(s-2)\dot{\alpha}(s-1)}
    =
    2i J^{++}_{(\alpha(s-2)\rho)\dot{\alpha}(s-1)}\,, \qquad  \bar{D}^+_{\dot{\rho}}
    \mathcal{J}^+_{\alpha(s-2)\dot{\alpha}(s-1)}
    =
    -2i \epsilon_{\dot{\rho}(\dot{\alpha}}
    J^{++}_{\alpha(s-2)\dot{\alpha}(s-2))}\,.
\end{equation}
The harmonic ${\cal D}^{++}$ conservation of the current superfields is fully dictated by the structure of the left hand sides of eqs. \eqref{EOM with currents}.
This can be easily checked using the zero curvature equations \eqref{zero-curv}, \eqref{zero-curv2}, the definitions \eqref{s-1}, \eqref{s-2} and
the analyticity property.

In complete analogy with the free theory (see Section \ref{component content of superfield equations}), we can study the component
contents of the aforementioned equations. We shall restrict our attention to the simplest cases of the spin $\mathbf{2}$  and spin $\mathbf{3}$  supermultiplets (in the spin $\mathbf{3}$ case
we will consider only the structure of current superfields). All the basic features  of the case of a generic ${\cal N}=2$  higher spin ${\bf s}$ match
those exhibited on these examples.

\subsection{Spin {\bf{2}}: $\mathcal{N}=2$ supercurrent of the free hypermultiplet} \label{hyper supercurrent}

The first non-trivial example of current superfields is supplied by the spin $\mathbf{2}$ case.
It is well known that the supergravity (spin $\mathbf{2}$) multiplet interacts with the ${\cal N}=1$
supercurrent \cite{Ogievetsky:1976qc}.
At the component level, the $\mathcal{N}=2$ supercurrent multiplet was introduced for the first time by
Sohnius \cite{Sohnius:1978pk}.
For further discussion and references, see, e.g., \cite{West:1997vm, Kuzenko:1999pi, Butter:2010sc,
West:1990tg}. To the best of our knowledge,  no detailed study of the current
superfields in the harmonic superspace approach for an arbitrary ${\bf s}$ was performed so far, with the exception of the case of ${\cal N}=2$
conformal supergravity (${\bf s} =2$) in \cite{Kuzenko:1999pi}, where, however, the component structure of the current superfields
was not studied.

Varying the cubic $(\mathbf{2}, \mathbf{\tfrac{1}{2}}, \mathbf{\tfrac{1}{2}})$ superfield coupling with respect to the unconstrained analytic potentials of the
linearized $\mathcal{N}=2$ supergravity,
 $$h^{++\alpha\dot{\alpha}}\,,\quad h^{++5}\,,\quad h^{++\alpha+}\,,\quad h^{++\dot{\alpha}+}$$ leads to the analytic current superfields:
\begin{equation}
    \begin{split}
     &J^{++}_{\alpha\dot{\alpha}} = - \frac{1}{2}
     q^{+a}  \partial_{\alpha\dot{\alpha}}  q^+_a\,,
     \quad
     J^{++} =  -
     \frac{1}{2} q^{+a} \partial_5 q^+_a \,,
     \\
     &J^+_\alpha = -\frac{1}{2} q^{+a} \partial^-_{\alpha} q^+_a\,, \quad
     J^+_{\dot{\alpha}} = -\frac{1}{2} q^{+a} \partial^-_{\dot{\alpha}} q^+_a\,.
     \end{split}
\end{equation}

So in the spin $\mathbf{2}$ case the superfield equations \eqref{EOM with currents} take the form:
\begin{equation}\label{EOM-spin 2}
    \begin{cases}
        (D^+)^4 G^{--\alpha \dot{\alpha} }=  \frac{1}{2}  \kappa_2 J^{++\alpha \dot{\alpha}}\,,
        \\
        (D^+)^4 G^{--5 } = \frac{1}{2}  \kappa_2 J^{++}\,,
        \\
        (\bar{D}^+)^2 D^{+\alpha} G^{--}_{\alpha \dot{\alpha}}
        -
        (D^+)^2 \bar{D}^+_{\dot{\alpha}} G^{--5}
        = 2i \kappa_2  \mathcal{J}^+_{ \dot{\alpha}}\,,
        \\
        (D^+)^2  \bar{D}^{+\dot{\alpha}} G^{--}_{\alpha \dot{\alpha}}
        +
        (\bar{D}^+)^2 D^+_{\alpha} G^{--5}
        =
        2i \kappa_2 \mathcal{J}^+_{\alpha },
    \end{cases}
\end{equation}
where
\begin{equation}
    ({\rm a})\;    \mathcal{J}^+_{\alpha} = J^+_\alpha - 2i \bar{\theta}^{-\dot{\mu}} J^{++}_{\alpha\dot{\mu}}
    +
    2i \theta^-_\alpha J^{++}, \quad ({\rm b})\; \mathcal{J}^+_{\dot{\alpha}}
    =
    J^+_{\dot{\alpha}}
    +
    2i \theta^{-\mu} J^{++}_{\mu\dot{\alpha}}
    +
    2i \bar{\theta}^-_{\dot{\alpha}} J^{++}.
\end{equation}}

Note that  all these  current superfields can be obtained  as proper projections of the single non-analytic ``master''  current superfield
$\mathcal{J}$ introduced (in the superconformal case) by S. Kuzenko
and S. Theisen in \cite{Kuzenko:1999pi}\footnote{ We are grateful to Sergei Kuzenko for bringing this interesting property to our attention.}:
\begin{equation}
    \begin{split}
    &\mathcal{J} := - \frac{1}{2} q^{+a} \mathcal{D}^{--} q^+_a,
    \qquad
    \mathcal{D}^{++} \mathcal{J} = 0.
    \end{split}
\end{equation}
Indeed, using the definitions \eqref{D-} and the analyticity of $q^{+ a}$, we obtain
\begin{equation}
   \mathcal{J}^+_\alpha =  D^+_\alpha \mathcal{J}\,,
    \qquad
    \mathcal{J}^+_{\dot{\alpha}} = -\bar{D}^+_{\dot{\alpha}} \mathcal{J}\,,
\end{equation}
\begin{equation}
   J^{++}_{\alpha\dot{\alpha}} = -\frac{i}{2} D^{+}_\alpha \bar{D}^+_{\dot{\alpha}} \mathcal{J}\,, \qquad
    J^{++} = \frac{i}{4} (D^+)^2 \mathcal{J}
    =
      - \frac{i}{4} (\bar D^+)^2 \mathcal{J}\,.
\end{equation}
The geometric meaning of $\mathcal{J}$ becomes clear in the superconformal case where it appears as a hypermultiplet current
superfield associated with the additional analytic gauge potential $H^{+4}$ \cite{Galperin:1987ek}.

All the superfield currents in eqs.  \eqref{EOM-spin 2} are killed by the harmonic derivative ${\cal D}^{++}$ on the hypermultiplet mass shell.
The first equation should also be compatible with the additional conservation condition $\partial_{\alpha\dot{\alpha}} J^{++\alpha\dot{\alpha}} =0$.
In contrast to the conservations law $\mathcal{D}^{++}J^{++\alpha\dot{\alpha}} = 0$, which is completely dictated by the structure of the left hand side
of this equation (it follows from the zero curvature conditions and the analyticity of the basic gauge potentials),
the extra condition just mentioned is satisfied only after elimination of the auxiliary fields in $G^{--\alpha \dot{\alpha} }$ in terms of the hypermultiplet fields
and taking into account the equations of motion for the latter (see eqs. \eqref{mathcal P} - \eqref{spin 23}).

Now we turn to the component analysis of the current superfields defined above.\\

\textbf{1. }For $s=2$ the current superfield \eqref{Supercurrent1} is written as:
\begin{equation}\label{Supercurrent1-spin2}
    J^{++}_{\alpha\dot{\alpha}}
    =
    - \frac{1}{2}
    q^{+a}  \partial_{\alpha\dot{\alpha}}  q^+_a
    =
    -
    \frac{1}{2} \tilde{q}^+ \partial_{\alpha\dot{\alpha}} q^+
    +
    \frac{1}{2} q^+ \partial_{\alpha\dot{\alpha}} \tilde{q}^+
    \,.
\end{equation}

In this case, as was already mentioned, besides the generic conservation law $\mathcal{D}^{++}J^{++}_{\alpha\dot{\alpha}}=0\,,$ there is one more conservation condition:
\begin{equation}\label{Another conservation}
    \partial^{\alpha\dot{\alpha}} J^{++}_{\alpha\dot{\alpha}}
    =
    -
    \frac{1}{2} \partial^{\alpha\dot{\alpha}} \tilde{q}^+ \partial_{\alpha\dot{\alpha}} q^+
    +
    \frac{1}{2} \partial^{\alpha\dot{\alpha}} q^+ \partial_{\alpha\dot{\alpha}} \tilde{q}^+
    -
     \tilde{q}^+ \Box q^+
    +
     q^+  \Box \tilde{q}^+
     =
     0\,,
\end{equation}
where we made use of the superfield Klein-Gordon equation \eqref{superfield-KG} for the hypermultiplet superfield.
This property indicates that the supercurrent $J^{++}_{\alpha\dot{\alpha}}$ is special. Indeed, all components of $J^{++}_{\alpha\dot{\alpha}}$
are conserved currents. We shall see that this superfield contains the conserved component currents which by Noether theorem are related to $\mathcal{N}=2$
global supersymmetry of the free hypermultiplet action \eqref{hyper}. Keeping this in mind,  we shall use the term ``\textit{supercurrent}'' for the current
superfield $J^{++}_{\alpha\dot{\alpha}}$. For the discussion of the general structure of $\mathcal{N}=2$ supercurrent see Appendix \ref{supercurrent}.

The supercurrent \eqref{Supercurrent1-spin2} is real:
\begin{equation}
    \widetilde{J^{++}_{\alpha\dot{\alpha}} } = J^{++}_{\alpha\dot{\alpha}}\,,
\end{equation}
due to the relation $\widetilde{(\tilde{q}^+)} = -q^+$. Using the on-shell $\theta$-expansion of the hypermultiplet, one can explicitly write down
its component currents in accordance with the general $\theta$ expansion in \eqref{supercurrent - superfield}:
\begin{equation}
    \begin{split}
    J^{++}_{\alpha\dot{\alpha}} =& R^{(ij)}_{\alpha\dot{\alpha}} u^+_i u^+_j
    +
    \theta^{+\beta} S^i_{\alpha\dot{\alpha} \beta} u^+_i
    +
    \bar{\theta}^{+\dot{\beta}} \bar{S}^i_{\alpha\dot{\alpha}\dot{\beta}} u^+_i
    +
    (\theta^+)^2 R_{\alpha\dot{\alpha}}
    +
    (\bar{\theta}^+)^2 \bar{R}_{\alpha\dot{\alpha}}
    \\&+
    2i \theta^+ \sigma^n \bar{\theta}^+ \sigma^m_{\alpha\dot{\alpha}} \left[T_{mn} + \partial_n R_{m}^{(ij)} u^+_{i} u^-_{j}\right]
    +
    \dots\,.
        \end{split}
\end{equation}
We obtain the following representation for the component currents in terms of the hypermultiplet components:

\begin{itemize}
    \item ${\rm SU}(2)$ $R$ symmetry current:
    \begin{equation}\label{R-symmetry current}
        R^{(ij)}_{\alpha\dot{\alpha}} = \frac{1}{2} \left(\bar{f}^{(i} \partial_{\alpha\dot{\alpha}} f^{j)} - f^{(i} \partial_{\alpha\dot{\alpha}} \bar{f}^{j)} \right),
    \end{equation}
    \begin{equation}
        \partial^{\alpha\dot{\alpha}} R^{(ij)}_{\alpha\dot{\alpha}} = 0\,;
    \end{equation}

    \item Fermionic current of $\mathcal{N}=2$ supersymmetry:
    \begin{equation}\label{S current}
        S^{\alpha\dot{\alpha}}_{i\mu}
        =
        \frac{1}{2}
        \left( \partial^{\alpha\dot{\alpha}} \psi_\mu \bar{f}_i - \psi_\mu \partial^{\alpha\dot{\alpha}}\bar{f}_i\right)
        -
         \frac{1}{2} \left(   \partial^{\alpha\dot{\alpha}} \chi_\mu f_i - \chi_\mu \partial^{\alpha\dot{\alpha}} f_i \right),
    \end{equation}
    \begin{equation}\label{ConsS}
        \partial_{\alpha\dot{\alpha}}S^{\alpha\dot{\alpha}}_{i\mu} = 0\,;
    \end{equation}
Modulo terms proportional to the fermionic equations of motion \eqref{hyper-equations}, this
current coincides with Noether $\mathcal{N}=2$ fermionic current \eqref{N=2 supercurrents}.

    \item The 
    current
    \begin{equation}
        R_{\alpha\dot{\alpha}} = \frac{1}{4} \left(\psi^\gamma \partial_{\alpha\dot{\alpha}} \chi_\gamma - \chi^\gamma \partial_{\alpha\dot{\alpha}} \psi_\gamma\right)\,,
        \qquad \partial^{\alpha\dot{\alpha}}R_{\alpha\dot{\alpha}} = 0\,,
    \end{equation}
    corresponds to the following rigid symmetry of the hypermultiplet action \eqref{hyper on shell}:
    \begin{equation}\label{Rsymm}
        \delta \psi_{\dot{\alpha}} = \lambda \partial_{\dot{\alpha}\beta} \chi^\beta\,,\qquad
        \delta \bar{\chi}_{\dot{\alpha}} = -\lambda \partial_{\dot{\alpha}\beta} \psi^\beta\,.
    \end{equation}
    On shell and after redefining the parameter $\lambda$ (for $m\neq 0$), these transformations are reduced to ${\rm U}(1)$ R-symmetry. Note that the physical component hypermultiplet action \eqref{hyper on shell} is invariant under
\eqref{Rsymm} off shell, in both $m=0$ and $m\neq 0$ cases.

    \item The energy-momentum tensor:
\begin{equation}\label{EM}
    \begin{split}
    T_{nm} =& \frac{1}{2} \partial_{(n} f^i \partial_{m)} \bar{f}_i
    - \frac{1}{4} \left( \bar{f}_i \partial_n \partial_m f^i + f^i \partial_m \partial_n \bar{f}_i \right)
    \\&
    -
    \frac{i}{8} \sigma_n^{\dot{\beta}\beta}
    \left(
    \chi_\beta \partial_m \bar{\chi}_{\dot{\beta}} +
    \bar{\chi}_{\dot{\beta}} \partial_m \chi_\beta
    +
    \psi_\beta \partial_m \bar{\psi}_{\dot{\beta}} +
    \bar{\psi}_{\dot{\beta}} \partial_m \psi_\beta
     \right).
    \end{split}
\end{equation}
\end{itemize}

The expression \eqref{EM} does not coincide with the canonical energy momentum tensor. Note that $T_{nm}$ is mass-independent.
On the hypermultiplet equations of motion \eqref{hyper-equations} it is conserved in the standard way:
\begin{equation}
    \partial^n T_{nm} = 0\,.
\end{equation}
The bosonic part of \eqref{EM},  $T^{(bos)}_{(nm)}$, differs from the analogous part of the canonical  energy momentum tensor $\mathcal{T}^{(bos)}_{(nm)}$:
\begin{equation}
    \mathcal{T}^{(bos)}_{(nm)} = \partial_{(n} f^i \partial_{m)} \bar{f}_i
    -
    \frac{1}{2} \eta_{nm} \left(\partial^k \bar{f}^i \partial_k f_i + m^2 \bar{f}^i f_i \right).
\end{equation}
The canonical energy momentum tensor $\mathcal{T}^{(bos)}_{mn}$ is mass-dependent.
Let us consider the difference between these two expressions:
\begin{equation}
    \begin{split}
    T^{(bos)}_{(nm)} - \mathcal{T}^{(bos)}_{(nm)}
    =&
    \frac{1}{2} \eta_{nm} \left(\partial^k \bar{f}^i \partial_k f_i + m^2 \bar{f}^i f_i \right)
    \\&-
    \frac{1}{4} \left( \bar{f}_i \partial_n \partial_m f^i + f^i \partial_m \partial_n \bar{f}_i \right)
    -
    \frac{1}{2} \partial_{(n} f^i \partial_{m)} \bar{f}_i\,.
        \end{split}
\end{equation}
On shell it can be rewritten as:
\begin{equation}\label{TrivBosTens}
    T^{(bos)}_{(nm)} - \mathcal{T}^{(bos)}_{(nm)}
    =
    \frac{1}{4} \left(\eta_{nm}\Box - \partial_n \partial_m\right) \left(\bar{f}^i f_i\right).
\end{equation}
So $T_{nm}$ differs from the canonical energy momentum tensor by
the trivially conserved ``current''.

The symmetric part of the fermionic term of the tensor \eqref{EM} coincides with the canonical fermionic energy momentum tensor. The antisymmetric
part reads:
\begin{equation}
    \mathcal{T}_{[nm]}^{(ferm)}
    =
    -
    \frac{i}{4} \sigma_{[n}^{\dot{\beta}\beta}
    \sigma_{m]}^{\rho\dot{\rho}}
    \left(
    \chi_\beta \partial_{\rho\dot{\rho}} \bar{\chi}_{\dot{\beta}} +
    \bar{\chi}_{\dot{\beta}} \partial_{\rho\dot{\rho}} \chi_\beta
    +
    \psi_\beta \partial_{\rho\dot{\rho}} \bar{\psi}_{\dot{\beta}} +
    \bar{\psi}_{\dot{\beta}} \partial_{\rho\dot{\rho}} \psi_\beta
    \right)\,.
\end{equation}
Using the relation:
\begin{equation}
    \sigma_{[n}^{\dot{\beta}\beta}
    \sigma_{m]}^{\rho\dot{\rho}}
    \sim
    \sigma^{(\beta\rho)}_{[nm]}\epsilon^{\dot{\beta}\dot{\rho}}
    +
    \tilde{\sigma}^{(\dot{\beta}\dot{\rho})}_{[nm]}
    \epsilon^{\beta\rho}
\end{equation}
we have on shell:
\begin{equation}
    \begin{split}
        \mathcal{T}_{[nm]}^{(ferm)}
        &\sim
        \sigma^{(\beta\rho)}_{[nm]}
        \left(
        \bar{\chi}^{\dot{\rho}} \partial_{\rho\dot{\rho}} \chi_\beta
        +
            \bar{\psi}^{\dot{\rho}} \partial_{\rho\dot{\rho}} \psi_\beta
        \right)
        +           \tilde{\sigma}^{(\dot{\beta}\dot{\rho})}_{[nm]}
        \left(
        \chi^\rho \partial_{\rho\dot{\rho}} \bar{\chi}_{\dot{\beta}}
        +
        \psi^\rho \partial_{\rho\dot{\rho}} \bar{\psi}_{\dot{\beta}}
        \right)
        \\&=
        \sigma^{(\beta\rho)}_{[nm]}
        \partial_{\rho\dot{\rho}}
        \left(
        \bar{\chi}^{\dot{\rho}}  \chi_\beta
        +
        \bar{\psi}^{\dot{\rho}}  \psi_\beta
        \right)
        +           \tilde{\sigma}^{(\dot{\beta}\dot{\rho})}_{[nm]}
         \partial_{\rho\dot{\rho}}
        \left(
       \chi^\rho \bar{\chi}_{\dot{\beta}}
        +
        \psi^\rho  \bar{\psi}_{\dot{\beta}}
        \right).
    \end{split}
\end{equation}
Exploiting further the relations:
\begin{subequations}
\begin{equation}
     (\tilde{\sigma}^{ns})^{\dot{\alpha}}_{\;\;\dot{\beta}}
     \sigma^m_{\alpha \dot{\alpha}}
    =
    -i (\eta^{ms}\sigma^n - \eta^{mn}\sigma^s)_{\alpha\dot{\beta}} + \epsilon^{nsmt}(\sigma_t)_{\alpha\dot{\beta}}\,,
\end{equation}
\begin{equation}
    (\sigma^{ns})_\alpha^{\;\;\beta} \sigma^m_{\beta\dot{\beta}}
    =
    i(\eta^{sm}\sigma^n - \eta^{mn}\sigma^s)_{\alpha\dot{\beta}} + \epsilon^{nsmt}(\sigma_t)_{\alpha\dot{\beta}}
\end{equation}
\end{subequations}
we finally obtain:
\begin{equation}
T_{[mn]}^{(ferm)}
\sim
\epsilon_{nmkt} \partial^k (\sigma^t)^\beta_{\dot{\rho}} \left(
\bar{\chi}^{\dot{\rho}}  \chi_\beta
+
\bar{\psi}^{\dot{\rho}}  \psi_\beta
\right)\,.
\end{equation}
So the on-shell antisymmetric part of the fermionic term in \eqref{EM} is conserved kinematically, without using equations of motion. It looks analogous to the trivially conserved bosonic
tensor \eqref{TrivBosTens}.

 We wish to point out once more that the spin ${\bf 2}$ supercurrent is distinguished among other hypermultiplet superfield currents in that, at the level of free fields,
it is just the hypermultiplet part of the full  supercurrent associated with the rigid ${\cal N}=2$ supersymmetry. This supercurrent contains
as well a part coming from the linearized ${\cal N}=2$ gauge superfield action\footnote{This second part in general involves contributions from all higher spin gauge superfield actions.}.
The rigid symmetry transformations of $q^{+}$ which are gauged in this case coincide in fact
with rigid ${\cal N}=2$ supersymmetry transformations (translations and supertranslations) and with those of $R$ symmetry, and they are same as rigid symmetries
of the gauge superfield actions. All other rigid higher spin symmetries
realized on $q^{+}$ (including the spin ${\bf 1}$ ones) have no realizations on the gauge superfields at the linearized level, so the relevant superfield currents
get contributions only from the hypermultiplet.

\medskip

\textbf{2.} In the ${\bf s}=2$ case, due to the relation \eqref{partial 5 and mass}, the current superfield \eqref{Supercurrent2} is non-vanishing only for $m\neq 0$:
 \begin{equation}\label{u(1) current spin 2}
    J^{++} =
    -
    \frac{1}{2} q^{+a} \partial_5 q^+_a
    =
    -\frac{m}{2} q^{+a} J q^+_a
    =
    im \tilde{q}^+ q^+\,.
\end{equation}
Up to the multiplier $(-m)$ it coincides with the spin $\mathbf{1}$ current superfield \eqref{electric current} introduced in Section \ref{sec:cubic Maxwell}.
The corresponding conserved current \eqref{el current} is ${\rm U}(1)$ PG-symmetry (or central charge) current of free hypermultiplet.

\medskip

\textbf{3.} The fermionic current superfield \eqref{Supercurrent4} for $s=\mathbf{2}$ case takes the form
\begin{equation}
        J^+_{\alpha} = -\frac{1}{2} q^{+a} \partial^-_{\alpha} q^+_a
        =
        - \frac{1}{2} \tilde{q}^+ \partial^-_{\alpha} q^+
        +
        \frac{1}{2} q^+ \partial^-_{\alpha} \tilde{q}^+\,.
\end{equation}
Since it satisfies the modified conservation law,
\begin{equation}\label{spin 2 conservation}
    \mathcal{D}^{++} J^+_{\alpha}
    =
    2i \bar{\theta}^{+\dot{\rho}}J^{++}_{\alpha\dot{\rho}} - 2i \theta^+_{\alpha} J^{++}\,,
\end{equation}
there are no conserved currents among its components. This is in agreement with  the fact that there are no physical fields
among the original components of the spinor potential $h^{++\alpha+}$
(they appear only after redefinitions of the auxiliary fields).

According to the general discussion in Appendix \ref{N=2 current superfield}, the conserved fermionic current defined by eq. \eqref{spin 2 conservation}
is given by the sum:
\begin{equation}
    \mathfrak{j}^i_{\rho\dot{\rho}\alpha}
    :=
    j^i_{\rho\dot{\rho}\alpha}
    -
    S^i_{\alpha\dot{\rho}\rho}
    +
    m
    \epsilon_{\rho\alpha} \bar{j}_{\dot{\rho}}^i\,.
\end{equation}
Here, $j^{i}_{\rho\dot{\rho}\sigma}$ comes from $J^+_{\alpha}$:
\begin{equation}
    J^+_{\alpha} = \dots + 2i \theta^{+\rho}\bar{\theta}^{+\dot{\rho}} j^i_{\rho\dot{\rho}\alpha} u^-_i
    +
    \dots\,,
\end{equation}
and it is expressed in terms of the on-shell hypermultiplet component fields (defined in \eqref{hyper q} and \eqref{hyper q tilde}) as:
\begin{equation}
    j^i_{\rho\dot{\rho} \alpha} =
    \psi_{(\alpha} \partial_{\rho)\dot{\rho}} \bar{f}^i
    -
    \chi_{(\alpha} \partial_{\rho)\dot{\rho}} f^i
    -
    \frac{1}{2} im \epsilon_{\rho\alpha} \left(\bar{\chi}_{\dot{\rho}}\bar{f}^i - \bar{\psi}_{\dot{\rho}} f^i\right).
\end{equation}
The current $S^i_{\alpha\dot{\rho}\rho}$ was defined in \eqref{S current} and $\bar{j}^i_{\dot{\rho}}$ in \eqref{pre S current}.

Recall the on-shell conservation law $\partial^{\alpha\dot{\rho}} S^i_{\alpha\dot{\rho}\rho}=0$ (eq. \eqref{ConsS}).

The final expression for the total fermionic current $\mathfrak{j}^i_{\rho\dot{\rho}\alpha}$ is:
\begin{equation}\label{j susy current}
    \begin{split}
    \mathfrak{j}^i_{\rho\dot{\rho}\alpha}
    =&\;\;\;\;
    \frac{1}{2}
    \left(
    \psi_\alpha \partial_{\rho\dot{\rho}} \bar{f}^i + 2 \psi_{\rho} \partial_{\alpha\dot{\rho}} \bar{f}^i
    -
    \partial_{\alpha\dot{\rho}} \psi_\rho \bar{f}^i
    \right)
    \\
    &
    -
    \frac{1}{2}
    \left(\chi_\alpha \partial_{\rho\dot{\rho}} f^i + 2\chi_\rho \partial_{\alpha\dot{\rho}} f^i
    - \partial_{\alpha\dot{\rho}}\chi_\rho f^i
    \right)
    \\
    &-
    \frac{3}{2} im \epsilon_{\rho\alpha} \left(\bar{\chi}_{\dot{\rho}}\bar{f}^i - \bar{\psi}_{\dot{\rho}} f^i\right).
    \end{split}
\end{equation}
Using the free hypermultiplet equations of motion \eqref{hyper-equations} one can check that this current
is conserved, $\partial^{\rho\dot{\rho}}\mathfrak{j}^i_{\rho\dot{\rho}\alpha} = 0$, in agreement with the general result \eqref{modefied conservation}.

\medskip

Note that the current constructed in this way does not coincide with the $\mathcal{N}=2$ supersymmetry fermionic current defined in eq. \eqref{N=2 supercurrents}. Let us
rewrite \eqref{j susy current} using the equations of motion for physical fields and the identities:
\begin{subequations}
    \begin{equation}
        \partial_{\rho\dot{\rho}} \psi_\alpha
        -
        \partial_{\alpha\dot{\rho}} \psi_\rho
        =
        \epsilon_{\rho\alpha} \partial_{\dot{\rho}}^\sigma\psi_\sigma
        =
        im \epsilon_{\rho\alpha} \bar{\chi}_{\dot{\rho}}\,,
    \end{equation}
    \begin{equation}
    \partial_{\rho\dot{\rho}} \chi_\alpha
    -
    \partial_{\alpha\dot{\rho}} \chi_\rho
    =
    \epsilon_{\rho\alpha} \partial_{\dot{\rho}}^\sigma\chi_\sigma
    =
    im \epsilon_{\rho\alpha} \bar{\psi}_{\dot{\rho}}
\end{equation}
\end{subequations}
We obtain
\begin{equation}\label{j current spin 32}
    \mathfrak{j}^i_{\rho\dot{\rho}\alpha}
    =
    -
    2S^i_{\rho\dot{\rho}\alpha}
    +
    \partial_{\alpha\dot{\rho}}
    \left(
     \psi_\rho \bar{f}^i
     -\chi_\rho  f^i \right)
    -
   \frac{1}{2} \partial_{\rho\dot{\rho}}   \left(
   \psi_\alpha \bar{f}^i
   -\chi_\alpha  f^i \right).
\end{equation}
So the current \eqref{j susy current} differs from Noether current for $\mathcal{N}=2$ supersymmetry by the expression (two last terms) which is conserved automatically,
without employing the equations of motion.

\medskip

There is also another way to rewrite the conserved $\mathcal{N}=2$ supersymmetry current of hypermultiplet:
\begin{equation}\label{S cal current}
    \mathcal{S}^i_{\rho\dot{\rho}\alpha} = \psi_\rho \partial_{\alpha\dot{\rho}} \bar{f}^i
    -
     \chi_\rho \partial_{\alpha\dot{\rho}} f^i
    -
    im \epsilon_{\rho\alpha} \left( \bar{\chi}_{\dot{\rho}} \bar{f}^i - \bar{\psi}_{\dot{\rho}} f^i \right),
    \qquad
    \partial^{\rho\dot{\rho}} \mathcal{S}^i_{\rho\dot{\rho}\alpha}  = 0.
\end{equation}
This current is related to $S^i_{\rho\dot{\rho}\alpha}$ and $\mathfrak{j}^i_{\rho\dot{\rho}\alpha}$ currents as:
\begin{equation}
    \mathcal{S}^i_{\rho\dot{\rho}\alpha} =
    \mathfrak{j}^i_{\rho\dot{\rho}\alpha}
    +
    S^i_{\rho\dot{\rho}\alpha}.
\end{equation}
Note that all the currents considered coincide with each other up to numerical factors and the trivially conserved terms.
Just the current $   \mathcal{S}^i_{\rho\dot{\rho}\alpha}$ current appears in the cubic $(\tfrac{3}{2}, \tfrac{1}{2}, 0)$ vertex in Section \ref{spin 2 coupling}.

\subsection{Component reduction of superfield equations of motions with source in the spin \textbf{2} case}
\label{spin 2 equations with matter}

Here we consider the modification of equations of motion for the fields $\Phi_{\rho\dot{\rho}}^{\alpha\dot{\alpha}}$, $\mathcal{P}^{\alpha\dot{\alpha}}$
and $\bar{\mathcal{P}}^{\alpha\dot{\alpha}}$ as compared to those found in the spin $\mathbf{2}$ case without taking account of sources. We shall see that the  auxiliary
$\mathcal{P}$-fields $\mathcal{P}^{\alpha\dot{\alpha}}$
and $\bar{\mathcal{P}}^{\alpha\dot{\alpha}}$, unlike the free case,  play a non-trivial role in getting the correct modified equations of motion
for the spin 2 fields \footnote{Note that in the first order in $\kappa$ the auxiliary fields do not contribute to the component Lagrangian
even in the interacting theory, as will be explained in the next Section.
    The reason is that the auxiliary fields $\mathcal{P}$ are vanishing on the free equations of motion. So in the interacting theory their contribution
    starts with the order $\sim \kappa^2$.}.

The relevant parts of the spin 2 current superfields (in the  bosonic sector) read:
\begin{subequations}
    \begin{equation}
        J^{++}_{\alpha\dot{\alpha}} = \dots - \frac{i}{2} \theta^{+\rho} \bar{\theta}^{+\dot{\rho}} \left(\partial_{\alpha\dot{\alpha}}\partial_{\rho\dot{\rho}}f^i \bar{f}_i - \partial_{\alpha\dot{\alpha}}f^i \partial_{\rho\dot{\rho}}\bar{f}_i
        - \partial_{\rho\dot{\rho}} f^i \partial_{\alpha\dot{\alpha}} \bar{f}_i
         + f^i \partial_{\alpha\dot{\alpha}}\partial_{\rho\dot{\rho}}\bar{f}_i \right)  + \dots\,,
    \end{equation}
    \begin{equation}
        J^+_\alpha = \dots - \frac{i}{2} \bar{\theta}^{+\dot{\alpha}} \left(\partial_{\alpha\dot{\alpha}}f^i\bar{f}_i  + f^i \partial_{\alpha\dot{\alpha}} \bar{f}_i \right) + \dots\,,
    \end{equation}
    \begin{equation}
        J^+_{\dot{\alpha}} = \dots +\frac{i}{2} \theta^{+\alpha} \left(f^i\partial_{\alpha\dot{\alpha}} \bar{f}_i + \partial_{\alpha\dot{\alpha}} f^i \bar{f}_i \right) + \dots\,.
    \end{equation}
\end{subequations}

So in the presence of sources the free equations from Section \ref{spin 2 equations} are  modified as:

\begin{equation}\label{mathcal P}
    \begin{cases}
        8 (\mathcal{P}_{\dot{\alpha}}^\mu - 3 \bar{\mathcal{P}}_{\dot{\alpha}}^\mu ) =  \kappa_2 \partial^\mu_{\dot{\alpha}} (f^i\bar{f}_i)\\
        8 ( \bar{\mathcal{P}}^{\dot{\rho}}_\alpha - 3 \mathcal{P}^{\dot{\rho}}_\alpha ) = \kappa_2 \partial_{\alpha}^{\dot{\rho}} (f^i\bar{f}_i)
    \end{cases}
    \quad \Rightarrow \quad \mathcal{P}_{\dot{\alpha}}^\mu = \bar{\mathcal{P}}_{\dot{\alpha}}^\mu  = -\frac{\kappa_2}{16} \partial_{\dot{\alpha}}^\mu \left(f^i \bar{f}_i\right).
\end{equation}
As a consequence, there appear important additional terms in the linearized Einstein equations:
\begin{eqnarray}
    && \mathcal{G}_{\rho\dot{\rho} \alpha\dot{\alpha}}^{}
    -
    \frac{1}{2} \partial^{}_{\rho\dot{\alpha}} \mathcal{P}_{\alpha\dot{\rho}}
    -
    \frac{1}{2} \partial_{\alpha\dot{\rho}} \bar{\mathcal{P}}^{}_{\rho\dot{\alpha}}
    -
    \frac{1}{2} \epsilon^{}_{\dot{\alpha}\dot{\rho}} \partial^{\dot{\mu}}_\rho \mathcal{P}_{\alpha\dot{\mu}}
    -
    \frac{1}{2} \epsilon_{\alpha\rho} \partial^\mu_{\dot{\rho}} \bar{\mathcal{P}}_{\mu\dot{\alpha}} \nonumber \\
    && = \, \frac{\kappa_2}{16}  \left(\partial_{\alpha\dot{\alpha}}\partial_{\rho\dot{\rho}}f^i \bar{f}_i
    -
     \partial_{\alpha\dot{\alpha}}f^i \partial_{\rho\dot{\rho}}\bar{f}_i
    -
   \partial_{\rho\dot{\rho}}   f^i  \partial_{\alpha\dot{\alpha}}  \bar{f}_i
    + f^i \partial_{\alpha\dot{\alpha}}\partial_{\rho\dot{\rho}}\bar{f}_i \right).
\end{eqnarray}
In this equation we made use of symmetry under the index permutation $(\alpha\rho)(\dot{\alpha}\dot{\rho}) \leftrightarrow (\rho\alpha)(\dot{\rho}\dot{\alpha})$.
Using eq. \eqref{mathcal P} and permuting the indices as $(\alpha\rho)(\dot{\alpha}\dot{\rho}) \leftrightarrow (\rho\alpha)(\dot{\rho}\dot{\alpha})$,  we obtain:
\begin{equation}\label{spin 23}
    \begin{split}
        \mathcal{G}_{\rho\dot{\rho}\alpha\dot{\alpha}} =&
        \, \frac{\kappa_2}{16}  \left(\partial_{\alpha\dot{\alpha}}\partial_{\rho\dot{\rho}}f^i \bar{f}_i
        -
         \partial_{\alpha\dot{\alpha}}f^i \partial_{\rho\dot{\rho}}\bar{f}_i
        -
        \partial_{\rho\dot{\rho}} f^i    \partial_{\alpha\dot{\alpha}} \bar{f}_i
        +
         f^i \partial_{\alpha\dot{\alpha}}\partial_{\rho\dot{\rho}}\bar{f}_i \right)
        \\&
        -
        \frac{\kappa_2}{16} \left(\partial_{\alpha\dot{\rho}} \partial_{\rho\dot{\alpha}} - \epsilon_{\alpha\rho} \epsilon_{\dot{\alpha}\dot{\rho}} \Box \right) \left(f^i \bar{f}_i\right).
    \end{split}
\end{equation}
This is just the linearized Einstein equations with a source. Thus we have shown that the auxiliary fields have a non-trivial impact on the eventual form
of the physical equations. In particular, the energy-momentum tensor
that stands in the right hand side of eq. \eqref{spin 23} differs from the tensor \eqref{EM} entering
$\mathcal{N}=2$ supercurrent \eqref{Supercurrent1-spin2}.
A similar situation takes place  for other higher spins,
but we will not present here the corresponding technical details.

Also, eq. \eqref{spin 23} ensures the on-shell compatibility of the first equation in the set \eqref{EOM-spin 2} (at least, in its bosonic sector)  with the source conservation
law $\partial_{\alpha\dot{\alpha}} J^{++\alpha\dot{\alpha}} = 0$  fulfilled on the free hypermultiplet
equations of motion.

Relevant to the consideration in this Section is also the component reduction of coupling in Section \ref{spin 2 coupling}, where we analyze the effect of elimination
of the hypermultiplet auxiliary fields and discuss the structure of the on-shell cubic vertex Lagrangian.

\subsection{Spin \textbf{3}}
\label{spin 3 supercurrent}

The spin  \textbf{3} theory is the first non-trivial new example.
Also the structure of the superfield currents  in the spin $\mathbf{3}$ case is of particular interest,
since it helps to clarify the role of the $\xi$-interaction.

Varying the cubic $(\mathbf{3}, \mathbf{\tfrac{1}{2}}, \mathbf{\tfrac{1}{2}})$ coupling with respect to the unconstrained analytical $\mathcal{N}=2$ spin $\mathbf{3}$
potentials $$h^{++(\alpha\beta)(\dot{\alpha}\dot{\beta})}\,,\quad h^{++5\alpha\dot{\alpha}}\,,\quad h^{++(\alpha\beta)\dot{\alpha}+}\,,\quad h^{++(\dot{\alpha}\dot{\beta})\alpha+}$$
yields the analytic current superfields:
\begin{subequations}\label{s=3 super current}
\begin{equation}
    J^{++}_{(\alpha\beta)(\dot{\alpha}\dot{\beta})}
    =
    - \frac{1}{2}
    q^{+a}  \partial_{(\alpha \dot{\alpha}} \partial_{\beta) \dot{\beta}} J q^+_a
    -
    \frac{1}{2} \xi  \,\partial_{(\alpha \dot{\alpha}} \partial_{\beta) \dot{\beta}} \left(q^{+a}  J  q^+_a \right),
\end{equation}
\begin{equation}
      J^{++}_{\alpha\dot{\alpha}}
    =
    -
    \frac{1}{2}
    q^{+a}\partial_{\alpha\dot{\alpha}} J \partial_5 q^+_a,
\end{equation}
\begin{equation}
    J^+_{(\alpha\beta)\dot{\alpha}} = -
    \frac{1}{2} q^{+a}\partial_{(\alpha\dot{\alpha}} \partial^-_{\beta)} J q^+_a - \frac{1}{2} \xi \partial_{(\alpha\dot{\alpha}} \partial^-_{\beta)} \left( q^{+a} J q^+_a \right),
\end{equation}
\begin{equation}
    J^+_{(\dot{\alpha}\dot{\beta})\alpha} = -
    \frac{1}{2} q^{+a}\partial_{\alpha(\dot{\alpha}} \partial^-_{\dot{\beta})} J q^+_a - \frac{1}{2} \xi \partial_{\alpha(\dot{\alpha}} \partial^-_{\dot{\beta})} \left( q^{+a} J q^+_a \right).
\end{equation}
\end{subequations}

Like in the ${\bf s}=2$ case, these current superfields can be related to a single ``master'' current superfield, this time the vector one,
\begin{equation}\label{s=3 precurrent}
    \mathcal{J}_{\alpha\dot{\alpha}} :=
    - \frac{1}{2} q^{+a} \mathcal{D}^{--} \partial_{\alpha\dot{\alpha}} J q^+_a
    - \frac{1}{2} \xi \,\mathcal{D}^{--} \partial_{\alpha\dot{\alpha}} \left( q^{+a}  J q^+_a \right),
\end{equation}
obeying, on the hypermultiplet mass shell,  the conservation law:
\begin{equation}\label{s=3 conservation law}
        \mathcal{D}^{++}    \mathcal{J}_{\alpha\dot{\alpha}} =  -\frac{1}{2} q^{+a} \partial_{\alpha\dot{\alpha}} J q^+_a - \xi \,
        \partial_{\alpha\dot{\alpha}}\left( q^{+a}  J q^+_a \right).
\end{equation}
We find
\begin{equation}
   \mathcal{J}^+_{(\alpha\beta)\dot{\alpha}} =  D^+_{(\alpha} \mathcal{J}_{\beta)\dot{\alpha}}\, ,
    \qquad
    \mathcal{J}^+_{(\dot{\alpha}\dot{\beta})\alpha} = -\bar{D}^+_{(\dot{\alpha}} \mathcal{J}_{\dot{\beta})\alpha}\,,
\end{equation}
\begin{equation}
    J^{++}_{(\alpha\beta)(\dot{\alpha}\dot{\beta})} = -\frac{i}{2} D^{+}_{(\alpha} \bar{D}^+_{(\dot{\alpha}} \mathcal{J}_{\beta)\dot{\beta})}\,,
\qquad J_{\alpha\dot{\alpha}}^{++} = \frac{i}{4}(D^+)^2 \mathcal{J}_{\alpha\dot{\alpha}}
    =
    -\frac{i}{4}(\bar D^+)^2 \mathcal{J}_{\alpha\dot{\alpha}}\,,
\end{equation}
where
\begin{subequations}
    \begin{equation}
        \mathcal{J}^+_{(\dot{\alpha}\dot{\beta})\alpha }
        =
        J^+_{(\dot{\alpha}\dot{\beta})\alpha }
        +
        2i \theta^{-\mu} J^{++}_{(\alpha\mu)(\dot{\alpha}\dot{\beta})}
        +
        2i \bar{\theta}^-_{(\dot{\alpha}}J^{++}_{\alpha\dot{\beta})},
    \end{equation}
    \begin{equation}
        \mathcal{J}^+_{(\alpha\beta) \dot{\alpha}}
        =
        J^+_{(\alpha\beta) \dot{\alpha}}
        -
        2i \bar{\theta}^{-\dot{\mu}} J^{++}_{(\alpha\beta)(\dot{\alpha}\dot{\mu})}
        +
        2i \theta^-_{(\alpha} J^{++}_{\beta)\dot{\alpha}}.
    \end{equation}
\end{subequations}
Since all the current superfields are obtained by action of analytic spinor derivatives on ${\cal J}^{\alpha\dot\alpha}$, and  the right hand side
of the harmonic constraint \eqref{s=3 conservation law}
is obviously analytic, these superfields  automatically satisfy the standard harmonic conservation laws, $\mathcal{D}^{++}J^{++}_{\alpha\dot\alpha} = 0\,,$ etc. \footnote{
This construction can straightforwardly be generalized to  an arbitrary ${\cal N}=2$ spin ${\bf s}$, with the appropriate non-analytic master supercurrent.}.

The current superfields \eqref{s=3 super current} provide sources in the relevant  superfield equations of motion:
\begin{equation}\label{EOM with currents spin 3}
    \begin{cases}
        (D^+)^4 G^{--(\alpha\beta) (\dot{\alpha}\dot{\beta}) }= -\frac{1}{2}  \kappa_3 J^{++(\alpha\beta) (\dot{\alpha}\dot{\beta}) }\,,
        \\
        (D^+)^4 G^{--5\alpha \dot{\alpha} } = -\frac{1}{2}  \kappa_3 J^{++\alpha \dot{\alpha}}\,,
        \\
        (\bar{D}^+)^2 D^{+\beta} G^{--}_{(\beta\alpha) (\dot{\alpha}\dot{\beta})}
        -
        (D^+)^2 \bar{D}^+_{(\dot{\alpha}} G^{--5}_{\alpha \dot{\beta})}
        = -2i \kappa_3 \mathcal{J}^+_{\alpha (\dot{\alpha}\dot{\beta})}\,,
        \\
        (D^+)^2  \bar{D}^{+\dot{\beta}} G^{--}_{(\alpha\beta) (\dot{\beta}\dot{\alpha})}
        +
        (\bar{D}^+)^2 D^+_{(\alpha} G^{--5}_{\beta) \dot{\alpha}}
        =
        -2i \kappa_3 \mathcal{J}^+_{(\alpha\beta) \dot{\alpha}}\,.
    \end{cases}
\end{equation}
The component content of spin \textbf{3} current superfields can be analyzed in full analogy with the spin \textbf{2} case.

\textbf{1.} The spin $\mathbf{3}$ current superfield has the form:
\begin{equation}\label{spin 3 superfield current}
    \begin{split}
J^{++}_{(\alpha\beta)(\dot{\alpha}\dot{\beta})} =&
- \frac{1}{2}
q^{+a}  \partial_{(\alpha \dot{\alpha}} \partial_{\beta) \dot{\beta}} J q^+_a
-
\frac{1}{2} \xi  \,\partial_{(\alpha \dot{\alpha}} \partial_{\beta) \dot{\beta}} \left(q^{+a}  J  q^+_a \right)
\\=&
- i \frac{1}{2} q^{+a} \partial_{(\alpha \dot{\alpha}} \partial_{\beta) \dot{\beta}} \tau^3_{ab} q^{+b}
-
i \frac{1}{2} \xi  \,\partial_{(\alpha \dot{\alpha}} \partial_{\beta) \dot{\beta}} \left( \tau^3_{ab} q^{+a}    q^{+b} \right).
    \end{split}
\end{equation}
To figure out its component structure, it is convenient to rewrite it in terms of $q^+$ and $\tilde{q}^{+}$:
\begin{equation}\label{spin 3 superfield exp2}
    J^{++}_{(\alpha\beta)(\dot{\alpha}\dot{\beta})} =  i \frac{1}{2} \left( \tilde{q}^{+} \partial_{(\alpha \dot{\alpha}} \partial_{\beta) \dot{\beta}}  q^{+}
    +
    q^{+} \partial_{(\alpha \dot{\alpha}} \partial_{\beta) \dot{\beta}}  \tilde{q}^{+} \right)
    +
    i  \xi \,\partial_{(\alpha \dot{\alpha}} \partial_{\beta) \dot{\beta}} \left( \tilde{q}^+ q^+\right).
\end{equation}
The divergence of the superfield \eqref{spin 3 superfield current} does not vanish,
\begin{equation}
    \partial^{\alpha\dot{\alpha}}J^{++}_{(\alpha\beta)(\dot{\alpha}\dot{\beta})} \neq 0\,.
\end{equation}
So, in contrast to the spin $\mathbf{2}$ case, the spin $\mathbf{3}$ current superfield reveals no any additional conservation
condition beyond ${\cal D}^{++}J^{++}_{(\alpha\beta)(\dot{\alpha}\dot{\beta})} =0\,$. We will be interested in the following components in
the $\theta$-expansion of the supefield \eqref{spin 3 superfield current}:
\begin{equation}\label{spin 3 superfield exp}
     J^{++}_{(\alpha\beta)(\dot{\alpha}\dot{\beta})}
     =
     \dots
     +
     \theta^{+\gamma} S^i_{(\alpha\beta)(\dot{\alpha}\dot{\beta})\gamma} u^+_i
     +
     \bar{\theta}^{+\dot{\gamma}} S^i_{(\alpha\beta)(\dot{\alpha}\dot{\beta})\dot{\gamma}} u^+_i
     +
     2i
     \theta^+ \sigma^n \bar{\theta}^+ j^n_{(\alpha\beta)(\dot{\alpha}\dot{\beta})}
     +
     \dots.
\end{equation}
We shall express them through the hypermultiplet fields from eq. \eqref{spin 3 superfield exp2}

\begin{itemize}
    \item The spinor component of \eqref{spin 3 superfield exp} reads:
    \begin{equation}\label{fermionic component spin 3}
        \begin{split}
        S^i_{(\alpha\beta)(\dot{\alpha}\dot{\beta})\rho}
        =&
        -
        \frac{i}{2}
        \left(\kappa_\rho \partial_{(\alpha \dot{\alpha}} \partial_{\beta) \dot{\beta}} f^i
        +
        \partial_{(\alpha \dot{\alpha}} \partial_{\beta) \dot{\beta}} \kappa_\rho  f^i \right)
        \\&-
        \frac{i}{2}
        \left(
        \psi_\rho \partial_{(\alpha \dot{\alpha}} \partial_{\beta) \dot{\beta}} \bar{f}^i
        +
        \partial_{(\alpha \dot{\alpha}} \partial_{\beta) \dot{\beta}} \psi_\rho  \bar{f}^i  \right)
        \\&+ i  \xi \,\partial_{(\alpha \dot{\alpha}} \partial_{\beta) \dot{\beta}} \left[ \bar{f}^i \psi_\rho + f^i \kappa_\rho \right].
            \end{split}
    \end{equation}
In contrast to the spin $\mathbf{2}$ case, this component does not satisfy any conservation law.
However, this does not mean that the spin $\tfrac{5}{2}$ fields interacts with a non-conserved current.
The corresponding conserved spin $\tfrac{5}{2}$ current consists of several constituents and will be explicitly presented below (eq. \eqref{5:2}).

    \item
The conserved spin 3 current component of the superfield \eqref{spin 3 superfield current} reads:
\begin{equation}\label{spin 3 bosonic current}
    \begin{split}
    j^n_{(\alpha\beta)(\dot{\alpha}\dot{\beta})} =& -i \frac{1}{4}
    \Bigr(
    f^i \partial_{(\alpha\dot{\alpha}} \partial_{\beta)\dot{\beta}}\partial^n \bar{f}_i
    -
    \partial_{(\alpha\dot{\alpha}} \partial_{\beta)\dot{\beta}}\partial^n f^i \bar{f}_i
    \\&\qquad -\,
    \partial^n f^i \partial_{(\alpha\dot{\alpha}} \partial_{\beta)\dot{\beta}} \bar{f} _i
    +
    \partial_{(\alpha\dot{\alpha}} \partial_{\beta)\dot{\beta}} f^i
    \partial^n \bar{f}_i
     \Bigr)
     \\& - i\frac{\xi}{2} \,\partial_{(\alpha \dot{\alpha}} \partial_{\beta) \dot{\beta}} \left( f^i \partial^n \bar{f}_i  - \partial^n f^i \bar{f}_i\right).
    \end{split}
\end{equation}
We observe that the $\xi$-term is just the derivative of the ${\rm U}(1)$ current.
The conservation law for the spin 3 current has the standard form:
\begin{equation}
    \partial_n j^n_{(\alpha\beta)(\dot{\alpha}\dot{\beta})} = 0\,.
\end{equation}

\end{itemize}

The current \eqref{spin 3 bosonic current} (with $\xi=0$) exactly coincides with the scalar field current elaborated in \cite{Berends:1985xx}.

\medskip

\textbf{2.} Spin 2 current superfield
\begin{equation} \label{Supercurrent 2 - spin 3}
    J^{++}_{\alpha\dot{\alpha}}
    =
    -
    \frac{1}{2}
    q^{+a}\partial_{\alpha\dot{\alpha}} J \partial_5 q^+_a
    =
    -m
    \frac{1}{2}
    q^{+a} \partial_{\alpha\dot{\alpha}} q^+_a
\end{equation}
 coincides (up to the multiplier $m$) with the $\mathcal{N}=2$ free hypermultiplet supercurrent \eqref{Supercurrent1-spin2} introduced in Section \ref{hyper supercurrent}.
 The corresponding spin 2 conserved current in the bosonic sector is:
 \begin{equation}\label{Supercurrent 2 - spin 3-component}
    j_{\alpha\beta\dot{\alpha}\dot{\beta}}
    =
    \frac{m}{4}
 \Big(
 \partial_{\alpha\dot{\alpha}} \partial_{\beta\dot{\beta}} f^i \bar{f}_i
 -
 \partial_{\alpha\dot{\alpha}}  f^i \partial_{\beta\dot{\beta}} \bar{f}_i
 -
 \partial_{\beta\dot{\beta}}   f^i  \partial_{\alpha\dot{\alpha}} \bar{f}_i
 +  f^i \partial_{\alpha\dot{\alpha}} \partial_{\beta\dot{\beta}}\bar{f}_i
 \Big).
 \end{equation}

\medskip

\textbf{3.}  The spin $\mathbf{3}$ fermionic current superfield \eqref{Supercurrent4} reads:
\begin{equation}\label{3FermCurr}
    \begin{split}
    J^+_{(\alpha\beta)\dot{\alpha}}
    =&
    -
    \frac{1}{2} q^{+a}\partial_{(\alpha\dot{\alpha}} \partial^-_{\beta)} J q^+_a - \frac{1}{2} \xi \partial_{(\alpha\dot{\alpha}} \partial^-_{\beta)} \left( q^{+a} J q^+_a \right)
    \\
    =&\;
    \frac{i}{2} \tilde{q}^+ \partial_{(\alpha\dot{\alpha}} \partial^-_{\beta)} q^+
    +
    \frac{i}{2} q^+ \partial_{(\alpha\dot{\alpha}} \partial^-_{\beta)} \tilde{q}^+
    -
    i \xi \partial_{(\alpha\dot{\alpha}} \partial^-_{\beta)}
    \left(\tilde{q}^+q^+\right).
    \end{split}
\end{equation}
As in the spin \textbf{2} case, it satisfies the modified conservation law \eqref{ModConser2}:
\begin{equation}
    \mathcal{D}^{++} J^+_{(\alpha\beta)\dot{\alpha}}
    =
    2i \bar{\theta}^{+\dot{\rho}} J^{++}_{(\alpha\beta)(\dot{\alpha}\dot{\rho})}
    -
    2i
    \theta^+_{(\alpha} J^{++}_{\beta)\dot{\alpha}}\,.
\end{equation}

The corresponding conserved spin $\tfrac{5}{2}$ fermionic current, according to the general discussion in Appendix \ref{N=2 current superfield},
is defined by the following combination of various spinorial components:
\begin{equation} \label{5:2}
    \mathfrak{j}^i_{\rho\dot{\rho}(\alpha\beta)\dot{\alpha}}
    :=
    j^i_{\rho\dot{\rho}(\alpha\beta)\dot{\alpha}}
    -
    S^i_{\rho\dot{\rho}(\alpha\beta)\dot{\alpha}}
    +
    m\epsilon_{\rho(\alpha} S^i_{\beta)\dot{\rho}\dot{\alpha}}\,.
\end{equation}
Here $j^i_{\rho\dot{\rho}(\alpha\beta)\dot{\alpha}}$ appears as a component of $J^+_{(\alpha\beta)\dot{\alpha}}$,
\begin{equation}
    J^+_{(\alpha\beta)\dot{\alpha}} = \dots + 2i \theta^{+\rho} \bar{\theta}^{+\dot{\rho}} j^i_{\rho\dot{\rho}(\alpha\beta)\dot{\alpha}}
    +
    \dots\,.
\end{equation}
Using the on-shell hypermultiplet superfield \eqref{hyper q} and the definition \eqref{3FermCurr} one can express it in terms of the free hypermultiplet fields:
\begin{equation}
    \begin{split}
    j^i_{\rho\dot{\rho}(\alpha\beta)\dot{\alpha}}
    =&
    \frac{i}{2} \left(\chi_\rho \partial_{(\alpha\dot{\alpha}}\partial_{\beta)\dot{\rho}} f^i - \partial_{\rho\dot{\rho}} f^i \partial_{(\alpha\dot{\alpha}} \chi_{\beta)}  \right)
    +
    \frac{i}{2} \left(\psi_\rho \partial_{(\alpha\dot{\alpha}}\partial_{\beta)\dot{\rho}} \bar{f}^i - \partial_{\rho\dot{\rho}} \bar{f}^i \partial_{(\alpha\dot{\alpha}} \psi_{\beta)} \right)
    \\& + i\xi \partial_{(\alpha\dot{\alpha}}
    \left( \chi_{\beta)} \partial_{\rho\dot{\rho}}
    f^i
    -
  \chi_\rho \partial_{\beta)\dot{\rho}} f^i \right)
    +
    i\xi \partial_{(\alpha\dot{\alpha}} \left( \psi_{\beta)} \partial_{\rho\dot{\rho}} \bar{f}^i
    -
    \psi_\rho \partial_{\beta)\dot{\rho}} \bar{f}^i \right).
    \end{split}
\end{equation}
The term $S^i_{\rho\dot{\rho}(\alpha\beta)\dot{\alpha}}$ in \eqref{5:2} is the fermionic component \eqref{fermionic component spin 3}
of the spin 3 current superfield $J^{++}_{(\alpha\beta)(\dot{\alpha}\dot{\beta})}$ \eqref{spin 3 superfield exp}.
The last component $S^i_{\beta\dot{\rho}\dot{\alpha}}$ in \eqref{5:2} is the $\mathcal{N}=2$ supersymmetry fermionic current defined in \eqref{S current}.

The total conserved fermionic current \eqref{5:2} (for simplicity, we present it for the $m=0$ case) is then given by the expression:
\begin{equation}\label{spin 5/2 current - component}
    \begin{split}
    \mathfrak{j}^i_{\rho\dot{\rho}(\alpha\beta)\dot{\alpha}}
    =&\;\;\;\;
    \frac{i}{2}
    \left(
    \partial_{(\alpha \dot{\alpha}} \partial_{\beta) \dot{\rho}} \chi_\rho  f^i
    +
    2\chi_\rho \partial_{(\alpha \dot{\alpha}} \partial_{\beta) \dot{\rho}} f^i
    -
     \partial_{(\alpha\dot{\alpha}} \chi_{\beta)} \partial_{\rho\dot{\rho}} f^i
    \right)
    \\&
    +
    \frac{i}{2}
    \left(
    \partial_{(\alpha \dot{\alpha}} \partial_{\beta) \dot{\rho}} \psi_\rho  \bar{f}^i
    +
    2\psi_\rho \partial_{(\alpha \dot{\alpha}} \partial_{\beta) \dot{\rho}} \bar{f}^i
    -  \partial_{(\alpha\dot{\alpha}} \psi_{\beta)} \partial_{\rho\dot{\rho}} \bar{f}^i \right)
    \\&
    + i\xi \partial_{(\alpha\dot{\alpha}}
    \left( \chi_{\beta)} \partial_{\rho\dot{\rho}}
    f^i
    -
   \chi_\rho \partial_{\beta)\dot{\rho}} f^i \right)
    +
    i\xi \partial_{(\alpha\dot{\alpha}} \left( \psi_{\beta)} \partial_{\rho\dot{\rho}} \bar{f}^i
    -
    \psi_\rho \partial_{\beta)\dot{\rho}} \bar{f}^i \right)
    \\&- i  \xi \,\partial_{(\alpha \dot{\alpha}} \partial_{\beta) \dot{\rho}} \left( \chi_\rho  f^i  + \psi_\rho \bar{f}^i \right).
\end{split}
\end{equation}

This current differs from the Noether spin $\tfrac{5}{2}$ current \eqref{spin 5/2 current}.
However, two conserved currents of the same spin can differ only by trivially conserved terms.
Let us explicitly give how these currents are related to each other.

The $\xi$-terms in the third and fourth lines of \eqref{spin 5/2 current - component}, modulo free massless equations of motions,
 \begin{equation}\label{EOM fermion}
     \partial_{\rho\dot{\rho}} \chi_\beta
     -
      \partial_{\beta\dot{\rho}} \chi_\rho = 0,
      \qquad
      \partial_{\rho\dot{\rho}} \psi_\beta
      -
      \partial_{\beta\dot{\rho}} \psi_\rho = 0,
 \end{equation}
can be reduced to the form:
\begin{equation}
    i\xi \partial_{(\alpha\dot{\alpha}} \partial_{\rho\dot{\rho}}
    \left( \chi_{\beta)}  f^i + \psi_{\beta)}     \bar{f}^i \right)
    - 2i  \xi \,\partial_{(\alpha \dot{\alpha}} \partial_{\beta) \dot{\rho}} \left( \chi_\rho  f^i  + \psi_\rho \bar{f}^i \right).
\end{equation}
It is straightforward to see that this $\xi$ contribution to $\mathfrak{j}$-current is trivially conserved.
 Moreover, it is a total derivative and so do not contribute to the corresponding conserved charge.

 Using the free equations \eqref{EOM fermion} one can further reduce the current $\mathfrak{j}$ to the form:
 \begin{equation}
    \begin{split}
 \mathfrak{j}^i_{\rho\dot{\rho}(\alpha\beta)\dot{\alpha}}
 =
& -2  j^i_{\rho\dot{\rho}(\alpha\beta)\dot{\alpha}}
 \\&+
 \frac{i}{2}
 \Big[
 2 \partial_{(\beta\dot{\rho}} \left( \chi_\rho \partial_{\alpha)\dot{\alpha}} f^i + \psi_\rho \partial_{\alpha)\dot{\alpha}} \bar{f}^i  \right)
 -
 \partial_{\rho\dot{\rho} } \left( \chi_{(\beta} \partial_{\alpha)\dot{\alpha}} f^i+ \psi_{(\beta}\partial_{\alpha)\dot{\alpha}} \bar{f}^i \right)
 \Big]
 \\&
 +
 \frac{i}{2}
 \partial_{\alpha\dot{\alpha}} \left( \chi_\beta \partial_{\rho\dot{\rho}} f^i -  \partial_{\rho\dot{\rho}}  \chi_\beta f^i  \right)
 +
 \frac{i}{2}
 \partial_{\alpha\dot{\alpha}} \left( \psi_\beta \partial_{\rho\dot{\rho}} \bar{f}^i -  \partial_{\rho\dot{\rho}}  \psi_\beta \bar{f}^i  \right).
 \end{split}
 \end{equation}
The second line in this expression is trivially conserved. The third line are derivatives of the independently conserved $\psi$ and $\chi$ parts
of the $\mathcal{N}=2$ supersymmetry fermionic current \eqref{S current}. Since these terms are total derivatives, they do not give contributions to the conserved charges.

Thus we see that, up to numerical factors, the trivially conserved current and derivatives of the supersymmetric current,
the current $\mathfrak{j}^i_{\rho\dot{\rho}(\alpha\beta)\dot{\alpha}}$ coincides with the canonical Noether spin $\tfrac{5}{2}$ current.

\medskip

The conserved spin $\tfrac{5}{2}$ current of hypermultiplet  can also be represented in the form:
\begin{equation}\label{S cal spin 52}
    \mathcal{S}^i_{\rho\dot{\rho}(\alpha\beta)\dot{\alpha}}
    :=
    i \chi_\rho
    \partial_{(\alpha\dot{\alpha}} \partial_{\beta)\dot{\rho}} f^i
    +
    i \psi_\rho
    \partial_{(\alpha\dot{\alpha}} \partial_{\beta)\dot{\rho}} \bar{f}^i,
    \qquad
    \partial^{\rho\dot{\rho}} \mathcal{S}^i_{\rho\dot{\rho}(\alpha\beta)\dot{\alpha}} = 0.
\end{equation}
The various spin $\tfrac{5}{2}$ currents discussed in this section are related by the equation:
\begin{equation}
        \mathcal{S}^i_{\rho\dot{\rho}(\alpha\beta)\dot{\alpha}} = \mathfrak{j}^i_{\rho\dot{\rho}(\alpha\beta)\dot{\alpha}}
        +
        j^i_{\rho\dot{\rho}(\alpha\beta)\dot{\alpha}}
\end{equation}
Thus, just as in the case of the   conserved  $\tfrac{3}{2}$ currents of rigid $\mathcal{N}=2$ supersymmetry, the various spin $\tfrac{5}{2}$ conserved currents
coincide up to numerical factors, trivially conserved terms and derivatives of the lower spin currents.
In Section \ref{spin 3 coupling} we shall see that the current  $\mathcal{S}^i_{\rho\dot{\rho}(\alpha\beta)\dot{\alpha}} $ reappears in the $(\tfrac{5}{2}, \tfrac{1}{2}, 0)$
cubic vertices.

\subsection{General  spin $\mathbf{s}$}

For an arbitrary spin, the calculations can be  performed quite analogously to the previous Subsections. As an example,
we quote here how the conserved bosonic spin s currents look for an arbitrary higher spin.

In the case of \textit{even spins }the current superfield \eqref{Supercurrent1} has the form:
\begin{equation}
    J^{++}_{\alpha(s-1)\dot{\alpha}(s-1)} =
    - \frac{1}{2}
    q^{+a}  \partial^{s-1}_{\alpha(s-1) \dot{\alpha}(s-1)}  q^+_a\,.
\end{equation}
The corresponding conserved spin s current in the bosonic sector is expressed as:
\begin{equation}\label{spin s even bosonic current}
    \begin{split}
    j^n_{\alpha(s-1)\dot{\alpha}(s-1)}
    =&
    - \frac{1}{4} \partial^n \bar{f}^i \partial^{s-1}_{\alpha(s-1) \dot{\alpha}(s-1)} f_i
    -
    \frac{1}{4}  \partial^{s-1}_{\alpha(s-1) \dot{\alpha}(s-1)} \bar{f}^i   \partial^nf_i
    \\&
    +\frac{1}{4}  \bar{f}^i \partial^{s-1}_{\alpha(s-1)  \dot{\alpha}(s-1)} \partial^n f_i
    +\frac{1}{4}  \partial^{s-1}_{\alpha(s-1) \partial^n   \dot{\alpha}(s-1)} \bar{f}^i f_i\,.
        \end{split}
\end{equation}

In the \textit{odd spin} case we have also a contribution from the $\xi$-term:
\begin{equation}
J^{++}_{\alpha(s-1)\dot{\alpha}(s-1)} =
- \frac{1}{2}
q^{+a}  \partial^{s-1}_{\alpha(s-1) \dot{\alpha}(s-1)} J q^+_a
-
\frac{1}{2} \xi  \, \partial^{s-1}_{\alpha(s-1) \dot{\alpha}(s-1)} \left(q^{+a}  J q^+_a \right).
\end{equation}
In this case the conserved current in the bosonic sector  reads:
\begin{equation}\label{spin s odd bosonic current}
    \begin{split}
        j^n_{\alpha(s-1)\dot{\alpha}(s-1)} =& -i \frac{1}{4}
        \Bigr(
        f^i \partial^{s-1}_{\alpha(s-1) \dot{\alpha}(s-1)} \partial^n \bar{f}_i
        -
        \partial^{s-1}_{\alpha(s-1) \dot{\alpha}(s-1)} \partial^n f^i \bar{f}_i
        \\&\qquad\qquad-
        \partial^n f^i \partial^{s-1}_{\alpha(s-1) \dot{\alpha}(s-1)} \bar{f} _i
        +
        \partial^{s-1}_{\alpha(s-1) \dot{\alpha}(s-1)} f^i
        \partial^n \bar{f}_i
        \Bigr)
        \\& - i\frac{\xi}{2} \,\partial^{s-1}_{\alpha(s-1) \dot{\alpha}(s-1)} \left( f^i \partial^n \bar{f}_i  - \partial^n f^i \bar{f}_i\right).
    \end{split}
\end{equation}
It is easy to check that the currents \eqref{spin s even bosonic current} and \eqref{spin s odd bosonic current} are conserved on the free hypermultiplet equations of motion \eqref{hyper-equations}:
\begin{equation}
    \partial_n j^n_{\alpha(s-1)\dot{\alpha}(s-1)} = 0\,.
\end{equation}

\section{Hypermultiplet equations of motion and the component reduction of cubic couplings}
\label{Hypermultiplet equations of motion and component reduction of cubic couplings}

In this Section, we shall study the component structure of $\mathcal{N}=2$ supersymmetric cubic interactions of hypermultiplet with higher spins.
One of our goals will be elucidation of the role of the $\xi$-interaction that appears for odd spins $\mathbf{s}= \mathbf{3}, \mathbf{5}, \mathbf{7}, \dots$.
We shall consider the simplest examples of the spin $\mathbf{1}$, $\mathbf{2}$ and $\mathbf{3}$ multiplets.

\subsection{Hypermultiplet equations of motion}\label{Hypermultiplet equations of motion}

The presence of the interaction of a hypermultiplet with fields of higher spins  modifies the equations of motion for the hypermultiplet.
This leads to a change in the on-shell component expansion of the hypermultiplet and, as a result, affects the component structure of the on-shell cubic vertex.
Varying the action \eqref{action-final} yields:
\begin{equation}\label{action-final-var}
    \begin{split}
        \delta_q S^{(s)}_{gauge}
        =&
        - \frac{1}{2} \int d\zeta^{(-4)}\;\Big[
        \delta q^{+a} \left(\mathcal{D}^{++} + \kappa_s \hat{\mathcal{H}}^{++}_{(s)}  (J)^{P(s)} +  \kappa_s \xi \Gamma^{++}_{(s)} (J)^{P(s)} \right) q^+_a
        \\&
        +
        q^{+a} \left(\mathcal{D}^{++} + \kappa_s \hat{\mathcal{H}}^{++}_{(s)}  (J)^{P(s)} +  \kappa_s \xi \Gamma^{++}_{(s)} (J)^{P(s)} \right) \delta q^+_a\Big].
    \end{split}
\end{equation}
After integration by parts, the variation becomes:
\begin{equation}
    \begin{split}
        \delta_q S^{(s)}_{gauge}
        =&
        - \frac{1}{2} \int d\zeta^{(-4)}\;
        \delta q^{+a} \Bigr[2 \mathcal{D}^{++} + \kappa_s \hat{\mathcal{H}}^{++}_{(s)}  (J)^{P(s)} +  2 P(s) \kappa_s \xi \Gamma^{++}_{(s)} (J)^{P(s)}
        \\&
        +(-1)^{s}\kappa_s \partial^{s-2}_{\alpha(s-2)\dot{\alpha}(s-2)} \big(\hat{\mathcal{H}}^{++\alpha(s-2)\dot{\alpha}(s-2)} + \Gamma^{++\alpha(s-2)\dot{\alpha}(s-2)} \big)  (J)^{P(s)}  \Bigr] q^+_a\,.
    \end{split}
\end{equation}
Here, all derivatives act freely to the right. So the hypermultiplet equation of motion on the higher spin gauge background reads:

\begin{equation}\label{hyper eom}
    \begin{split}
        &\Bigr[2 \mathcal{D}^{++} + \kappa_s \hat{\mathcal{H}}^{++}_{(s)}  (J)^{P(s)} +  2 P(s) \kappa_s \xi \Gamma^{++}_{(s)} (J)^{P(s)}
        \\&
        +(-1)^{s+P(s)}\kappa_s \partial^{s-2}_{\alpha(s-2)\dot{\alpha}(s-2)} \big(\hat{\mathcal{H}}^{++\alpha(s-2)\dot{\alpha}(s-2)} + \Gamma^{++\alpha(s-2)\dot{\alpha}(s-2)} \big)  (J)^{P(s)}  \Bigr] q^+_a = 0\,.
    \end{split}
\end{equation}

As a characteristic example, consider the case of spin $\mathbf{2}$, for which the equation \eqref{hyper eom} is greatly simplified:
\begin{equation}\label{hyper eom spin 2}
    \Bigr( \mathcal{D}^{++} + \kappa_2 \hat{\mathcal{H}}^{++}_{(s=2)}
    + \frac{\kappa_2}{2} \Gamma^{++}_{(s=2)}   \Bigr) q^+_a = 0\,.
\end{equation}
It is seen that the new terms appearing in this equation include a coupling constant. Thus, in the presence of coupling with $\mathcal{N}=2$ gauge superfields,  the on-shell component content
of the hypermultiplet differs from  that in the non-interacting case \eqref{hyper q} by $\kappa_2$ corrections. This will play an important role
in performing the component reduction of the cubic vertex.

In the spin $\mathbf{3}$ case the superfield equation is more complicated :
\begin{equation}\label{hyper eom spin 3}
    \begin{split}
    \Bigr[ \mathcal{D}^{++} +  \kappa_3 \hat{\mathcal{H}}^{++}_{(s=3)}  J +& \frac{\kappa_3}{2}  \left( \partial_{\alpha\dot{\alpha}} \hat{\mathcal{H}}^{++\alpha\dot{\alpha}} \right) J
    \\&+  \kappa_3  \left(\xi + \frac{1}{2} \right) \Gamma^{++}_{(s=3)} J
    +\frac{\kappa_3}{2}   \Gamma^{++\alpha\dot{\alpha}} \partial_{\alpha\dot{\alpha}}  J  \Bigr]\; q^+_a = 0\,.
        \end{split}
\end{equation}

Now we shall proceed to a detailed discussion of these two simplest cases, with an emphasis on the role of the auxiliary fields. As a preliminary step,
we once again first consider the simplest example of the spin ${\bf 1}$ ${\cal N}=2$ multiplet.

\subsection{Spin \textbf{1} coupling}

The equation of motion for the hypermultiplet in the spin $\mathbf{1}$ background has the form:
\begin{equation}\label{eom spin 1}
    \left(\mathcal{D}^{++} + i \kappa_1 h^{++} \right) q^+ = 0\,.
\end{equation}
To analyze the component structure of the cubic vertex, one needs to know how the non-dynamical part of the constraint \eqref{eom spin 1}  modifies
the free on-shell   expression \eqref{hyper q} for the hypermultiplet superfield due to the presence of interaction with the spin $\mathbf{1}$ multiplet.
Before proceeding to calculations, it is useful to note that only those auxiliary fields which are non-zero in the free $q^+$ theory can contribute to a cubic vertex.
In particular, this explains why the auxiliary fields of the gauge spin $\mathbf{1}$ multiplet do not contribute.

In the gauge sector, the spin $\mathbf{1}$ analytic potential has the form:
\begin{equation}
    h^{++}_A = -2i \theta^{+\rho} \bar{\theta}^{+\dot{\rho}} A_{\rho\dot{\rho}}.
\end{equation}
Eliminating the auxiliary fields on such a background by the non-dynamical part of \eqref{eom spin 1} (terms up to $\sim (\theta^+)^2$)
leads to the following modified on-shell hypermultiplet superfield:
 \begin{equation}\label{hyper q spin 1}
    q^+_{(s=1)} = f^i u^+_i
    +
    \theta^{+\alpha} \psi_\alpha
    +
    \bar{\theta}^+_{\dot{\alpha}} \bar{\kappa}^{+\dot{\alpha}}
    +
    (\theta^{\hat{+}})^2 mf^i u^-_i
    +
    2i \theta^+ \sigma^m \bar{\theta}^+ \nabla^{(s=1)}_m f^i u^-_i\,,
 \end{equation}
where
\begin{equation}
    \nabla^{(s=1)}_m f^i := \left(\partial_m + i\kappa_1 A_m \right) f^i.
\end{equation}
Thus we observe the appearance  of $\kappa_1$-correction to the free on-shell hypermultiplet \eqref{hyper q}.

It will be useful to divide the on-shell hypermultiplet superfield as:
\begin{equation}
    q^+_{(s=1)} = q^+_{(0)} + \kappa_1 q^{+}_{(\kappa_1)},
\end{equation}
where the zeroth-order term $q_{(0)}$ is still given by the expression \eqref{hyper q}.

In this notations,  the action to the leading order in $\kappa_1$  reads:
\begin{equation}
    \begin{split}\label{Cubic1}
    S_{gauge}^{(s=1)} =& - \int d\zeta^{(-4)}\; \tilde{q}^+ \left(\mathcal{D}^{++} + i \kappa_1 h^{++} \right) q^+
    \\
    =& - \int d\zeta^{(-4)}\; \tilde{q}_{(0)}^+ \mathcal{D}^{++}  q^+_{(0)}
    \\&- \kappa_1 \int d\zeta^{(-4)}\; \left( \tilde{q}_{(0)}^+ \mathcal{D}^{++}  q_{(\kappa_1)}^+
    +
    \tilde{q}_{(\kappa_1)}^+ \mathcal{D}^{++}  q_{(0)}^+ \right)
     \\  &- i \kappa_1 \int d\zeta^{(-4)}\; h^{++} \tilde{q}_{(0)}^+  q^+_{(0)}.
    \end{split}
\end{equation}

Using  the on-shell hypermultiplet expression \eqref{hyper q spin 1} one can easily calculate the bosonic contribution from the first two lines:
\begin{equation}\label{spin 1 free part}
    \begin{split}
    &-\int d\zeta^{(-4)} \; \tilde{q}^+_{(s=1)} \mathcal{D}^{++} q^+_{(s=1)}
    \\& \Rightarrow
    \int d^4x \; \left(-\eta^{nm}
    \nabla^{(s=1)}_n f^i \nabla^{(s=1)}_m \bar{f}_i
    +
    \partial^m f^i \nabla^{(s=1)}_m \bar{f}_i
    +
    \nabla^{(s=1)}_m f^i \partial^m \bar{f}_i
    - m^2 f^i \bar{f}_i  \right)
    \\&=
    \int d^4x\;
    \left(
    \partial^m f^i \partial_m \bar{f}_i - m^2 f^i \bar{f}_i+ \kappa_1^2 A^m A_m f^i \bar{f}_i \right)
    \\&=
    \int d^4x\;
    \left(\partial^m f^i \partial_m \bar{f}_i
    - m^2 f^i \bar{f}_i \right)
    +
    \mathcal{O}(\kappa_1^2).
    \end{split}
\end{equation}
Thus these terms do not contribute to the cubic component interaction  (but contribute to quartic $(1,1,0,0)$ vertex!).  So we reduce the cubic gauge coupling in \eqref{Cubic1}
to the form:
\begin{equation}
    S_{int}^{(s=1)}
    =
   - i \kappa_1 \int d\zeta^{(-4)}\; h^{++} \tilde{q}_{(0)}^+  q^+_{(0)}
    =
    \kappa_1 \int d\zeta^{(-4)} h^{++} J^{++}\,.
\end{equation}
Here we used the expression  \eqref{electric current} for the spin $\mathbf{1}$ current superfield. We need to leave in it only the bosonic term relevant to our study:
\begin{equation}
    J^{++} = - i \tilde{q}_{(0)}^+ q^+_{(0)} = ... + \theta^{+\rho}\bar{\theta}^{+\dot{\rho}} \left( \partial_{\rho\dot{\rho}} f^i \bar{f}_i - f^i \partial_{\rho\dot{\rho}} \bar{f}_i \right).
\end{equation}

As a net result, the gauge field coupling in the first order in $\kappa_1$ is found to be:
\begin{equation}\label{Minspin1}
    S_{int}^{(s=1)} = - \frac{i\kappa_1}{2} \int d^4x\; A^{\rho\dot{\rho}} \left( \partial_{\rho\dot{\rho}} f^i \bar{f}_i - f^i \partial_{\rho\dot{\rho}} \bar{f}_i \right).
\end{equation}
This is the standard minimal Noether interaction of the spin 1 field with the ${\rm U}(1)$ current\footnote{As distinct from the cases of ${\bf s}\geq 3$, in the spin ${\bf 1}$ and
${\bf 2}$ cases one can restore the gauge field - hypermultiplet couplings to all orders in $\kappa_{(s)}$ since the relevant theories are well known in full generality: these are ${\cal N}=2$
supersymmetric Maxwell theory and ${\cal N}=2$ Einstein supergravity.  For instance, the evident completion of the minimal bosonic coupling  \eqref{Minspin1} reads
$$S^{(s=1)}_{int}{}' = S^{(s=1)}_{int} + \kappa_1^2\int d^4x\, (A^mA_m) (f^i\bar{f}_i)\,.$$  However, keeping in mind  the construction of the minimal vertices for the non-trivial
${\bf s}\geq 3$ cases as our main goal here, we limit our attention altogether only to such minimal vertices.}.

\subsection{Spin \textbf{2} coupling}
\label{spin 2 coupling}

Like in the previous example,
we consider the equations of motion \eqref{hyper eom spin 2} for a hypermultiplet interacting with the spin $\mathbf{2}$ gauge superfield.
Once again, the interaction modifies the component expansion of the hypermultiplet as compared to the free case \eqref{hyper q}.

We will consider the cubic interaction to the first order in the coupling constant $\kappa_2$:
\begin{equation}
    q^+_{(s=2)} = q^+_{(0)} + \kappa_2 q^+_{(\kappa_2)} + \mathcal{O}(\kappa_2^2)\,,
\end{equation}
where $q^+_{(0)}$ was determined in \eqref{hyper q} and $q^+_{(\kappa_2)}$ is  a correction induced by the interaction.

The component cubic coupling then comes from the $\kappa_2$-terms of the hypermultiplet action:
\begin{equation}\label{hyper spin 2 action}
    \begin{split}
    S^{(s=2)}_{gauge}
    =&
    - \frac{1}{2} \int d\zeta^{(-4)}\;
    q^{+a} \left(\mathcal{D}^{++} + \kappa_2 \hat{\mathcal{H}}^{++}_{(s=2)}  \right) q^+_{a}
    \\=&
    - \frac{1}{2} \int d\zeta^{(-4)}\;
    q^{+a}_{(0)} \mathcal{D}^{++}  q^+_{(0)a}
    - \kappa_2 \int d\zeta^{(-4)}\; q^{+a}_{(\kappa_2)} \mathcal{D}^{++}  q^+_{(0)a}
    \\&-   \frac{\kappa_2}{2} \int d\zeta^{(-4)}\;
    q^{+a}_{(0)}   \hat{\mathcal{H}}^{++}_{(s=2)}   q^+_{(0)a} + \mathcal{O}(\kappa^2_2)\,,
    \end{split}
\end{equation}
where we sill retained the free action for $q^{+a}_{(0)}$.

Below we shall show that in the sectors of spins 2 and 1, after eliminating the auxiliary fields, the cubic vertex in the first order in $\kappa_2$ is obtained as:
\begin{equation}\label{Cubic2}
    \begin{split}
    S^{(s=2)}_{int} =&-\frac{\kappa_2}{2} \int d\zeta^{(-4)}\;
    q^{+a}_{(0)}   \hat{\mathcal{H}}^{++}_{(s=2)}   q^+_{(0)a}
    \\=&-\frac{\kappa_2}{2} \int d\zeta^{(-4)}\; \left[ \tilde{q}_{(0)}^+ \hat{\mathcal{H}}^{++}_{(s=2)} q_{(0)}^+ -  q_{(0)}^+ \hat{\mathcal{H}}^{++}_{(s=2)} \tilde{q}_{(0)}^+ \right],
        \end{split}
\end{equation}
Thus we have written the minimal cubic interaction in terms of the on-shell free hypermultiplet $q^{+a}_{(0)}$. We now turn to considering the component structure
of the bosonic sector of the spin $\mathbf{2}$ theory, also in the first order in $\kappa_2$. For this we should make use of the representation \eqref{hyper q}
and \eqref{hyper q tilde} for $q^{+a}_{(0)}$.

\medskip

\textbf{\underline{\textit{Spin 2 sector}}}

\medskip

The spin 2 sector of the gauge potentials is spanned by the following terms in the relevant $\theta$-expansions:
\begin{subequations}\label{spin 2 spin 2 sector}
\begin{equation}
    h^{++\alpha\dot{\alpha}}_{(\Phi, \mathcal{P})} = - 2i \theta^{+\rho}\bar{\theta}^{+\dot{\rho}} \Phi_{\rho\dot{\rho}}^{\alpha\dot{\alpha}}\,,
\end{equation}
\begin{equation}
    h^{++\alpha+}_{(\Phi, \mathcal{P})} = -2i (\theta^+)^2 \bar{\theta}^{+\dot{\beta}} B_{\dot{\beta}}^\alpha -2i (\theta^+)^2 \bar{\theta}^{+\dot{\beta}} \mathcal{P}_{\dot{\beta}}^\alpha\,,
\end{equation}
\begin{equation}
    h^{++\dot{\alpha}+}_{(\Phi, \mathcal{P})} = - 2i (\bar{\theta}^+)^2 \theta^{+\beta}B_\beta^{\dot{\alpha}} - 2i (\bar{\theta}^+)^2 \theta^{+\beta}\bar{\mathcal{P}}_\beta^{\dot{\alpha}} \,.
\end{equation}
\end{subequations}

After elimination of the auxiliary fields by the non-dynamical part of the constraint \eqref{hyper eom spin 2} one obtains the following expression for the on-shell hypermultiplet on
the spin 2 background \eqref{spin 2 spin 2 sector}:
\begin{equation}
    q^+_{(s=2)}
    =
    f^i u^+_i
    +
    \theta^{+\alpha} \psi_\alpha
    +
    \bar{\theta}^+_{\dot{\alpha}} \bar{\kappa}^{+\dot{\alpha}}
    +
    (\theta^{\hat{+}})^2 mf^i u^-_i
    +
    2i \theta^{+\rho}\bar{\theta}^{+\dot{\rho}} \nabla^{(s=2)}_{\rho\dot{\rho}} f^i u^-_i,
\end{equation}
where we have introduced the derivative covariant under the linearized gauge spin 2 transformations:
\begin{equation}
    \nabla^{(s=2)}_{\rho\dot{\rho}} f^i
    :=
    \left(\partial_{\rho\dot{\rho}} + \kappa_2 \Phi^{\alpha\dot{\alpha}}_{\rho\dot{\rho}} \partial_{\alpha\dot{\alpha}}
    +\frac{\kappa_2}{2}
        \left[\partial_{\alpha\dot{\alpha}} \Phi_{\rho\dot{\rho}}^{\alpha\dot{\alpha}}
        +
        4B_{\rho\dot{\rho}}
        +
        2\mathcal{P}_{\rho\dot{\rho}}
        +
        2\bar{\mathcal{P}}_{\rho\dot{\rho}}
        \right]\right) f^i.
\end{equation}
One can unfold the second line of \eqref{hyper spin 2 action}, in full analogy with the spin $\mathbf{1}$ case:
\begin{equation}
    \begin{split}
        &-\int d\zeta^{(-4)} \; \tilde{q}^+_{(s=2)} \mathcal{D}^{++} q^+_{(s=2)}
        \\&=
        \int d^4x \; \left(-\eta^{nm}
        \nabla^{(s=2)}_n f^i \nabla^{(s=2)}_m \bar{f}_i
        +
        \partial^m f^i \nabla^{(s=2)}_m \bar{f}_i
        +
        \nabla^{(s=2)}_m f^i \partial^m \bar{f}_i
        - m^2 f^i \bar{f}_i\ \right)
        \\&=
        \int d^4x\;
        \left(\partial^m f^i \partial_m \bar{f}_i
        - m^2 f^i \bar{f}_i\right)
        +
        \mathcal{O}(\kappa_2^2).
    \end{split}
\end{equation}
So this line does not contribute to the cubic coupling.

Substituting the expressions \eqref{spin 2 spin 2 sector} into \eqref{Cubic2} and using the $\theta$ -expansions \eqref{hyper q} and \eqref{hyper q tilde},
we obtain in the bosonic sector:
\begin{equation}
    \begin{split}
    S^{(s=2)}_{int} =&-\frac{\kappa_2}{4} \int d^4x\; \Phi^{\alpha\beta\dot{\alpha}\dot{\beta}} \left(
    \partial_{\alpha\dot{\alpha}} \partial_{\beta\dot{\beta}} f^i \bar{f}_i
    -
     \partial_{\alpha\dot{\alpha}}  f^i \partial_{\beta\dot{\beta}} \bar{f}_i
    -
     \partial_{\beta\dot{\beta}}   f^i \partial_{\alpha\dot{\alpha}} \bar{f}_i
    +  f^i \partial_{\alpha\dot{\alpha}} \partial_{\beta\dot{\beta}}\bar{f}_i  \right)
    \\&\;+ \frac{\kappa_2}{2} \int d^4x \left(2B^{\alpha\dot{\alpha}} + \mathcal{P}^{\alpha\dot{\alpha}} + \bar{\mathcal{P}}^{\alpha\dot{\alpha}}\right) \partial_{\alpha\dot{\alpha}} \left(f^i\bar{f}_i\right).
        \end{split}
\end{equation}
The second line contains an interaction with the linearized scalar curvature due to the relation $\partial_{\alpha\dot{\alpha}}B^{\alpha\dot{\alpha}}= -2 \mathcal{R}$.

After using the expression \eqref{B field} for $B^{\alpha\dot{\alpha}}$ and eliminating the auxiliary fields $\mathcal{P}^{\alpha\dot{\alpha}}$
and $\bar{\mathcal{P}}^{\alpha\dot{\alpha}}$
by eqs. \eqref{P fields}, the full spin 2 action involving the cubic coupling to scalars  takes the
form:
\begin{equation}\label{spin 2-2 full action}
    \begin{split}
    S^{(s=2)}_{_{(\Phi)}} + S_{int}^{(s=2)}
    =&
    \int d^4x\;
    2\Phi^{\alpha\beta\dot{\alpha}\dot{\beta}} \mathcal{G}_{\alpha\beta\dot{\alpha}\dot{\beta}}
    \\&
    - \frac{\kappa_2}{4} \int d^4x\; \Phi^{\alpha\beta\dot{\alpha}\dot{\beta}} \Big[
    \partial_{\alpha\dot{\alpha}} \partial_{\beta\dot{\beta}} f^i \bar{f}_i
    -
     \partial_{\alpha\dot{\alpha}}  f^i \partial_{\beta\dot{\beta}} \bar{f}_i
     -
    \partial_{\beta\dot{\beta}}   f^i  \partial_{\alpha\dot{\alpha}} \bar{f}_i
      +  f^i \partial_{\alpha\dot{\alpha}} \partial_{\beta\dot{\beta}}\bar{f}_i
      \\&+
    \left( \partial_{\alpha\dot{\beta}} \partial_{\beta\dot{\alpha}} - \epsilon_{\alpha\beta} \epsilon_{\dot{\alpha}\dot{\beta}} \Box \right) \left(f^i \bar{f}_i\right)\Big] + \mathcal{O}(\kappa_2^2) \,.
        \end{split}
\end{equation}
The equation of motion for $\Phi^{\alpha\beta\dot{\alpha}\dot{\beta}}$ yields:
\begin{equation}\label{eom spin 22}
    \begin{split}
    \mathcal{G}_{\alpha\beta\dot{\alpha}\dot{\beta}} =&
    \frac{\kappa_2}{16}  \Big[
    \partial_{\alpha\dot{\alpha}} \partial_{\beta\dot{\beta}} f^i \bar{f}_i
    -
     \partial_{\alpha\dot{\alpha}}  f^i \partial_{\beta\dot{\beta}} \bar{f}_i
    -
      \partial_{\beta\dot{\beta}} f^i  \partial_{\alpha\dot{\alpha}} \bar{f}_i
    +  f^i \partial_{\alpha\dot{\alpha}} \partial_{\beta\dot{\beta}}\bar{f}_i
        \\&-
    \left( \partial_{\alpha\dot{\beta}} \partial_{\beta\dot{\alpha}} - \epsilon_{\alpha\beta} \epsilon_{\dot{\alpha}\dot{\beta}} \Box \right) \left(f^i \bar{f}_i\right)\Big],
        \end{split}
\end{equation}
which coincides with \eqref{spin 23}. The interaction of spin 2 with a scalar field can now be read off from eq. \eqref{spin 2-2 full action} and cast in
the generic Noether form \eqref{Noether form interaction}:
\begin{equation}
    S^{(s=2)}_{int} = \kappa_2 \int d^4x\, \Phi^{\alpha\dot{\alpha}\beta\dot{\beta}} Y_{\alpha\dot{\alpha}\beta\dot{\beta}}\,.
\end{equation}
The corresponding $Y$ current is:
\begin{equation}\label{Y spin 2}
    \begin{split}
    Y_{\alpha\dot{\alpha}\beta\dot{\beta}}
    =
   - \frac{1}{4}
    \Big[
    &\partial_{\alpha\dot{\alpha}} \partial_{\beta\dot{\beta}} f^i \bar{f}_i
    -
    \partial_{\alpha\dot{\alpha}}  f^i \partial_{\beta\dot{\beta}} \bar{f}_i
    -
    \partial_{\beta\dot{\beta}}   f^i  \partial_{\alpha\dot{\alpha}} \bar{f}_i
    +  f^i \partial_{\alpha\dot{\alpha}} \partial_{\beta\dot{\beta}}\bar{f}_i
    \\&-
    \left( \partial_{\alpha\dot{\beta}} \partial_{\beta\dot{\alpha}} - \epsilon_{\alpha\beta} \epsilon_{\dot{\alpha}\dot{\beta}} \Box \right) \left(f^i \bar{f}_i\right)\Big].
    \end{split}
\end{equation}
Up to the last term (automatically satisfying the conservation law), the $Y$-current obtained coincides with the energy-momentum tensor \eqref{EM}.

In Appendix \ref{Minimal gravity coupling}, using the proper field redefinitions,  we prove that the coupling constructed is equivalent to the
minimal linearized gravity coupling of scalar fields. It is also possible to remove the last line in \eqref{spin 2-2 full action} (and the last line in the corresponding
$Y$ current \eqref{Y spin 2}) through a redefinition of the spin 2 field, as described in Appendix \ref{Redefinition of spin 2 gauge field}.
After such a redefinition we reproduce the standard linearized spin 2 coupling to a complex scalar, as it is given, e.g., in \cite{Bekaert:2009ud}.

\medskip

\textbf{\underline{\textit{Spin 1 sector}}}

\medskip

After elimination of the auxiliary fields,  we are left with the following spin 1 sector:
\begin{equation}
    h^{++5}_{(C)} = -2i \theta^{+\rho} \bar{\theta}^{+\dot{\rho}} C_{\rho\dot{\rho}}\,,
\end{equation}
\begin{equation}
    h^{++\alpha+}_{(C)}  = i (\bar{\theta}^+)^2 \theta^+_\nu \partial^{(\alpha}_{\dot{\rho}} C^{\dot{\rho} \nu)}\,,
\end{equation}
\begin{equation}
    h^{++\dot{\alpha}+}_{(C)}  = -i (\theta^+)^2 \bar{\theta}^+_{\dot{\nu}} \partial^{(\dot{\alpha}}_\rho C^{\rho \dot{\nu})}\,.
\end{equation}
In this spin 1 sector of $\mathcal{N}=2$ spin $\mathbf{2}$ theory, the interaction term \eqref{hyper eom spin 2} deforms the free on-shell hypermultiplet \eqref{hyper q} as:
 \begin{equation}\label{hyper q spin 21}
    q^+_{(s=2)} = f^i u^+_i
    +
    \theta^{+\alpha} \psi_\alpha
    +
    \bar{\theta}^+_{\dot{\alpha}} \bar{\kappa}^{+\dot{\alpha}}
    +
    (\theta^{\hat{+}})^2 mf^i u^-_i
    +
    2i \theta^+ \sigma^n \bar{\theta}^+ \nabla^{(s=2)}_n f^i u^-_i.
\end{equation}
Here we have defined the spin 1 covariant derivative in the spin $\mathbf{2}$ ${\cal N}=2$ theory as:
\begin{equation}
    \nabla^{(s=2)}_n f^i := \left(\partial_n + im\kappa_2 C_n \right) f^i.
\end{equation}
Using the same argument as in \eqref{spin 1 free part},
 we conclude that term
 \begin{equation*}
    - \kappa_2 \int d\zeta^{(-4)}\; q^{+a}_{(\kappa_2)} \mathcal{D}^{++}  q^+_{(0)a}
 \end{equation*}
does not contribute to $(1,0,0)$  vertex. There is only one contribution to cubical $(1,0,0)$ vertex
 and it is given by \eqref{Cubic2}.

The actual contribution to the interaction in the first order in $\kappa_2$ comes only from the prepotential $h^{++5}$, $\hat{\mathcal{H}}^{++}_{(s=2)} \rightarrow h^{++5}\partial_5$:
\begin{equation}
    \begin{split}
        S^{(s=2)}_{int} =& - \frac{\kappa_2}{2} \int d\zeta^{(-4)}\; \left[ \tilde{q}_{(0)}^+ \hat{\mathcal{H}}^{++}_{(s=2)} q_{(0)}^+ -  q_{(0)}^+ \hat{\mathcal{H}}^{++}_{(s=2)} \tilde{q}_{(0)}^+ \right]
        \\ \Rightarrow &-
        \frac{im\kappa_2}{2} \int d^4x\; C^{\rho\dot{\rho}} \left( f^i \partial_{\rho\dot{\rho}} \bar{f}_i - \partial_{\rho\dot{\rho}} f^i  \bar{f}_i \right).
    \end{split}
\end{equation}
So, the resulting spin 1 coupling is proportional to the mass $m$ of the hypermultiplet and has the  standard Noether form \eqref{Noether form interaction}:
\begin{equation}
    S^{(s=2)}_{int} = \kappa_2 \int d^4x\, C^{\rho\dot{\rho}} Y_{\rho\dot{\rho}},
\end{equation}
where :
\begin{equation}
    Y_{\rho\dot{\rho}}  = -\frac{im}{2} \left( f^i \partial_{\rho\dot{\rho}} \bar{f}_i - \partial_{\rho\dot{\rho}} f^i  \bar{f}_i \right).
\end{equation}
This current coincides with the appropriate conserved bosonic component of the current superfield \eqref{u(1) current spin 2}.

\medskip
\textbf{\underline{\textit{Spin $3/2$ sector}}}
\medskip

A similar analysis can also be carried out for the spinor sector. The corresponding cubic interactions also has the Noether form \eqref{Noether form interaction}:
\begin{equation}
    S^{(s=2)}_{int} = \kappa_2 \int d^4x \, \left(\psi^{\alpha\beta \dot{\alpha}i} Y_{\alpha\beta \dot{\alpha}i}
    +
    \bar{\psi}^{\alpha\dot{\alpha}\dot{\beta}i}
    \bar{Y}_{\alpha\dot{\alpha}\dot{\beta}i}\right).
\end{equation}
The corresponding spin $3/2$ currents coincides with the conserved  fermionic current of $\mathcal{N}=2$ supersymmetry \eqref{S cal current}, up to terms which are
trivially conserved without making use of the equations of motion:
\begin{equation}
        \begin{split}
            Y^i_{\rho\dot{\rho}\alpha} =
            -
            \frac{1}{4} \mathcal{S}^i_{\rho\dot{\rho}\alpha}
            =
            -
            \frac{1}{4}
            \Big[\psi_\rho \partial_{\alpha\dot{\rho}} \bar{f}^i
            -
            \chi_\rho \partial_{\alpha\dot{\rho}} f^i
            -
            im \epsilon_{\rho\alpha} \left( \bar{\chi}_{\dot{\rho}} \bar{f}^i - \bar{\psi}_{\dot{\rho}} f^i \right)\Big].
        \end{split}
\end{equation}
The overall coefficient is uniquely fixed by inspecting the interaction of the symmetric part $\psi^{(\alpha\rho)\dot{\rho}i}$ of the spin $\tfrac{3}{2}$
gauge field with the hypermultiplet fields. The contribution to this part of vertex comes from a single term in the WZ gauge of the  analytic
potential $h^{++\alpha\dot{\alpha}}$.

The terms by which the various conserved spin $\tfrac{3}{2}$ currents defined in Section \ref{hyper supercurrent}
differ from each other yield fake cubic $(\frac32, \frac12, 0)$ interactions.
Such terms can be eliminated by means of some field redefinition, according to the general prescription of Appendix \ref{Fields redefinitions}.
We do not provide the relevant details here.

\subsection{Spin \textbf{3} coupling}
\label{spin 3 coupling}

Like in the spin $\mathbf{2}$ case, in order to obtain the cubic coupling to the leading order in $\kappa_3$, one needs to partly solve the equation of motion for the hypermultiplet
in the spin $\mathbf{3}$ background:
\begin{equation}\label{eom - spin 3}
    \begin{split}
    \Big[\mathcal{D}^{++} +  \kappa_3 \hat{\mathcal{H}}^{++}_{(s=3)}  J &+ \frac{\kappa_3}{2}  \left( \partial_{\alpha\dot{\alpha}} \hat{\mathcal{H}}^{++\alpha\dot{\alpha}} \right) J
    \\&+  \kappa_3  \left(\xi + \frac{1}{2} \right) \Gamma^{++}_{(s=3)} J
    +\frac{\kappa_3}{2}   \Gamma^{++\alpha\dot{\alpha}} \partial_{\alpha\dot{\alpha}}  J  \Big]\, q^+_a = 0\,.
        \end{split}
\end{equation}

We represent the  solution as a perturbative series in $\kappa_3$:
\begin{equation}
    q^+ = q^{+}_{(0)} + \kappa_3 q^+_{(\kappa_3)} + \mathcal{O}(\kappa_3^2)\,,
\end{equation}
where $q^+_{0}$ is described by the free hypermultiplet action and is given in \eqref{hyper q}.
The action, up to the first order in $\kappa_3$, is written as:
\begin{equation}\label{action-final-spin3}
    \begin{split}
    S^{(s=3)}_{gauge}
    =&
    - \frac{1}{2} \int d\zeta^{(-4)}\;
    q^{+a} \left(\mathcal{D}^{++} + \kappa_3\hat{\mathcal{H}}^{++}_{(s=3)}  J + \kappa_3 \xi \Gamma^{++}_{(s=3)} J \right) q^+_a
    \\=&
    - \frac{1}{2} \int d\zeta^{(-4)}\;
    q^{+a}_{(0)} \mathcal{D}^{++}  q^+_{(0)a}
    \\&
    - \kappa_3 \int d\zeta^{(-4)}\; q^{+a}_{(\kappa_3)} \mathcal{D}^{++}  q^+_{(0)a}
    \\& - \frac{\kappa_3}{2} \int d\zeta^{(-4)}\;
    q^{+a}_{(0)} \left( \hat{\mathcal{H}}^{++}_{(s=3)}  J +  \xi \Gamma^{++}_{(s=3)} J \right) q^+_{(0)a}
    + \mathcal{O}(\kappa_3^2)\,.
        \end{split}
\end{equation}
Below we will show that after elimination of the auxiliary fields the third line does not contribute to the cubic vertex, and the latter takes the form:
\begin{equation}
    \label{spin 3 on-shell superfield}
    \begin{split}
        S^{(s=3)}_{int} =
        &- \frac{\kappa_3}{2} \int d\zeta^{(-4)}\;
        q^{+a}_{(0)} \left( \hat{\mathcal{H}}^{++}_{(s=3)}  J +  \xi \Gamma^{++}_{(s=3)} J \right) q^+_{(0)a}
        \\=&
        \;\frac{i\kappa_3}{2} \int d\zeta^{(-4)}\; \left[  q_{(0)}^+ \hat{\mathcal{H}}^{++}_{(s=3)} \tilde{q}_{(0)}^+
        +
           \tilde{q}_{(0)}^+ \hat{\mathcal{H}}^{++}_{(s=3)} q_{(0)}^+  \right]
        \\
        & +
        \xi \cdot i\kappa_3 \int d\zeta^{(-4)}\; \Gamma^{++}_{(s=3)} \; \tilde{q}^+_{(0)} q^+_{(0)}\,.
    \end{split}
\end{equation}
Another convenient form of the cubic interaction related to \eqref{spin 3 on-shell superfield} by a total derivative is as follows :
\begin{equation}
    \begin{split}
        S^{(s=3)}_{int} =&-
         \frac{i\kappa_3}{2} \int d\zeta^{(-4)}\; \left[ \partial_{\alpha\dot{\alpha}}\tilde{q}_{(0)}^+ \hat{\mathcal{H}}^{++\alpha\dot{\alpha}}_{(s=3)} q_{(0)}^+
         +  \partial_{\alpha\dot{\alpha}} q_{(0)}^+ \hat{\mathcal{H}}^{++\alpha\dot{\alpha}}_{(s=3)} \tilde{q}_{(0)}^+ \right]
         \\
         &-
         \frac{i\kappa_3}{2}
         \int d\zeta^{(-4)}\; \left[ \partial_{\alpha\dot{\alpha}}\tilde{q}_{(0)}^+ \Gamma^{++\alpha\dot{\alpha}}_{(s=3)} q_{(0)}^+ +  \partial_{\alpha\dot{\alpha}}
         q_{(0)}^+ \Gamma^{++\alpha\dot{\alpha}}_{(s=3)} \tilde{q}_{(0)}^+ \right]
         \\
         &+
         \xi \cdot i\kappa_3 \int d\zeta^{(-4)}\; \Gamma^{++}_{(s=3)} \; \tilde{q}^+_{(0)} q^+_{(0)}\,.
    \end{split}
\end{equation}
Using these expressions for the cubic vertex, it is easy to find the component structure of the $({3}, {0}, {0})$ interaction.

\medskip

\textbf{\underline{\textit{Spin 3 sector}}}

\medskip
Since the spin ${\bf 3}$ coupling is an essentially new subject as compared to the previous cases, we consider it with more intermediate details.

The  spin 3 sector of the gauge $\mathcal{N}=2$ spin $\mathbf{3}$ potentials is encompassed by the following contributions:
\begin{subequations}\label{spin 3 spin 3 sector}
    \begin{equation}
        h^{++(\alpha\beta)(\dot{\alpha}\dot{\beta})} = - 2i \theta^{+\rho} \bar{\theta}^{+\dot{\rho}} \Phi^{(\alpha\beta)(\dot{\alpha}\dot{\beta})}_{\rho\dot{\rho}}\,,
    \end{equation}
    \begin{equation}
        h^{++(\alpha\beta)\dot{\alpha}+} = -2i (\theta^+)^2 \bar{\theta}^{+\dot{\beta}} B^{(\alpha\beta)\dot{\alpha}}_{\dot{\beta}}\,,
    \end{equation}
    \begin{equation}
        h^{++(\dot{\alpha}\dot{\beta})\alpha+}
        =
        -2i (\bar{\theta}^+)^2 \theta^{+\beta} B_\beta^{(\dot{\alpha}\dot{\beta})\alpha}\,.
    \end{equation}
\end{subequations}
Here we used the notations from Appendix \ref{spin 3 bosonic sector}:
\begin{equation}
    \Phi_{(\alpha\beta)\gamma(\dot{\alpha}\dot{\beta})\dot{\gamma}} = \Phi_{(\alpha\beta\gamma)(\dot{\alpha}\dot{\beta}\dot{\gamma})}
    +
    \Phi_{(\alpha(\dot{\alpha}} \epsilon_{\beta)\gamma} \epsilon_{\dot{\beta})\dot{\gamma}}\,,
\end{equation}
\begin{eqnarray}
&& B_{(\alpha\beta)\dot{\alpha}\dot{\beta}} = - \frac{1}{2} \Big[ \partial^{\gamma\dot{\gamma}} \Phi_{(\alpha\beta\gamma)(\dot{\alpha}\dot{\beta}\dot{\gamma})} - \partial_{(\alpha(\dot{\alpha}} \Phi_{\beta)\dot{\beta})}
    - \partial_{(\alpha\dot{\alpha}} \Phi_{\beta)\dot{\beta}} \Big], \\
&& B_{\alpha\beta(\dot{\alpha}\dot{\beta})} = - \frac{1}{2} \Big[ \partial^{\gamma\dot{\gamma}} \Phi_{(\alpha\beta\gamma)(\dot{\alpha}\dot{\beta}\dot{\gamma})} - \partial_{(\alpha(\dot{\alpha}} \Phi_{\beta)\dot{\beta})}
    - \partial_{\alpha(\dot{\alpha}} \Phi_{\beta\dot{\beta})} \Big].
\end{eqnarray}
Using these fields and hypermultiplet equation in the spin 3 background \eqref{hyper eom spin 3}, we obtain for the on-shell hypermultiplet:
\begin{equation}\label{hyper q spin 3}
    q^+_{(s=3)} =
        f^i u^+_i
    +
    \theta^{+\alpha} \psi_\alpha
    +
    \bar{\theta}^+_{\dot{\alpha}} \bar{\kappa}^{+\dot{\alpha}}
    +
    (\theta^{\hat{+}})^2 mf^i u^-_i
    +
    2i \theta^{+\rho}\bar{\theta}^{+\dot{\rho}} \nabla^{(s=3)}_{\rho\dot{\rho}} f^i u^-_i.
\end{equation}
Here we introduced the spin 3 gauge-covariant derivative:
\begin{multline}
    \nabla^{(s=3)}_{\rho\dot{\rho}} f^i
    =
    \Big\{\partial_{\rho\dot{\rho}}     -i\kappa_3  \Phi_{\rho\dot{\rho}}^{(\alpha\beta)(\dot{\alpha}\dot{\beta})} \partial_{\alpha\dot{\alpha}} \partial_{\beta\dot{\beta}}
    -i\frac{\kappa_3}{2}  \left(\partial_{\alpha\dot{\alpha}}\Phi_{\rho\dot{\rho}}^{(\alpha\beta)(\dot{\alpha}\dot{\beta})} \right) \partial_{\beta\dot{\beta}}
    \\ -i\frac{\kappa_3}{2} (1+2\xi)
        \left[ \partial_{\alpha\dot{\alpha}} \partial_{\beta\dot{\beta}} \Phi^{(\alpha\beta)(\dot{\alpha}\dot{\beta})}_{\rho\dot{\rho}}
        +
        2 \partial^{\alpha\dot{\alpha}}B^{}_{(\alpha\rho) \dot{\rho}\dot{\alpha}}
        +
        2\partial^{\alpha\dot{\alpha}}B^{}_{\rho \alpha(\dot{\alpha}\dot{\rho})}
        \right]
    \\  -i\frac{\kappa_3}{2}
        \left[ \partial_{\beta\dot{\beta}} \Phi^{(\alpha\beta)(\dot{\alpha}\dot{\beta})}_{\rho\dot{\rho}}
        +
        2B^{(\alpha\dot{\alpha}}_{\rho) \dot{\rho}}
        +
        2B^{\alpha(\dot{\alpha}}_{\rho \dot{\rho})}
        \right] \partial_{\alpha\dot{\alpha}}
     \Big\} f^i.
\end{multline}

Using the expression \eqref{hyper q spin 3}, one can easily unfold the second and third lines of \eqref{action-final-spin3} in full analogy with the spin $\mathbf{1}$ and spin $\mathbf{2}$ cases:
\begin{equation}
    \begin{split}
        &-\int d\zeta^{(-4)} \; \tilde{q}^+_{(s=3)} \mathcal{D}^{++} q^+_{(s=3)}
        \\&=
        \int d^4x \; \left(-\eta^{nm}
        \nabla^{(s=3)}_n f^i \nabla^{(s=3}_m \bar{f}_i
        +
        \partial^m f^i \nabla^{(s=3)}_m \bar{f}_i
        +
        \nabla^{(s=3)}_m f^i \partial^m \bar{f}_i
        - m^2 f^i \bar{f}_i \right)
        \\&=
        \int d^4x\;
        \left(\partial^m f^i \partial_m \bar{f}_i
        - m^2 f^i \bar{f}_i\right)
        +
        \mathcal{O}(\kappa_3^2).
    \end{split}
\end{equation}
So these terms of \eqref{action-final-spin3} doe not contribute to the first-order cubic gauge $(3,0,0)$ coupling.

Substituting the expressions \eqref{spin 3 spin 3 sector} into the action \eqref{spin 3 on-shell superfield},
we obtain the component bosonic core of the $(3,0,0)$ cubic coupling as:
\begin{equation}\label{spin 3 action xi term}
    \begin{split}
    S^{(s=3)}_{int}=&-\frac{i\kappa_3}{4}\; \int d^4x\; \Phi^{\rho\dot{\rho}(\alpha\beta)(\dot{\alpha}\dot{\beta})}\Big(  f^i \partial_{\alpha\dot{\alpha}} \partial_{\beta\dot{\beta}} \partial_{\rho\dot{\rho}}  \bar{f}_i
    -  \partial_{\alpha\dot{\alpha}} \partial_{\beta\dot{\beta}} \partial_{\rho\dot{\rho}} f^i   \bar{f}_i
    \\&\qquad\qquad\;\;\;\;
    -
    \partial_{\rho\dot{\rho}} f^i \partial_{\alpha\dot{\alpha}} \partial_{\beta\dot{\beta}} \bar{f}_i
    +
    \partial_{\alpha\dot{\alpha}} \partial_{\beta\dot{\beta}}   f^i \partial_{\rho\dot{\rho}} \bar{f}_i
    \Big)
    \\&
    + \frac{i\kappa_3}{2} \int d^4x\; \Big(B^{(\alpha\beta)\dot{\alpha}\dot{\beta}} + B^{(\dot{\alpha}\dot{\beta})\alpha\beta}\Big)
    \Big(f^i \partial_{\alpha\dot{\alpha}} \partial_{\beta\dot{\beta}} \bar{f}_i
    -  \partial_{\alpha\dot{\alpha}} \partial_{\beta\dot{\beta}} f^i  \bar{f}_i \Big)
    \\&
     -
    \frac{i \xi \kappa_3}{2}\; \int d^4x\;
    \partial_{\alpha\dot{\alpha}}\Big[ \partial_{\beta\dot{\beta}} \Phi^{\rho\dot{\rho}(\alpha\beta)(\dot{\alpha}\dot{\beta})} + 2\Big( B^{(\alpha\beta)\dot{\alpha}\dot{\beta}}
    + B^{(\dot{\alpha}\dot{\beta})\alpha\beta}\Big)\Big] \Big(f^i\partial_{\rho\dot{\rho}} \bar{f}_i - \partial_{\rho\dot{\rho}} f^i \bar{f}_i \Big).
    \end{split}
\end{equation}
We now turn to the analysis of various terms in this expression. Some of such terms can be removed by a field redefinition, essentially simplifying the resulting vertex.

\textbf{1.} One of the basic relations we use is the following:
\begin{eqnarray}\label{BcalR}
&& \partial^{\alpha\dot{\alpha}} B_{(\alpha\beta)\dot{\alpha}\dot{\beta}} =
    \partial^{\alpha\dot{\alpha}} B_{\alpha\beta(\dot{\alpha}\dot{\beta})} = - \frac{1}{2} \mathcal{R}_{\beta\dot{\beta}} \nonumber \\
&& =\, - \frac{1}{2}\Big( \partial^{\alpha\dot{\alpha}}\partial^{\gamma\dot{\gamma}} \Phi_{(\alpha\beta\gamma)(\dot{\alpha}\dot{\beta}\dot{\gamma})}
    -
    \frac{1}{4} \partial_{\beta\dot{\beta}} \partial^{\alpha\dot{\alpha}} \Phi_{\alpha\dot{\alpha}} - \frac{5}{2} \Box \Phi_{\beta\dot{\beta}} \Big).
\end{eqnarray}
This is the linearized tensor (spin 3 ``scalar curvature'') defined in \eqref{spin 3 curvature 1}. So the last term in \eqref{spin 3 action xi term} is reduced to
\begin{equation}
   i\xi\kappa_3 \int d^4x\; \mathcal{R}^{\beta\dot{\beta}}  \Big( f^i \partial_{\beta\dot{\beta}} \bar{f}_i
    -
    \partial_{\beta\dot{\beta}}f^i \bar{f}_i  \Big).
\end{equation}
As explained in Appendix \ref{spin 3 redef gauge}, it  can be eliminated by a field redefinition \footnote{One can also note that on the free equation of motion for the spin 3 gauge field
one can eliminate this term thanks to the relation \eqref{R alpha dot alpha}. }.

\textbf{2.} The second line in \eqref{spin 3 action xi term} can also be partially rewritten in the more convenient form, integrating by parts and once again using  \eqref{BcalR}:
\begin{equation}\label{spin3II}
    \begin{split}
    &\int d^4x \;B^{(\dot{\alpha}\dot{\beta})(\alpha\beta)} \Big( f^i \partial_{\alpha\dot{\alpha}} \partial_{\beta\dot{\beta}} \bar{f}_i -  \partial_{\alpha\dot{\alpha}} \partial_{\beta\dot{\beta}} f^i  \bar{f}_i \Big)
     \\&\qquad =
     \frac{1}{2}  \; \int d^4x\; \left(\mathcal{R}^{\alpha\dot{\alpha}} - \frac{1}{4}\partial^{\alpha\dot{\alpha}} \partial_{\beta\dot{\beta}} \Phi^{\beta\dot{\beta}} + \frac{1}{2} \Box \Phi^{\alpha\dot{\alpha}} \right)
     \Big(f^i \partial_{\alpha\dot{\alpha}} \bar{f}_i - \partial_{\alpha\dot{\alpha}} f^i \bar{f}_i \Big).
    \end{split}
\end{equation}
This term can be eliminated through a field redefinition described in Appendices \ref{spin 3 redef scalar} and \ref{spin 3 redef gauge}.

\textbf{3.} Now it becomes possible to clarify the role of the $\xi$-interaction. The remaining $\xi$-term with $\Phi$ in the last line of \eqref{spin 3 action xi term}
can be reduced to the form:
\begin{equation}
    \begin{split}
        &\int d^4x\;
        \partial_{\alpha\dot{\alpha}} \partial_{\beta\dot{\beta}} \Phi^{\rho\dot{\rho}(\alpha\beta)(\dot{\alpha}\dot{\beta})} \Big(f^i\partial_{\rho\dot{\rho}} \bar{f}_i - \partial_{\rho\dot{\rho}} f^i \bar{f}_i \Big)
        \\&=
        \int d^4x\;
        \Big[
        \partial_{\alpha\dot{\alpha}} \partial_{\beta\dot{\beta}} \Phi^{(\alpha\beta\rho)(\dot{\alpha}\dot{\beta}\dot{\rho})}
        +
        \frac{1}{2}
        \partial^{\rho\dot{\rho}} \partial_{\alpha\dot{\alpha}} \Phi^{\alpha\dot{\alpha}}
        +
        \frac{1}{2} \partial^{\rho\dot{\alpha}} \partial^{\alpha\dot{\rho}} \Phi_{\alpha\dot{\alpha}}
        \Big]
         \Big(f^i\partial_{\rho\dot{\rho}} \bar{f}_i - \partial_{\rho\dot{\rho}} f^i \bar{f}_i \Big).
    \end{split}
\end{equation}
The first and second terms in this expression can be eliminated  in the way  described in Appendices \ref{spin 3 redef scalar} and \ref{spin 3 redef gauge}. Let us consider the third term:
\begin{equation}
    \begin{split}
    &\int d^4x\; \Phi_{\alpha\dot{\alpha}}\, \partial^{\rho\dot{\alpha}} \partial^{\alpha\dot{\rho}} \, \Big(f^i\partial_{\rho\dot{\rho}} \bar{f}_i - \partial_{\rho\dot{\rho}} f^i \bar{f}_i \Big)
    \\&= \int d^4x\;  \Phi^{\alpha\dot{\alpha}}
    \Box \Big(f^i\partial_{\alpha\dot{\alpha}} \bar{f}_i - \partial_{\alpha\dot{\alpha}} f^i \bar{f}_i \Big)
    +
    2\int d^4x\; \partial_{\alpha\dot{\alpha}} \Phi^{\alpha\dot{\alpha}}
    \Big( \Box f^i \bar{f}_i -  f^i \Box \bar{f}_i \Big).
        \end{split}
\end{equation}
The first piece here can be eliminated  by the scalar field redefinition described in Appendix \ref{spin 3 redef scalar}.
On shell, using the equation $(\Box + m^2) f^i = 0$, one can also eliminate the second piece.

\medskip

As the result of our analysis, we conclude that the $\xi$-term can be completely removed on shell  through various field redefinitions.
Finally, the  $(3,0,0)$ coupling is reduced to the form:
\begin{equation}\label{(300) vertex}
        \begin{split}
            S^{(s=3)}_{int}=&-\frac{i\kappa_3}{4}\; \int d^4x\; \Phi^{(\alpha\beta)\rho(\dot{\alpha}\dot{\beta})\dot{\rho}}
            \Bigr(  f^i \partial_{\alpha\dot{\alpha}} \partial_{\beta\dot{\beta}} \partial_{\rho\dot{\rho}}  \bar{f}_i
            -  \partial_{\alpha\dot{\alpha}} \partial_{\beta\dot{\beta}} \partial_{\rho\dot{\rho}} f^i   \bar{f}_i
            \\&\qquad\qquad\qquad\qquad\qquad\qquad\quad\;\;\;\;
            -
            \partial_{\rho\dot{\rho}} f^i \partial_{\alpha\dot{\alpha}} \partial_{\beta\dot{\beta}} \bar{f}_i
            +
            \partial_{\alpha\dot{\alpha}} \partial_{\beta\dot{\beta}}   f^i \partial_{\rho\dot{\rho}} \bar{f}_i
            \Bigr).
        \end{split}
\end{equation}
It has the standard Noether form \eqref{Noether form interaction}:
\begin{equation}
    S^{(s=3)}_{int}
    =
    \kappa_3 \int d^4x\,
    \Phi^{(\alpha\beta)\rho(\dot{\alpha}\dot{\beta})\dot{\rho}}
    Y_{(\alpha\beta)\rho(\dot{\alpha}\dot{\beta})\dot{\rho}},
\end{equation}
where the corresponding conserved current (in massless case) is:
\begin{equation}
    Y_{(\alpha\beta)\rho(\dot{\alpha}\dot{\beta})\dot{\rho}}
    =
    -\frac{i}{4}
    \Bigr(  f^i \partial_{(\alpha\dot{\alpha}} \partial_{\beta)\dot{\beta}} \partial_{\rho\dot{\rho}}  \bar{f}_i
    -  \partial_{(\alpha\dot{\alpha}} \partial_{\beta)\dot{\beta}} \partial_{\rho\dot{\rho}} f^i   \bar{f}_i
    -
    \partial_{\rho\dot{\rho}} f^i \partial_{(\alpha\dot{\alpha}} \partial_{\beta)\dot{\beta}} \bar{f}_i
    +
    \partial_{(\alpha\dot{\alpha}} \partial_{\beta)\dot{\beta}}   f^i \partial_{\rho\dot{\rho}} \bar{f}_i
    \Bigr).
\end{equation}
This current coincides with the conserved component \eqref{spin 3 bosonic current} of the spin $\mathbf{3}$ current superfield \eqref{spin 3 superfield current} for the $\xi=0$ case.

\medskip
It is useful to further transform the action \eqref{(300) vertex} with the help of the relation (valid up to total derivatives):
\begin{eqnarray}
&&\int d^4x\;
    \partial_{\alpha\dot{\alpha}} \partial_{\beta\dot{\beta}} \Phi^{\rho\dot{\rho}(\alpha\beta)(\dot{\alpha}\dot{\beta})} \Big(f^i\partial_{\rho\dot{\rho}} \bar{f}_i - \partial_{\rho\dot{\rho}} f^i \bar{f}_i \Big) \nonumber \\
&&\Rightarrow\, \int d^4x\;
 \Phi^{\rho\dot{\rho}(\alpha\beta)(\dot{\alpha}\dot{\beta})} \partial_{\alpha\dot{\alpha}} \partial_{\beta\dot{\beta}} \Big( f^i\partial_{\rho\dot{\rho}} \bar{f}_i - \partial_{\rho\dot{\rho}} f^i \bar{f}_i \Big) \nonumber \\
 && =\, \int d^4x\; \Phi^{\rho\dot{\rho}(\alpha\beta)(\dot{\alpha}\dot{\beta})}\Bigr(  f^i \partial_{\alpha\dot{\alpha}} \partial_{\beta\dot{\beta}} \partial_{\rho\dot{\rho}}  \bar{f}_i
 -  \partial_{\alpha\dot{\alpha}} \partial_{\beta\dot{\beta}} \partial_{\rho\dot{\rho}} f^i   \bar{f}_i \nonumber \\
 && \qquad\qquad\qquad\qquad\qquad
 -
 \partial_{\rho\dot{\rho}} f^i \partial_{\alpha\dot{\alpha}} \partial_{\beta\dot{\beta}} \bar{f}_i
 +
 \partial_{\alpha\dot{\alpha}} \partial_{\beta\dot{\beta}}   f^i \partial_{\rho\dot{\rho}} \bar{f}_i
 \Bigr) \nonumber \\
 && \qquad+ 2 \int d^4x \; \Phi^{\rho\dot{\rho}(\alpha\beta)(\dot{\alpha}\dot{\beta})} \Big( \partial_{\alpha\dot{\alpha}} f^i  \partial_{\beta\dot{\beta}}\partial_{\rho\dot{\rho}} \bar{f}_i
 - \partial_{\alpha\dot{\alpha}} \partial_{\rho\dot{\rho}} f^i \partial_{\beta\dot{\beta}} \bar{f}_i \Big).
 \end{eqnarray}
On the one hand, this term as a whole can be omitted by properly redefining the fields and using the equations of motion.
On the other hand, using this relation, one can cast the cubic $(3,0,0)$ vertex in the form:
\begin{equation}
    \begin{split}
        S^{(s=3)}_{int}=&\;\frac{i\kappa_3}{2}\; \int d^4x\; \Phi^{(\alpha\beta)\rho(\dot{\alpha}\dot{\beta})\dot{\rho}}
        \left( \partial_{\alpha\dot{\alpha}} f^i  \partial_{\beta\dot{\beta}}\partial_{\rho\dot{\rho}} \bar{f}_i - \partial_{\alpha\dot{\alpha}} \partial_{\rho\dot{\rho}} f^i \partial_{\beta\dot{\beta}} \bar{f}_i  \right)
        \\=&\; \frac{i\kappa_3}{2}\; \int d^4x\; \Biggr[ \Phi^{(\alpha\beta\rho)(\dot{\alpha}\dot{\beta}\dot{\rho})}
        \left( \partial_{\alpha\dot{\alpha}} f^i  \partial_{\beta\dot{\beta}}\partial_{\rho\dot{\rho}} \bar{f}_i - \partial_{\alpha\dot{\alpha}} \partial_{\rho\dot{\rho}} f^i \partial_{\beta\dot{\beta}} \bar{f}_i  \right)
        \\& \qquad\qquad\qquad\qquad+ \Phi^{\alpha\dot{\alpha}} \left(\partial_{\alpha\dot{\alpha}}f^i \Box \bar{f}_i - \Box f^i \partial_{\alpha\dot{\alpha}}\bar{f}_i\right)
        \\& \qquad\qquad\qquad\qquad+
         \frac{1}{4} \Phi^{\alpha\dot{\alpha}} \left(\partial^{\beta\dot{\beta}} f^i \partial_{\alpha\dot{\alpha}} \partial_{\beta\dot{\beta}} \bar{f}_i
        -
         \partial_{\alpha\dot{\alpha}} \partial_{\beta\dot{\beta}}f^i \partial^{\beta\dot{\beta}} \bar{f}_i  \right) \Biggr].
    \end{split}
\end{equation}
This vertex exactly coincides with the (3,0,0) vertex given in \cite{Zinoviev:2010cr}.

\medskip

\textbf{\underline{\textit{Spin 2 sector}}}

\medskip

The spin 2 sector is spanned by the following pieces in the analytic potentials:
\begin{subequations}
\begin{equation}\label{AA}
    h^{++5\alpha\dot{\alpha}} = -2i \theta^{+\rho} \bar{\theta}^{+\dot{\rho}} C_{\rho\dot{\rho}}^{\alpha\dot{\alpha}}\,,
\end{equation}
\begin{equation}\label{BB}
    h^{++\dot{\alpha}(\alpha\beta)+} = i (\bar{\theta}^+)^2 \theta^+_\rho H^{\dot{\alpha}\rho (\alpha\beta)}\,,
\end{equation}
\begin{equation}\label{CC}
    h^{++\alpha(\dot{\alpha}\dot{\beta})} = i (\theta^+)^2 \bar{\theta}^+_{\dot{\rho}} \bar{H}^{\alpha\dot{\rho} (\dot{\alpha}\dot{\beta})}\,.
\end{equation}
\end{subequations}
Here we used the notations \eqref{H1}, \eqref{H2} and \eqref{H3}.

The relevant contribution to the on-shell hypermultiplet is:
\begin{equation}
    \begin{split}
    q^{+}_{(s=3)}
    =&
        f^i u^+_i
        +
        \theta^{+\alpha} \psi_\alpha
        +
        \bar{\theta}^+_{\dot{\alpha}} \bar{\kappa}^{+\dot{\alpha}}
        +
        (\theta^{+})^2 \left[mf^i u^-_i + \kappa_3 M^- \right]
        \\&-
        (\bar{\theta}^{+})^2 \left[ mf^i u^-_i + \kappa_3N^- \right]
        +
        2i \theta^{+\rho}\bar{\theta}^{+\dot{\rho}} \nabla^{(s=3)}_{\rho\dot{\rho}} f^i u^-_i.
        \end{split}
\end{equation}
Here we introduced the spin 2 covariant derivative in the spin $\mathbf{3}$ ${\cal N}=2$ theory:
\begin{equation}
    \nabla^{(s=3)}_{\rho\dot{\rho}} f^i
    =
    \Bigr(\partial_{\rho\dot{\rho}}     +m\kappa_3  C_{\rho\dot{\rho}}^{\alpha\dot{\alpha}} \partial_{\alpha\dot{\alpha}}  F^+
+\frac{1}{2}m\kappa_3  (\partial_{\alpha\dot{\alpha}} C_{\rho\dot{\rho}}^{\alpha\dot{\alpha}})\Bigr)
    f^i\,,
\end{equation}
as well as, defined:
\begin{equation}
    N^- = - M^- = - \xi  \mathcal{R} f^i u^-_i
    +
        \frac{3}{2} \partial_{\alpha\dot{\alpha}} (H^{\alpha\dot{\alpha}}  f^i u^-_i).
\end{equation}
Here we used the notation $\mathcal{R}$ \eqref{scalar curvature}, with replacing $\Phi\to C$.

Now it is direct to show that these terms do not contribute to the cubic $(2,0,0)$ vertex:
\begin{equation}
    -\int d\zeta^{(-4)}\, \tilde{q}^{+}_{(s=3)}\mathcal{D}^{++} q^+_{(s=3)}
    =
    \int d^4x\;
    \left(\partial^m f^i \partial_m \bar{f}_i - m^2 f^i \bar{f}_i\right)
    +
    \mathcal{O}(\kappa_3^2).
\end{equation}

Thus the minimal coupling $(2,0,0)$ in the spin 2 sector (up to a total derivative and discarding fermionic terms) is generated as:
\begin{equation}
    \begin{split}
    S_{int-min}^{(s=3)} = & \frac{i\kappa_3}{2} \int d\zeta^{(-4)}\; \left(  q_{(0)}^+ \hat{\mathcal{H}}^{++}_{(s=3)} \tilde{q}_{(0)}^+
    +
    \tilde{q}_{(0)}^+ \hat{\mathcal{H}}^{++}_{(s=3)} q_{(0)}^+  \right)
    \\=&
    -
    \frac{\kappa_3m}{4}
    \int d^4x\;
    C^{\alpha\beta\dot{\alpha}\dot{\beta}} \left(
    \partial_{\alpha\dot{\alpha}} \partial_{\beta\dot{\beta}} f^i \bar{f}_i
    -
     \partial_{\alpha\dot{\alpha}}  f^i \partial_{\beta\dot{\beta}} \bar{f}_i
     -
    \partial_{\beta\dot{\beta}} f^i
     \partial_{\alpha\dot{\alpha}}  \bar{f}_i
    +  f^i \partial_{\alpha\dot{\alpha}} \partial_{\beta\dot{\beta}}\bar{f}_i  \right)
    \\&+
    \frac{\kappa_3m}{4}
    \int d^4x\;
    \underbrace{\left(
    \partial_{\alpha\dot{\alpha}} \bar{H}^{\alpha(\dot{\alpha}\dot{\beta})}_{\dot{\beta}}
    -
    \partial_{\alpha\dot{\alpha}}H^{\dot{\alpha}(\alpha\beta)}_\beta
    \right) }_{8\mathcal{R}} f^i\bar{f}_i\,.
        \end{split}
\end{equation}
The superfield form of $\Gamma^{++}_{(s=3)}$ on the background \eqref{BB} and \eqref{CC} is reduced to
\begin{equation}
    \begin{split}
    \Gamma^{++}_{(s=3)} =& \partial_{\alpha\dot{\alpha}} \left(- \partial^-_\beta h^{++(\alpha\beta)\dot{\alpha}} - \partial^-_{\dot{\beta}} h^{++(\dot{\alpha}\dot{\beta})\alpha}\right)
    \\=& i (\bar{\theta}^+)^2 \partial_{\alpha\dot{\alpha}}H^{\dot{\alpha}(\alpha\beta)}_\beta
    +
     i (\theta^+)^2 \partial_{\alpha\dot{\alpha}}\bar{H}^{\alpha(\dot{\alpha}\dot{\beta})}_{\dot{\beta}}
     \\=& - 4 i (\bar{\theta}^+)^2  \mathcal{R}
     +
     4i (\theta^+)^2 \mathcal{R}
     = 4i (\theta^{\hat{+}})^2 \mathcal{R}\,.
        \end{split}
\end{equation}
From this expression one can easily deduce the contribution of $\xi$-term to the spin 2 sector:
\begin{equation}
    \begin{split}
S_{int-\xi}^{(s=3)} \sim \int d^4xd^4\theta^+\;    \Gamma^{++}_{s=3} \; \tilde{q}^+_{(0)} q^+_{(0)}
    &=
     4i \int d^4xd^4\theta^+\;  (\theta^{\hat{+}})^2 \mathcal{R}
     \cdot m\, (\theta^{\hat{+}})^2 f^i \bar{f}_i
     \\&=
     -8i \int d^4x\;  m\, \mathcal{R}\, f^i \bar{f}_i\,.
        \end{split}
\end{equation}

Putting together all the above expressions, we can write the total  $(2,0,0)$ vertex in the spin 2 sector of $\mathcal{N}=2$ spin $\mathbf{3}$ theory:
\begin{equation}\label{Spin32}
    \begin{split}
    S_{int}^{(s=3)}
    =&
    -\frac{\kappa_3m}{4}
    \int d^4x\;
    C^{\alpha\beta\dot{\alpha}\dot{\beta}} \left(
    \partial_{\alpha\dot{\alpha}} \partial_{\beta\dot{\beta}} f^i \bar{f}_i
    -
    2 \partial_{\alpha\dot{\alpha}}  f^i \partial_{\beta\dot{\beta}} \bar{f}_i
    +  f^i \partial_{\alpha\dot{\alpha}} \partial_{\beta\dot{\beta}}\bar{f}_i  \right)
    \\&+
    2\kappa_3m
    \int d^4x\;
    \mathcal{R} f^i\bar{f}_i
    -
    8\kappa_3m \cdot \xi
    \int d^4x\;
    \mathcal{R} f^i\bar{f}_i\,.
    \end{split}
\end{equation}
Comparing it with the results of Appendix \ref{Minimal gravity coupling}, we conclude that \eqref{Spin32} contains a non-minimal interaction of the scalar field
with the spin 2 field (2nd  line) and this deviation from the minimal interaction includes, in particular, the term proportional to the parameter $\xi$.
However, using the results of Appendix \ref{Redefinition of spin 2 gauge field}, one can remove on shell both non-minimal terms in \eqref{Spin32} by a proper field redefinition.

Thus we conclude that in the cubic bosonic vertices which come from the superfield cubic vertex involving the spin ${\bf 3}$ ${\cal N}=2$ gauge superfield,
the dependence on $\xi$ can be fully eliminated on shell.

Finally, the spin 2 coupling to scalars can be rewritten in the  Noether form \eqref{Noether form interaction}:
\begin{equation}\label{Spin32-1}
    \begin{split}
        S_{int}^{(s=3)}
        =&
        -\frac{\kappa_3m}{4}
        \int d^4x\;
        C^{\alpha\beta\dot{\alpha}\dot{\beta}} \left(
        \partial_{\alpha\dot{\alpha}} \partial_{\beta\dot{\beta}} f^i \bar{f}_i
        -
         \partial_{\alpha\dot{\alpha}}  f^i \partial_{\beta\dot{\beta}} \bar{f}_i
        -
         \partial_{\beta\dot{\beta}}  f^i   \partial_{\alpha\dot{\alpha}}  \bar{f}_i
        +  f^i \partial_{\alpha\dot{\alpha}} \partial_{\beta\dot{\beta}}\bar{f}_i  \right),
    \end{split}
\end{equation}
with
\begin{equation}
    Y_{\alpha\beta\dot{\alpha}\dot{\beta}}
    =
   - \frac{m}{4}
    \left(
    \partial_{\alpha\dot{\alpha}} \partial_{\beta\dot{\beta}} f^i \bar{f}_i
    -
    \partial_{\alpha\dot{\alpha}}  f^i \partial_{\beta\dot{\beta}} \bar{f}_i
    -
    \partial_{\beta\dot{\beta}}  f^i   \partial_{\alpha\dot{\alpha}}  \bar{f}_i
    +  f^i \partial_{\alpha\dot{\alpha}} \partial_{\beta\dot{\beta}}\bar{f}_i  \right).
\end{equation}
This current coincides with the conserved current \eqref{Supercurrent 2 - spin 3-component} (modulo the mass coefficient) entering the current superfield \eqref{Supercurrent 2 - spin 3}.

\medskip
\textbf{\underline{\textit{Spin $5/2$ sector}}}
\medskip

The analysis of the spinor sector can be performed in exactly the same way and here we do not give the relevant details.
The interaction has also the generic Noether form \eqref{Noether form interaction}:
\begin{equation}
    S_{int}^{(s=3)}
    =
    \kappa_3
    \int d^4x\,
    \left(
    \psi^{(\alpha\beta)(\dot{\alpha}\dot{\beta})\gamma i} Y_{(\alpha\beta)(\dot{\alpha}\dot{\beta})\gamma i}
    +
    \bar{\psi}^{(\alpha\beta)(\dot{\alpha}\dot{\beta})\dot{\gamma} i} \bar{Y}_{(\alpha\beta)(\dot{\alpha}\dot{\beta})\dot{\gamma} i}
    \right),
\end{equation}
 where the corresponding fermionic spin $\tfrac{5}{2}$ $Y$-current (in massless case), up to numerical factor, coincides with the
 conserved spin $\tfrac{5}{2}$ $\mathcal{S}$-current, defined in \eqref{S cal spin 52}:
\begin{equation}
    \begin{split}
    Y^i_{(\alpha\beta)(\dot{\alpha}\dot{\rho})\rho }
    =
    -
    \frac{1}{4}
    \mathcal{S}^i_{(\alpha\beta)(\dot{\alpha}\dot{\rho})\rho }
    =
    -\frac{i}{4}
    \chi_\rho \partial_{(\alpha\dot{\alpha}} \partial_{\beta)\dot{\rho}} \bar{f}^i
    -
    \frac{i}{4}
     \psi_\rho \partial_{(\alpha\dot{\alpha}} \partial_{\beta)\dot{\rho}} f^i.
    \end{split}
\end{equation}
The overall coefficient is fixed by inspecting the contribution of symmetric part
of the spin $\tfrac{5}{2}$ gauge field $\psi^{(\alpha\beta\gamma)(\dot{\alpha}\dot{\beta})i}$. It solely comes from the analytic
potential $h^{++(\alpha\beta)(\dot{\alpha}\dot{\beta})}$.

Other spin $\tfrac{5}{2}$ currents defined in Section \ref{spin 3 supercurrent} differ by terms which are conserved automatically,
without use of the equations of motion, and by terms which are derivatives of the conserved spin $\tfrac{3}{2}$ currents.
This type of terms results in fake cubic $(\tfrac{5}{2}, \tfrac{1}{2}, 0)$ vertices which can be eliminated by tricks much similar to
those employed in Appendix \ref{Fields redefinitions} for bosonic vertices. We do not give here any details.

\section{Summary and outlook}
\label{Summary and outlook}

Here we briefly summarize the basic results and discuss some further problems.

\begin{enumerate}
    \item[\bf{1.}] In Section \ref{equations of motion}, we have derived superfield equations of
    motion for an arbitrary $\mathcal{N}=2$ spin $\mathbf{s}$. They have the universal form:
    \begin{equation}\label{final EOM free}
        \begin{cases}
            (D^+)^2 \bar{D}^{+\dot{\beta}} G^{--}_{\alpha(s-1) (\dot{\beta}\dot{\alpha}(s-2))} + (\bar{D}^+)^2 D^+_{(\alpha} G^{--}_{\alpha(s-2) )  \dot{\alpha}(s-2) } = 0\,,\\
            (\bar{D}^+)^2 D^{+\beta} G^{--}_{(\beta\alpha(s-2))\dot{\alpha}(s-1)}  -  (D^+)^2 \bar{D}^+_{(\dot{\alpha}} G^{--}_{\alpha(s-2)  \dot{\alpha}(s-2))} = 0\,.
        \end{cases}
    \end{equation}
We have derived these equations by varying with respect  either to the analytic potentials $h^{++\dots}(\zeta)$ or to unconstrained
Mezincescu-type prepotentials $\Psi^{-\dots}(\zeta, \theta^-)$. These prepotentials are defined as\footnote{Here we assume the complete symmetry between dotted and undotted spinor indices.}
    \begin{equation}
        \begin{split}
    \hat{\mathcal{H}}^{++}_{(s)} =&\; h^{++\alpha(s-2)\dot{\alpha}(s-2)M}\partial_M \partial^{s-2}_{\alpha(s-2)\dot{\alpha}(s-2)}
    \\:=&\; (D^+)^4 \left(\Psi^{-\alpha(s-2)\dot{\alpha}(s-2)\hat{\mu}} D^-_{\hat{\mu}}\right) \partial^{s-2}_{\alpha(s-2)\dot{\alpha}(s-2)}
         \end{split}
\end{equation}
and provide alternative representations for $\mathcal{N}=2$ integer spin $\mathbf{s}$ supermultiplets.
We have also tracked in detail how these superfield equations yield the ordinary equations of
motion in the bosonic sector.

\item[\bf{2.}] In the presence of cubic couplings \eqref{action-final} of hypermultiplet to
$\mathcal{N}=2$ spin $\mathbf{s}$ supermultiplets the equations \eqref{final EOM free} acquire the superfield sources constructed as bi-linear combinations of the hypermultiplet superfield:
\begin{equation}\label{final EOM source}
    \begin{cases}
        (D^+)^2 \bar{D}^{+\dot{\beta}} G^{--}_{\alpha(s-1) (\dot{\beta}\dot{\alpha}(s-2))} + (\bar{D}^+)^2 D^+_{(\alpha} G^{--}_{\alpha(s-2) )  \dot{\alpha}(s-2) } = i  \kappa_s (-1)^s
        \mathcal{J}^+_{\dot{\alpha}(s-2)\alpha(s-1)} \,,\\
        (\bar{D}^+)^2 D^{+\beta} G^{--}_{(\beta\alpha(s-2))\dot{\alpha}(s-1)}  -  (D^+)^2 \bar{D}^+_{(\dot{\alpha}} G^{--}_{\alpha(s-2)  \dot{\alpha}(s-2))} = i  \kappa_s (-1)^s \mathcal{J}^+_{\alpha(s-2)\dot{\alpha}(s-1)}\,.
    \end{cases}
\end{equation}
The conserved current superfieds $\mathcal{J}^+_{\dots}$  are
defined as
\begin{equation}
    \mathcal{J}^{+}_{\alpha(s-2)\dot{\alpha}(s-2)\hat{\mu}}   =
    -
    \frac{1}{2}
    q^{+a}   D^-_{\hat{\mu}} \partial^{s-2}_{\alpha(s-2)\dot{\alpha}(s-2)} (J)^{P(s)} q^+_a
    -
    \frac{1}{2}
    \xi  D^-_{\hat{\mu}} \partial^{s-2}_{\alpha(s-2)\dot{\alpha}(s-2)} \left[ q^{+a} (J)^{P(s)}  q^+_a \right].
\end{equation}
We have analyzed the  bosonic component structure of these currents and found the explicit expressions for the components
in terms of the on-shell hypermultiplet fields.

\item[\bf{3.}] In Section \ref{Hypermultiplet equations of motion and component reduction of cubic couplings} we have derived the equations of motion for the hypermultiplet coupled to $\mathcal{N}=2$ spin $\mathbf{s}$ gauge supermultiplet. We analyzed the component contents of the spin $\mathbf{2}$ and spin
$\mathbf{3}$ models in the bosonic sectors and found the full agreement with the standard results for the corresponding cubic vertices. In particular, we demonstrated that on shell and after proper field redefinitions, one
can eliminate the $\xi$-interactions which are present in the cubic vertices involving odd spin ${\cal N}=2$ gauge superfields. It is still an open question whether it is possible to eliminate such $\xi$ interaction at the
full off-shell superfield level. We have explicitly presented, in the leading order in the coupling constants, the on-shell component form of the superfield cubic vertices involving the spin ${\bf 2}$ and ${\bf 3}$ gauge
supermultiplets.

\end{enumerate}

Some further possible extensions of the results obtained can be summarized as follows:

\begin{itemize}

    \item One of the urgent problems is a generalization to the superconformal case.
    It is likely that in the superconformal case the relevant structures cannot significantly differ from the non-conformal case, as implied by
    $\mathcal{N}=2$ conformal supergravity in the HSS approach. Indeed, $\mathcal{N}=2$ Weyl multiplet
    \cite{Galperin:1987ek}
    is described in a way very similar to the formulation of $\mathcal{N}=2$ Einstein supergravity \cite{Galperin:1987em}; one just needs
    to further covariantize the analyticity-preserving harmonic derivative by an extra vielbein associated with local ${\rm SU}(2)_{aut}$ symmetry.
    The concrete generalization of such a picture to higher spins will of course require some efforts. Once this is done, a generalization of $\mathcal{N}=2$ higher spin theory
    to superconformally flat backgrounds, including the most interesting (A)dS$_d$ option, will hopefully be performed rather straightforwardly.

\item The task of generalizing the  results obtained to other non-trivial supergravity backgrounds is indeed of significant interest.
In particular, such theories in (A)dS$_d$ could display a number of features characteristic of the fully consistent higher spin theories.
How these structures are changed and constrained in the presence of $\mathcal{N}=2$ supersymmetry is an important open question.

    It is worth noting that such theories have already been discussed in ref. \cite{Kuzenko:2021pqm}
    (see also \cite{Buchbinder:2018nkp} and a recent paper \cite{KPR}), without use of the harmonic superspace approach. These  theories seemingly
    correspond to the choice of special gauge for the analytic potentials.
    We hope to address these questions elsewhere \footnote{Let us point out once more that in all these papers the current superfields
    were constructed within the on-shell description of the hypermultiplet in terms of the properly constrained conventional ${\cal N}=2$ (or ${\cal N}=1$) superfields.
    The basic distinction of our approach is that we use the off-shell description of both higher spin gauge ${\cal N}=2$ multiplets and the matter hypermultiplets
    in terms of unconstrained harmonic analytic superfields.}.

\item Quartic interactions (see, e.g. \cite{Karapetyan:2021wdc} and the references therein)
also constitute a very interesting area for application of the
strategy adopted here and in our previous work
\cite{Buchbinder:2022kzl}. This circle of problems would involve, in
particular, finding out nonlinear corrections to the linearized
higher spin $\mathcal{N}=2$ superfield actions. Note that
the question of the structure of quartic higher spin interactions
may be closely related to the generic problem of locality in higher
spin theory (see a recent discussion of this issue in
\cite{Vasiliev}, \cite{Didenko1}).
\end{itemize}

\medskip

\acknowledgments

We thank X. Bekaert, V.A. Krykhtin, T.V. Snegirev, M. Tsulaia and
Y.M. Zinoviev for useful discussions and S.M. Kuzenko for valuable comments.
Work of I.B. was supported in part by the Ministry of Education of the Russian
Federation, project QZOY-2023-0003.
Work of N.Z. was partially supported by the grant 22-1-1-42-2 from the Foundation for the Advancement
of Theoretical Physics and Mathematics ``BASIS''.

\appendix

\section{Component reduction of bosonic gauge sector} \label{Component reduction of bosonic gauge sector}

In this Appendix  we present details of the component structure of the theories of ${\cal N}=2$ spins $\mathbf{2}$ and $\mathbf{3}$. We will use the results
of our previous work \cite{Buchbinder:2021ite}.

\subsection{${\cal N}=2$ spin $\mathbf{2}$ theory: bosonic sector}
\label{spin 2 theory -- components}

We start with collecting all the necessary details related to the bosonic sector of the spin $\mathbf{2}$ ${\cal N}=2$ theory. It consists of the spin 2 and spin 1 sectors.
Particular attention is paid to
the role of auxiliary fields in deriving the correct component equations from the superfield equations.

\underline{\textbf{\textit{Spin 2 sector}}}

\medskip

The bosonic spin 2 sector involves three relevant fields: the gauge spin 2 field  $\Phi^{\alpha\dot{\alpha}}_{\beta\dot{\beta}}$ and auxiliary fields $P_{\alpha\dot{\alpha}}$, $\bar{P}_{\alpha\dot{\alpha}}$.
In the spinor notation, the irreducible parts of the spin 2 field are defined by the decomposition:
\begin{equation}\label{Spin2Dec}
    \Phi_{\alpha\beta\dot{\alpha}\dot{\beta}} = \Phi_{(\alpha\beta)(\dot{\alpha}\dot{\beta})}
    +
    \epsilon_{\alpha\beta} \epsilon_{\dot{\alpha}\dot{\beta}} \Phi\,.
\end{equation}
The spin 2 gauge transformations with parameters $a_{\alpha\dot{\alpha}}$ have the form:
\begin{equation}\label{spin 2 gauge transformations}
    \delta \Phi_{\alpha\beta\dot{\alpha}\dot{\beta}}
    =
    \frac{1}{2} \left(\partial_{\alpha\dot{\alpha}} a_{\beta\dot{\beta}} + \partial_{\beta\dot{\beta}} a_{\alpha\dot{\alpha}} \right)\,\qquad \Leftrightarrow \qquad
    \begin{cases}
        \delta \Phi_{(\alpha\beta)(\dot{\alpha}\dot{\beta})} =
        \partial_{(\alpha(\dot{\alpha}} a_{\beta)\dot{\beta})}\,,\\
        \delta \Phi = \frac{1}{4}\partial_{\alpha\dot{\alpha}} a^{\alpha\dot{\alpha}} \,.
    \end{cases}
\end{equation}
These transformations define the residual bosonic symmetries of the Wess-Zumino gauge \eqref{WZ3 s=2} and fully agree with the spin 2 gauge transformations in Fronsdal theory.

The $\Phi, P$ sector of the superfield $G^{++\alpha\dot{\alpha}}$ is as follows:
\begin{equation}\label{G++ P}
    G^{++\alpha\dot{\alpha}}_{(\Phi,P)} =
    -
    2i \theta^{+\beta} \bar{\theta}^{+\dot{\beta}} \Phi_{\beta\dot{\beta}}^{\alpha\dot{\alpha}}
    +
    4 (\theta^+)^2 \bar{\theta}^{+\dot{\beta}} \bar{\theta}^{-\dot{\alpha}} P_{\dot{\beta}}^\alpha
    -
    4 (\bar{\theta}^+)^2 \theta^{+\beta} \theta^{-\alpha} \bar{P}_\beta^{\dot{\alpha}}\,.
\end{equation}
The corresponding solution of the zero-curvature equation \eqref{zero-curv} is given by:
\begin{equation}\label{G-- solution}
    \begin{split}
        G^{--\alpha\dot{\alpha}}_{(\Phi,P)} =&
        -
        2i \theta^{-\beta} \bar{\theta}^{-\dot{\beta}} \Phi_{\beta\dot{\beta}}^{\alpha\dot{\alpha}}
        \\&
        +
        2 (\theta^-)^2 \bar{\theta}^{-(\dot{\rho}} \bar{\theta}^{+\dot{\beta})} \partial_{\dot{\rho}}^\beta \Phi_{\beta\dot{\beta}}^{\alpha\dot{\alpha}}
        -
        2 (\bar{\theta}^-)^2 \theta^{-(\rho} \bar{\theta}^{+\beta)} \partial_{\rho}^{\dot{\beta}} \Phi_{\beta\dot{\beta}}^{\alpha\dot{\alpha}}
        \\&+ 2i (\theta^-)^2 (\bar{\theta}^-)^2 \theta^{+\rho} \bar{\theta}^{+\dot{\rho}} \left(\Box \Phi^{\alpha\dot{\alpha}}_{\rho\dot{\rho}}
        - \frac{1}{2} \partial_{\rho\dot{\rho}} \partial^{\beta\dot{\beta}} \Phi_{\beta\dot{\beta}}^{\alpha\dot{\alpha}} \right)
        \\&+
        4 (\theta^-)^2 \bar{\theta}^{+(\dot{\alpha}} \bar{\theta}^{-\dot{\beta})} P_{\dot{\beta}}^\alpha
        -
        2 (\theta^-)^2 (\bar{\theta}^+ \bar{\theta}^-) \bar{P}^{\alpha\dot{\alpha}}
        \\&-
        4 (\bar{\theta}^-)^2 \theta^{+(\alpha} \theta^{-\beta)} \bar{P}_\beta^{\dot{\alpha}}
        -
        2 (\bar{\theta}^-)^2 (\theta^+ \theta^-) P^{\alpha\dot{\alpha}}
        \\&
        + 4i (\theta^-)^2 (\bar{\theta}^-)^2 \theta^{+\rho} \bar{\theta}^{+(\dot{\alpha}} \partial^{\dot{\rho})}_\rho P^\alpha_{\dot{\rho}}
        -
        4i (\theta^-)^2 (\bar{\theta}^-)^2 \bar{\theta}^{+\dot{\rho}} \theta^{+(\alpha} \partial^{\rho)}_{\dot{\rho}} \bar{P}^{\dot{\alpha}}_{\rho}\,.
    \end{split}
\end{equation}

Passing to the auxiliary fields $\mathcal{P}_{\mu\dot{\mu}}, \bar{\mathcal{P}}_{\mu\dot{\mu}}$ which are invariant under the gauge transformation \eqref{spin 2 gauge transformations}
is accomplished by the following shifts \cite{Buchbinder:2021ite}
\begin{equation}\label{p-field}
    P_{\mu\dot{\mu}} = B_{\mu\dot{\mu}} + \mathcal{P}_{\mu\dot{\mu}}\,,
\end{equation}
\begin{equation}\label{p-bar-field}
    \bar{P}_{\mu\dot{\mu}} = B_{\mu\dot{\mu}} + \bar{\mathcal{P}}_{\mu\dot{\mu}}\,,
\end{equation}
where
\begin{equation}\label{B field}
    B_{\mu\dot{\mu}}
    =
    \frac{1}{4} \left(3 \partial_{\mu\dot{\mu}} \Phi - \partial^{\rho\dot{\rho}} \Phi_{(\mu\rho)(\dot{\mu}\dot{\rho})}\right)
    =
    \frac{1}{4} \left( \partial_{\mu\dot{\mu}} \Phi - \partial^{\rho\dot{\rho}} \Phi_{\mu\rho\dot{\mu}\dot{\rho}}\right)
\end{equation}
is the spin 2-dependent part. Using these expressions,
one can express the spin 2 part of the coefficient of the monomial $2i(\theta^-)^2(\bar{\theta}^-)^2$ in \eqref{G-- solution}
through the linearized Einstein tensor $\mathcal{G}_{\rho\dot{\rho}}^{\alpha\dot{\alpha}}$:
\begin{eqnarray}
 \left(\Box \Phi^{\alpha\dot{\alpha}}_{\rho\dot{\rho}} - \frac{1}{2} \partial_{\rho\dot{\rho}} \partial^{\beta\dot{\beta}} \Phi_{\beta\dot{\beta}}^{\alpha\dot{\alpha}} \right)
    +
    \partial^{\dot{\alpha}}_\rho B_{\dot{\rho}}^\alpha
    +
    \partial^\alpha_{\dot{\rho}} B_\rho^{\dot{\alpha}}
    +
    \delta^{\dot{\alpha}}_{\dot{\rho}} \partial^{\dot{\mu}}_\rho B_{\dot{\mu}}^\alpha
    +
    \delta^\alpha_\rho \partial^\mu_{\dot{\rho}} B^{\dot{\alpha}}_\mu =
    -2 \mathcal{G}_{\rho\dot{\rho}}^{\alpha\dot{\alpha}}\,.
\end{eqnarray}
Here
\begin{equation}\label{lin-ein}
    \mathcal{G}_{\alpha\beta\dot{\alpha}\dot{\beta}}
    =
    \mathcal{R}_{(\alpha\beta)(\dot{\alpha}\dot{\beta})} - \varepsilon_{\alpha\beta} \varepsilon_{\dot{\alpha}\dot{\beta}} \mathcal{R}\,,
\end{equation}
and
\begin{equation}\label{Ricchi-lin}
    \mathcal{R}_{(\alpha\beta)(\dot{\alpha}\dot{\beta})} =
    \frac{1}{2} \partial_{(\alpha(\dot{\alpha}} \partial^{\sigma\dot{\sigma}} \Phi_{\beta)\sigma\dot{\beta})\dot{\sigma}}
    -
    \frac{1}{2} \Box \Phi_{(\alpha\beta)(\dot{\alpha}\dot{\beta})}
    -
    \frac{1}{2} \partial_{(\alpha(\dot{\alpha}} \partial_{\beta)\dot{\beta})} \Phi \, ,
\end{equation}
\begin{equation}\label{scalar curvature}
    \mathcal{R} = \frac{1}{8} \partial^{\alpha\dot{\alpha}} \partial^{\beta\dot{\beta}}  \Phi_{(\alpha\beta) (\dot{\alpha}\dot{\beta})}
    - \frac{3}{4} \Box \Phi \,.
\end{equation}
In the first term of the r.h.s. of eq. \eqref{Ricchi-lin}, to simplify the notation,  we omitted symmetrization of indices in $\Phi_{(\beta\sigma)(\dot\beta\dot\sigma)}$.
All these objects are linearized tensors which are invariant with respect to the gauge transformations \eqref{spin 2 gauge transformations}.
It is useful to note that the linearized Einstein tensor \eqref{lin-ein} by construction satisfies the transversality condition:
\begin{equation}\label{Bianchi spin 2}
    \partial^{\alpha\dot{\alpha}}  \mathcal{G}_{\alpha\beta\dot{\alpha}\dot{\beta}} = 0\,.
\end{equation}

With the above relations at hand, the complete expression for the highest component of $G^{--\alpha\dot{\alpha}}_{(\Phi,P)}$ in \eqref{G-- solution} reads:
\begin{equation}\label{spin 2 highest component}
    G^{--\alpha\dot{\alpha}}_{(\Phi,P)} = \dots
    + 4i (\theta^-)^2 (\bar{\theta}^-)^2 \left[- \theta^{+\rho}\bar{\theta}^{+\dot{\rho}} \mathcal{G}^{\alpha\dot{\alpha}}_{\rho\dot{\rho}} + \theta^{+\rho} \bar{\theta}^{+(\dot{\alpha}} \partial^{\dot{\rho})}_\rho \mathcal{P}^\alpha_{\dot{\rho}}
    -
    \bar{\theta}^{+\dot{\rho}} \theta^{+(\alpha} \partial^{\rho)}_{\dot{\rho}} \bar{\mathcal{P}}^{\dot{\alpha}}_{\rho}\right].
\end{equation}
We used this result in the analysis of the superfield equations of motion in Sections \ref{spin 2 equations} and \ref{spin 2 equations with matter}.
It is worth recalling  that this leading component is a linearized gauge invariant.

One more object in the spin 2 sector is the superfield $G^{++5}$:
\begin{equation}\label{G++5 P}
    G_{(\Phi, P)}^{++5} = -4 (\theta^+)^2  \bar{\theta}^{+\dot{\rho}} \theta^-_\mu P^\mu_{\dot{\rho}} - 4 (\bar{\theta}^+)^2 \theta^{+\mu} \bar{\theta}^-_{\dot{\rho}} \bar{P}_\mu^{\dot{\rho}}\,.
\end{equation}
The corresponding solution of the zero-curvature equation \eqref{zero-curv2} reads:

\begin{equation}\label{G--5 P}
    \begin{split}
        G_{(\Phi, P)}^{--5} & = 4 (\theta^-)^2 \bar{\theta}^{-\dot{\rho}} \theta^+_\mu P^\mu_{\dot{\rho}}
        +
        2i (\theta^+)^2 (\theta^-)^2 (\bar{\theta}^-)^2 \partial_{\rho\dot{\rho}} P^{\rho\dot{\rho}}
        \\&\,+
        4 (\bar{\theta}^-)^2 \theta^{-\mu} \bar{\theta}^+_{\dot{\rho}} \bar{P}_\mu^{\dot{\dot{\rho}}}
        -
        2i (\bar{\theta}^+)^2 (\theta^-)^2 (\bar{\theta}^-)^2 \partial_{\rho\dot{\rho}} \bar{P}^{\rho\dot{\rho}}\,.
    \end{split}
\end{equation}

Using \eqref{G++ P}, \eqref{G++5 P} and the zero-curvature solutions \eqref{G-- solution}, \eqref{G--5 P}), we deduce from the general superfield action
\eqref{Action} the following  spin $\mathbf{2}$ component action:
\begin{equation}
    \begin{split}
        S^{(s=2)}_{(\Phi, P)} = -\int d^4x\;
        \Bigr[ &\Phi^{\alpha\beta\dot{\alpha}\dot{\beta}} \left(\Box \Phi_{\alpha\beta\dot{\alpha}\dot{\beta}} - \frac{1}{2} \partial_{\alpha\dot{\alpha}} \partial_{\beta\dot{\beta}} \Phi \right)
        \\&
        +
        8 (P^{\alpha\dot{\alpha}} + \bar{P}^{\alpha\dot{\alpha}}) B_{\alpha\dot{\alpha}}
        -
        6 P^{\alpha\dot{\alpha}} P_{\alpha\dot{\alpha}}
        -
        6\bar{P}^{\alpha\dot{\alpha}} \bar{P}_{\alpha\dot{\alpha}}
        +
        4
        P^{\alpha\dot{\alpha}} \bar{P}_{\alpha\dot{\alpha}}
        \Bigr].
    \end{split}
\end{equation}
Now, singling out the gravity-dependent parts from  $P$-fields as in \eqref{p-field} and \eqref{p-bar-field}, one can expose the pure spin $2$
part of this Lagrangian.
 In particular, using the expression \eqref{B field} one finds
\begin{equation}
    8 \int d^4x\; B^{\alpha\dot{\alpha}}B_{\alpha\dot{\alpha}}
    =
    16
    \int d^4x\;
    \Phi \mathcal{R}
    +
    \int d^4x\; \Phi^{\alpha\beta\dot{\alpha}\dot{\beta}} \left(\frac{3}{2}\partial_{\alpha\dot{\alpha}}\partial_{\beta\dot{\beta}} \Phi
    - \frac{1}{2}   \partial_{\alpha\dot{\alpha}} \partial^{\rho\dot{\rho}} \Phi_{(\rho\beta)(\dot{\rho}\dot{\beta})}\right).
\end{equation}
The final form of the action is as follows:
\begin{equation}\label{component spin 2}
    \begin{split}
        S^{(s=2)}_{(\Phi, P)} = \int d^4x\;
        \Bigr[ &2\Phi^{\alpha\beta\dot{\alpha}\dot{\beta}} \mathcal{G}_{\alpha\beta\dot{\alpha}\dot{\beta}}
        +
        6 \mathcal{P}^{\alpha\dot{\alpha}} \mathcal{P}_{\alpha\dot{\alpha}}
        +
        6\bar{\mathcal{P}}^{\alpha\dot{\alpha}} \bar{\mathcal{P}}_{\alpha\dot{\alpha}}
        -
        4
        \mathcal{P}^{\alpha\dot{\alpha}} \bar{\mathcal{P}}_{\alpha\dot{\alpha}}
        \Bigr].
    \end{split}
\end{equation}
Here, the first term is just the linearized action of the spin 2 field (the Pauli-Fierz action).
In the free spin $\mathbf{2}$ theory, the algebraic equations of motion for the  auxiliary $\mathcal{P}$-fields are trivial:
\begin{equation}\label{P fields}
    \mathcal{P}_{\alpha\dot{\alpha}} = \bar{\mathcal{P}}_{\alpha\dot{\alpha}} = 0\,,
\end{equation}
and after elimination of these fields  we are left with the linearized spin 2 theory:
\begin{equation}\label{Pauli_Fiertz}
    \begin{split}
        S^{(s=2)}_{(\Phi)} &= 2 \int d^4x\;
        \Phi^{\alpha\beta\dot{\alpha}\dot{\beta}} \mathcal{G}_{\alpha\beta\dot{\alpha}\dot{\beta}}
        \\&=
        -\int d^4x\; \Big[\Phi^{(\alpha\beta)(\dot{\alpha}\dot{\beta})} \Box
        \Phi_{(\alpha\beta)(\dot{\alpha}\dot{\beta})} -
        \Phi^{(\alpha\beta)(\dot{\alpha}\dot{\beta})}
        \partial_{\alpha\dot{\alpha}} \partial^{\rho\dot{\rho}}
        \Phi_{(\rho\beta)(\dot{\rho}\dot{\beta})}
        \\&\;\;\;\;\;\;\;\;\;\;\; \qquad\quad        + 2\, \Phi
        \partial^{\alpha\dot{\alpha}} \partial^{\beta\dot{\beta}}
        \Phi_{(\alpha\beta)(\dot{\alpha}\dot{\beta})}
        - 6 \Phi \Box \Phi \Big].
    \end{split}
\end{equation}
In the theory including an interaction with matter the auxiliary fields will no longer vanish and play an important role in forming the final matter-modified
gauge field action. We discuss this in detail  below.

\medskip

\underline{\textbf{\textit{Spin 1 sector}}}

\medskip

The bosonic spin 1 sector of ${\cal N}=2$ spin $\mathbf{2}$ theory involves three relevant fields: the gauge field $C_{\alpha\dot{\alpha}}$ and
the auxiliary fields $T^{(\alpha\beta)}$, $T^{(\dot{\alpha}\dot{\beta})}$. The gauge transformations of $C_{\alpha\dot{\alpha}}$ have the standard form:
\begin{equation}
    \delta C_{\alpha\dot{\alpha}} = \partial_{\alpha\dot{\alpha}} b \,.
\end{equation}

The fields $C$ and $T$ make the following contributions to the superfields $G^{++\alpha\dot{\alpha}}$ and $G^{++5}$:
\begin{equation}
    G^{++\alpha\dot{\alpha}}_{(C, T)} = 2i (\bar{\theta}^+)^2 \theta^+_\nu \bar{\theta}^{-\dot{\alpha}} T^{(\alpha\nu)}
    -
    2i (\theta^+)^2 \bar{\theta}^+_{\dot{\nu}} \theta^{-\alpha} \bar{T}^{(\dot{\alpha}\dot{\nu})}\,,
\end{equation}
\begin{equation}
    G^{++5}_{(C, T)} = -2i \theta^{+\rho}\bar{\theta}^{+\dot{\rho}} C_{\rho\dot{\rho}}
    -
    2i (\bar{\theta}^+)^2  \theta^+_\nu \theta^-_\alpha T^{(\alpha\nu)}
    -
    2i (\theta^+)^2 \bar{\theta}^+_{\dot{\nu}} \bar{\theta}^-_{\dot{\alpha}} \bar{T}^{(\dot{\alpha}\dot{\nu})}\,.
\end{equation}
The corresponding solutions of the zero-curvature equations \eqref{zero-curv} and \eqref{zero-curv2} read:

\begin{equation}\label{spin 1-1}
    \begin{split}
        G^{--\alpha\dot{\alpha}}_{(C, T)}
        =&
        -2i (\bar{\theta}^-)^2 \theta^-_\nu \bar{\theta}^{+\dot{\alpha}} T^{(\alpha\nu)} +
        2i (\theta^-)^2 \bar{\theta}^-_{\dot{\nu}} \theta^{+\alpha} \bar{T}^{(\dot{\alpha}\dot{\nu})} \\&+
        (\bar{\theta}^+)^2 (\bar{\theta}^-)^2 (\theta^-)^2 \partial^{\dot{\alpha}}_\nu T^{(\alpha\nu)} -  (\theta^+)^2 (\theta^-)^2 ( \bar{\theta}^-)^2 \partial^\alpha_{\dot{\nu}} \bar{T}^{(\dot{\alpha}\dot{\nu})} \,,
    \end{split}
\end{equation}

\begin{equation}\label{spin 1-2}
    \begin{split}
        G^{--5}_{(C, T)} =& -2i \theta^{+\rho}\bar{\theta}^{+\dot{\rho}} C_{\rho\dot{\rho}}
        \\&
        +
        2 (\theta^-)^2 \bar{\theta}^{-(\dot{\rho}} \bar{\theta}^{+\dot{\beta})} \partial_{\dot{\rho}}^\beta C_{\beta\dot{\beta}}
        -
        2 (\bar{\theta}^-)^2 \theta^{-(\rho} \theta^{+\beta)}
        \partial_\rho^{\dot{\beta}} C_{\beta\dot{\beta}}
        \\&
        +
        2i (\theta^-)^2 (\bar{\theta}^-)^2 \theta^{+\rho} \bar{\theta}^{+\dot{\rho}} \left[ \Box C_{\rho\dot{\rho}} - \frac{1}{2} \partial_{\rho\dot{\rho}} \partial^{\beta\dot{\beta}} C_{\beta\dot{\beta}} \right]
        \\&
        -
        2i (\bar{\theta}^-)^2  \theta^+_\nu \theta^-_\alpha T^{(\alpha\nu)}
        -
        2i (\theta^-)^2 \bar{\theta}^+_{\dot{\nu}} \bar{\theta}^-_{\dot{\alpha}} \bar{T}^{(\dot{\alpha}\dot{\nu})}
        \\&
        +
        2 (\theta^-)^2 (\bar{\theta}^-)^2 \bar{\theta}^{+\dot{\rho}} \theta^+_\mu \partial_{\rho\dot{\rho}} T^{(\mu\rho)}
        +
        2 (\theta^-)^2 (\bar{\theta}^-)^2
        \theta^{+\rho} \bar{\theta}^+_{\dot{\mu}} \partial_{\rho\dot{\rho}} \bar{T}^{(\dot{\mu}\dot{\rho})}\,.
    \end{split}
\end{equation}
The spin 1 sector of the total action is expressed as:

\begin{equation}
    \begin{split}
        S^{(s=2)}_{(C, T)} = - \int d^4x \; \Bigr[ &C^{\rho\dot{\rho}} \left(\Box C_{\rho\dot{\rho}} - \frac{1}{2}\partial_{\rho\dot{\rho}} \partial^{\beta\dot{\beta}}C_{\beta\dot{\beta}}  \right)
        \\&
        +
        2i C^{\dot{\rho}}_\mu \partial_{\rho\dot{\rho}} T^{(\mu\rho)}
        -
        2i C^\rho_{\dot{\mu}} \partial_{\rho\dot{\rho}} \bar{T}^{(\dot{\mu}\dot{\rho})}
        -
        T^{(\mu\nu)} T_{(\mu\nu)}
        -
        \bar{T}^{(\dot{\mu}\dot{\nu})} \bar{T}_{(\dot{\mu}\dot{\nu})} \Bigr].
    \end{split}
\end{equation}

The equations of motion for the auxiliary fields yield:
\begin{equation}\label{spin 1 aux}
    T_{(\mu\nu)} = -i \partial_{(\mu\dot{\rho}} C^{\dot{\rho}}_{\nu)}\,,
    \qquad\qquad
    \bar{T}_{(\dot{\mu}\dot{\nu})} = i \partial_{(\dot{\mu}\rho} C^{\rho}_{\dot{\nu})}\,.
\end{equation}
After elimination of the auxiliary fields we reproduce the standard Maxwell action in the spinor formalism:
\begin{equation}\label{Maxwell action -- spin2}
    \begin{split}
        S^{(s=2)}_{(C)} &=  \int d^4x \;  C^{\rho\dot{\rho}} \left(\Box C_{\rho\dot{\rho}} - \frac{1}{2}\partial_{\rho\dot{\rho}} \partial^{\beta\dot{\beta}}C_{\beta\dot{\beta}}  \right)
        \\&= - \frac{1}{2} \int d^4x\; \partial^{\beta\dot{\beta}} C^{\rho\dot{\rho}}
        \left(\partial_{\beta\dot{\beta}} C_{\rho\dot{\rho}}
        -
        \partial_{\rho\dot{\rho}} C_{\beta\dot{\beta}}
        \right).
    \end{split}
\end{equation}
We see that the role of the tensorial auxiliary fields $T_{(\mu\nu)}, \bar{T}_{(\dot\mu\dot\nu)}$ in the present case is just
to ensure the correct sign of the final spin ${\bf 1}$ gauge action.

\subsection{${\cal N}=2$ spin $\mathbf{3}$ theory: bosonic sector}
\label{spin 3 bosonic sector}

Here we perform the corresponding bosonic component reduction of the spin $\mathbf{3}$ ${\cal N}=2$ theory \cite{Buchbinder:2021ite}. As distinct from the spin $\mathbf{2}$ case, here we do not discuss
the auxiliary fields $\mathcal{P}^{(\alpha\beta)\dot{\alpha}\dot{\beta}}$, $\bar{\mathcal{P}}^{(\dot{\alpha}\dot{\beta})\alpha\beta}$ since they give a trivial contribution in the free spin $\mathbf{3}$ theory.
Like in the spin $\mathbf{2}$ case, they still play an essential role in the interacting theory, when passing to the superfield equations with sources.
The bosonic gauge field sector of the ${\cal N}=2$ spin $\mathbf{3}$ theory involves the spin 3 and spin 2  fields.

\medskip

\underline{\textbf{\textit{Spin 3 sector}}}

\medskip
The gauge fields of the  spin 3 sector are collected in:
\begin{equation}\label{spin 3 field}
    \Phi_{(\alpha\beta)\gamma (\dot{\alpha}\dot{\beta})\dot{\gamma}}
    =
    \Phi_{(\alpha\beta\gamma) (\dot{\alpha}\dot{\beta}\dot{\gamma})}
    +
    \Phi_{(\alpha(\dot{\alpha}} \epsilon_{\beta)\gamma} \epsilon_{\dot{\beta})\dot{\gamma}}\,.
\end{equation}
They are subject to gauge transformations with the gauge parameters $a_{(\alpha\beta)(\dot{\alpha}\dot{\beta})}$:
\begin{equation}\label{spin 3 gauge transformations}
    \delta  \Phi_{(\alpha\beta\gamma) (\dot{\alpha}\dot{\beta}\dot{\gamma})}  = \partial_{(\alpha(\dot{\alpha}} a_{\beta\gamma)\dot{\beta}\dot{\gamma})}\,,\qquad\qquad
    \delta \Phi_{\alpha\dot{\alpha}} = \frac{4}{9} \partial^{\beta\dot{\beta}} a_{(\alpha\beta)(\dot{\alpha}\dot{\beta})}\,.
\end{equation}
For the component contents of relevant $G^{++\dots}$-superfields we have:
\begin{equation}
    \begin{split}
        G_{(\Phi, B)}^{++(\alpha\beta)(\dot{\alpha}\dot{\beta})} =&  -2i \theta^{+\gamma} \bar{\theta}^{+\dot{\gamma}} \Phi_{\gamma\dot{\gamma}}^{(\alpha\beta)(\dot{\alpha}\dot{\beta})}
        \\&+
        4 (\theta^+)^2 \bar{\theta}^{+\dot{\gamma}} \bar{\theta}^{-(\dot{\alpha}} B^{\dot{\beta})(\alpha\beta)}_{\dot{\gamma}}
        -\,4 (\bar{\theta}^+)^2 \theta^{+\gamma} \theta^{-(\alpha} \bar{B}^{\beta)(\dot{\alpha}\dot{\beta})}_{\gamma}\,,
    \end{split}
\end{equation}
\begin{equation}
    G_{(\Phi, B)}^{++5\alpha\dot{\alpha}} = -
    2 (\theta^+)^2 \bar{\theta}^{+\dot{\rho}} \theta^-_{\mu}
    B^{(\mu\alpha)\dot{\alpha}}_{\dot{\rho}}
    -
    2 (\bar{\theta}^+)^2 \theta^{+\beta} \bar{\theta}^-_{\dot{\rho}} \bar{B}^{\alpha(\dot{\rho}\dot{\alpha})}_{\beta}\,.
\end{equation}
Here we employed the notations:
\begin{equation}\label{A34}
    B_{(\alpha\beta)\dot{\alpha}\dot{\beta}} = - \frac{1}{2} \left[ \partial^{\gamma\dot{\gamma}} \Phi_{(\alpha\beta\gamma)(\dot{\alpha}\dot{\beta}\dot{\gamma})} - \partial_{(\alpha(\dot{\alpha}} \Phi_{\beta)\dot{\beta})}
    - \partial_{(\alpha\dot{\alpha}} \Phi_{\beta)\dot{\beta}} \right],
\end{equation}
\begin{equation}\label{A35}
    B_{\alpha\beta(\dot{\alpha}\dot{\beta})} = - \frac{1}{2} \left[ \partial^{\gamma\dot{\gamma}} \Phi_{(\alpha\beta\gamma)(\dot{\alpha}\dot{\beta}\dot{\gamma})} - \partial_{(\alpha(\dot{\alpha}} \Phi_{\beta)\dot{\beta})}
    - \partial_{\alpha(\dot{\alpha}} \Phi_{\beta\dot{\beta})} \right].
\end{equation}
Note useful identities for the derivatives of \eqref{A34} and \eqref{A35}:
\begin{equation}\label{BR}
    \partial_{\rho\dot{\rho}} B^{(\rho\alpha)\dot{\rho}\dot{\alpha}} = - \frac{1}{2} \mathcal{R}^{\alpha\dot{\alpha}}\,,
    \qquad
    \partial_{\rho\dot{\rho}} \bar{B}^{\rho\alpha(\dot{\rho\dot{\alpha}})} = - \frac{1}{2} \mathcal{R}^{\alpha\dot{\alpha}}\,.
\end{equation}

The corresponding solutions of the zero-curvature equations \eqref{zero-curv} and \eqref{zero-curv2} read:
\begin{equation}\label{G--31}
    \begin{split}
        G_{(\Phi, B)}^{--(\alpha\beta)(\dot{\alpha}\dot{\beta})}
        =&
        -2i \theta^{-\beta} \bar{\theta}^{-\dot{\beta}} \Phi_{\beta\dot{\beta}}^{(\alpha\beta)(\dot{\alpha}\dot{\beta})}
        \\&+
        2 (\theta^-)^2 \bar{\theta}^{-(\dot{\rho}} \bar{\theta}^{+\dot{\beta})}
        \partial^\beta_{\dot{\rho}} \Phi_{\beta\dot{\beta}}^{(\alpha\beta)(\dot{\alpha}\dot{\beta})}
        -
        2 (\bar{\theta}^-)^2 \theta^{-(\rho} \theta^{+\beta)}
        \partial_{\rho}^{\dot{\beta}} \Phi_{\beta\dot{\beta}}^{(\alpha\beta)(\dot{\alpha}\dot{\beta})}
        \\&
        +\, 4(\theta^-)^2 \bar{\theta}^{+(\dot{\alpha}} \bar{\theta}^{-\dot{\rho}} B_{\dot{\rho}}^{\dot{\beta})(\alpha\beta)}
        -
        4 (\bar{\theta}^-)^2 \theta^{+\rho} \theta^{-(\alpha} \bar{B}^{\beta)(\dot{\alpha}\dot{\beta})}_{\rho}
        \\&
        - 3i(\theta^-)^2 (\bar{\theta}^-)^2 \theta^{+\rho} \bar{\theta}^{+\dot{\rho}}  \mathcal{G}_{\rho\dot{\rho}}^{(\alpha\beta)(\dot{\alpha}\dot{\beta})}\,,
    \end{split}
\end{equation}
\begin{equation}\label{G--32}
    \begin{split}
        G_{(\Phi, B)}^{--5\alpha\dot{\alpha}} =&
        2 (\theta^-)^2 \bar{\theta}^{-\dot{\rho}} \theta^+_{\mu}
        B^{(\mu\alpha)\dot{\alpha}}_{\dot{\rho}}
        +
        i (\theta^+)^2 (\bar{\theta}^-)^2 (\theta^-)^2  \partial_{\rho\dot{\rho}} B^{(\rho\alpha)\dot{\rho}\dot{\alpha}} \\&
        + 2\, (\bar{\theta}^-)^2 \theta^{-\beta} \bar{\theta}^+_{\dot{\rho}} \bar{B}_\beta^{\alpha(\dot{\alpha}\dot{\rho})}
        -
        i (\bar{\theta}^+)^2 (\bar{\theta}^-)^2 (\theta^-)^2 \partial_{\rho\dot{\rho}} \bar{B}^{\rho\alpha(\dot{\rho}\dot{\alpha})}\,.
    \end{split}
\end{equation}
In eq. \p{G--31} the following expression for the linearized Fronsdal tensor was used:
\bea\label{Eintein3}
\mathcal{G}_{(\alpha_1\alpha_2)\alpha_3 (\dot{\alpha}_1  \dot{\alpha}_2) \dot{\alpha}_3}
=
\mathcal{R}_{(\alpha_1\alpha_2\alpha_3) (\dot{\alpha}_1  \dot{\alpha}_2 \dot{\alpha}_3)}
- \frac{4}{9}
\mathcal{R}_{(\alpha_1 (\dot{\alpha}_1} \epsilon_{\dot{\alpha}_2) \dot{\beta}} \epsilon_{\alpha_2) \beta}\,,
\eea
where the linearized spin 3 curvatures are defined as:
\begin{equation}\label{spin 3 curvature 1}
    \mathcal{R}_{\alpha_1\dot{\alpha}_1}
    =
    \partial^{\alpha_2\dot{\alpha}_2} \partial^{\alpha_3\dot{\alpha}_3} \Phi_{(\alpha_1\alpha_2\alpha_3)( \dot{\alpha}_1\dot{\alpha}_2\dot{\alpha}_3)}
    -
    \frac{1}{4}\partial_{\alpha_1\dot{\alpha}_1} \partial^{\alpha_2\dot{\alpha}_2} \Phi_{\alpha_2\dot{\alpha}_2}
    -
    \frac{5}{2} \Box \Phi_{\alpha_1\dot{\alpha}_1}\,,
\end{equation}
\begin{eqnarray}
    \mathcal{R}_{(\alpha_1\alpha_2\alpha_3)( \dot{\alpha}_1\dot{\alpha}_2\dot{\alpha}_3)}
    &=&
    \partial_{(\alpha_1(\dot{\alpha}_1}\partial^{\rho\dot{\rho}} \Phi_{\alpha_2\alpha_3)\rho \dot{\alpha}_2\dot{\alpha}_3)\dot{\rho}}
    -
    \frac{2}{3}
    \Box \Phi_{(\alpha_1\alpha_2\alpha_3)( \dot{\alpha}_1\dot{\alpha}_2\dot{\alpha}_3)} \nn
    &&-\,
    \partial_{(\alpha_1(\dot{\alpha}_1}
    \partial_{\alpha_2\dot{\alpha}_2}
    \Phi_{\alpha_3)\dot{\alpha}_3)}\,.\label{SymmCurvspin3}
\end{eqnarray}
In the first term in eq. \eqref{SymmCurvspin3} we omitted the symmetrization of the indices in the involved field. The curvatures $\mathcal{R}_{\alpha\dot{\alpha}}$ and $\mathcal{R}_{(\alpha_1\alpha_2\alpha_3)( \dot{\alpha}_1\dot{\alpha}_2\dot{\alpha}_3)}$
are invariant under the spin 3 gauge transformations \eqref{spin 3 gauge transformations}. The  linearized Fronsdal tensor satisfies the spin-3 transversality condition:
\begin{equation}
    \partial^{\alpha_3\dot{\alpha}_3} \mathcal{G}_{(\alpha_1\alpha_2)\alpha_3 (\dot{\alpha}_1  \dot{\alpha}_2) \dot{\alpha}_3} = 0\,.
\end{equation}

The resulting action for the spin 3 gauge field is written as:
\begin{equation}\label{spin 3 action}
    \begin{split}
        & S^{(s= 3)}_{(\Phi)}
        = - \frac{3}{2} \int d^4x\; \Phi^{(\alpha_1\alpha_2)\alpha_3 (\dot{\alpha}_1\dot{\alpha}_2)\dot{\alpha}_3} \mathcal{G}_{(\alpha_1\alpha_2)\alpha_3 (\dot{\alpha}_1  \dot{\alpha}_2) \dot{\alpha}_3}
        \\&\;\;\;\;\;\;\;\;\;\;=   \int d^4x\; \Big[\Phi^{(\alpha_1\alpha_2\alpha_3)( \dot{\alpha}_1\dot{\alpha}_2\dot{\alpha}_3)}
        \Box \Phi_{(\alpha_1\alpha_2\alpha_3)( \dot{\alpha}_1\dot{\alpha}_2\dot{\alpha}_3)} \\
        & \;\;\;\;\;\;\;\;\;\qquad\qquad\qquad- \,\frac{3}{2} \Phi^{(\alpha_1\alpha_2\alpha_3)( \dot{\alpha}_1\dot{\alpha}_2\dot{\alpha}_3)}
        \partial_{\alpha_1\dot{\alpha}_1} \partial^{\rho\dot{\rho}} \Phi_{(\rho\alpha_2\alpha_3)( \dot{\rho}\dot{\alpha}_2\dot{\alpha}_3)} \\
        &\;\;\;\;\;\;\;\;\;\qquad\qquad\qquad+ \,3 \Phi^{(\alpha_1\alpha_2\alpha_3)( \dot{\alpha}_1\dot{\alpha}_2\dot{\alpha}_3)} \partial_{\alpha_1\dot{\alpha}_1} \partial_{\alpha_2\dot{\alpha}_2}
        \Phi_{\alpha_3\dot{\alpha}_3}
        - \frac{15}{4} \Phi^{\alpha\dot{\alpha}} \Box  \Phi_{\alpha\dot{\alpha}} \\
        & \;\;\;\;\;\;\;\;\;\qquad\qquad\qquad+\,
        \frac{3}{8} \partial_{\alpha_1\dot{\alpha}_1} \Phi^{\alpha_1\dot{\alpha}_1} \partial_{\alpha_2\dot{\alpha}_2} \Phi^{\alpha_2\dot{\alpha}_2}\Big].
    \end{split}
\end{equation}
This is the spin 3 Fronsdal action in spinor notation. It is invariant under the gauge transformations \eqref{spin 3 gauge transformations}.

\medskip

\underline{\textbf{\textit{Spin 2 sector}}}

\medskip

The spin 2 sector of the spin $\mathbf{3}$ ${\cal N}=2$ multiplet involves the field $C^{\alpha\dot{\alpha}}_{\beta\dot{\beta}}$
\begin{equation}\label{spin 3-2 sector}
    C_{\alpha\beta\dot{\alpha}\dot{\beta}} = C_{(\alpha\beta)(\dot{\alpha}\dot{\beta})}
    +
    \epsilon_{\alpha\beta} \epsilon_{\dot{\alpha}\dot{\beta}} C\,,
\end{equation}
with the gauge freedom:
\begin{equation}
    \delta C_{\alpha\beta\dot{\alpha}\dot{\beta}} = \frac{1}{2} \left(\partial_{\alpha\dot{\alpha}} b_{\beta\dot{\beta}} + \partial_{\beta\dot{\beta}} b_{\alpha\dot{\alpha}} \right).
\end{equation}
Here $b_{\alpha\dot{\alpha}}$ is the spin 2 gauge parameter.

The relevant pieces of $G^{++\dots}$ superfields are expressed as:
\begin{equation}\label{Spin3G}
    G_{(C)}^{++(\alpha\beta)(\dot{\alpha}\dot{\beta})}
    =
    - 2 (\bar{\theta}^+)^2 \theta^+_\rho \bar{\theta}^{-(\dot{\alpha}} H^{\dot{\beta}) \rho(\alpha\beta)}
    -
    2 (\theta^+)^2 \bar{\theta}^+_{\dot{\rho}} \theta^{-(\alpha}
    \bar{H}^{\beta)\dot{\rho}(\dot{\alpha}\dot{\beta})}\,,
\end{equation}
\begin{equation}\label{Spin3G5}
    G^{++5\alpha\dot{\alpha}}_{(C)}
    =
    -2 i \theta^{+\rho} \bar{\theta}^{+\dot{\rho}} C_{\rho\dot{\rho}}^{\alpha\dot{\alpha}}
    -
    2 (\theta^+)^2 \bar{\theta}^{+}_{\dot{\rho}} \bar{\theta}^{-}_{\dot{\mu}} \bar{H}_{}^{\alpha\dot{\rho}(\dot{\mu}\dot{\alpha})}
    +
    2 (\bar{\theta}^+)^2 \theta^{+}_{\rho}\theta^{-}_{\mu} H_{}^{\dot{\alpha}\rho(\alpha\mu)}
    \,.
\end{equation}
Here the following notations were used:
\begin{equation}\label{H1}
    \bar{H}^{\alpha\dot{\rho}(\dot{\alpha}\dot{\mu})}
    :=
    \bar{H}^{\alpha(\dot{\rho}\dot{\alpha}\dot{\mu})}
    +
    \epsilon^{\dot{\rho}(\dot{\alpha}} \bar{H}^{\alpha\dot{\mu})}\,,
    \;\;\;\;\;\;\;\;\;
    H^{\dot{\alpha}\rho(\mu\alpha)}
    :=
    H^{\dot{\alpha}(\rho\mu\alpha)}
    +
    \epsilon^{\rho(\alpha} H^{\dot{\alpha}\mu) }\,,
\end{equation}
\begin{equation}\label{H2}
    \bar{H}^{\alpha(\dot{\rho}\dot{\alpha}\dot{\mu})}
    =
    \partial_\beta^{(\dot{\rho}}C^{\dot{\alpha}\dot{\mu})(\alpha\beta)}\,,
    \;\;\;\;\;\;\;\;
    H^{\dot{\alpha}(\rho\mu\alpha)}
    =
    \partial^{(\alpha}_{\dot{\beta}} C^{\mu\rho)(\dot{\alpha}\dot{\beta})}\,,
\end{equation}
\begin{equation}\label{H3}
    H^{\mu\dot{\mu}}
    =
    \partial^{\mu\dot{\mu}} C
    -
    \frac{1}{3} \partial_{\rho\dot{\rho}} C^{(\mu\rho)(\dot{\mu}\dot{\rho})}
    = \bar{H}^{\mu\dot{\mu}}\,.
\end{equation}
These objects satisfy the relations:
\begin{equation}\label{HR}
    \partial_{\rho}^{(\dot{\alpha}} H^{\dot{\beta})\rho(\alpha\beta)} = - \frac{2}{3} \mathcal{R}^{(\alpha\beta)(\dot{\alpha}\dot{\beta})}\,,\qquad
    \partial^{(\alpha}_{\dot{\rho}} \bar{H}^{\beta)\dot{\rho}(\dot{\alpha}\dot{\beta})}
    =
    - \frac{2}{3} \mathcal{R}^{(\alpha\beta)(\dot{\alpha}\dot{\beta})}\,.
\end{equation}
Here we used $\mathcal{R}^{(\alpha\beta)(\dot{\alpha}\dot{\beta})}$ defined in \eqref{Ricchi-lin}, with the replacement $\Phi \to C$\,.

The negatively charged partners of  \eqref{Spin3G} and \eqref{Spin3G5} can be obtained as solutions of the zero-curvature equations \eqref{zero-curv} and \eqref{zero-curv2}:
\begin{equation}\label{G-- 321}
    \begin{split}
        G_{(C)}^{--(\alpha\beta)(\dot{\alpha}\dot{\beta})}
        =&\;
        2 (\bar{\theta}^-)^2 \theta^-_\rho \bar{\theta}^{+(\dot{\alpha}} H^{\dot{\beta}) \rho(\alpha\beta)}
        +
        2 (\theta^-)^2 \bar{\theta}^-_{\dot{\rho}} \theta^{+(\alpha}
        \bar{H}^{\beta)\dot{\rho}(\dot{\alpha}\dot{\beta})}
        \\&+
        i (\bar{\theta}^+)^2 (\theta^-)^2 (\bar{\theta}^-)^2 \partial_{\rho}^{(\dot{\alpha}} H^{\dot{\beta})\rho(\alpha\beta)}
        +
        i (\theta^+)^2 (\theta^-)^2 (\bar{\theta}^-)^2 \partial^{(\alpha}_{\dot{\rho}} H^{\beta)\dot{\rho}(\dot{\alpha}\dot{\beta})}
        \,,
    \end{split}
\end{equation}
\begin{multline}\label{spin 3-2 solution}
    G_{(C)}^{--5\alpha\dot{\alpha}} = -2 i \theta^{-\rho} \bar{\theta}^{-\dot{\rho}} C_{\rho\dot{\rho}}^{\alpha\dot{\alpha}}
    +2
    (\theta^-)^2 \bar{\theta}^{-(\dot{\rho}} \bar{\theta}^{+\dot{\beta})}
    \partial^\beta_{\dot{\rho}} C_{\beta\dot{\beta}}^{\alpha\dot{\alpha}}
    -2
    (\bar{\theta}^-)^2 \theta^{-(\rho} \theta^{+\beta)}
    \partial_{\rho}^{\dot{\beta}} C_{\beta\dot{\beta}}^{\alpha\dot{\alpha}}
    \\
    +
    2 (\theta^-)^2 \bar{\theta}^{+}_{\dot{\rho}} \bar{\theta}^{-}_{\dot{\mu}} \bar{H}_{}^{\alpha\dot{\rho}(\dot{\mu}\dot{\alpha})}
    +2 (\bar{\theta}^-)^2 \theta^{+}_{\rho} \theta^{-}_{\mu} H_{}^{\dot{\alpha}\rho(\mu\alpha)}
    +4
    i(\theta^-)^2 (\bar{\theta}^-)^2 \theta^{+}_{\rho} \bar{\theta}^{+}_{\dot{\rho}}
    \mathcal{G}^{\alpha\rho\dot{\alpha}\dot{\rho}}\,.
\end{multline}
Here, $\mathcal{G}^{\alpha\rho\dot{\alpha}\dot{\rho}}$ is the linearized form of Einstein tensor introduced in \p{lin-ein}.

Substituting all these expressions in the superfield action \eqref{Action} for $\mathbf{s}= 3$, we
obtain the spin 2 action of the $\mathcal{N}=2$ spin $\mathbf{3}$ theory
\begin{multline}\label{spin 2 action -- spin 3}
    S^{(s=3)}_{(C)}
    =
    4\int d^4x \; C^{\alpha\beta\dot{\alpha}\dot{\beta}}
    \mathcal{G}_{\alpha\beta\dot{\alpha}\dot{\beta}}
    =
    -2 \int d^4x \; \Big[C^{(\alpha\beta)(\dot{\alpha}\dot{\beta})} \Box
    C_{(\alpha\beta)(\dot{\alpha}\dot{\beta})}
    \\-
    C^{(\alpha\beta)(\dot{\alpha}\dot{\beta})}
    \partial_{\alpha\dot{\alpha}} \partial^{\rho\dot{\rho}}
    C_{(\rho\beta)(\dot{\rho}\dot{\beta})}  + 2\, C
    \partial^{\alpha\dot{\alpha}} \partial^{\beta\dot{\beta}}
    C_{(\alpha\beta)(\dot{\alpha}\dot{\beta})}
    - 6 C \Box C \Big].
\end{multline}
It is precisely the linearized Pauli-Fierz action for spin 2.

It is  worth to note that in this case the auxiliary fields play no any role in forming the correct final action, in contradistinction to the spin 1 sector
of the spin $\mathbf{2}$ ${\cal N}=2$ supermultiplet (recall eq. \eqref{spin 1 aux}).

\section{Unconstrained $\mathcal{N}=2$ higher spin prepotentials and an alternative derivation of superfield equations}
\label{Pre-prepotentials}

In this Appendix we introduce the Mezincescu-type $\mathcal{N}=2$ higher spin prepotentials and describe an alternative derivation of the superfield equations of motion.

\subsection{Arbitrary $\mathcal{N}=2$ higher spin: unconstrained  prepotential}
\label{Arbitrary higher spin: unconstrained  pre-prepotential}

In order to define prepotentials for $\mathcal{N}=2$ higher spins, it is necessary to compactly collect all the  analytic potentials
$h^{++\dots}(\zeta)$ into a single object. Beforehand, we need to introduce some appropriate formalism.

In the previous paper \cite{Buchbinder:2022kzl} we introduced analytic differential operators, constructed in terms of the analytic potentials $h^{++\dots}(\zeta)$:
\begin{equation}\label{operator-in-1}
    \hat{\mathcal{H}}^{++\alpha(s-2)\dot{\alpha}(s-2)} : = h^{++\alpha(s-2)\dot{\alpha}(s-2)M} \partial_M\,,
\end{equation}
\begin{equation}\label{analyt-constr}
    [D^+_{\hat{\mu}},  \hat{\mathcal{H}}^{++\alpha(s-2)\dot{\alpha}(s-2)}  ] = 0\,.
\end{equation}
One can also construct higher order analytic differential operators
\begin{equation}
    \hat{\mathcal{H}}^{++}_{(s)} :=  \hat{\mathcal{H}}^{++\alpha(s-2)\dot{\alpha}(s-2)} \partial^{s-2}_{\alpha(s-2)\dot{\alpha}(s-2)}\,,
\end{equation}
which are invariant under $\mathcal{N}=2$ supersymmetry  and encode gauge transformations for an arbitrary integer spin $\bf{s}$ $\mathcal{N}=2$ supermultiplet:
\begin{equation}\label{transf-H}
    \delta_\lambda \mathcal{H}^{++}_{(s)} = [\mathcal{D}^{++}, \hat{\Lambda}_{(s)}]\,.
\end{equation}

Here we used the notations
\begin{equation}
    \hat{\Lambda}_{(s)} :=  \hat{\Lambda}^{\alpha(s-2)\dot{\alpha}(s-2)} \partial^{s-2}_{\alpha(s-2)\dot{\alpha}(s-2)}\,,
\end{equation}
\begin{equation}\label{LAMBDA s-2}
    \hat{\Lambda}^{\alpha(s-2)\dot{\alpha}(s-2)} : = \lambda^{\alpha(s-2)\dot{\alpha}(s-2)M} \partial_M\,.
\end{equation}
Note that the presence of the derivative $\partial_5$ in \eqref{operator-in-1}  plays an important role: it ensures the invariance of $\hat{\mathcal{H}}^{++}_{(s)}$ with respect to $\mathcal{N}=2$ supersymmetry transformations.
The operators defined above are crucial for building interactions of higher spins with the hypermultiplet, as discussed in Section \ref{hs-couplings}. Recall that
the spinor indices hidden in the multi-index $M$ in both operators \eqref{operator-in-1} and \eqref{LAMBDA s-2} are assumed to be fully symmetrized with other indices of the same type.

The analyticity constraints \eqref{analyt-constr} for the supermultiplet of spin ${\bf s}$ can be solved in terms of unconstrained
fermionic prepotentials $\Psi^{-\alpha(s-2)\dot{\alpha}(s-2)\hat{\mu}}(Z)$ as:
\begin{equation}\label{constr-sol}
    \begin{split}
        \hat{\mathcal{H}}^{++}_{(s)}
        &=
        \hat{\mathcal{H}}^{++\alpha(s-2)\dot{\alpha}(s-2)} \partial^{s-2}_{\alpha(s-2)\dot{\alpha}(s-2)}
        \\&= (D^+)^4 \left(\Psi^{-\alpha(s-2)\dot{\alpha}(s-2)\hat{\mu}} D^-_{\hat{\mu}}\right) \partial^{s-2}_{\alpha(s-2)\dot{\alpha}(s-2)} \,,
    \end{split}
\end{equation}
where we have assumed that $\Psi^{-\alpha(s-2)\dot{\alpha}(s-2)\hat{\mu}}$ are symmetric with respect to all indices of the same type as in \eqref{operator-in-1} and \eqref{LAMBDA s-2}.
Here we have also used the covariant spinor derivatives $D^-_{\hat{\mu}}$ defined in \eqref{D-}.

The relation \eqref{constr-sol} is a higher spin generalization of the Mezincescu prepotential relation \eqref{M-superfield} of  ${\cal N}=2$ Maxwell theory.
For the spin $\mathbf{2}$ case (the linearized $\mathcal{N}=2$ Einstein supergravity) it was generalized in \cite{Gates:1981qq} and further used in ref. \cite{Zupnik:1998td}.

Since the operator $\hat{\mathcal{H}}^{++}_{(s)}$  is invariant with respect to supersymmetry transformations \eqref{susy},
the prepotentials introduced  above are $\mathcal{N}=2$ scalars by construction:
\begin{equation}\label{susy PSI}
    \delta_\epsilon \Psi^{-\alpha(s-2)\dot{\alpha}(s-2)\hat{\mu}} = 0\,.
\end{equation}

Using \eqref{constr-sol}, one can deduce the  explicit expressions for the potentials $h^{++\dots}(\zeta)$ through the unconstrained prepotentials $\Psi^{-\alpha(s-2)\dot{\alpha}(s-2)\hat{\mu}}(Z)$:
\begin{equation}\label{h-fields}
    \begin{cases}
        h^{++(\alpha(s-2)\beta)(\dot{\alpha}(s-2)\dot{\beta})} =
        2i \Bigr[ (D^+)^2 \bar{D}^{+(\dot{\beta}} \Psi^{- (\alpha(s-2)\beta) \dot{\alpha}(s-2)) }
        +
        (\bar{D}^+)^2 D^{+(\beta} \bar{\Psi}^{-\alpha(s-2))(\dot{\alpha}(s-2)\dot{\beta})}\Bigr]_{\theta^{-}=0}\,,\\
        h^{++\alpha(s-1)\dot{\alpha}(s-2)+} =  -(D^+)^4 \Psi^{-\alpha(s-1)\dot{\alpha}(s-2)}\,,
        \\
        h^{++\alpha(s-2)\dot{\alpha}(s-1)+} =  -(D^+)^4 \bar{\Psi}^{-\alpha(s-2)\dot{\alpha}(s-1)}\,,
        \\
        h^{++5\alpha(s-2)\dot{\alpha}(s-2)} =
        2i \Bigr[ (\bar{D}^+)^2 D^+_\beta \Psi^{-(\beta\alpha(s-2))\dot{\alpha}(s-2)} - (D^+)^2 \bar{D}^+_{\dot{\beta}} \bar{\Psi}^{-\alpha(s-2)(\dot{\beta}\dot{\alpha}(s-2))} \Bigr]_{\theta^{-}=0}.\\
    \end{cases}
\end{equation}
The notation ``$\theta^{-}=0$'' means the restriction to the $\theta^-$-independent parts.

The prepotentials  give an alternative superfield representation for the higher spin degrees of freedom, but at cost of losing analyticity and enlarging the gauge group.
We will not present these gauge transformations here, though they can easily be written.

One can express $G^{++}$-superfields through $\Psi^{-\dots}$. In order to accomplish  this, it is useful and instructive to rewrite
the differential operator \eqref{operator-in} in the basis of the $\mathcal{N}=2$ covariant
derivatives, $D_M =(\partial_{\alpha\dot{\alpha}}, D^{-}_{\hat{\alpha}}, \partial_5)$:
\begin{equation}\label{cov}
    \hat{\mathcal{H}}^{++\alpha(s-2)\dot{\alpha}(s-2)}  = G^{++\alpha(s-2)\dot{\alpha}(s-2)M} D_M\,.
\end{equation}
Derivatives $D_M$ commute with $\mathcal{N}=2$ supersymmetry, so the coefficients $G^{++\dots}$ are invariant under supersymmetry by construction, though they are not analytic.
Their explicit expressions through the $h^{++\dots}$-potentials is given by eqs. \eqref{s-1} and \eqref{s-2}.

From eqs. \eqref{constr-sol} and \eqref{cov} one can directly express $G^{++\dots}$ through prepotentials:
\begin{equation}
    G^{++(\alpha(s-2)\beta)(\dot{\alpha}(s-2)\dot{\beta})} = 4i (D^+)^2 \bar{D}^{+(\dot{\beta}} \Psi^{-(\alpha(s-2)\beta)\dot{\alpha}(s-2))}
    +
    4i  (\bar{D}^+)^2 D^{+(\beta} \bar{\Psi}^{-\alpha(s-2))(\dot{\alpha}(s-1)\dot{\beta})}\,,
\end{equation}
\begin{equation}
    G^{++5\alpha(s-2)\dot{\alpha}(s-2)} =
    4i (\bar{D}^+)^2 D^+_\beta \Psi^{-(\beta\alpha(s-2))\dot{\alpha}(s-2)}
    -
    4i (D^+)^2 \bar{D}^+_{\dot{\beta}} \bar{\Psi}^{-\alpha(s-2)(\dot{\beta}\dot{\alpha}(s-2))}\,.
\end{equation}
Using these formulas, one can vary $G$-superfields with respect to the prepotentials $\Psi^-$. Varying the action \eqref{action++++} in this way yields:
\begin{equation}\label{Action-var}
    \begin{split}
        \delta_\Psi S_{(s)} =& 8i (-1)^{s+1}  \int d^4x d^8\theta du\; \Bigr\{ (D^+)^2 \bar{D}^{+(\dot{\beta}} \delta \Psi^{-\alpha(s-1)\dot{\alpha}(s-2))}  G^{--}_{\alpha(s-1)(\dot{\beta}\dot{\alpha}(s-2))}
        \\&
        \quad\quad\quad\quad\quad\quad\quad\quad\quad\quad\quad
        +
        (\bar{D}^+)^2 D^+_\beta \delta \Psi^{-(\beta\alpha(s-2))\dot{\alpha}(s-2)} G^{--5}_{\alpha(s-2)\dot{\alpha}(s-2)} \Bigr\},
    \end{split}
\end{equation}
\begin{equation}\label{Action-var-bar}
    \begin{split}
        \delta_{\bar{\Psi}} S_{(s)} =& 8i (-1)^{s+1}  \int d^4x d^8\theta du\; \Bigr\{ (\bar{D}^+)^2 D^{+(\beta} \delta \bar{\Psi}^{-\alpha(s-2))\dot{\alpha}(s-1)} G^{--}_{(\beta\alpha(s-2))\dot{\alpha}(s-1)}
        \\&
        \quad\quad\quad\quad\quad\quad\quad\quad\quad\quad\quad
        -
        (D^+)^2 \bar{D}^+_{\dot{\beta}} \delta \bar{\Psi}^{-\alpha(s-2)(\dot{\beta}\dot{\alpha}(s-2))} G^{--5}_{\alpha(s-2)\dot{\alpha}(s-2)} \Bigr\}.
    \end{split}
\end{equation}
Thus the variations of the actions were reduced to the variations of prepotentials which are unconstrained harmonic superfields.
After integration by parts, we deduce the equations of motion:
\begin{equation}\label{spin s EOM}
    \begin{cases}
        (D^+)^2 \bar{D}^{+\dot{\beta}} G^{--}_{(\beta\alpha(s-2)) (\dot{\beta}\dot{\alpha}(s-2))} + (\bar{D}^+)^2 D^+_{(\beta} G^{--5}_{\alpha(s-2) )  \dot{\alpha}(s-2) } = 0\,,\\
        (\bar{D}^+)^2 D^{+\beta} G^{--}_{(\beta\alpha(s-2))(\dot{\beta}\dot{\alpha}(s-2))}  -  (D^+)^2 \bar{D}^+_{(\dot{\beta}} G^{--5}_{\alpha(s-2)  \dot{\alpha}(s-2))} = 0\,.
    \end{cases}
\end{equation}
These equations coincide with eqs. \eqref{spinor1} and \eqref{spinor2} obtained earlier by
varying with respect to the analytic potentials $h^{++\dots}$.

\subsection{Current superfields for $\Psi^{-\dots}$ prepotentials}\label{non analytical supercurrents}

Varying the cubic coupling \eqref{action-final} with respect to the Mezincescu-type $\Psi^{-\dots}$ prepotential yields:
\begin{equation}
    \begin{split}
        \delta_{\Psi} S^{(s)}_{gauge}
        =&
        - \frac{\kappa_s}{2}  \int d\zeta^{(-4)}\;
        \Biggr\{  q^{+a}   (D^+)^4 \left(\delta\Psi^{-\alpha(s-2)\dot{\alpha}(s-2)\hat{\mu}} D^-_{\hat{\mu}}\right) \partial^{s-2}_{\alpha(s-2)\dot{\alpha}(s-2)} (J)^{P(s)} q^+_a
        \\& \qquad\qquad
        +
        \xi (D^+)^4 \left(\delta\Psi^{-\alpha(s-2)\dot{\alpha}(s-2)\hat{\mu}} D^-_{\hat{\mu}}\right) \partial^{s-2}_{\alpha(s-2)\dot{\alpha}(s-2)}\left[ q^{+a} (J)^{P(s)}  q^+_a \right]  \Biggr\}
        \\=&
        - \frac{\kappa_s}{2}  \int d\zeta^{(-4)}\,  (D^+)^4\;
        \Biggr\{  q^{+a}   \left(\delta\Psi^{-\alpha(s-2)\dot{\alpha}(s-2)\hat{\mu}} D^-_{\hat{\mu}}\right) \partial^{s-2}_{\alpha(s-2)\dot{\alpha}(s-2)} (J)^{P(s)} q^+_a
        \\& \qquad\qquad\qquad
        +
        \xi  \left(\delta\Psi^{-\alpha(s-2)\dot{\alpha}(s-2)\hat{\mu}} D^-_{\hat{\mu}}\right) \partial^{s-2}_{\alpha(s-2)\dot{\alpha}(s-2)}\left[ q^{+a} (J)^{P(s)}  q^+_a \right]  \Biggr\}.
    \end{split}
\end{equation}
Here we used the representation \eqref{constr-sol}. In the last equation one can restore the full harmonic superspace integration measure.

So we find that the superfield currents conjugated to the prepotentials $\Psi^{-\dots}$ are given by the expression:
\begin{equation}\label{pre-currents}
    \begin{split}
        \mathcal{J}^{+}_{\alpha(s-2)\dot{\alpha}(s-2)\hat{\mu}} : =&\; -\frac{1}{\kappa_s} \frac{\delta S^{(s)}_{gauge}}{\delta \Psi^{-\alpha(s-2)\dot{\alpha}(s-2)\hat{\mu}}}
        \\=&\;
        \frac{1}{2}
        q^{+a}   D^-_{\hat{\mu}} \partial^{s-2}_{\alpha(s-2)\dot{\alpha}(s-2)} (J)^{P(s)} q^+_a
        +
        \frac{1}{2}
        \xi  D^-_{\hat{\mu}} \left[ q^{+a} (J)^{P(s)}  q^+_a \right]\,.
    \end{split}
\end{equation}
These currents coincide with the supersymmetry-invariant spinor current superfields
\eqref{covariant current superfield} and \eqref{covariant current superfield 2} found earlier. The conservation law has the simple form:
\begin{equation}
    \mathcal{D}^{++}\mathcal{J}^{+}_{\alpha(s-2)\dot{\alpha}(s-2)\hat{\mu}} = 0\,.
\end{equation}
It is a direct consequence of the analyticity of the hypermultiplet  and its equation of motion.

It is also important to note that the supercurrents constructed in this way are manifestly invariant under ${\cal N}=2$ supersymmetry:
\begin{equation}\label{susy non analytic current}
    \delta_\epsilon \mathcal{J}^{+}_{\alpha(s-2)\dot{\alpha}(s-2)\hat{\mu}} = 0\,.
\end{equation}
In all these relations it is assumed that the index $\hat\mu (\mu, \dot\mu)$ is fully symmetrized with all other
spinorial indices of the same nature.

\section{Current superfields} \label{current}

Composite operators in a supersymmetric theory are members of some supermultiplets. This means the existence
of a superfield that contains composite operators as its components related to each other by supersymmetry transformations. We will be interested in superfields containing conserved currents,
since they appear as sources in the equations of motion for fields of higher spins.
Moreover, conserved currents furnish an important class of local operators and can be used to study the dynamics of supersymmetric field theories, supersymmetry breaking
and coupling to gauge theories (including higher spin theories) \cite{Dumitrescu:2011zz}. In this Appendix we will discuss general properties of supercurrents and present
analytic harmonic superfields which capture $\mathcal{N}=2$ conserved current supermultiplet\footnote{Among the first papers where supercurrents were defined and discussed
we mention, e.g., works \cite{Ferrara:1974pz, Ogievetsky:1976qc}.}.

\medskip

There are two essentially different types of the current supermultiplets:

\begin{itemize}
    \item {Symmetry current supermultiplet (for example, multiplets involving currents of internal symmetry)};
    \item {Supercurrent} (it contains a fermionic current generating rigid supersymmetry together with the stress tensor associated with space-time translations).

\end{itemize}

The structure of $\mathcal{N}=1$ current superfields is well known, see, e.g., \cite{Dumitrescu:2011zz, Bertolini}. The construction of higher spin conserved
$\mathcal{N}=1$ current superfields was presented
in refs. \cite{Howe:1981qj, HK1, Buchbinder:2018wzq, Buchbinder:2018nkp, Buchbinder:2018gle, Gates:2019cnl, Hutomo:2020wca}. We shall describe in Section \ref{N=2 current superfield}
$\mathcal{N}=2$ harmonic superfields that accommodate various current supermultiplets. In particular, we shall present harmonic superfields containing higher spin currents.

The structure of $\mathcal{N}=1$ supercurrents is  also well known, see for example \cite{BK, Kuzenko:2010am, Bertolini}. In $\mathcal{N}=2$ theories the structure of supercurrents
was elaborated in \cite{Butter:2010sc}. However, to the best of our knowledge, the supercurrent structure in harmonic superspace was not systematically discussed so far.

\subsection{$\mathcal{N}=2$ current superfields}\label{N=2 current superfield}

\textbf{1.} We consider the real (with respect to the tilde conjugation, $\widetilde{J^{++A}} = J^{++A} $) analytic superfield $J^{++A}$,
where $A$ is some multi-index of the form $A = \alpha(n)\dot{\alpha}(n)$.
We impose the ``harmonic conservation law'' on the superfield $J^{++A}$ \footnote{This conservation law formally coincides with the harmonic constraint $\mathcal{D}^{++} L^{++} = 0$
defining $\mathcal{N}=2$ tensor multiplet $L^{++}(\zeta)$ off shell. For details see Section 6.2 of \cite{18}. }:
\begin{equation}\label{conservation}
    \mathcal{D}^{++} J^{++A} = \partial^{++} J^{++A} -2i \theta^{+\alpha} \bar{\theta}^{+\dot{\alpha}} \partial_{\alpha\dot{\alpha}} J^{++A}  = 0\,.
\end{equation}
Here we assumed that the supercurrent does not depend on $x^5$.

The general solution of the constraint $\mathcal{D}^{++} J^{++A} = 0$ is given by:
\begin{equation}\label{current - exp}
    \begin{split}
        J^{++A}(x, \theta^+, u) =& j^{(ij) A}(x) u^+_i u^+_j  + \theta^{+\alpha} j^{i A}_{\alpha} (x) u^+_i
        +
        \bar{\theta}^{+\dot{\alpha}} \bar{j}^{i A}_{\dot{\alpha}} (x) u^+_i
        \\&+
        (\theta^{+})^2 j^A(x) +
        (\bar{\theta}^+)^2 \bar{j}^A(x)
        \\& +
        2i \theta^+ \sigma^n \bar{\theta}^+ \left[ j_n^A (x) + \partial_n j^{(ij) A}(x) u^+_i u^-_j \right]
        \\&
        - i
        (\bar{\theta}^+)^2\theta^{+\alpha} \partial^{\dot{\alpha}}_\alpha \bar{j}^{i A}_{\dot{\alpha}} u^-_i
        - i
        (\theta^+)^2 \bar{\theta}^{+\dot{\alpha}} \partial_{\dot{\alpha}}^\alpha j^{iA}_{\alpha} u^-_i
        \\&
        -
        (\theta^+)^2 (\bar{\theta}^+)^2 \Box j^{(ij)A} u^-_i u^-_j\,.
    \end{split}
\end{equation}
As independent components, it involves the  real field $j^{(ij)A}(x)$, the  complex field $j^{iA}_{\alpha}(x)$ and its complex conjugate $\bar{j}^{iA}_{\dot{\alpha}}(x)$, the complex field $j^A(x)$ and
the real field $j_n^A(x)$. Equation \eqref{conservation} implies the  current conservation condition for the latter:
\begin{equation}
    \partial^n j_n^A = 0\,.
\end{equation}

This allows us to identify the harmonic superfield $J^{++A}$ with the supermultiplet containing the conserved current and the equation $\mathcal{D}^{++}J^{++A} = 0$ with the relevant superfield conservation condition.
All bosonic higher spin current superfields considered in Section \ref{Hyper-current-superfields} have such a structure.

\textbf{2.} Fermionic higher spin current superfields are also analytic, but, in contrast to bosonic higher spin current superfields, they satisfy different conditions \eqref{ModConserv}:
\begin{equation}\label{current superfield mod}
    \begin{cases}
    \mathcal{D}^{++} J^{+}_{\alpha(s-2)\dot{\alpha}(s-1)} =
    -2i \theta^{+\rho} J^{++}_{(\alpha(s-2)\rho)\dot{\alpha}(s-1)}
    -
    2i \bar{\theta}^+_{(\dot{\alpha}} J^{++}_{\alpha(s-2)\dot{\alpha}(s-2))}\,,
    \\
    \mathcal{D}^{++}  J^{++}_{\alpha(s-1)\dot{\alpha}(s-1)} = 0\,,
    \\
        \mathcal{D}^{++}  J^{++}_{\alpha(s-2)\dot{\alpha}(s-2)} = 0\,.
    \end{cases}
\end{equation}
Using the general solution  \eqref{current - exp} of the second and third equations in \eqref{current superfield mod} and substituting it in the first equation o,
one can obtain, for the coefficient in  the $\theta^{+\rho}\bar{\theta}^{+\dot{\rho}}$ piece of the fermionic current superfield,
\begin{equation}
    J^+_{\alpha(s-2)\dot{\alpha}(s-1)} = \dots + 2i \theta^{+\rho} \bar{\theta}^{+\dot{\rho}} j^i_{\rho\dot{\rho} \alpha(s-2)\dot{\alpha}(s-1)} u^-_i + \dots\,,
\end{equation}
the following conservation condition:
\begin{equation}
    \partial^{\rho\dot{\rho}} \left(j^i_{\rho\dot{\rho} \alpha(s-2)(\dot{\alpha}(s-2)\dot{\sigma})} - j^i_{\dot{\rho}(\alpha(s-2)\rho)(\dot{\alpha}(s-2)\dot{\sigma})}
    +
    \epsilon_{(\dot{\sigma}\dot{\rho}} j^i_{\rho\alpha(s-2)\dot{\alpha}(s-2))} \right) = 0\,.
\end{equation}
Here, the fields $j^i_{\dot{\rho}\alpha(s-1)\dot{\alpha}(s-1)}$ and $j^i_{\rho\alpha(s-2)\dot{\alpha}(s-2)}$ are the appropriate fermionic components of the bosonic current superfields
$J^{++}_{\alpha(s-1)\dot{\alpha}(s-1)}$ and $J^{++}_{\alpha(s-2)\dot{\alpha}(s-2)}$, respectively.

So we conclude that the fermionic current, defined as
\begin{equation}
    \mathfrak{j}^i_{{\rho\dot{\rho} \alpha(s-2)(\dot{\alpha}(s-2)\dot{\sigma})} } := j^i_{\rho\dot{\rho} \alpha(s-2)(\dot{\alpha}(s-2)\dot{\sigma})} - j^i_{\dot{\rho}(\alpha(s-2)\rho)(\dot{\alpha}(s-2)\dot{\sigma})}
    -
    \epsilon_{\dot{\rho}(\dot{\sigma}} j^i_{\rho\alpha(s-2)\dot{\alpha}(s-2))}
\end{equation}
is conserved as a consequence of the superfield equations \eqref{current superfield mod}.

\textbf{3.} An  analogous consideration for the complex conjugate current superfield \eqref{ModConser2},
\begin{equation}
    \label{conservation mod 2}
    \begin{cases}
    \mathcal{D}^{++} J^{+}_{\alpha(s-1)\dot{\alpha}(s-2)}
    =
    2i \bar{\theta}^{+\dot{\rho}} J^{++}_{\alpha(s-1)(\dot{\alpha}(s-2)\dot{\rho})}
    -
    2i
    \theta^+_{(\alpha} J^{++}_{\alpha(s-2))\dot{\alpha}(s-2)}\,,
    \\
    \mathcal{D}^{++} J^{++}_{\alpha(s-1)\dot{\alpha}(s-1)} = 0\,,
    \\
        \mathcal{D}^{++} J^{++}_{\alpha(s-2)\dot{\alpha}(s-2)} = 0\,,
    \end{cases}
\end{equation}
evidently implies that the corresponding fermionic component,
\begin{equation}
    J^+_{\alpha(s-1)\dot{\alpha}(s-2)} = \dots +2i \theta^{+\rho} \bar{\theta}^{+\dot{\rho}}
    j^i_{\rho\dot{\rho}\alpha(s-1)\dot{\alpha}(s-2)} + \dots\,,
\end{equation}
obeys the conservation law:
\begin{equation}\label{modefied conservation}
    \partial^{\rho\dot{\rho}} \left(j^i_{\rho\dot{\rho}(\alpha(s-2)\sigma)\dot{\alpha}(s-2)}
    -
    j^i_{\rho(\alpha(s-2)\sigma)(\dot{\alpha}(s-2)\dot{\rho})}
    +
    \epsilon_{\rho(\sigma} j^i_{\dot{\rho}\alpha(s-2)) \dot{\alpha}(s-2)}\right) = 0\,.
\end{equation}
Here, the fields $j^i_{\rho\alpha(s-1)\dot{\alpha}(s-1)}$ and $j^i_{\dot{\rho}\alpha(s-2)\dot{\alpha}(s-2)}$ are proper
fermionic components of the bosonic current superfields
$J^{++}_{\alpha(s-1)\dot{\alpha}(s-1)}$ and $J^{++}_{\alpha(s-2)\dot{\alpha}(s-2)}$, respectively.
Thus the relevant conserved fermionic current is defined by:
\begin{equation}
    \mathfrak{j}^i_{\rho\dot{\rho}(\alpha(s-2)\sigma)\dot{\alpha}(s-2)}
    :=
    j^i_{\rho\dot{\rho}(\alpha(s-2)\sigma)\dot{\alpha}(s-2)}
    -
    j^i_{\rho(\alpha(s-2)\sigma)(\dot{\alpha}(s-2)\dot{\rho})}
    +
    \epsilon_{\rho(\sigma} j^i_{\dot{\rho}\alpha(s-2)) \dot{\alpha}(s-2)}\,.
\end{equation}

In Sections \ref{hyper supercurrent} and \ref{spin 3 supercurrent} we presented the explicit expressions of these current superfields in terms of the free hypermultiplet
in the spin $\mathbf{2}$ (see also Appendix \ref{supercurrent}) and the spin $\mathbf{3}$ cases.

In Appendix \ref{Noether} the conservation conditions \eqref{conservation}, \eqref{current superfield mod} and \eqref{conservation mod 2} will be independently derived
by making use of $\mathcal{N}=2$ superfield Noether theorem for rigid symmetries of the free
hypermultiplet.

\subsection{$\mathcal{N}=2$ supercurrent}\label{supercurrent}

Let us now consider the special supercurrent $J^{++\alpha\dot{\alpha}}$ having an additional conservation law:
\begin{equation}
    \begin{cases}
        \mathcal{D}^{++} J^{++\alpha\dot{\alpha}} = 0\,,\\
        \partial_{\alpha\dot{\alpha}} J^{++\alpha\dot{\alpha}} = 0\,.
    \end{cases}
\end{equation}

The second condition implies the conservation of all the component currents in its relevant $\theta$-expansion \eqref{current - exp}:
\begin{equation}
    \begin{cases}
        \partial_{\alpha\dot{\alpha}} j^{(ij)\alpha\dot{\alpha}} = 0\,,\\
        \partial_{\alpha\dot{\alpha}} j^{i\alpha\dot{\alpha}}_{\beta} = 0\,,\\
        \partial_{\alpha\dot{\alpha}} j^{\alpha\dot{\alpha}} =0\,,\\
        \partial_{m} j_n^m = 0\,.
    \end{cases}
\end{equation}

These currents can be identified with the currents inherent to any $\mathcal{N}=2$ super Poincar\'e invariant theory: $j_n^m$ with the energy-momentum tensor,
$j^{\alpha\dot{\alpha}}$ with ${\rm U}(1)$ R-symmetry current, $j^{i\alpha\dot{\alpha}}_{\beta}$ with the doublet of supersymmetry fermionic currents and $j^{(ij)\alpha\dot{\alpha}}$ with the ${\rm SU}(2)$ R-symmetry current.

In what follows we will use the notation:
\begin{equation}\label{supercurrent - superfield}
    \begin{split}
        J^{++}_{\alpha\dot{\alpha}}(x, \theta^+, u) =& R^{(ij)}_{\alpha\dot{\alpha}}(x) u^+_i u^+_j  + \theta^{+\beta} S^{i }_{\alpha\dot{\alpha} \beta} (x) u^+_i
        +
        \bar{\theta}^{+\dot{\beta}} \bar{S}^{i }_{\alpha\dot{\alpha}\dot{\beta}} (x) u^+_i
        \\&+
        (\theta^{+})^2 R_{\alpha\dot{\alpha}}(x) +
        (\bar{\theta}^+)^2 \bar{R}_{\alpha\dot{\alpha}}(x)
        \\& +
        2i \theta^+ \sigma^n \bar{\theta}^+ \left[ T_{n\, \alpha\dot{\alpha}} (x) + \partial_n R^{(ij)}_{\alpha\dot{\alpha}}(x) u^+_i u^-_j \right]
        \\&
        - i
        (\bar{\theta}^+)^2\theta^{+\beta} \partial^{\dot{\beta}}_\beta \bar{S}^{i }_{\alpha\dot{\alpha} \dot{\beta}} u^-_i
        - i
        (\theta^+)^2 \bar{\theta}^{+\dot{\beta}} \partial_{\dot{\beta}}^\beta S^{i}_{\alpha\dot{\alpha}\beta} u^-_i
        \\&
        -
        (\theta^+)^2 (\bar{\theta}^+)^2 \Box R^{(ij)}_{\alpha\dot{\alpha}} u^-_i u^-_j\,.
    \end{split}
\end{equation}
For example, the analytic current superfield \eqref{Supercurrent1-spin2} constructed out of the hypermultiplet superfield $q^{+}$,  has such a structure
and so can be identified with the $\mathcal{N}=2$ supercurrent superfield of hypermultiplet, as was already mentioned in Section \ref{hyper supercurrent}.

\section{$\mathcal{N}=2$ Noether theorem and rigid symmetries of hypermultiplet} \label{Noether}

In this Appendix, we briefly review the issue of higher-derivative rigid symmetries of the free hypermultiplet action defined in \cite{Buchbinder:2022kzl}
and relate these symmetries to the conserved current superfields.

\subsection{Rigid symmetries}
\label{Rigid symmetries}

The rigid symmetries of hypermultiplet considered in \cite{Buchbinder:2022kzl} are realized by the following transformations\footnote{Here $s\geq 2$.
For $s=1$, the symmetry transformation is:
$$
\delta_{rig}^{(s=1)} q^{+a} = - \lambda J q^{+a}.
$$
Unlike these transformations, all higher spin ($s\geq 2$) transformations contain derivatives.}:
\begin{equation}\label{global-sym}
    \delta^{(s)}_{rig} q^{+a} = -\hat{\Lambda}_{rig}^{\alpha(s-2)\dot{\alpha}(s-2)}\partial^{s-2}_{\alpha(s-2)\dot{\alpha}(s-2)} (J)^{P(s)} q^{+a}\,,
\end{equation}
where the differential operator $\hat{\Lambda}_{rig}^{\alpha(s-2)\dot{\alpha}(s-2)}$ is defined as
\begin{equation}\label{Lambda rigid}
    \hat{\Lambda}_{rig}^{\alpha(s-2)\dot{\alpha}(s-2)} := \Lambda^{\alpha(s-2)\dot{\alpha}(s-2)M} \partial_M,
    \qquad M = (\alpha\dot{\alpha}, \alpha, \dot{\alpha}, 5),
\end{equation}
with the parameters of the transformations as coefficients:
\begin{subequations}\label{LambdA}
    \begin{equation}
        \Lambda^{(\alpha(s-2)\alpha)(\dot{\alpha}(s-2)\dot{\alpha})} =
        \lambda^{\alpha(s-1)\dot{\alpha}(s-1)} - 2i \lambda^{\alpha(s-1)(\dot{\alpha}(s-2)-} \bar{\theta}^{+\dot{\alpha})}
        -
        2i
        \theta^{+(\alpha} \lambda^{\alpha(s-2)) \dot{\alpha}(s-1)-}\,,
    \end{equation}
    \begin{equation}
        \Lambda^{\alpha(s-2)\dot{\alpha}(s-2)5} =
        \lambda^{\alpha(s-2)\dot{\alpha}(s-2)5}
        +
        2i \lambda^{(\alpha(s-2)\beta)\dot{\alpha}(s-2)-} \theta^+_{{\beta}} + 2i \lambda^{\alpha(s-2)(\dot{\alpha}(s-2)\dot\beta) -} \bar\theta^+_{{\dot\beta}}\,,
    \end{equation}
    \begin{equation}
        \Lambda^{\alpha(s-1)\dot{\alpha}(s-2)+}
        =
        \lambda^{\alpha(s-1)\dot{\alpha}(s-2)+}\,,
        \;\;\;\;\;
        \lambda^{\alpha(s-1)\dot{\alpha}(s-2)\pm}
        =
        \lambda^{\alpha(s-1)\dot{\alpha}(s-2)i} u_i^\pm\,,
    \end{equation}
    \begin{equation}
        \Lambda^{\alpha(s-2) \dot{\alpha}(s-1)+}=
        \lambda^{\alpha(s-2) \dot{\alpha}(s-1)+}\,,
        \;\;\;\;\;
        \lambda^{\alpha(s-2) \dot{\alpha}(s-1)\pm}=
        \lambda^{\alpha(s-2) \dot{\alpha}(s-1)i} u_i^\pm\,.
    \end{equation}
\end{subequations}
In \eqref{Lambda rigid} the same convention concerning the spinor indices inside the index $M$ is assumed as in the differential operators \eqref{operator-in}.
All the parameters $\lambda^{\alpha(s-1)\dot{\alpha}(s-1)}\,, \lambda^{\alpha(s-2)\dot{\alpha}(s-2)5}\,,$  $\lambda^{\alpha(s-1)\dot{\alpha}(s-2)i},
\,\lambda^{\alpha(s-2) \dot{\alpha}(s-1)i}$ in the operator $\hat{\Lambda}_{rig}^{\alpha(s-2)\dot{\alpha}(s-2)}$ are coordinate-independent.
The analytic-superspace localization of these parameters produces the spin $s$, spin $s-1$ and
spin $s-\tfrac{1}{2}$ gauge transformations, correspondingly.
The analytic operator \eqref{Lambda rigid}
satisfies  the condition:
\begin{equation}
    [\mathcal{D}^{++}, \hat{\Lambda}_{rig}^{\alpha(s-2)\dot{\alpha}(s-2)}] \partial^{s-2}_{\alpha(s-2)\dot{\alpha}(s-2)} = 0\,.
\end{equation}
This property is crucial for the invariance of the free hypermultiplet action \eqref{hyper} under \eqref{global-sym}.

Let us consider the algebra of these symmetry transformations. For simplicity, we shall limit our study to the massless case
({\it i.e.}, with  $\partial_5 q^{+a} = 0$).
The commutation relation of two rigid symmetries \ref{global-sym} has the following generic form:
\begin{equation}
    \begin{split}
    [\delta^{(s)}_{\lambda_1}, \delta^{(s)}_{{\lambda}_2} ] q^{+a}
    =&
    \Lambda_1^{\alpha(s-2)\dot{\alpha}(s-2)N} \partial_N
    \Lambda_2^{\beta(s-2)\dot{\beta}(s-2)M}
    \partial_M
    \partial^{s-2}_{\alpha(s-2)\dot{\alpha}(s-2)}
    \partial^{s-2}_{\beta(s-2)\dot{\beta}(s-2)}
    \\&-
    \Lambda_2^{\alpha(s-2)\dot{\alpha}(s-2)M} \partial_M
    \Lambda_1^{\beta(s-2)\dot{\beta}(s-2)N} \partial_N
    \partial^{s-2}_{\alpha(s-2)\dot{\alpha}(s-2)}
    \partial^{s-2}_{\beta(s-2)\dot{\beta}(s-2)}.
    \end{split}
\end{equation}
Only transformations with fermionic parameters can give a nontrivial contribution to this commutator.
Keeping this in mind, we stick to the rigid spinor parameters $\lambda_1^{\alpha(s-1)\dot{\alpha}(s-2)i}$,
$\lambda_2^{\alpha(s-2)\dot{\alpha}(s-1)i}$. A simple computation yields:
\begin{equation}\label{algebra general}
    \begin{split}
    [\delta^{(s)}_{\lambda_1}, \delta^{(s)}_{\lambda_2}]
    q^{+a}
    =
    &2i
    \left(\lambda_1^{\alpha(s-1)\dot{\alpha}(s-2)+} \lambda_2^{\beta(s-2)\dot{\beta}(s-1)-}
    -
    \lambda_1^{\alpha(s-1)\dot{\alpha}(s-2)-} \lambda_2^{\beta(s-2)\dot{\beta}(s-1)+}
    \right)
    \\&
    \times
    \partial^{s-2}_{\alpha(s-2)\dot{\alpha}(s-2)}
    \partial^{s-2}_{\beta(s-2)\dot{\beta}(s-2)}
    \partial_{\alpha\dot{\beta}} q^{+a} + {\rm c.c.}
    \end{split}
\end{equation}
Using the completeness relation
\begin{equation}
    u^+_i u^-_j - u^-_i u^+_j = \epsilon_{ij}\,,
\end{equation}
we obtain:
\begin{equation}\label{superfield commutator}
    [\delta^{(s)}_{\lambda_1}, \delta^{(s)}_{\lambda_2}]
    q^{+a}
    =
    2i
    \lambda_1^{\alpha(s-1)\dot{\alpha}(s-2)j} \left(\lambda_2^{\beta(s-2)\dot{\beta}(s-1)}\right)_j
    \partial^{s-2}_{\alpha(s-2)\dot{\alpha}(s-2)}
    \partial^{s-2}_{\beta(s-2)\dot{\beta}(s-2)}
    \partial_{\alpha\dot{\beta}} q^{+a} + {\rm c.c.}
\end{equation}

 For $s=2$ case, we reproduce the standard result that commutator of two rigid supersymmetries
 leads to translations.
 In fact, the transformations \eqref{global-sym} are closed only in the $s=2$ case, while for $s>2$
 new transformations are
 inevitably generated (as is expected for higher-spin algebras).

Let us adduce more details for the simplest non-trivial case of $s=3$:
\begin{equation}\label{superfield commutator spin 3}
    [\delta^{(3)}_{\lambda_1}, \delta^{(3)}_{\lambda_2}]
    q^{+a}
    =
    2i
    \lambda_1^{(\alpha\beta)\dot{\alpha}j} \left(\lambda_2^{\gamma(\dot{\beta}\dot{\gamma})}\right)_j
    \partial^{}_{\alpha\dot{\alpha}}
    \partial^{}_{\gamma\dot{\gamma}}
    \partial_{\beta\dot{\beta}} q^{+a} + {\rm c.c.}
\end{equation}
The transformation in the right hand side consists of the two irreducible pieces:

1. Totally symmetric one:
\begin{equation}
        [\delta^{(3)}_{\lambda_1}, \delta^{(3)}_{\lambda_2}]_{sym}
    q^{+a}
    =
    \lambda^{(\alpha\beta\gamma)(\dot{\alpha}\dot{\beta}\dot{\gamma})}
    \partial^{}_{\alpha\dot{\alpha}}
    \partial^{}_{\gamma\dot{\gamma}}
    \partial_{\beta\dot{\beta}} q^{+a},
\end{equation}
where:
\begin{equation}
    \lambda^{(\alpha\beta\gamma)(\dot{\alpha}\dot{\beta}\dot{\gamma})}
    =
        2i
    \lambda_1^{(\alpha\beta)\dot{\alpha}j} \left(\lambda_2^{\gamma(\dot{\beta}\dot{\gamma})}\right)_j +{\rm symm}(\alpha, \beta, \gamma) +  {\rm c.c.}\,.
\end{equation}

2. Antisymmetric in $[\alpha\gamma]$, $[\dot{\alpha}\dot{\gamma}]$:

\begin{equation}
    [\delta^{(3)}_{\lambda_1}, \delta^{(3)}_{\lambda_2}]_{antisym}
    q^{+a}
    =
    \lambda^{\beta\dot{\beta}}
    \partial_{\beta\dot{\beta}} \Box q^{+a},
\end{equation}
where:
\begin{equation}
        \lambda^{\beta\dot{\beta}}
        =
            -i
        \left(\lambda_1\right)^{(\alpha\beta)j}_{\dot{\gamma}} \left(\lambda_2\right)^{(\dot{\beta}\dot{\gamma})}_{\alpha j}\,
        +
        {\rm c.c.}
\end{equation}
where we used the relation:
\begin{equation}
    \partial_{[\alpha\dot{\alpha}} \partial_{\gamma]\dot{\gamma}}
    =
    \frac{1}{2} \epsilon_{\alpha\gamma} \epsilon_{\dot{\alpha}\dot{\gamma}} \Box.
\end{equation}
Thus the commutator of two rigid spin $\tfrac{3}{2}$ supersymmetries generates a rigid bosonic spin $s=4$ transformation associated with the
parameter $\lambda^{(\alpha\beta\gamma)(\dot{\alpha}\dot{\beta}\dot{\gamma})}$ and
the spin $s=2$-like transformation with a vector parameter and one $\Box$ operator acting on $q^{+a}$.
This clearly demonstrates that the algebra of transformations \eqref{Rigid symmetries} is not closed for
 $s\geq 3$ \footnote{It is a crucial difference of the higher spin $3/2$
supersymmetry (and its higher-rank cousins)
from those closed $3/2$ superalgebras (and their higher-rank analogs) which were considered, e.g.,
in \cite{32SUSY}.}.

\medskip

In the general case, we see that the commutator of two symmetry transformations with spin $(s-\tfrac{3}{2})$ parameters
results in a bosonic symmetry transformation with a spin $2s-3$ parameter $(s \geq 2)$:
\begin{equation}
    \lambda^{(\alpha(s-1)\beta(s-2))(\dot{\alpha}(s-2)\dot{\beta}(s-1))}
    =
    2i \lambda_1^{(\alpha(s-1)(\dot{\alpha}(s-2)j} \left(\lambda_2^{\beta(s-2))\dot{\beta}(s-1))}\right)_j + {\rm symm}(\alpha(s-1), \beta(s-2)) +  {\rm c.c.},
\end{equation}
plus some extra parts (for $(s \geq 3)$) which include powers of $\Box$ acting on $q^{+a}$ and generalize those found in the $s=3$ example.

\medskip

As a simple generalization, one can also calculate the bracket of two different fermionic symmetries:
\begin{equation}
    \begin{split}
        [\delta^{(s_1)}_{\lambda_1}, \delta^{(s_2)}_{\lambda_2}]
        q^{+a}
        =
        &2i
        \left(\lambda_1^{\alpha(s_1-1)\dot{\alpha}(s_1-2)+} \lambda_2^{\beta(s_2-2)\dot{\beta}(s_2-1)-}
        -
        \lambda_1^{\alpha(s_1-1)\dot{\alpha}(s_1-2)-} \lambda_2^{\beta(s_2-2)\dot{\beta}(s_2-1)+}
        \right)
        \\&
        \times
        \partial^{s_1-2}_{\alpha(s_1-2)\dot{\alpha}(s_1-2)}
        \partial^{s_2-2}_{\beta(s_2-2)\dot{\beta}(s_2-2)}
        \partial_{\alpha\dot{\beta}}  \left(J\right)^{P(s_1)+ P(s_2)}q^{+a}
        \\=
        &
        2i\,
        \lambda_1^{\alpha(s_1-1)\dot{\alpha}(s_1-2)j} \left(\lambda_2^{\beta(s_2-2)\dot{\beta}(s_2-1)}\right)_j
        \\ &\times
        \partial^{s_1-2}_{\alpha(s_1-2)\dot{\alpha}(s_1-2)}
        \partial^{s_2-2}_{\beta(s_2-2)\dot{\beta}(s_2-2)}
        \partial_{\alpha\dot{\beta}}  \left(J\right)^{P(s_1+ s_2)}q^{+a}.
    \end{split}
\end{equation}
It leads to the bosonic rigid spin  $s=s_1+s_2-2$ symmetry transformation with the spin $s=s_1+s_2-3$ constant parameter and some
additional symmetries involving powers of $\Box$ acting on $q^{+a}$.  For example, in the simplest nontrivial case ($s_1=2,\; s_2=3$)
the bracket  of rigid spin $\tfrac12$ ${\cal N}=2$ supersymmetry with a rigid spin $\tfrac{3}{2}$ supersymmetry  in the right hand side yields, as a leading term,
the bosonic spin 3 symmetry transformation with the rank 2 symmetric traceless tensor constant parameter.

Here we considered the global symmetries of a free hypermultiplet. If we switch to the corresponding gauge transformations,
then a similar reasoning leads to the conclusion that in the right hand sides of the commutation relations,
gauge transformations of an increasing higher spins will appear. This implies
that in the hypothetical complete consistent interacting theory of ${\cal N}=2$ higher spins, the entire infinite towers
of all higher spins should simultaneously participate, in agreement with a general belief.

\subsection{Fermionic  symmetry with spin $\tfrac{3}{2}$ parameter}
\label{5/2 symmetry}

It is instructive to see how the above transformations look at the component level.
Let us consider the $s = 3$ rigid transformations.
It is easy to verify that the bosonic symmetry is the well-known spin $3$ symmetry of the free bosonic action \cite{Berends:1985xx}. We shall pay our attention to
the less trivial fermionic rigid symmetries:
\begin{equation}
    \begin{split}
    \delta^{(3)}_{rig} q^{+a} =& \quad 2i \left( \lambda^{(\alpha\beta)(\dot{\alpha}-} \bar{\theta}^{+\dot{\beta})} + \theta^{+(\alpha}\lambda^{\beta)(\dot{\alpha}\dot{\beta})-} \right) \partial_{\alpha\dot{\alpha}} \partial_{\beta\dot{\beta}} i (\tau_3)^a_{\;b} q^{+b}
    \\&-2i  \left( \lambda^{(\alpha\beta)\dot{\alpha}-}\theta^+_\beta + \lambda^{\alpha(\dot{\alpha}\dot{\beta})-}\bar{\theta}^+_{\dot{\beta}}\right) \partial_{\alpha\dot{\alpha}} \partial_5 i (\tau_3)^a_{\;b} q^{+b}
    \\&
    + \left(\lambda^{(\alpha\beta)\dot{\alpha}+}\partial^-_{\beta} + \lambda^{(\dot{\alpha}\dot{\beta})\alpha+} \partial^-_{\dot{\alpha}} \right)
    \partial_{\alpha\dot{\beta}}  i (\tau_3)^a_{\;b} q^{+b}\,.
    \end{split}
\end{equation}
They are characterized by the constant spin $\tfrac{3}{2}$ Grassmann parameters $\lambda^{(\alpha\beta)\dot{\alpha}i}$ and $\bar{\lambda}^{(\dot{\alpha}\dot{\beta})\alpha i}$. These transformations are realized on the on-shell hypermultiplet fields as:
\begin{equation}
    \begin{cases}
        \delta f^i = i\lambda^{(\alpha\beta)\dot{\alpha}i} \partial_{\alpha\dot{\alpha}} \chi_\beta
        +
        i \bar{\lambda}^{(\dot{\alpha}\dot{\beta})\alpha i} \partial_{\alpha\dot{\alpha}}\bar{\psi}_{\dot{\beta}}\,,
        \\
        \delta \bar{f}_i
 = - i \lambda_i^{(\alpha\beta)\dot{\alpha}} \partial_{\alpha\dot{\alpha}} \psi_\beta
 -
 i
 \bar{\lambda}_i^{(\dot{\alpha}\dot{\beta})\alpha} \partial_{\alpha\dot{\alpha}} \bar{\chi}
_{\dot{\beta}} \,,\\
         \delta \psi_\alpha = -2\bar{\lambda}^{\beta(\dot{\alpha}\dot{\beta})i} \partial_{\alpha\dot{\alpha}}\partial_{\beta\dot{\beta}} \bar{f}_i - 2 im \lambda_\alpha^{\beta\dot{\beta}i} \partial_{\beta\dot{\beta}} \bar{f}_i
         \,,\\
         \delta \bar{\psi}_{\dot{\alpha}}
         =
          2 \lambda^{(\alpha\beta)\dot{\beta}i} \partial_{\alpha\dot{\alpha}} \partial_{\beta\dot{\beta}} f_i
         -
         2i m \bar{\lambda}_{\dot{\alpha}}^{\beta\dot{\beta}i} \partial_{\beta\dot{\beta}} f_i
         \,,\\
         \delta \chi_\alpha =
         - 2 \bar{\lambda}^{\beta(\dot{\alpha}\dot{\beta})i} \partial_{\alpha\dot{\alpha}} \partial_{\beta\dot{\beta}} f_i
         - 2i m \lambda_\alpha^{\beta\dot{\beta} i} \partial_{\beta\dot{\beta}} f_i
         \,,\\
        \delta \bar{\chi}_{\dot{\alpha}}
        =
        2 \lambda^{(\alpha\beta)\dot{\beta}i} \partial_{\alpha\dot{\alpha}} \partial_{\beta\dot{\beta}} \bar{f}_i
        -
        2i m \bar{\lambda}_{\dot{\alpha}}^{\dot{\beta}\beta i} \partial_{\beta\dot{\beta}} \bar{f}_i\,.
    \end{cases}
\end{equation}
The resulting transformations look similar to $\mathcal{N}=2$ supersymmetry transformations
\eqref{susy hyp} of the free hypermultiplet, but involve more derivatives.
One can easily check that  on-shell hypermultiplet action \eqref{hyper on shell} is invariant under
these transformations (up to total derivatives).
So we observe that among symmetries of free hypermultiplet \eqref{global-sym} there is a kind of
$\mathcal{N}=2$ ``higher spin supersymmetry'' transformations.
The corresponding conserved Noether current is given by the expression
\begin{equation}\label{spin 5/2 current}
    \begin{split}
    J^{\alpha\dot{\alpha}}_{(\mu\nu) \dot{\mu} i}
    =
    &\frac{i}{2} \left[\partial_{(\mu\dot{\mu}} \chi_{\nu)} \partial^{\alpha\dot{\alpha}}\bar{f}_i
    -
    \partial^{\alpha\dot{\alpha}}\partial_{(\mu\dot{\mu}} \chi_{\nu)} \bar{f}_i\right]
    +
    \frac{i}{2}
    \left[ \partial_{\mu\dot{\mu}}\psi_\beta \partial^{\alpha\dot{\alpha}} f_i
    -
    \partial^{\alpha\dot{\alpha}}\partial_{\mu\dot{\mu}}\psi_\beta  f_i
    \right]
    \\
    &- \delta_{(\mu}^\alpha \Big[ \partial_{\nu)\dot{\mu}} \bar{f}_i \left(i \partial^{\dot{\alpha}\rho} \chi_\rho + m \bar{\psi}^{\dot{\alpha}}\right)
    +
    \partial_{\nu) \dot{\mu}} f_i \left(i \partial^{\dot{\alpha}\rho} \phi_\rho + m \bar{\chi}^{\dot{\alpha}}\right) \Big]
    \\
    &
    +
    \delta^\alpha_{(\nu} \delta^{\dot{\alpha}}_{\dot{\mu}}
    \Bigr[ i f_i \left(\Box \psi_\mu + m^2 \psi_\mu\right)
    +
    i
    \bar{f}_i \left(\Box \chi_\mu + m^2 \chi_\mu\right) \Bigr],
    \end{split}
\end{equation}
\begin{equation}
    \partial_{\alpha\dot{\alpha}} J^{\alpha\dot{\alpha}}_{(\mu\nu) \dot{\mu} i}  = 0\,.
\end{equation}
Note that the second and third lines of \eqref{spin 5/2 current} are vanishing on the free hypermultiplet equations of motion \eqref{hyper-equations}.

One can calculate Lie bracket of such transformations, for example:
\begin{equation}
    \begin{split}
    [\delta_{\lambda_1}, \delta_{\bar{\lambda}_2}] f^i
    =&
    -2i
    \left(\lambda_1^{(\alpha\beta)\dot{\alpha} j} \bar{\lambda}_2^{(\dot{\rho}\dot{\sigma})\rho i}
    -
    \lambda_1^{(\alpha\beta)\dot{\alpha} i} \bar{\lambda}_2^{(\dot{\rho}\dot{\sigma})\rho j}\right)
    \partial_{\alpha\dot{\alpha}}\partial_{\beta\dot{\sigma}}\partial_{\rho\dot{\rho}} f_j
    \\=&
    -2i
    \lambda_1^{(\alpha\beta)\dot{\alpha} j} \left(\bar{\lambda}_2^{(\dot{\rho}\dot{\sigma})\rho }\right)_j
    \partial_{\alpha\dot{\alpha}}\partial_{\beta\dot{\sigma}}\partial_{\rho\dot{\rho}} f^i
    .
    \end{split}
\end{equation}
So in the full agreement with the discussion of off-shell $s=3$ transformation \eqref{superfield commutator spin 3} in the previous Subsection \ref{Rigid symmetries},
we observe that the commutator of two  spin $\tfrac{3}{2}$ ``supersymmetries'' yields spin $s=4$ symmetry with a spin 3 parameter
and an extra  $\Box$ term.

\subsection{Noether theorem}

In this subsection we derive the current superfields corresponding to the  rigid symmetries \eqref{global-sym} of the free hypermultiplet action.

We will use Noether theorem in the form which can be simply extended to the superspace setting. If some continuous transformations with parameters $a^{A}$ ($A$ is an index enumerating symmetries)
form a rigid symmetry, then the variation of the action with the coordinate-dependent parameters $a^{A}(x)$ has the universal form:
\begin{equation}\label{NoetherBos}
    \delta S = \int d^4x \; \partial^\mu a^{A}(x) J^A_\mu(x)\,.
\end{equation}
If $a^{A}(x) = const$ we obtain that $\delta S = 0$ in agreement with the assumption that the given transformation is a symmetry of the action.
On the equation of motion, any field variation is vanishing:
\begin{equation}
    \delta S = 0\,,
\end{equation}
and we conclude, that on equations of motions, due  to an arbitrariness of the parameters $a^{A}(x)$, the current in \eqref{NoetherBos} is subject
to the conservation law:
\begin{equation}
    \partial^\mu J^A_\mu(x) = 0 \,.
\end{equation}

Now we shall apply this general procedure to rigid symmetries of the free hypermultiplet. Varying the action \eqref{hyper} by the transformations \eqref{global-sym} with arbitrary analytic parameters, which constitute a local version of the rigid symmetry
transformations from the previous subsection, we obtain:
\begin{equation}\label{action variation}
    \begin{split}
    \delta S =&  \frac{1}{2} \int d\zeta^{(-4)} \; \Bigr[ \lambda^{\alpha(s-2)\dot{\alpha}(s-2)M} \left(\partial_M  \partial^{s-2}_{\alpha(s-2)\dot{\alpha}(s-2)} (J)^{P(s)} q^{+a} \right) \mathcal{D}^{++} q^+_a
    \\&\qquad \qquad \qquad +
     q^{+a}  \left( \mathcal{D}^{++} \lambda^{\alpha(s-2)\dot{\alpha}(s-2)M} \partial_M  \partial^{s-2}_{\alpha(s-2)\dot{\alpha}(s-2)} (J)^{P(s)} q^+_a\right) \Bigr].
        \end{split}
\end{equation}
One can rewrite the second term using the relation:
\begin{equation}
    \begin{split}
     \Bigr( \mathcal{D}^{++} \lambda^{\alpha(s-2)\dot{\alpha}(s-2)M} \partial_M &  \partial^{s-2}_{\alpha(s-2)\dot{\alpha}(s-2)} (J)^{P(s)} q^+_a\Bigr)
     \\
     =&
     \left( \mathcal{D}^{++} \lambda^{\alpha(s-2)\dot{\alpha}(s-2)M}\right) \partial_M  \partial^{s-2}_{\alpha(s-2)\dot{\alpha}(s-2)} (J)^{P(s)} q^+_a
     \\&+
     \lambda^{\alpha(s-2)\dot{\alpha}(s-2)M}  \left[\mathcal{D}^{++},\partial_M \right] \partial^{s-2}_{\alpha(s-2)\dot{\alpha}(s-2)} (J)^{P(s)} q^+_a
     \\&+
      \lambda^{\alpha(s-2)\dot{\alpha}(s-2)M} \partial_M  \partial^{s-2}_{\alpha(s-2)\dot{\alpha}(s-2)} (J)^{P(s)} \mathcal{D}^{++} q^+_a\,.
        \end{split}
\end{equation}
To obtain the conserved currents it is convenient to make use of the free equations of motion to eliminate terms involving $\mathcal{D}^{++}q^+_a$ which vanish on the hypermultiplet mass shell.
It is also useful to separate, in the variation \eqref{action variation}, the contributions with different index structures:
\begin{equation}
    \delta S \Bigr|_{M = \alpha\dot{\alpha}} = \frac{1}{2} \int d\zeta^{(-4)}\;
     \left( \mathcal{D}^{++} \lambda^{\alpha(s-1)\dot{\alpha}(s-1)}\right) q^{+a}   \partial^{s-1}_{\alpha(s-1)\dot{\alpha}(s-1)} (J)^{P(s)} q^+_a\,,
\end{equation}
\begin{equation}
    \delta S \Bigr|_{M = 5} = \frac{1}{2} \int d\zeta^{(-4)}\;
    \left( \mathcal{D}^{++} \lambda^{\alpha(s-2)\dot{\alpha}(s-2)}\right) q^{+a}   \partial^{s-2}_{\alpha(s-2)\dot{\alpha}(s-2)} \partial_5 (J)^{P(s)} q^+_a\,,
\end{equation}
\begin{equation}
    \begin{split}
    \delta S \Bigr|_{M = \alpha} = \frac{1}{2} \int d\zeta^{(-4)}\;
    &\left( \mathcal{D}^{++} \lambda^{\alpha(s-1)\dot{\alpha}(s-2)+}\right) q^{+a}  \partial^-_{\alpha} \partial^{s-2}_{\alpha(s-2)\dot{\alpha}(s-2)} (J)^{P(s)} q^+_a
    \\& + \lambda^{\alpha(s-1)\dot{\alpha}(s-2)} q^{+a} [
    \mathcal{D}^{++}, \partial^-_{\alpha} ] \partial^{s-2}_{\alpha(s-2)\dot{\alpha}(s-2)} (J)^{P(s)} q^+_a\,,
    \end{split}
\end{equation}
\begin{equation}
    \begin{split}
    \delta S \Bigr|_{M = \dot{\alpha}} = \frac{1}{2} \int d\zeta^{(-4)}\;
    &\left( \mathcal{D}^{++} \lambda^{\alpha(s-2)\dot{\alpha}(s-1)+}\right) q^{+a}  \partial^-_{\dot{\alpha}} \partial^{s-2}_{\alpha(s-2)\dot{\alpha}(s-2)} (J)^{P(s)} q^+_a
    \\&
    +
     \lambda^{\alpha(s-2)\dot{\alpha}(s-1)+} q^{+a}  [ \mathcal{D}^{++}, \partial^-_{\dot{\alpha}}] \partial^{s-2}_{\alpha(s-2)\dot{\alpha}(s-2)} (J)^{P(s)} q^+_a\,.
        \end{split}
\end{equation}
One immediately observes that, if the parameters are rigid as in \eqref{LambdA}, {\it i.e.}, satisfy the condition $[\mathcal{D}^{++},\lambda^M\partial_M] = 0$,
all these variations vanish in full agreement with the general statement of Noether theorem. In this way we derive the current superfield conservation laws:
\begin{equation}\label{conser 1}
    \mathcal{D}^{++} \left(q^{+a}   \partial^{s-1}_{\alpha(s-1)\dot{\alpha}(s-1)} (J)^{P(s)} q^+_a\right) = 0\,,
\end{equation}
\begin{equation}\label{conser 2}
    \mathcal{D}^{++} \left(q^{+a}   \partial^{s-2}_{\alpha(s-2)\dot{\alpha}(s-2)} \partial_5 (J)^{P(s)} q^+_a\right) = 0\,,
\end{equation}
\begin{equation}\label{conser 3}
    \mathcal{D}^{++} \left(q^{+a}  \partial^-_{\alpha} \partial^{s-2}_{\alpha(s-2)\dot{\alpha}(s-2)} (J)^{P(s)} q^+_a\right)
    =
    q^{+a} [
    \mathcal{D}^{++}, \partial^-_{\alpha} ] \partial^{s-2}_{\alpha(s-2)\dot{\alpha}(s-2)} (J)^{P(s)} q^+_a\,,
\end{equation}
\begin{equation}\label{conser 4}
    \mathcal{D}^{++} \left(q^{+a}  \partial^-_{\dot{\alpha}} \partial^{s-2}_{\alpha(s-2)\dot{\alpha}(s-2)} (J)^{P(s)} q^+_a\right)
    =
    q^{+a}  [ \mathcal{D}^{++}, \partial^-_{\dot{\alpha}}] \partial^{s-2}_{\alpha(s-2)\dot{\alpha}(s-2)} (J)^{P(s)} q^+_a\,.
\end{equation}
The first two equations \eqref{conser 1} and \eqref{conser 2} directly give the laws of conservation of current superfields
\eqref{Supercurrent1} and \eqref{Supercurrent2}.

Taking into account the equations
\begin{subequations}
\begin{equation}
    [\mathcal{D}^{++}, \partial^-_\alpha] = - \partial^+_\alpha
    +
    2i \bar{\theta}^{+\dot{\alpha}} \partial_{\alpha\dot{\alpha}}
    +
    2i \theta^+_{\alpha} \partial_5,
\end{equation}
\begin{equation}
    [\mathcal{D}^{++}, \partial^-_{\dot{\alpha}}] = - \partial^+_{\dot{\alpha}}
    -
    2i \theta^{+\alpha} \partial_{\alpha\dot{\alpha}}
    +
    2i \bar{\theta}^+_{\dot{\alpha}} \partial_5,
\end{equation}
\end{subequations}
we can immediately identify eqs. \eqref{conser 3} and \eqref{conser 4} with the modified conservation laws \eqref{ModConserv} and \eqref{ModConser2}.

So we conclude that the above equations fully reproduce the current superfield conservation conditions from Section \ref{Construction of current superfields}
(we restricted our consideration to the $\xi=0$ case, because $\xi\neq 0$ gives a
trivial contribution to the current superfield) and give the explicit form of the current superfields. Thus we have established the direct connection between
the rigid symmetries \eqref{global-sym} and the conserved current superfields. This Noether theorem can be considered as a generalization of Noether theorem  in ordinary
superspace \cite{Magro:2001aj} to the HSS case.

In full agreement with the results of Appendix \ref{N=2 current superfield}, the superfield conditions derived above
lead to the conservation of the bosonic spin $s$ current, the  doublet of fermionic spin $s-\tfrac{1}{2}$ currents and the bosonic spin $s-1$ current.
These currents correspond to the global spin $\mathbf{s}$ symmetry  \eqref{global-sym} of the free hypermultiplet.

In conclusion of this Section, note that the conservation law \eqref{Another conservation}
\begin{equation}
    \partial_{\alpha\dot{\alpha}} J^{++\alpha\dot{\alpha}} = 0
\end{equation}
does not directly follow from Noether theorem for rigid symmetries with the structure \eqref{global-sym}.
Presumably, it is related to some other symmetry of the free hypermultiplet. It would be interesting
to explicitly identify this extra bosonic symmetry.

\section{Field redefinitions}
\label{Fields redefinitions}

In this Appendix we describe various field redefinitions capable to simplify the structure of the linearized cubic bosonic vertices in the spin 2 and spin 3 cases.

Before proceeding to concrete examples, it is important to note a simple observation that if a cubic vertex is equal to zero on free equations of motion,
then such a vertex can be removed by appropriate redefinition of fields. This observation is based on the property that, when redefining fields:
\begin{equation}
    \phi \to \phi + F(\phi)\,,
\end{equation}
the action is changing as
\begin{equation}
    S[\phi] \to S[\phi] + \int d^4x \frac{\delta S[\phi]}{\delta \phi} F(\phi).
\end{equation}
The variation of the action on the equations of motion is zero, which leads to the vanishing of the vertex on the equations of motion.
Such interactions are called "fake interactions", see e.g. \cite{Ponomarev:2022vjb}.

\subsection{Minimal gravity coupling of scalar fields and scalar field redefinitions} \label{Minimal gravity coupling}

For comparing the results  of Section \ref{spin 2 coupling} with those known in literature, it is more convenient  to pass from the spinor notation for spin 2 fields to the more accustomed vector notation:
\begin{equation}\label{Phi-field}
    \Phi^{\alpha\dot{\alpha}\beta\dot{\beta}} := \sigma^{\alpha\dot{\alpha}}_n \sigma^{\beta\dot{\beta}}_m h^{(nm)}\,,
\end{equation}
\begin{equation}
    \mathcal{R} = \frac{1}{2}R = \frac{1}{2} \left(\partial^n \partial^m - \eta^{nm}\Box\right)  h_{mn}\,.
\end{equation}
With this notation, the scalar part of on-shell  hypermultiplet action \eqref{hyper on shell} and the (2,0,0) coupling derived in \eqref{spin 2-2 full action} looks as:
\begin{equation}
    \begin{split}
        S_{free} + S^{(s=2)}_{int} =
        & \int d^4x\, \left( \partial_n f^i \partial^n \bar{f}_i - m^2 f^i \bar{f}_i \right)
        \\&- \kappa_2 \int d^4x\; h^{(mn)} \left(
        \partial_m \partial_n f^i \bar{f}_i
        -
        2 \partial_m  f^i \partial_n \bar{f}_i
        +  f^i \partial_m \partial_n\bar{f}_i  \right)
        \\&+\kappa_2 \int d^4x \; R \left(f^i\bar{f}_i\right) + \mathcal{O}(\kappa_2^2)\,.
    \end{split}
\end{equation}
It will be convenient to integrate by parts in the second line to bring the interacting action in the form:
\begin{equation}\label{inter-norm}
    S^{(s=2)}_{int} =  \kappa_2 \int d^4x\; \left( 4 h^{(mn)} \partial_m  f^i \partial_n \bar{f}_i - \Box h f^i \bar{f}_i \right) + \mathcal{O}(\kappa_2^2).
\end{equation}
On the other hand, the standard (non-linear) minimal $(2,0,0)$ interaction has a different form:

\begin{equation}\label{minimal}
    S^{(s=2)}_{min} = \int d^4 x \sqrt{-g}\; \left(g^{mn} \partial_m f^i \partial_n \bar{f}_i - m^2 f^i \bar{f}_i \right).
\end{equation}

To compare it with our result, we need to consider the linearized cubic vertices that follow from this Lagrangian.
To do this, it is necessary to establish a relationship between the metric $g_{mn}$ and the fields \eqref{Phi-field} used for spin 2.
According to \cite{Galperin:1987ek, Galperin:1987em} the full analytic $\mathcal{N}=2$ supergravity potential  $H^{++m}$ in Wess-Zumino gauge has the form:
\begin{equation}
    H^{++m} = -2i \theta^+ \sigma^a \bar{\theta}^+ e_a^m + \dots,
\end{equation}
where $e_a^m$ is the  ``inverse vierbein'', i.e.:
\begin{equation}
g_{mn} = \eta_{ab} e^{a}_m e^{b}_n.
\end{equation}

Using the relation
 \begin{equation}
    H^{++m} = -2i \theta^+ \sigma^m \bar{\theta}^+
    +
    \kappa_2 h^{++\alpha\dot{\alpha}}\sigma^m_{\alpha\dot{\alpha}},
 \end{equation}
the spin $\mathbf{2}$ Wess-Zumino gauge \eqref{WZ3 s=2} and the relation \eqref{Phi-field}, we obtain:
\begin{equation}
    e_a^m = \delta_a^m + 2 \kappa_2 h_a^m.
\end{equation}
So we have:
\begin{equation}
    g_{mn} = \eta_{mn} - 4 \kappa_2 h_{mn} + \mathcal{O}(h^2)\,,
    \qquad
    \sqrt{-g} = 1 - 2\kappa_2 h + \mathcal{O}(\kappa_2^2)\,,
\end{equation}
and obtain that the action \eqref{minimal}, to the linear order in the spin 2 field $h^{mn}$, reads:
\begin{equation}\label{int-min}
    \begin{split}
        S^{(s=2)}_{int-min} =& \int d^4 x \; \left(\eta^{mn} \partial_m f^i \partial_n \bar{f}_i - m^2 f^i \bar{f}_i \right)
        \\&
        +4
       \kappa_2 \int d^4x \; h^{mn} \partial_m f^i \partial_n \bar{f}_i
        -
        2\kappa_2
        \int d^4x \; h \left(\eta^{mn} \partial_m f^i \partial_n \bar{f}_i - m^2 f^i \bar{f}_i \right) + \mathcal{O}(\kappa_2^2)\,.
    \end{split}
\end{equation}

Now we redefine the scalar field as
\begin{equation}\label{field redef}
    f^i = \left(-g\right)^{-\frac14} \mathfrak{f}^i = \mathfrak{f}^i + \kappa_2 h  \mathfrak{f}^i + \mathcal{O}(\kappa_2^2)\,.
\end{equation}
Then the free kinetic term of $f^i$ is transformed to
\begin{equation}
    \begin{split}
        S_{free} =&
        \int d^4 x \; \left(\eta^{mn} \partial_m f^i \partial_n \bar{f}_i - m^2 f^i \bar{f}_i \right)
        \\= &
        \int d^4 x \; \left(\eta^{mn} \partial_m \mathfrak{f}^i \partial_n \bar{\mathfrak{f}}_i - m^2 \mathfrak{f}^i \bar{\mathfrak{f}}_i \right)
        \\& +2 \kappa_2 \int d^4 x \;h \left(\eta^{mn} \partial_m \mathfrak{f}^i \partial_n \bar{\mathfrak{f}}_i - m^2 \mathfrak{f}^i \bar{\mathfrak{f}}_i \right)
        -\kappa_2 \int d^4x \; \Box h \left(\mathfrak{f}^i\bar{\mathfrak{f}}_i\right) + \mathcal{O}(\kappa_2^2)\,,
    \end{split}
\end{equation}
which modifies the linearized cubic interaction. The action \eqref{int-min} takes the form:
\begin{equation}\label{int-min-mod}
    \begin{split}
        S^{(s=2)}_{int-min} =& \int d^4 x \; \left(\eta^{mn} \partial_m \mathfrak{f}^i \partial_n \bar{\mathfrak{f}}_i - m^2 \mathfrak{f}^i \bar{\mathfrak{f}}_i \right)
        \\&
       +
       4\kappa_2 \int d^4x \; h^{mn} \partial_m \mathfrak{f}^i \partial_n \bar{\mathfrak{f}}_i
        -
        \kappa_2 \int d^4x \; \Box h \left(\mathfrak{f}^i\bar{\mathfrak{f}}_i\right)
        + \mathcal{O}(\kappa_2^2)\,.
    \end{split}
\end{equation}

Thus the interaction of spin 2 with scalars \eqref{inter-norm} found in Section \ref{spin 2 coupling}, actually coincides with the minimal \eqref{minimal}
 up to the field redefinitions \eqref{field redef}.

\subsection{Redefinition of the spin 2 gauge field and $(2,0,0)$ coupling}
\label{Redefinition of spin 2 gauge field}

In this Subsection we study how a redefinition of the spin 2 gauge field can affect the structure of the (2, 0, 0) vertex.

Recall the Pauli-Fierz action \eqref{Pauli_Fiertz} in the spinor notations:
\begin{equation}
    \begin{split}
        S_{PF} (\Phi^{(\alpha\beta)(\dot{\alpha}\dot{\beta})}, \Upsilon) &=
        -\int d^4x\; \Big[\Phi^{(\alpha\beta)(\dot{\alpha}\dot{\beta})} \Box
        \Phi_{(\alpha\beta)(\dot{\alpha}\dot{\beta})} -
        \Phi^{(\alpha\beta)(\dot{\alpha}\dot{\beta})}
        \partial_{\alpha\dot{\alpha}} \partial^{\rho\dot{\rho}}
        \Phi_{(\rho\beta)(\dot{\rho}\dot{\beta})}
        \\&\;\;\;\;\;\;\;\;\;\;\;         + 2\, \Phi
        \partial^{\alpha\dot{\alpha}} \partial^{\beta\dot{\beta}}
        \Phi_{(\alpha\beta)(\dot{\alpha}\dot{\beta})}
        - 6 \Phi \Box \Phi \Big].
    \end{split}
\end{equation}

Redefining the field $\Phi$,
\begin{equation}\label{redefinition spin 2}
    \Phi  \,\Rightarrow \,\Phi + a f^i\bar{f}_i\,,
\end{equation}
leads to:
\begin{equation}
    \begin{split}
    S_{PF} (\Phi^{(\alpha\beta)(\dot{\alpha}\dot{\beta})}, \Phi)\,
    \Rightarrow\, &
    S_{PF} (\Phi^{(\alpha\beta)(\dot{\alpha}\dot{\beta})}, \Phi)
    \\&- a \int d^4x\; \left[2 \Phi^{(\alpha\beta)(\dot{\alpha}\dot{\beta})} \partial_{\alpha\dot{\alpha}}\partial_{\beta\dot{\beta}} \left(f^i\bar{f}_i\right)
    - 12 \Phi \Box   \left(f^i\bar{f}_i\right)  \right]+\mathcal{O}(a^2).
    \end{split}
\end{equation}

After integration by parts one can rewrite the second line in terms of scalar curvature $\mathcal{R}$ introduced in \eqref{scalar curvature}:
\begin{equation}
    S_{PF} (\Phi^{(\alpha\beta)(\dot{\alpha}\dot{\beta})}, \Phi)\;
\Rightarrow \;
    S_{PF} (\Phi^{(\alpha\beta)(\dot{\alpha}\dot{\beta})}, \Phi)
    -
    16 a \int d^4x\; \mathcal{R} \left(f^i\bar{f}_i\right) +\mathcal{O}(a^2)\,.
\end{equation}
Thus we have shown that at the linearized level, the $(2,0,0)$ vertex of form $\mathcal{R}\bar{f} f$ can be eliminated
by the redefinition \eqref{redefinition spin 2} of the spin 2 field trace. This property was used, e.g.,  in the discussion in the end of Section \ref{spin 3 coupling}.

\subsection{Redefinition of scalar fields and $(3,0,0)$ coupling}
\label{spin 3 redef scalar}

\quad\;\;\textbf{1.} In a direct analogy with \eqref{field redef} it is possible to redefine the complex scalar fields as:
\begin{equation}\label{redef spin 3}
    f^i = \mathfrak{f}^i - i b \left(\partial_{\alpha\dot{\alpha}} \Phi^{\alpha\dot{\alpha}}\right) \mathfrak{f}^i\,,\qquad
    \bar{f}_i = \bar{\mathfrak{f}}_i + i b\left(\partial_{\alpha\dot{\alpha}} \Phi^{\alpha\dot{\alpha}}\right) \bar{\mathfrak{f}}_i\,.
\end{equation}
Here $b$ is some real parameter and the field $\Phi^{\alpha\dot{\alpha}}$ was defined in \eqref{spin 3 field}. Under this redefinition the free scalar Lagrangian changes as:
\begin{equation}\label{redef spin 3 result}
    S_{free} (f^i) = S_{free} (\mathfrak{f} ^i)
    -
    \frac{i b}{2} \Phi^{\alpha\dot{\alpha}} \partial_{\alpha\dot{\alpha}} \partial_{\beta\dot{\beta}} \left(\mathfrak{f}^i \partial^{\beta\dot{\beta}} \bar{\mathfrak{f}}_i
    -
    \partial^{\beta\dot{\beta}} \mathfrak{f}^i  \bar{\mathfrak{f}}_i \right)
    +
    \mathcal{O}(b^2)\,.
\end{equation}
So, making the redefinition \eqref{redef spin 3}, one can eliminate the cubic vertex of the form appearing, e.g., in \eqref{spin3II}.

\medskip
\textbf{2.} There is one more admissible scalar fields redefinition:
\begin{equation}\label{redef spin 3-2}
    f^i = \mathfrak{f}^i - i c \, \Phi^{\alpha\dot{\alpha}} \partial_{\alpha\dot{\alpha}} \mathfrak{f}^i\,,\qquad
    \bar{f}_i = \bar{\mathfrak{f}}_i + i c \, \Phi^{\alpha\dot{\alpha}} \partial_{\alpha\dot{\alpha}} \bar{\mathfrak{f}}_i\,.
\end{equation}
This redefinition leads to:
\begin{equation}
    \begin{split}
    S_{free}(f^i) =&\; S_{free}(\mathfrak{f}^i) - \frac{i c}{2} \Phi^{\beta\dot{\beta}} \left( \partial_{\alpha\dot{\alpha}} \partial_{\beta\dot{\beta}} \mathfrak{f}^i \partial^{\alpha\dot{\alpha}} \bar{\mathfrak{f}}_i
    -
    \partial_{\alpha\dot{\alpha}} \mathfrak{f}^i \partial^{\alpha\dot{\alpha}} \partial_{\beta\dot{\beta}} \bar{\mathfrak{f}}_i \right)
    \\&+ i c\; m^2 \Phi^{\beta\dot{\beta}} \left(\partial_{\beta\dot{\beta}} \mathfrak{f}^i \bar{\mathfrak{f}}_i
    -  \mathfrak{f}^i \partial_{\beta\dot{\beta}} \bar{\mathfrak{f}}_i  \right)+\mathcal{O}(c^2).
        \end{split}
\end{equation}
Further manipulations lead to the expression:
\begin{equation}
    \begin{split}
        S_{free}(f^i) =&\; S_{free}(\mathfrak{f}^i) - \frac{i c}{2} \int d^4x\, \Phi^{\beta\dot{\beta}} \Box \left( \partial_{\beta\dot{\beta}} \mathfrak{f}^i  \bar{\mathfrak{f}}_i
        -
         \mathfrak{f}^i \partial_{\beta\dot{\beta}} \bar{\mathfrak{f}}_i \right)
        \\&+ \frac{i c}{2}\; \int d^4x\, \Phi^{\beta\dot{\beta}} \Bigr\{\partial_{\beta\dot{\beta}} (\Box+m^2) \mathfrak{f}^i \bar{\mathfrak{f}}_i
        +
        \partial_{\beta\dot{\beta}} \mathfrak{f}^i (\Box+m^2) \bar{\mathfrak{f}}_i
         \\&\qquad\qquad\qquad\qquad-
         (\Box+m^2)\mathfrak{f}^i \partial_{\beta\dot{\beta}} \bar{\mathfrak{f}}_i
         -
         \mathfrak{f}^i \partial_{\beta\dot{\beta}} (\Box+m^2) \bar{\mathfrak{f}}_i \Bigr\} +\mathcal{O}(c^2).
    \end{split}
\end{equation}
So on the free equations of motions for $\mathfrak{f}^i$ one can eliminate the  cubic $(3,0,0)$ coupling of the form:
\begin{equation}
    \int d^4x\, \Phi^{\beta\dot{\beta}}\, \Box \, \left(  \partial_{\beta\dot{\beta}} \mathfrak{f}^i  \bar{\mathfrak{f}}_i
    -
    \mathfrak{f}^i \partial_{\beta\dot{\beta}} \bar{\mathfrak{f}}_i \right).
\end{equation}

\subsection{Redefinition of spin 3 gauge field and $(3,0,0)$ coupling}
\label{spin 3 redef gauge}

The spin 3 action \eqref{spin 3 action} has the form:
\begin{eqnarray}
&& S^{(s= 3)} \Big(\Phi^{(\alpha_1\alpha_2\alpha_3)( \dot{\alpha}_1\dot{\alpha}_2\dot{\alpha}_3)}, \Phi^{\alpha\dot{\alpha}}\Big) =   \int d^4x\,
\Big[\Phi^{(\alpha_1\alpha_2\alpha_3)( \dot{\alpha}_1\dot{\alpha}_2\dot{\alpha}_3)}
        \Box \Phi_{(\alpha_1\alpha_2\alpha_3)( \dot{\alpha}_1\dot{\alpha}_2\dot{\alpha}_3)} \nonumber \\
&&- \,\frac{3}{2} \Phi^{(\alpha_1\alpha_2\alpha_3)( \dot{\alpha}_1\dot{\alpha}_2\dot{\alpha}_3)}
        \partial_{\alpha_1\dot{\alpha}_1} \partial^{\rho\dot{\rho}} \Phi_{(\rho\alpha_2\alpha_3)( \dot{\rho}\dot{\alpha}_2\dot{\alpha}_3)}
        + 3 \Phi^{(\alpha_1\alpha_2\alpha_3)( \dot{\alpha}_1\dot{\alpha}_2\dot{\alpha}_3)} \partial_{\alpha_1\dot{\alpha}_1} \partial_{\alpha_2\dot{\alpha}_2}
        \Phi_{\alpha_3\dot{\alpha}_3} \nonumber \\
&&\,        - \frac{15}{4} \Phi^{\alpha\dot{\alpha}} \Box  \Phi_{\alpha\dot{\alpha}}
        +
        \frac{3}{8} \partial_{\alpha_1\dot{\alpha}_1} \Phi^{\alpha_1\dot{\alpha}_1} \partial_{\alpha_2\dot{\alpha}_2} \Phi^{\alpha_2\dot{\alpha}_2}\Big].
\end{eqnarray}

After the field redefinition,
\begin{equation}\label{redef spin 3 - 2}
    \Phi^{\alpha\dot{\alpha}} \Rightarrow  \Phi^{\alpha\dot{\alpha}} + i  u \left(f^i \partial^{\alpha\dot{\alpha}} \bar{f}_i - \partial^{\alpha\dot{\alpha}}f^i  \bar{f}_i \right),
\end{equation}
the action is redefined as:
\begin{equation}\label{TransfSpin3}
    \begin{split}
    S^{(s= 3)} \Big(&\Phi^{(\alpha_1\alpha_2\alpha_3)( \dot{\alpha}_1\dot{\alpha}_2\dot{\alpha}_3)}, \Phi^{\alpha\dot{\alpha}}\Big)
    \Rightarrow\,
    S^{(s= 3)} \Big(\Phi^{(\alpha_1\alpha_2\alpha_3)( \dot{\alpha}_1\dot{\alpha}_2\dot{\alpha}_3)}, \Phi^{\alpha\dot{\alpha}}\Big)
    \\& + i u \int d^4x\; \Bigr[
    3 \Phi^{(\alpha_1\alpha_2\alpha_3)( \dot{\alpha}_1\dot{\alpha}_2\dot{\alpha}_3)} \partial_{\alpha_1\dot{\alpha}_1} \partial_{\alpha_2\dot{\alpha}_2}
\left(f^i \partial_{\alpha_3\dot{\alpha}_3} \bar{f}_i - \partial_{\alpha_3\dot{\alpha}_3}f^i  \bar{f}_i \right)
    \\&\qquad\qquad\qquad- \frac{15}{2} \Phi^{\alpha\dot{\alpha}} \Box  \left(f^i \partial^{\alpha\dot{\alpha}} \bar{f}_i - \partial^{\alpha\dot{\alpha}}f^i  \bar{f}_i \right)
    \\&\qquad\qquad\qquad
    -\frac{3}{4}  \Phi^{\alpha_1\dot{\alpha}_1} \partial_{\alpha_1\dot{\alpha}_1} \partial_{\alpha_2\dot{\alpha}_2} \left(f^i \partial^{\alpha_2\dot{\alpha}_2} \bar{f}_i - \partial^{\alpha_2\dot{\alpha}_2}f^i  \bar{f}_i \right)
    \Bigr] + \mathcal{O}(u^2)\,.
        \end{split}
\end{equation}
Using the definition of the linearized spin 3 curvature \eqref{spin 3 curvature 1} and integrating by parts, one can rewrite the transformation \eqref{TransfSpin3} as
\begin{equation}
        \begin{split}
            S^{(s= 3)} \Big(&\Phi^{(\alpha_1\alpha_2\alpha_3)( \dot{\alpha}_1\dot{\alpha}_2\dot{\alpha}_3)}, \Phi^{\alpha\dot{\alpha}}\Big)\;
            \Rightarrow
            \;
            S^{(s= 3)} \Big(\Phi^{(\alpha_1\alpha_2\alpha_3)( \dot{\alpha}_1\dot{\alpha}_2\dot{\alpha}_3)}, \Phi^{\alpha\dot{\alpha}}\Big)
            \\& + i u \int d^4x\; \Bigr[
            3 \mathcal{R}^{\alpha\dot{\alpha}}
            \left(f^i \partial_{\alpha\dot{\alpha}} \bar{f}_i - \partial_{\alpha\dot{\alpha}}f^i  \bar{f}_i \right)
            \Bigr] + \mathcal{O}(u^2)\,.
        \end{split}
\end{equation}
So, the field redefinition \eqref{redef spin 3 - 2} ensures eliminating the $(3,0,0)$ cubic couplings of the form:
\begin{equation}
     i  \int d^4x\;
     \mathcal{R}^{\alpha\dot{\alpha}}
    \left(f^i \partial_{\alpha\dot{\alpha}} \bar{f}_i - \partial_{\alpha\dot{\alpha}}f^i  \bar{f}_i \right).
\end{equation}

\section{Gauge-fixing}
\label{Gauge fixing}

The aim of this Section is to rewrite the superfield equations of motions for gauge superfields in terms of unconstrained analytic  potentials $h^{++\dots}$ subject to the proper manifestly
supersymmetric gauge conditions. Equations \eqref{EOM} are written in terms of
the superfields $G^{--\dots}$ defined through the zero curvature equations \eqref{zero-curv}, \eqref{zero-curv2} and related to $h^{++\dots}$ by eqs. \eqref{s-1} and \eqref{s-2}.
It would be very convenient to write the equations directly through the analytic gauge potentials.

\subsection{Spin $\mathbf{1}$ theory}

$\mathcal{N}=2$ Maxwell multiplet is described by the analytic superfield $h^{++}(\zeta)$ with the gauge transformation:
\begin{equation}
    \delta h^{++} = \mathcal{D}^{++}\lambda\,.
\end{equation}
and the equation of motion
\begin{equation}\label{MaxwellEOM2}
    (D^+)^4 h^{--} = 0\,,
\end{equation}
where $h^{--}$ is a solution of the zero-curvature equation,
\begin{equation}
    \mathcal{D}^{++} h^{--} = \mathcal{D}^{--} h^{++}\,.
\end{equation}

Let us  consider the superfield transformation law of $\mathcal{D}^{++} h^{++}$:
\begin{equation}\label{Zero-curvature-Maxwell}
    \delta \left( \mathcal{D}^{++} h^{++} \right) =  \mathcal{D}^{++}\mathcal{D}^{++} \lambda\,.
\end{equation}
This law prompts that, in addition to the Wess-Zumino gauge
\eqref{Maxwell WZ}, one can gauge away one more component from
$\mathcal{D}^{++}h^{++}$. Indeed, keeping in mind that
\begin{equation}
    \left(\mathcal{D}^{++} h^{++} \right)_{gauge} = - 2 (\theta^+)^2 (\bar{\theta}^+)^2 \partial_m A^m\,,
\end{equation}
and making use of the residual gauge freedom,
\begin{equation}
    \lambda(\zeta) = a(x)\,,
\end{equation}
we can choose Lorentz gauge,
\begin{equation}
    \delta \left(\partial_m A^m\right) = \Box a \qquad \Rightarrow \qquad \partial_m A^m = 0\,,
\end{equation}
that further restricts the residual gauge freedom to
\begin{equation}
    \lambda(\zeta) = a(x)\,, \qquad \Box a(x) = 0\,.
\end{equation}
Gauge-fixing $\Big(\mathcal{D}^{++}h^{++}\Big)_{gauge} = 0$ is just the $\mathcal{N}=2$ superfield generalization of Lorentz gauge in Maxwell theory.

It is possible from the very beginning to fix the gauge in the manifestly supersymmetric way as
\begin{equation}\label{spin1Gauge}
    \begin{split}
        \mathcal{D}^{++}h^{++} = 0\;\;\;\;\; &\Rightarrow \;\;\;\;\; h^{++} = h^{(ij)}(Z) u^+_i u^+_j \;\;\;\;\; \\& \Rightarrow \;\;\;\;\; h^{--} = h^{(ij)}(Z) u^-_i u^-_j = \frac{1}{2} \mathcal{D}^{--} \mathcal{D}^{--} h^{++}\,.
    \end{split}
\end{equation}
Here we used the cental basis in HSS and, at the  last step, rewrote the result of solving the harmonic zero-curvature condition in the covariant basis-independent form.

It is straightforward to check that, taking into account the analyticity of $h^{++}$, we can transform the equation of motion \eqref{MaxwellEOM2} to the form
\begin{equation}
    (D^{+})^4 h^{--} = \frac12 (D^{+})^4(D^{--})^2h^{++} \sim \Box  h^{++}\,.
\end{equation}

So we conclude that $\mathcal{N}=2$ superspin \textbf{1} can be described by a pair of equations:
\begin{equation}
    \begin{cases}
        \Box h^{++} = 0\,,\\
        \mathcal{D}^{++}h^{++} = 0\,.
    \end{cases}
\end{equation}
The first equation is the massless Klein-Gordon equation, the second one is the gauge-fixing condition.

\subsection{Spin $\mathbf{2}$ theory}

The spin $\mathbf{2}$ equations \eqref{EOM spin 2} are more complicated:
\begin{equation}
    \begin{cases}
        (D^+)^4 G^{--\alpha \dot{\alpha} }= 0\,,
        \\
        (D^+)^4 G^{--5} = 0\,,
        \\
        (\bar{D}^+)^2 D^{+\alpha} G^{--}_{\alpha \dot{\alpha}}
        -
        (D^+)^2 \bar{D}^+_{\dot{\alpha}} G^{--5}_{} = 0\,,
        \\
        (D^+)^2  \bar{D}^{+\dot{\alpha}} G^{--}_{\alpha \dot{\alpha}}
        +
        (\bar{D}^+)^2 D^+_{\alpha} G^{--5}_{}
        =
        0\,.
    \end{cases}
\end{equation}

It is convenient to choose the covariant gauge in full analogy to the spin $\mathbf{1}$ case:
\begin{subequations}
    \begin{equation}
        \mathcal{D}^{++} G^{++\alpha\dot{\alpha}} =0 \,,
    \end{equation}
    \begin{equation}
        \mathcal{D}^{++} G^{++5} =0 \,.
    \end{equation}
\end{subequations}
Note, that this gauge-fixing condition explicitly involves the spinor coordinates $\theta^{+ \hat{\alpha}}$. In terms of the analytic potentials these conditions amount to the set:
\begin{subequations}
    \begin{equation}
        \mathcal{D}^{++} h^{++\alpha+} = 0\,,\qquad  \mathcal{D}^{++} h^{++\dot{\alpha}+} = 0\,, \label{Gauge21}
    \end{equation}
    \begin{equation}
        \mathcal{D}^{++} h^{++\alpha\dot{\alpha}} + 2i h^{++\alpha+} \bar{\theta}^{+\dot{\alpha}} - 2i h^{++\dot{\alpha}+} \theta^{+\alpha} = 0\,,
    \end{equation}
    \begin{equation}
        \mathcal{D}^{++} h^{++5} - 2i h^{++\alpha+} \theta^+_\alpha - 2i h^{++\dot{\alpha}+} \bar{\theta}^+_{\dot{\alpha}}= 0\,.\label{Gauge2Last}
    \end{equation}
\end{subequations}
This gauge is just the superfield form of de-Donder gauge:
\begin{equation}
    \partial_m \bar{h}^{mn} = 0\,, \qquad \text{where} \qquad
    \bar{h}^{mn} := h^{mn} - \frac{1}{2} \eta^{mn} h\,.
\end{equation}

In this gauge, one can resolve the zero-curvature conditions in analogy with \eqref{spin1Gauge}:
\begin{equation}
    G^{--\alpha\dot{\alpha}} = \frac{1}{2} (\mathcal{D}^{--})^2 G^{++\alpha\dot{\alpha}}\,,
\end{equation}
\begin{equation}
    G^{--5} = \frac{1}{2} (\mathcal{D}^{--})^2 G^{++5}\,.
\end{equation}
Using these solutions, one arrives at the equations of motions  written in terms of the analytic potentials only:
\begin{equation}\label{EOM2}
    \begin{cases}
        \Box h^{++\alpha \dot{\alpha} } - 4 \partial^{\mu(\dot{\alpha}}  \partial^-_{\mu} h^{++\alpha\dot)+} + 4 \partial^{(\alpha\dot{\mu}} \bar{\partial}^-_{\dot{\mu}} h^{++\dot\alpha)+} = 0,\\
        \Box h^{++\alpha+} = 0,\\
        \Box h^{++\dot\alpha+} = 0\,,\\
        \Box h^{++ 5} + 2 \partial^{\mu\dot{\mu}}  \partial^-_{\dot{\mu}} h^{+++}_\mu - 2 \partial^{\mu\dot{\mu}} \partial^-_{\mu} h^{+++}_{\dot{\mu}} = 0\,.
    \end{cases}
\end{equation}
We see that, in addition to the $\Box$ terms, the first and fourth equations contain additional  terms linear in the derivatives.
The first equation, in particular, yields:
\begin{equation}
    \Box \bar{h}^{mn} = 0\,.
\end{equation}

Thus the equations of motion for the spin ${\bf 2}$ case amount to the Lorentz-type covariant gauge conditions \eqref{Gauge21} - \eqref{Gauge2Last}
and eqs. \eqref{EOM2}, all written in terms of analytic potentials.
\subsection{General spin $\mathbf{s}$ theory}
One can easily repeat the same steps as in spin $\mathbf{2}$ case for the generic spin ${\bf s}$ superfield equation \eqref{EOM}:
\begin{equation}
    (D^+)^4 G^{--\alpha(s-1) \dot{\alpha}(s-1) }= 0\,.
\end{equation}
It is possible to show that this equation is equivalent to the system of equations:
\begin{equation}
 \begin{cases}
        \Box G^{++\alpha(s-1) \dot{\alpha}(s-1) } + i \partial^{\mu\dot{\mu}} \left( D^-_{\mu} \bar{D}^+_{\dot{\mu}} - \bar{D}^-_{\dot{\mu}} D^+_\mu  \right) G^{++\alpha(s-1) \dot{\alpha}(s-1) } = 0\,,\\
        D^{++} G^{++\alpha(s-1) \dot{\alpha}(s-1) } = 0\,.
    \end{cases}
\end{equation}
Using the relation (\ref{s-1}), this system can be represented as a system of equations for the analytic potentials.

The gauge conditions in terms of the analytic potentials read:
\begin{equation}
    \begin{cases}
        D^{++}h^{++\alpha(s-1)\dot\alpha(s-1)} + 2i
        \big[h^{++\alpha(s-1)(\dot\alpha(s-2)+}\bar\theta^{+\dot\alpha_{s-1})}
        +
        h^{++\dot\alpha(s-1)(\alpha(s-2)+}\,\theta^{+\alpha_{s-1})} \big] = 0\,,\\
        D^{++}h^{++\alpha(s-1)\dot\alpha(s-2)+} = 0\,,\\
        D^{++}h^{++\dot\alpha(s-1)\alpha(s-2)+} = 0\,.
    \end{cases}
\end{equation}
The superfield equations of motion are written as
\begin{equation}
    \begin{cases}
        \Box h^{++\alpha(s-1) \dot{\alpha}(s-1) } - 4 \partial^{\mu(\dot{\alpha}(s-1)}  \partial^-_{\mu} h^{++\alpha(s-1)\dot\alpha(s-2))+} + 4 \partial^{(\alpha(s-1)\dot{\mu}} \bar{\partial}^-_{\dot{\mu}} h^{++\dot\alpha(s-1)\alpha(s-2))+} = 0\,,\\
        \Box h^{++\alpha(s-1)\dot\alpha(s-2)+} = 0\,,\\
        \Box h^{++\dot\alpha(s-1)\alpha(s-2)+} = 0\,.
    \end{cases}
\end{equation}

The equation
\begin{equation}
    (D^+)^4 G^{--5\alpha(s-2) \dot{\alpha}(s-2) }= 0
\end{equation}
can also be restructured  in the same way and it amounts to the gauge condition
\begin{equation}
    \mathcal{D}^{++} h^{++5\alpha(s-2)\dot{\alpha}(s-2)}
    -
    2i h^{++(\alpha(s-2)\rho)\dot{\alpha}(s-2)} \theta^+_\rho
    -
    2i h^{++\alpha(s-2)(\dot{\alpha}(s-2)\dot{\rho})} \bar{\theta}^+_{\dot{\rho}} = 0
\end{equation}
and the following equation for the analytic potentials:
\begin{equation}
    \Box h^{++5\alpha(s-2) \dot{\alpha}(s-2) } + 2 \partial^{\mu\dot{\mu}}  \partial^-_{\dot{\mu}} h^{++\alpha(s-2)\dot\alpha(s-2)+}_\mu
- 2 \partial^{\mu\dot{\mu}} \partial^-_{\mu} h^{++\alpha(s-2)\dot\alpha(s-2)+}_{\dot{\mu}} = 0\,.
\end{equation}

\medskip

The results presented in this Appendix should be considered as first steps in the study of representations of $\mathcal{N}=2$ supersymmetry algebra
on harmonic superfields from the group-theory point of view. More complete analysis of the group-theoretical origin of these equations will
be a subject of separate publication.

\end{document}